%% file: paper_body.tex
\ifdefined\pdfminorversion
\fi

\makeatletter
\let\qpgft@kernel@label\label
\makeatother
\documentclass[
aps,
prx,
showpacs,
preprintnumbers,
twocolumn,
superscriptaddress,
longbibliography,
nofootinbib,
10pt,
]{revtex4-2}
\makeatletter
\IfFormatAtLeastTF{2023-06-01}{\let\label\qpgft@kernel@label}{}
\makeatother

\usepackage[T1]{fontenc}
\usepackage{newtxtext}
\usepackage{newtxmath}

\usepackage{enumitem}

\usepackage{physics}
\usepackage{amsmath}
\usepackage{bbold}
\DeclareFontFamily{U}{bbold}{}
\DeclareFontShape{U}{bbold}{m}{n}{
  <-5> bbold5
  <5><6><7><8><9> gen * bbold
  <5.5> bbold5
  <10><10.95> bbold10
  <12><14.4> bbold12
  <17.28><20.74><24.88> bbold17
}{}
\usepackage{mathtools}
\usepackage{dsfont}
\usepackage{braket}
\usepackage{array}[=2016-10-06]
\usepackage{centernot}

\usepackage{placeins}

\usepackage[dvipsnames]{xcolor}

\definecolor{nqdcolor}{rgb}{0.5586, 0.0586, 0.4219}

\usepackage{tikz}

\usepackage{hyperref}
\hypersetup{colorlinks,citecolor=nqdcolor,linkcolor=nqdcolor,urlcolor=nqdcolor}

\usepackage{makecell}
\usepackage{multirow}
\usepackage{colortbl}
\makeatletter
\providecommand\insert@pcolumn{\insert@column}
\makeatother

\usepackage{graphicx}
\usepackage{algpseudocode}
\makeatletter
\newcounter{algorithm}
\renewcommand\thealgorithm{\arabic{algorithm}}
\def\fps@algorithm{htbp}
\def\ftype@algorithm{8}
\def\ext@algorithm{loa}
\def\fnum@algorithm{ALGO.~\thealgorithm}
\expandafter\newbox\csname fbox@8\endcsname
\newenvironment{algorithm}[1][htbp]{%
  \@float{algorithm}[#1]%
  \hrule height.8pt depth0pt\kern2pt
}{%
  \end@float
}
\makeatother
\usepackage[all]{hypcap}
\graphicspath{{figs/}}

\input{preamble/macros}

\makeatletter
\@ifundefined{ifmainmode}{\newif\ifmainmode\mainmodetrue}{}%
\@ifundefined{ifsuppmode}{\newif\ifsuppmode\suppmodetrue}{}%
\@ifundefined{ifcombinedmode}{\newif\ifcombinedmode\combinedmodetrue}{}%
\makeatother

\makeatletter
\patchcmd{\@bibdataout@aps}{author="08"}{author="48"}{}{}
\patchcmd{\@bibdataout@aps}{author="08"}{author="48"}{}{}
\makeatother
\makeatletter
\def\qpgft@checkpagefloat\@elt#1#2\@nil{%
 \@if@notdblfloat{#1}{}{\check@deferlist@stuck\do@startpage}%
}
\patchcmd{\do@startpage}
 {\check@deferlist@stuck\do@startpage}
 {\ifx\@deferlist\@empty
    \check@deferlist@stuck\do@startpage
  \else
    \expandafter\qpgft@checkpagefloat\@deferlist\@nil
  \fi}
 {}{}
\makeatother

\begin{document}

\author{Chengzhi Ye\orcid{0009-0004-0190-1625}}
\email{chengzhi.ye@ens-lyon.fr}
\affiliation{Max Planck Institute for the Physics of Complex Systems, N\"{o}thnitzer Str.~38, 01187 Dresden, Germany}
\affiliation{\'{E}cole Normale Sup\'{e}rieure de Lyon, CNRS, Laboratoire de Physique, 69342 Lyon, France}
\author{Marin Bukov\orcid{0000-0002-3688-9599}}
\email{mgbukov@pks.mpg.de}
\affiliation{Max Planck Institute for the Physics of Complex Systems, N\"{o}thnitzer Str.~38, 01187 Dresden, Germany}

\clearpage
\title{Geometric Theory of Quasiperiodically Driven Systems}

\ifmainmode

\begin{abstract}
    Quasiperiodically driven systems give rise to a rich set of physical phenomena in contemporary quantum physics, with applications ranging from nonequilibrium dynamics to Hamiltonian engineering and control. However, unlike periodically driven (Floquet) systems, their study is hindered by the absence of a generic theory to describe their evolution using an effective time-independent Hamiltonian and its eigenstates. Based on counterdiabatic driving, we construct a general framework to identify and determine the dynamically invariant spaces of arbitrary quasiperiodic drives (i.e., the generalization of Floquet states). We decompose their unitary evolution using the exact effective average-energy operator \AE\ and the corresponding gauge potential, which we compute via a variational principle in frequency-lattice operator space. We then formulate a four-fold taxonomy of quasiperiodic drives based on the geometry and topology of their dynamically invariant spaces into (non)abelian and (un)twisted classes. The abelian untwisted class admits a direct extension of Floquet's theorem to incommensurate quasiperiodic drives; going beyond, we derive a nonabelian generalization of Floquet theory. Finally, we propose a concept of ergodicity whose onset is marked by coalescing dynamically invariant eigenspaces, in direct analogy to the emergence of ergodicity in a mixed classical phase space. The effective mathematical framework, variational principle, and nonabelian ergodicity we put forward apply beyond quasiperiodic drives.
\end{abstract}

\maketitle

\section{Introduction}

Out-of-equilibrium quantum dynamics stands at the focal point of contemporary research in atomic, molecular and optical physics, condensed matter physics, and quantum information science. Since, for generic time-dependent Hamiltonians, the Schr\"odinger equation admits no closed-form solutions, theoretical progress has relied on protocols in which the time dependence possesses additional structure: sudden quenches~\cite{polkovnikov2011colloqium}, slow adiabatic ramps and their counterdiabatic shortcuts~\cite{kolodrubetz2017geometry,guerry-odelin2019shortcuts}, and periodic drives~\cite{Bukov_2015general_HFE}. For the latter, Floquet theory decomposes the time evolution into a periodic micromotion operator and a time-independent effective Hamiltonian, defining quasienergies and Floquet states. This theoretical framework underlies research in quantum simulation~\cite{Goldman2014FloquetGaugeFields,Goldman2016Topo,eckardt2017colloquium,Weitenberg2021FloquetGases,gavensky2025streda}, studies of quantum chaos and thermalization~\cite{dalessio2016quantum,ho2023quantum}, and nonequilibrium quantum matter~\cite{GomezLeon2013FloquetBlochBandTheory,Oka2019FloquetMaterials}. The immense success of this approach naturally leads to the question of which concepts survive when time-periodicity no longer holds.

Quasiperiodic drives, characterized by multiple incommensurate frequencies $\omegavec$ [i.e., $\kvec\cdot\omegavec\neq 0$ for all $\kvec {\in} \mathbb{Z}^D{\setminus}\{0\}$], trace trajectories that cover the torus $\T^D{=}[0,2\pi)^D$ ergodically and provide a natural generalization beyond the Floquet paradigm ($D{=}1$), see Fig.~\ref{fig:octopusQPCD}(c,d). Over the last decade, previous work has focused on topological transport and response in insulators~\cite{martin2017topological,crowley2019topological,crowley2020half,vuina2023giant,kaushik2026topological,long2022coupled} and superconductors~\cite{peng2018time}, on frequency conversion~\cite{nathan2019topological,korber2020interacting,boyers2020exploring,malz2021topological,korber2022topological,long2022boosting}, and on energy pumping~\cite{kolodrubetz2018topological,qi2021universal}. Studies of thermalization and energy-space localization~\cite{chabe2008experimental,nathan2021quasiperiodic,long2022many} have also attracted considerable attention in both integrable~\cite{nandy2017aperiodically,maity2019fibonacci} and critical systems~\cite{lapierre2020fine}, as well as in generic interacting models~\cite{dumitrescu2018logarithmically,else2020long,gallone2024prethermalization,he2023quasi,zhou2026emergent}. In condensed matter and statistical physics, the enlarged parameter space of quasiperiodic drives has emerged as an engineering playground for new dynamical phases of matter, including topological phases~\cite{long2021nonadiabatic,else2020long,friedman2022topological,dumitrescu2022dynamical} and discrete time quasicrystals~\cite{giergiel2019discrete,zhao2019floquet,else2020long,marripour2025time,marripour2026emergence,luo2025discrete,he2025experimental,moon2025experimental,zhu2026observation}. Many of these phenomena have been realized experimentally in cold atoms~\cite{chabe2008experimental}, superconducting qubits~\cite{malz2021topological}, trapped ions~\cite{dumitrescu2022dynamical}, nitrogen-vacancy (NV) centers~\cite{boyers2020exploring,he2023quasi,he2025experimental,moon2025experimental}, and Rydberg atoms~\cite{zhu2026observation}. All these studies rely, whether explicitly or implicitly, on some notion of a stationary state and a quasienergy for the quasiperiodic drive; however, for $D{>}1$ neither object has a generically valid definition, and the works above obtain them, where needed, only under model-specific assumptions.

Indeed, despite their apparent similarity with Floquet systems ($D{=}1$), quasiperiodic drives with $D{>}1$ lack a self-contained description in terms of effective operators such as the micromotion and quasienergy. Analytically, many-mode Floquet theory~\cite{ho1983semiclassical} has limitations arising from the nature of the quasienergy spectrum~\cite{luck1988response,blekher1992floquet} that manifest as small $\kvec\cdot\omegavec$ denominators in generalized inverse-frequency expansions~\cite{verdeny2016quasi}, closely related to the existence of Anderson localization on the frequency lattice~\cite{casati1989anderson}. Numerically, Diophantine approximation has been widely used to construct sequences that approximate a given quasiperiodic drive using a family of periodic drives~\cite{crowley2019topological,verdeny2016quasi}; however, the sequence of Floquet states need not have a well-defined limit or may converge to incorrect states~\cite{martin2017topological}. Similarly, exact diagonalization on the frequency lattice \cite{verdeny2016quasi,crowley2019topological} suffers from uncontrolled truncation since the quasienergy spectrum is invariant under shifts by $\kvec\cdot\omegavec$ which are dense on the energy axis, see Fig.~\ref{fig:octopusQPCD}(e); the procedure is intrinsically ill-posed for topological stationary states that cannot be defined globally on $\T^D$~\cite{crowley2019topological,crowley2020half,long2022coupled}, and for nonabelian stationary-state spaces that cannot be reduced to single wavefunctions~\cite{eliasson2002ergodic}. These issues create a formidable bottleneck to unleashing the full potential of quasiperiodically driven systems, and point to the need to develop a reliable framework for their investigation.

\begin{figure*}[t!]
    \centering
    \includegraphics[width=\textwidth]{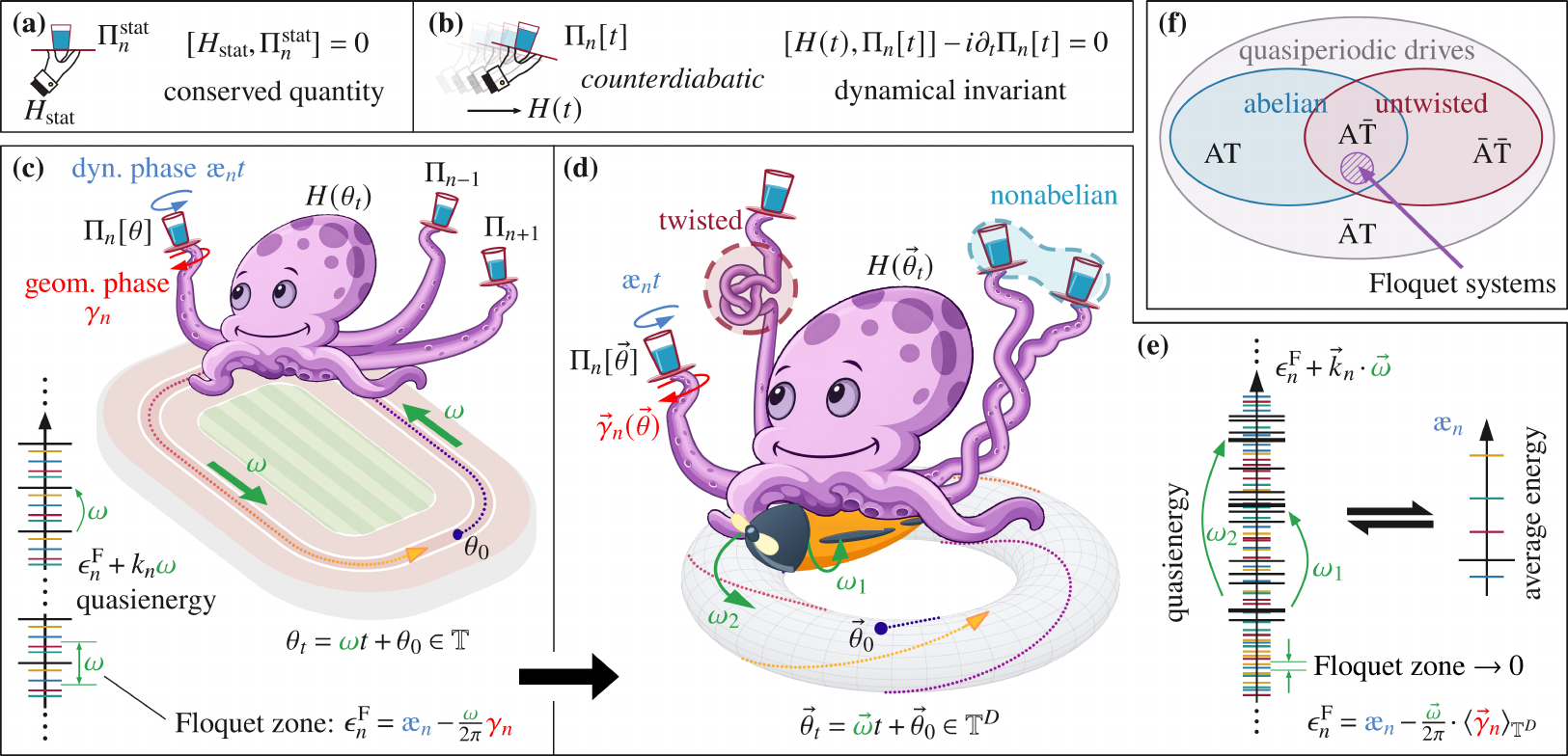}
    \caption{ \textbf{Geometric theory of quasiperiodic drives from counterdiabatic driving.} (a) In static systems, conserved quantities correspond to symmetry operators commuting with the time-independent Hamiltonian $H_\mathrm{stat}$, and are built out of its eigenprojectors $\Pi^{\rm stat}_n$. (b) In dynamical systems described by a time-dependent Hamiltonian $H(t)$, applying Noether's theorem identifies the so-called dynamically invariant projectors $\Pi_n[t]$ which commute with the Hamiltonian up to a time derivative. States encoded in $\Pi_n[t]$ are called stationary; they are protected against excitations during evolution generated by $H(t)$ by a form of counterdiabatic driving w.r.t.~the gauge potential $[i\partial_t\Pi_n[t],\Pi_n[t]]$. (c) A Floquet system can be thought of as a waiter-octopus running in a cycle at frequency $\omega$, representing the time-periodic Hamiltonian $H(\theta_t)$ with $\theta_t=\omega t + \theta_0$; its tentacles embody dynamically invariant spaces $\Pi_n=|n^\F\rangle\langle n^\F|$, whose Floquet states $|n^\F\rangle$ accumulate only dynamical $\mae_n t$ and geometric $\gamma_n$ phases. These phases sum up to define the quasienergies $\epsilon^\F_n$ which are invariant w.r.t.~discrete shifts by integer multiples of the drive frequency $\omega$, giving rise to a finite Floquet zone of size $\omega$. (d) Quasiperiodically driven systems, $H(\thetavec_t)$, follow trajectories on a torus $\mathbb{T}^D$ parametrized by time $\thetavec_t=\omegavec t + \thetavec_0$; here $D=2$ with frequency vector $\omegavec=(\omega_1,\omega_2)$; knotted and intertwined tentacles, respectively, represent twisted (topological) and nonabelian dynamically invariant spaces $\Pi_n$. Conventional Floquet theory breaks down due to these topological and dynamical geometric obstructions. (e) Incommensurate frequencies give rise to quasienergies $\epsilon^\F_n$ whose folding by $\kvec_n\cdot\omegavec$ is dense on the energy axis, and the Floquet zone vanishes. We develop an alternative description using geometric phase $\gamma$ and average energy $\mae$. (f) Quasiperiodic drives can be classified as abelian/nonabelian (A/\=A) and twisted/untwisted (T/\=T); the abelian untwisted (A\= T) class admits an exact Floquet decomposition for the unitary evolution even for incommensurate drives with no period. }
    \label{fig:octopusQPCD}
\end{figure*}

\subsection{Summary of results}

We consider generic quasiperiodically driven systems governed by a smooth Hamiltonian $H(\thetavec_t)$ with driving phase $\thetavec_t=\omegavec t+\thetavec_0$ on the torus $\T^D$. Our main result is an exact decomposition of their evolution operator,
\begin{eqnarray}
\label{eq:main}
    U(t,0) &=& \mathcal{T}_s\mathrm e^{-i \int_{0}^t\mathrm ds\, H(\thetavec_s)}  \\
    &=& W^\mAE(t;0)\,\mathrm{e}^{-i t\mAE[\thetavec_0]}, \nonumber\\
    W^\mAE(t;0)&=&\mathcal{T}_s\mathrm e^{-i \int_{0}^t\mathrm ds\, A^\mAE(\thetavec_s)}, \nonumber
\end{eqnarray}
into a Wilson line $W^\mAE(t;0)$ generated by a gauge potential $A^{\mAE}$ and a dynamical phase generated by a time-independent average-energy operator $\mAE$, both globally smooth on $\T^D$ for every smooth drive and every incommensurate $\omegavec$, with no further assumption. The similarity of the decomposition to Geometric Floquet theory~\cite{schindler2025geometric} is formal; as we demonstrate, Eq.~\eqref{eq:main} contains a strictly richer set of physical phenomena without Floquet counterparts.

The eigenprojectors $\Pi_n[\thetavec]$ of the average energy $\mAE[\thetavec]$ define dynamically invariant spaces; the states they contain are stationary\footnote{i.e., states which accumulate only a phase under the exact $U(t,0)$, for one-dimensional invariant spaces.}, and generalize the Floquet states. These stationary states can be dressed by the quasiperiodic drive in the same way a Floquet drive modifies the properties of static states; this opens an extension of Hamiltonian engineering to quasiperiodic drives. Unlike Floquet states, the dynamically invariant eigenprojectors of quasiperiodic drives are not exact eigenstates of the evolution operator $U(t,0)$ at any $t$; by continuity, however, they are approximate eigenspaces at a sequence of near-return times $T^{(i)}$, because they evolve back arbitrarily close to themselves at these instants. Under the exact quasiperiodic drive, their time evolution reduces to collecting a generalized geometric phase (a.k.a.~holonomy) described by the Wilson line operator $W^\mAE$ generated by the average-energy gauge potential $A^\mAE$, and a dynamical phase generated by the average energy operator $\mAE$, see Fig.~\ref{fig:octopusQPCD}(d). The average-energy decomposition thus represents an instance of exact counterdiabatic (CD) driving, illustrated in Fig.~\ref{fig:octopusQPCD}(a,b).

The dynamical geometric and topological properties of the drive-induced stationary-state projectors $\Pi_n[\thetavec]$ imply a four-fold classification of quasiperiodically driven systems, see Fig.~\ref{fig:octopusQPCD}(f):
\begin{itemize}
    \item[A\=T:] \textit{Abelian untwisted} drives%
    \footnote{ We use the term \textit{twisted} to denote a topologically nontrivial wavefunction fiber bundle over the torus $\T^D$ (i.e., admitting no globally smooth frame); the term \textit{topological} can be ambiguous in the condensed matter literature. } are defined by possessing topologically trivial, one-dimensional stationary-state invariant spaces. We find this class contains all Floquet systems and forms a strictly larger set, and provide unambiguous definitions of quasienergy, Floquet states, and micromotion for incommensurate drives, together with numerical algorithms for computing them.
\end{itemize}
Beyond the A\=T class, conventional Floquet theory fails; the corresponding drive classes therefore provide a playground to exhibit phenomena without Floquet counterparts:
\begin{itemize}
    \item[AT:] \textit{Abelian twisted} drives have topologically nontrivial one-dimensional dynamically invariant spaces [Fig.~\ref{fig:octopusQPCD}(d)]. The topological obstruction forbids globally defined stationary eigenkets on the entire torus $\T^D$~\cite{crowley2019topological}; as a result, the micromotion is not globally well-defined, and the quasienergy operator no longer captures the evolution despite being a dynamical invariant. The Floquet decomposition for the unitary $U(t,0)$ breaks down.

    \item[\=A\=T:] \textit{Nonabelian untwisted} drives have dynamically invariant spaces characterized by trivial topology and nonabelian geometry. The notion of individual stationary-state eigenstates breaks down for \=A\=T drives [Fig.~\ref{fig:octopusQPCD}(d)], since nonabelian invariant spaces are irreducible along the quasiperiodic trajectory $\thetavec_t$. Hence, $\mAE$-basis transport via the Wilson line is inherently nonabelian.

    \item[\=AT:] Finally, the \textit{nonabelian twisted} class combines the features of the AT and \=A\=T classes; hence, it is the most general class among the four, reducing to any of the other three in special cases, see Fig.~\ref{fig:octopusQPCD}(f).
\end{itemize}

For almost all incommensurate frequency vectors, we also derive generalizations of Floquet theory for nonabelian and twisted drives with micromotion and quasienergy operators defined almost everywhere on $\T^D$ [Appendix~\ref{sec:SuppDynamicsReduction}]. Remarkably, unlike its Floquet counterpart, the average-energy decomposition is always well-defined for any quasiperiodic drive, and the unitary evolution operator~\eqref{eq:main} has a single time-ordered exponential, the Wilson line. This makes it a well-behaved object for numerical simulations.

We compute the average-energy decomposition variationally. Once a sufficiently accurate approximation for the AGP is obtained, we compute the average energy operator as $\mAE=H-A^\mAE$; in general, its spectral decomposition gives directly the dynamically invariant spaces corresponding to the generalized stationary states $\Pi_n[\thetavec_t]$ of the quasiperiodic drive $\thetavec_t$. We apply our variational principle to a quasiperiodically driven two-qubit toy model with independently tunable parameters controlling abelianness and twistedness. This allows us to demonstrate all four classes (A\=T, AT, \=A\=T, \=AT). We also benchmark the accuracy of the variational approximation using an analytically solvable version of this toy model.

Investigating the ultimate limits of the average energy decomposition, we propose the notion of \textit{nonabelian ergodicity}. We examine the evolution at times $T^{(i)}$ when the drive trajectory returns arbitrarily close to its initial value; at these near-return times, each invariant space $\Pi_n[\thetavec_t]$ almost returns to itself, while its states can acquire phases or undergo nonabelian rotations. Nonabelian ergodicity means that each space is explored densely by the near-return evolution. As a model parameter is tuned, the smallest dynamically invariant spaces can merge into larger ones to be explored, as it happens, e.g., when a quasiperiodic drive transitions from an abelian to a nonabelian class. This resembles the onset of classical ergodicity: the projectors $\Pi_n$ play a role analogous to action variables labeling invariant tori, whose breakdown connects previously separated regions of phase space. Once all eigenspaces have merged, the only globally smooth quasiperiodic dynamical invariants left are multiples of the identity; thus, $\mAE\propto\mathds{1}_{N\times N}$ and the average-energy decomposition no longer separates distinct stationary-state sectors of the dynamics; it encodes only an overall dynamical phase. Hence, complete nonabelian ergodicity corresponds to near-return evolution exploring all unitary rotations of the Hilbert space allowed by symmetry constraints.

\subsection{Relation to previous work}

The study of quasiperiodically driven systems has a long-standing history, with questions about their effective description and classification raised earlier. For instance, Ref.~\cite{crowley2019topological} first proposed a topological classification of quasiperiodic drives and related the topological obstruction to delocalization of the Floquet states on the frequency lattice; here, we show explicitly that the dynamically invariant projectors remain localized since they are insensitive to the topological singularity in the wavefunction phase. This observation enables us to investigate higher-rank irreducible invariant spaces, and identifies frequency-lattice \textit{operator} space as the setting for stable numerical algorithms. In the mathematical literature, the existence of a Floquet-type decomposition is the question of reducibility of the quasiperiodic cocycle generated by $H(\thetavec_t)$~\cite{eliasson2002ergodic,karaliolios2018continuous}; the relation of our invariant projectors, which exist irrespective of whether the cocycle is reducible, to that body of work is discussed in Appendix~\ref{app:subsec:math_lit}. To the best of our knowledge, no numerically stable procedure, free of dense quasienergy folding [Fig.~\ref{fig:octopusQPCD}(e)], is currently available for computing decomposition~\eqref{eq:main} of generic drives. Moreover, the mathematics literature often discusses the spectrum (pure point vs.~continuous) of the frequency-lattice Schr\"odinger operator $\mathcal{K}$~\cite{blekher1992floquet}, an object that our geometric theory elegantly circumvents. Instead, analyzing the dynamical geometric and topological properties of the stationary-state projectors allows us to formulate a more general, nonabelian version of Floquet's theorem that goes significantly beyond the existing literature as it applies to nonabelian and twisted drives, for almost all frequency vectors $\omegavec$. Finally, the notion of nonabelian ergodicity we propose is distinct from unitary ergodicity or Hilbert-space ergodicity~\cite{pilatowsky2024hilbert}: unlike Hilbert-space or unitary ergodicity, which follow a single state or the evolution operator at all times, nonabelian ergodicity probes the evolution at near-return times where similar dynamics recur, distinguishing internal nonabelian mixing from the full motion of invariant spaces over the drive manifold.

Against this background, our work achieves four key contributions: (i) Motivated by counterdiabatic driving [Fig.~\ref{fig:octopusQPCD}(b)], we formulate an effective description for the dynamics of quasiperiodically driven systems using the average-energy operator $\mAE[\thetavec]$ and its gauge potential $A^\mAE(\thetavec)$; unlike the quasienergy and the micromotion, these objects are well-behaved (i.e., smooth on $\T^D$) for any smooth quasiperiodic drive. (ii) These properties, together with the counterdiabatic relation between them, allow us to identify the appropriate space and formulate a variational principle to compute the average energy. Importantly, our variational principle applies to any quasiperiodic drive. The identification of a procedure to compute stationary states (the counterpart of Floquet states) thus brings quasiperiodically driven systems on par with Floquet systems in the pursuit of new physical phenomena. (iii) In this respect, our work provides an algorithmic procedure to directly classify quasiperiodically driven systems according to the geometric and topological properties of their dynamically invariant spaces; we determine the class (A\=T, AT, \=A\=T, \=AT) by diagonalizing $\mAE$. Crucially, this is a classification of the eigenspaces of the effective average energy operator $\mAE$, not the instantaneous drive Hamiltonian $H$. (iv) Moreover, by identifying the merging of these invariant spaces at critical values of the system parameters, we propose a new notion of ergodicity whose onset is determined by the dynamical nonabelian properties of the effective eigenprojectors. Remarkably, our theory is not limited to quasiperiodic drives, and readily extends to more general time-dependencies~\cite{wu2026geometric}; it therefore provides a new framework to understand and investigate nonequilibrium systems.

\subsection{Structure of the paper}

This paper is organized as follows. In Sec.~\ref{sec:GFT}, we derive the universal average-energy decomposition of quasiperiodic drives, and then the dynamically invariant projectors, which are the central objects of this study. Section~\ref{sec:classification} identifies four drive classes from their abelianness and twistedness; we illustrate this four-fold classification with a coupled two-qubit model. In Sec.~\ref{sec:num_algos}, we explain why Diophantine approximations and frequency lattice diagonalization can fail, and develop a variational method to compute the average-energy operator and its dynamically invariant eigenprojectors. Section~\ref{sec:ergodicity} connects the merging of invariant spaces to nonabelian ergodicity in the near-return dynamics. We conclude with a discussion and outlook in Sec.~\ref{sec:outro}. In addition, we derive a generalized Floquet theory for all four quasiperiodic classes in Appendix~\ref{sec:SuppDynamicsReduction}; complementary numerical results and mathematical details about the classification, the ubiquity of these drives, and different methods can also be found in Appendices~\ref{app:omega-irreducibility}, \ref{sec:LandscapeQP}, \ref{app:sec:model}, and \ref{sec:Diophantine}.

\section{\label{sec:GFT}Geometric Decomposition and Effective Evolution}

Before we investigate the generalization of Geometric Floquet theory to quasiperiodically driven systems, let us briefly recap the key ideas behind this equivalent formulation of Floquet theory, illustrated in Fig.~\ref{fig:octopusQPCD}(c). Instead of the usual explicit time dependence, we adjust the notation to foreshadow a generalization of the formalism to quasiperiodic drives.

\subsection{\label{sec:GFT-floquet}Geometric Floquet theory}

Periodic drives are defined by a trajectory $\theta_t=\theta(t)=\omega t+\theta_0$ on the circle $\T=[0,2\pi)$, parametrized by time $t$ and wrapping at a single constant frequency $\omega$; we assume the starting time is $t=0$ and hence the initial phase is $\theta_0$. With a slight abuse of notation, we will interchangeably use $\theta_t$ and $t$ below. We also use the real time variable $t$ to parametrize physical time evolution, whereas operators defined globally on the circle $\T$ are parametrized by $\theta\in\T$; we often drop the $t$-dependence to make the notation lighter.

Consider a Hamiltonian $H(\theta_t)$, with $H$ a smooth operator-valued function on the circle $\mathbb T=[0,2\pi)$ evaluated along the trajectory $\theta_t$. Geometric Floquet theory~\cite{schindler2025geometric} allows us to state Floquet's theorem infinitesimally as three equivalent decompositions:
\begin{subequations}
\label{eq:GFT_decomp}
\begin{alignat}{2}
    H(\theta_t) &= H^\F[\theta_t] + A^\F(\theta_t)   \quad &&\makebox[3.5cm][l]{\text{(Floquet/q'energy gauge)}} \label{eq:HF_decomp} \\
    &= H^\K(\theta_t) + A^\K(\theta_t)   \quad &&\makebox[3.5cm][l]{\text{(Kato gauge)}} \label{eq:HK_decomp} \\
    &= \mAE[\theta_t] + A^\mAE(\theta_t) \quad &&\makebox[3.5cm][l]{\text{(average-energy gauge).}} \label{eq:AE_decomp}
\end{alignat}
\end{subequations}
In the Floquet gauge of Eq.~\eqref{eq:HF_decomp}, $H^\F[\theta_0]=\sum_{n=1}^N \epsilon^\F_n\Pi_n[\theta_0]$ is the quasienergy operator with time-independent quasienergies $\epsilon^\F_n$ and Floquet eigenprojectors $\Pi_n[\theta_0]=\ket{n^\F[\theta_0]}\bra{n^\F[\theta_0]}$. The square bracket $[\theta_0]$ denotes a parametric dependence on the starting phase of the drive. We assume the Floquet states are nondegenerate; for the degenerate case, see~\cite{zenodo}. The micromotion $P(\theta_t;\theta_0)=\mathcal T_s \mathrm e^{-i\int_0^t \mathrm ds\,A^{\F}(\theta_s)}$ is generated by the Floquet gauge potential $A^\F(\theta) = H(\theta)-H^\F[\theta]$, and depends on the starting phase $\theta_0$ via the trajectory $\theta_t$. The Floquet decomposition Eq.~\eqref{eq:HF_decomp} is defined by the periodic gauge $P(\theta+2\pi;\theta_0) = P(\theta;\theta_0)$, a partial gauge-fixing condition with residual gauge freedom leading to $\epsilon^\F_n \to \epsilon^\F_n + k_n \omega$ for some integer $k_n\in\mathbb{Z}$ \cite{schindler2025geometric}. The relation $H^\F[\theta] = P(\theta;\theta_0)H^\F[\theta_0]P(\theta;\theta_0)^\dagger$ defines the Floquet Hamiltonian corresponding to the starting phase $\theta$.

The equivalent Kato decomposition, Eq.~\eqref{eq:HK_decomp}, instead uses the Kato Hamiltonian $H^\K(\theta) {=} \sum_n E^\K_n(\theta)\Pi_n[\theta]$, where the Kato energies are given by $E^\K_n(\theta) {=} \mathrm{tr}\left(\Pi_n[\theta] H(\theta)\right)$, and $\Pi_n[\theta]$ are again the Floquet eigenprojectors. The Kato adiabatic gauge potential (AGP) is $A^\K(\theta) = -1/2\sum_n [\Pi_n[\theta], i\omega \partial_\theta\Pi_n[\theta]]$. The Kato decomposition makes use of the instantaneous parallel transport gauge, defined by the instantaneous orthogonality conditions $\Pi_n[\theta]A^{\K}(\theta)\Pi_n[\theta]=0$ for all $n$; it then follows that $\mathrm{tr}\left( H^\K(\theta)A^\K(\theta) \right)=\sum_nE_n^{\K}\text{tr}(\Pi_nA^{\K})=0$ at any point on the circle, $\theta\in\T$, and hence for every time $t$. Unlike the Floquet AGP $A^\F$, the Kato AGP $A^\K$ generates the geometric phases of the Floquet states when integrated over the circle.

Finally, Eq.~\eqref{eq:AE_decomp} defines the average-energy decomposition with the average-energy operator $\mAE[\theta]{=}\sum_n \mae_n \Pi_n[\theta]$ whose eigenvalues are the $\theta$-independent  average energies $\mae_n {=} (2\pi)^{-1}\int_0^{2\pi}\mathrm d\theta E^\K_n(\theta) = (2\pi)^{-1} \int_0^{2\pi}\mathrm d\theta \, \mathrm{tr}\left(\Pi_n[\theta] H(\theta)\right)$, and with the same Floquet eigenprojectors $\Pi_n[\theta]$. The average-energy gauge potential reads as $A^\mAE(\theta) = A^\K(\theta) + \sum_n (E^\K_n(\theta) - \mae_n)\Pi_n[\theta]$. The average-energy decomposition imposes a weaker parallel transport gauge, valid on average over the circle: $(2\pi)^{-1}\int_0^{2\pi}\mathrm d\theta\, \mathrm{tr}\!\left(\Pi_n[\theta]A^\mAE(\theta)\Pi_n[\theta]\right)=0$. This combines the best of the periodic and parallel transport gauges: the average energies $\mae_n$ are constant over the circle like the quasienergies $\epsilon^\F$, whereas $A^\mAE$ generates the geometric phases over the circle much like $A^\K$.

To see this explicitly, let us look at the unitary evolution generated by the three decompositions, which can accordingly be written in three equivalent ways:
\begin{subequations}
\label{eq:U_evo}
\begin{align}
    U(t,0) &= \mathcal{T}_s\mathrm e^{-i \int_{0}^t\mathrm ds\,  A^\F(\theta_s)}\; \mathrm{e}^{-i t H^\F[\theta_0]}
    \label{eq:U_F} \\
    &= \mathcal{T}_s\mathrm e^{-i \int_{0}^t\mathrm ds\,  A^\K(\theta_s)}\; \mathrm{e}^{-i \int_{0}^t\mathrm ds\, \sum_n E^\K_n(\theta_s) \Pi_n[\theta_0]}
    \label{eq:U_K} \\
    &= \mathcal{T}_s\mathrm e^{-i \int_{0}^t\mathrm ds\,  A^\mAE(\theta_s)}\; \mathrm{e}^{-i t \mAE[\theta_0]}.
    \label{eq:U_AE}
\end{align}
\end{subequations}
Unlike the micromotion, which returns to the identity operator after one cycle, the Kato and average-energy AGPs both generate the geometric phases $\gamma_n$ of the Floquet states over one drive cycle, via $\mathrm e^{i\gamma_n}=\langle n^{\F}[\theta_0]|\,\mathcal T_s \mathrm e^{-i\int_0^T \mathrm ds\,A^{\K}(\theta_s)}\,|n^{\F}[\theta_0]\rangle$. In turn, the average energy itself carries the physical meaning of a dynamical `phase'. We therefore have\footnote{The equality between Eqs.~\eqref{eq:UF_Kato} and~\eqref{eq:UF_AE} relies on the relation $\int_0^T(E_n^{\K}-\mae_n)\mathrm dt=0$, i.e., on the Floquet eigenspaces being abelian, see Sec.~\ref{sec:classification}.}
\begin{subequations}
\label{eq:UF}
\begin{eqnarray}
    U(T,0) &=& \mathrm{e}^{-i T H^\F[\theta_0]} \\
    \label{eq:UF_Kato}
    &=& \mathcal{T}_s\mathrm e^{-i \int_{0}^T\mathrm ds\,  A^\K(\theta_s)}\; \mathrm{e}^{-i T\mAE[\theta_0]} \\
    \label{eq:UF_AE}
    &=& \mathcal{T}_s\mathrm e^{-i \int_{0}^T\mathrm ds\,  A^\mAE(\theta_s)}\; \mathrm{e}^{-i T\mAE[\theta_0]}.
\end{eqnarray}
\end{subequations}
Because the Floquet Hamiltonian $H^\F$, the Kato Hamiltonian $H^\K$, and the average energy operator $\mAE$ share the eigenprojectors $\Pi_n[\theta]$, we obtain the following useful relation~\cite{fainshtein1978some,grifoni1998contemporary,le2022missing}
\begin{eqnarray}
\label{eq:epsF-gamma-ae}
    \epsilon^\F_n &=& \left(\mae_n - T^{-1}\gamma_n\right)  \text{mod}\ \omega \nonumber\\
    &=& \left(E^\K_n(\theta) - a^\F_n(\theta)\right)  \text{mod}\ \omega  ,
\end{eqnarray}
where $a^\F_n(\theta) = \langle n^\F[\theta]|A^\F(\theta)|n^\F[\theta]\rangle$ is the Floquet-Berry connection which obeys the gauge transformation, $a^\F_n(\theta)\longrightarrow a^\F_n(\theta) - k_n\omega$; the second equality follows from projecting Eq.~\eqref{eq:HF_decomp} onto the Floquet eigenspace $\Pi_n[\theta]$.

Unlike the Kato spectrum $E^\K_n(\theta)$, the quasienergy $\epsilon^\F_n$ and average energy $\mae_n$ spectra are constant over the circle (i.e., do not depend on the initial phase $\theta_0$). Moreover, the projectors $\Pi_n[\theta]$ are built out of Floquet eigenstates which are stationary states of the dynamics generated by the lab-frame Hamiltonian $H(\theta)$. In general, Hermitian operators obeying such a structure are known as (Lewis-Riesenfeld) dynamical invariants, and are defined by the equation
\begin{equation}
\label{eq:L}
    \LL(O[\theta])=0,\qquad \LL(\cdot) := [H(\theta),\cdot]- i\omega\partial_\theta(\cdot) ,
\end{equation}
where the superoperator $\LL$ is the Floquet-Liouvillian. Dynamical invariants serve as generalizations of symmetries in dynamical systems [Fig.~\ref{fig:octopusQPCD}(a,b)]. Generalized conserved quantities are linear combinations of dynamical invariants which, in the case of many-body systems, obey spatial locality in addition. To clearly designate them, we use the square bracket notation $O[\theta]$. Thus, the quasienergy operator $H^\F[\theta]$ and the average-energy operator $\mAE[\theta]$ are dynamical invariants, whereas the Kato Hamiltonian $H^\K(\theta)$ is not; it is for this reason that its dynamical phase in Eq.~\eqref{eq:U_K} involves a proper time integral.

Observe that, for any static operator $O$, $U(t,0) O U(t,0)^\dagger$ defines trivially a dynamical invariant. However, the quasienergy and the average energy are special because they are invariants periodic with the same period as the drive; i.e., they are smooth global functions on the circle $\T$. We therefore restrict the Liouvillian $\LL$ to operators on $\T$. One can convince oneself that, for nondegenerate quasienergies, the Floquet eigenprojectors $\Pi_n[\theta]$ span the space of all periodic dynamical invariants for periodically driven systems. Floquet's theorem is then equivalent [up to gauge fixing, cf.~Eq.~\eqref{eq:GFT_decomp}] to the linear-algebra statement that the number of linearly independent directions in the space of dynamical invariants equals the dimension $N$ of the Hilbert space, i.e., $\dim\ker\LL=N$, which holds exactly when the quasienergies are nondegenerate.

\subsection{Generalization to Quasiperiodically Driven Systems}

We now turn to quasiperiodically driven systems, where $\thetavec_t=\omegavec t+\thetavec_0\in\T^D$ is the drive phase for $D$ incommensurate frequencies $\omegavec$. Quasiperiodic drives lead to trajectories that fill the torus $\T^D$ densely in time; hence, they never close and do not possess a finite period. This raises the question about the existence of an effective time-independent generator of dynamics, different from the drive $H(\thetavec)$ but akin to the Floquet Hamiltonian for $D{=}1$. In Sec.~\ref{sec:classification}, we show that the micromotion-quasienergy paradigm fails to generalize to quasiperiodically driven systems when the drive induces topological or nonabelian obstructions, and argue that the average energy operator provides a natural description for the effective dynamics.

First, notice that the notion of dynamical invariants is general, and hence it also applies to quasiperiodic drives. In particular, we call $O[\thetavec]$ an invariant under the dynamics generated by $H(\thetavec)$, if it obeys the generalization of Eq.~\eqref{eq:L}:
\begin{equation}
\label{eq:Lqp}
    \LL(O[\thetavec])=0,\qquad \LL(\cdot) := [H(\thetavec),\cdot]- i\omega^\mu\partial_\mu(\cdot) ,
\end{equation}
where $\partial_\mu = \partial_{\theta_\mu}$, $\mu=1,\dots,D$. Repeated indices are summed over. Note that all dynamical invariants implicitly depend on the frequency vector%
\footnote{ This feature will give rise to a property called dynamical irreducibility that we introduce in Sec.~\ref{sec:classification}. } $\omegavec$ through the Liouvillian $\LL$, now acting on smooth operator-valued functions on the torus $\T^D$.
As in the Floquet case, we are interested in the space of globally smooth dynamical invariants on $\T^D$, which we identify with $\ker\LL$. Let $\{\Pi_n[\thetavec]\}$ denote a projector-valued orthonormal basis for the vector space $\ker\LL$,%
\footnote{ $\ker\LL$ is a finite-dimensional unital *-algebra, i.e., it is closed under (i) addition, by linearity; (ii) multiplication: $A,B\in\ker\LL$, $\LL(AB)=A\LL(B)+\LL(A)B=0$; (iii) Hermitian conjugation: $\LL(O^\dagger) = -[\LL(O)]^\dagger=0$; and (iv) contains the identity, $\LL(\mathds{1}_{N\times N})=0$. We assume that this algebra is commutative; the noncommutative case leading to degeneracy is discussed in~\cite{zenodo}. } with $\Pi_m[\thetavec]\Pi_n[\thetavec]=\delta_{mn}\Pi_n[\thetavec]$ and $\sum_n \Pi_n[\thetavec]=\mathds{1}_{N\times N}$ pointwise for every $\thetavec\in\T^D$; we relate the physical meaning of the size of this basis set to a new notion of ergodicity in Sec.~\ref{sec:ergodicity}.

Given the lab-frame Hamiltonian $H(\thetavec)$, we can generalize the Kato Hamiltonian and gauge potential to quasiperiodic drives as
\begin{eqnarray}
\label{eq:Kato_objs}
    H^\K(\thetavec) &=& \sum_n \Pi_n[\thetavec]H(\thetavec)\Pi_n[\thetavec], \\
    A^\K(\thetavec) &=& \omega^\mu A^\K_\mu(\thetavec),\quad A^\K_\mu = -\frac{1}{2}\sum_n [\Pi_n[\thetavec],\, i\partial_\mu\Pi_n[\thetavec] ]. \nonumber
\end{eqnarray}
The Kato gauge is defined by the condition $\Pi_n[\thetavec]A^{\K}_\mu(\thetavec)\Pi_n[\thetavec]{=}0$ for all $n$. As before, these definitions guarantee the orthogonality $\text{tr}\left( H^\K(\thetavec) A^\K(\thetavec) \right){=}0$ for all $\thetavec\in\T^D$. We can also generalize the Kato energies as $E^\K_n(\thetavec) {=} \text{tr}\left(\Pi_n[\thetavec]H(\thetavec)\right)/r_n$, with $r_n{=}\text{rank}(\Pi_n){=}\text{tr}\,(\Pi_n)$ constant over $\T^D$ by continuity. However, note that the Kato Hamiltonian $H^{\rm K}{=}\sum_n\Pi_nH\Pi_n$ reduces to $\sum_n E^{\K}_n\Pi_n$ only for rank-one spaces, $r_n{=}1$; for higher-rank spaces, $r_n{>}1$, it retains the traceless block $\Pi_nH\Pi_n{-}E^{\K}_n\Pi_n$, see Appendix~\ref{app:KatoQP}.

Therefore, in complete analogy with Geometric Floquet theory, we change the gauge to define the average energy and its AGP
\begin{eqnarray}
\label{eq:AE_objs}
    \mAE[\thetavec] &=& \sum_n \mae_n\, \Pi_n[\thetavec], \nonumber\\
    A^\mAE(\thetavec) &=& H(\thetavec)  - \sum_n \mae_n\,\Pi_n[\thetavec]\, .
\end{eqnarray}
Once again, the average energy $\mAE[\thetavec]$ is a dynamical invariant and its eigenvalues $\mae_n = (2\pi)^{-D}\int_{\T^D}\mathrm d^D\theta\, E^\K_n(\thetavec)$ are constant over the torus $\T^D$. Notice that, similar to Floquet systems, we can readily use the average energy to unambiguously define a lowest (ground-state) manifold, i.e., the $\mae_n$ spectrum comes out sorted~\cite{schindler2025geometric}. In Sec.~\ref{sec:num_algos} we discuss numerical techniques to compute the average energy $\mAE[\thetavec]$ and its gauge potential $A^\mAE(\thetavec)$.

By construction, the lab-frame drive Hamiltonian obeys the exact geometric decompositions
\begin{subequations}
\label{eq:qpGFT_decomp}
\begin{align}
    \!\!\! H(\thetavec) &= H^\K(\thetavec) + A^\K(\thetavec) \qquad &&\text{(Kato gauge)}
    \label{eq:qpKato_decomp}
    \\
    &= \mAE[\thetavec] + A^\mAE(\thetavec) \qquad &&\text{(average-energy gauge)}\, .
    \label{eq:qpAE_decomp}
\end{align}
\end{subequations}
We assume that the invariant projectors $\Pi_n$ all have minimal ranks $r_n$.

Since the geometric decompositions implement exact counterdiabatic driving for the invariant projectors $\Pi_n[\thetavec]$, it follows directly that along the quasiperiodic trajectory $\thetavec_t=\omegavec t+\thetavec_0$, the evolution factorizes exactly as
\begin{eqnarray}
\label{eq:U_qp}
    U(t,0) &=& \mathcal{T}_s\mathrm e^{-i \int_{0}^t\mathrm ds\,  A^\mAE(\thetavec_s)}\; \mathrm{e}^{-i t\mAE[\thetavec_0]}, \nonumber \\
    &=& W^\mAE(t;0)\,\mathrm{e}^{-i t\mAE[\thetavec_0]}\, ,
\end{eqnarray}
which defines the Wilson line $W^{\mAE}(t;0)$. A key contribution of this work is that the unitary evolution decomposition~\eqref{eq:U_qp} is generic and applies to all quasiperiodically driven systems. The corresponding Kato equation acquires a modification in general, and is discussed in Appendix~\ref{app:KatoQP}. By contrast, the concepts of a quasienergy operator $H^\F$ and the corresponding micromotion $P$ cannot be unconditionally generalized to all quasiperiodically driven systems, see Secs.~\ref{subsec:AN}-\ref{subsec:NT} and Appendix~\ref{app:sec:4classes}.

The physical interpretation of the geometric decomposition~\eqref{eq:U_qp} for the evolution carries over from Floquet systems: the average-energy AGP generates the geometric component of the evolution, together with the zero-average dynamical part, through the Wilson line operator $W^\mAE(t;0)$, whereas $\mAE$ generates the dynamical phase. Unlike Floquet systems, however, there is no well-defined notion of stroboscopic dynamics, in general. Related to this, whereas $\mAE[\thetavec]$ and $A^\mAE(\thetavec)$ are globally defined smooth operators on the torus $\T^D$, the Wilson line $W^\mAE(t;0)$ is not: it depends on the entire path $\{\thetavec_s\}_{s\in[0,t]}$, not only on its endpoints, and accumulates the geometric phases. Nonetheless, the concept of a dynamically invariant space, i.e., the image of the stationary projector $\Pi_n[\thetavec]$ of the dynamics generated by $H(\thetavec)$ (and, by construction, an eigenspace of $\mAE$), carries over to quasiperiodic drives; however, the underlying physics can be much richer, as we now discuss.

\begin{table*}[t!]
\centering
\setlength{\tabcolsep}{4pt}
\renewcommand{\arraystretch}{1.45}
\begin{tabular}{>{\centering\arraybackslash}p{3.0cm}|p{7.0cm}|p{7.0cm}}
\hline
\shortstack[c]{\vspace{0.01cm}\\\textbf{Classification of}\\ \textbf{quasiperiodic drives}} & \multicolumn{1}{c|}{\shortstack[c]{\textbf{untwisted (\=T)}\\(topologically trivial)}} & \multicolumn{1}{c}{\shortstack[c]{\textbf{twisted (T)}\\(topological)}} \\
\hline
\vspace*{1.1cm} \textbf{abelian (A)}, $r_n=1$
& Globally smooth micromotion $P(\thetavec;\thetavec_0)$ exists and, together with the quasienergy operator $H^\F[\thetavec_0]$, reconstructs the evolution $U(t,0)$ in Eq.~\eqref{eq:qp-Floquet_decomp}:
\[
U(t,0) = P(\thetavec_t;\thetavec_0)\, \mathrm e^{-i t H^\F[\thetavec_0]}\,,
\]
where $H^\F[\thetavec]\!=\!\sum_n\epsilon_n^\F \Pi_n[\thetavec]$ is formed from the dynamically invariant eigenprojectors $\Pi_n$ and the quasienergy $\epsilon_n^\F\!=\!\mae_n\!-\omega^\mu/(2\pi)\!\left\langle\gamma_{n,\mu}\right\rangle\!_{\T^D}$. & At least one invariant space does not admit a globally smooth Floquet ket-vector due to the topological obstruction. The lack of a globally smooth micromotion $P$ breaks conventional Floquet theory for this class.

The average-energy decomposition in Eq.~\eqref{eq:U_qp} remains valid. A generalized Floquet decomposition also exists by defining an almost-everywhere smooth micromotion, see Appendix~\ref{app:sec:twisted_abelian}. \\ \hline
\vspace*{0.8cm} \textbf{nonabelian (\=A)}, $r_n>1$ \textit{and} \textbf{dynamically irreducible}
& At least one invariant space $\Pi_n$ has geometric and dynamical phases promoted to irreducible matrices; their nonabelian time-ordered average is no longer the torus-averaged scalar $\epsilon^\F$, resulting in the breakdown of conventional Floquet theory.

The average-energy decomposition in Eq.~\eqref{eq:U_qp} remains valid. A nonabelian generalized Floquet decomposition also exists by extracting the nonabelian dynamics explicitly, see Appendix~\ref{app:sec:untwisted_nonabelian}. & No globally smooth micromotion $P$ exists. Nonabelian geometry and dynamics are not captured by a scalar quasienergy $\epsilon^\F$. Both topological and nonabelian obstructions forbid conventional Floquet theory.

The average-energy decomposition in Eq.~\eqref{eq:U_qp} still holds. A nonabelian generalized Floquet decomposition universal for all four classes also exists, see Appendix~\ref{app:sec:twisted_nonabelian}. \\ \hline
\end{tabular}
\caption{\textbf{Classification of incommensurate quasiperiodic drives and failure of Floquet theory due to topological and dynamical geometric obstructions.} Topological twistedness and dynamically irreducible abelianness divide quasiperiodic drives into four classes, see Fig.~\ref{fig:octopusQPCD}(f). Only abelian untwisted (A\=T) quasiperiodic drives admit a conventional Floquet decomposition of the evolution operator $U(t,0)$. The average energy decomposition~\eqref{eq:U_qp} holds universally for all four classes. Generalized Floquet decompositions also exist for each class, see Table \ref{table:table} and Appendix~\ref{app:sec:4classes}. }
\label{table:Floquet4classes}
\end{table*}

\section{\label{sec:classification}Dynamical Geometric-Topological Classification of Quasiperiodic Drives}

As in Floquet systems, the infinitesimal geometric decomposition in Eq.~\eqref{eq:qpGFT_decomp} allows us to transfer concepts from static systems to classify the properties of stationary states in quasiperiodically driven systems. However, we emphasize up front that the invariant projectors we discuss, $\Pi_n[\thetavec]$, are defined only in the presence of the drive and are hence inherently nonequilibrium. Below, we first introduce the geometric and topological notions required to define the classification, and then investigate a concrete example.

In the language of quantum geometry, each invariant projector $\Pi_n[\thetavec]$ of rank $r_n=\text{rank}(\Pi_n)<N$ defines a $U(r_n)$ principal bundle over the torus $\T^D$. The geometric properties of this bundle can be derived from the quantum geometric tensor
\begin{equation}
\begin{aligned}
\label{eq:QGT}
    Q_{n,\mu\nu}(\thetavec) &= \Pi_n[\thetavec](\partial_\mu\Pi_n[\thetavec])(\partial_\nu\Pi_n[\thetavec])\Pi_n[\thetavec]\\&=\Pi_n[\thetavec]A^\K_\mu(\thetavec)A^\K_\nu(\thetavec)\Pi_n[\thetavec].
    \end{aligned}
\end{equation}
For instance, the Fubini-Study metric and Berry curvature are extracted from the Hermitian and anti-Hermitian parts of $Q_{n,\mu\nu} = G_{n,\mu\nu} - i F_{n,\mu\nu}/2$, respectively, with $Q_{n,\mu\nu}$ viewed as an operator\footnote{We deliberately work with the Hilbert-space operators, which are globally smooth on the torus $\T^D$, rather than with the corresponding $r_n\times r_n$ matrices defined on $\im(\Pi_n)$; the latter require fixing a frame which cannot be globally smooth when the bundle is twisted.} on the Hilbert space for each fixed $(n,\mu\nu)$:
\begin{subequations}
\label{eq:FS_metric-Berry_curv}
\begin{eqnarray}
    G_{n,\mu\nu}(\thetavec) &=& \frac{1}{2}\Pi_n[\thetavec]\left\{\partial_\mu\Pi_n[\thetavec],\partial_\nu\Pi_n[\thetavec]\right\}\Pi_n[\thetavec]
    \label{eq:FS_metric}, \nonumber\\
    &=&\frac{1}{2}\Pi_n[\thetavec]\left\{A^\K_\mu(\thetavec),A^\K_\nu(\thetavec)\right\}\Pi_n[\thetavec]\\
    F_{n,\mu\nu}(\thetavec) &=& i \Pi_n[\thetavec]\left[\partial_\mu\Pi_n[\thetavec],\partial_\nu\Pi_n[\thetavec]\right]\Pi_n[\thetavec] \nonumber\\
    &=& i\Pi_n[\thetavec]\left[A^\K_\mu,A^\K_\nu\right]\Pi_n[\thetavec],
    \label{eq:Berry_curv}
\end{eqnarray}
\end{subequations}
where $\{\cdot,\cdot\}$ is the anticommutator.

\textbf{Irreducibility of invariant spaces.---}%
For projectors $\Pi_n$ of rank $r_n>1$, the $\U(r_n)$-bundle describing the space $\im\Pi_n$ is called nonabelian. Nonabelian spaces can either be geometrically reducible, in which case the holonomy group is contained in a proper subgroup of $\U(r_n)$ acting reducibly on the fiber (e.g., $\U(2)\supset \U(1)\times \U(1)$), or geometrically irreducible, see Appendix~\ref{app:omega-irreducibility}. Geometric reducibility is an inherently global property of the connection (i.e., of the eigenprojector $\Pi_n[\thetavec]$), irrespective of spectral degeneracies in its parent Hamiltonian; hence, nonabelian connections arising from spectral degeneracies can still yield reducible fiber bundles. In static systems, one can directly compute the eigenprojectors and analyze their reducibility properties. Reducibility is central for quantum geometry, because it determines whether the effective quantum geometric tensor $Q_{n,\mu\nu}$, an operator of rank at most $r_n$, block-diagonalizes into two independent blocks of rank $r'_n$ and $r''_n$, respectively, with $r_n=r'_n+r''_n$.

Geometric reducibility is a global property applicable to static subspaces. However, each quasiperiodic drive singles out a specific direction $\omegavec$ on the torus $\T^D$. Therefore, the relevant notion of reducibility for these dynamical systems is defined along the direction $\omegavec$. We call an invariant space $\Pi_n$ \textit{dynamically reducible} if there exists a globally smooth subprojector $\Phi_n$ of strictly smaller rank, $0<\text{rank}(\Phi_n)<r_n$, which is itself a dynamical invariant:
\begin{equation}
\label{eq:AaeTransportEquation}
    \LL \Phi_n[\thetavec] = 0\,\quad \Leftrightarrow \quad i\omega^\mu \partial_\mu\Phi_n[\thetavec] - [A^\mAE(\thetavec),\Phi_n[\thetavec]] =0
\end{equation}
i.e., it solves the transport equation of $A^\mAE$; otherwise it is called \textit{dynamically irreducible}. In contrast to geometric irreducibility, dynamical irreducibility refers to minimal dynamically invariant spaces; it is central to the identification of effective decompositions of the evolution operator, cf.~Eq.~\eqref{eq:qpGFT_decomp}, because it determines whether effective generators, such as the Kato Hamiltonian $H^\K$, contain an irreducible traceless component that couples the degrees of freedom inside the degenerate nonabelian invariant space.
We emphasize that these two notions of reducibility are distinct for $D{\geq}2$;%
\footnote{ In Floquet systems, $D{=}1$, the notions of geometric and dynamical reducibility trivialize. } neither implies the other: an invariant space can be reducible in one sense and irreducible in the other, as discussed in Appendix~\ref{app:omega-irreducibility}. In this work, we use exclusively dynamical irreducibility to define nonabelian invariant spaces.

The dynamically invariant projectors $\Pi_n$ that span $\ker\LL$ inherit an $\omegavec$-dependence from the Liouvillian $\LL$. In practice, the spectral projectors $\Pi_n$ obtained from diagonalizing the average energy operator $\mAE$ are already dynamically irreducible, except for rare cases of accidental degeneracies in the $\mae$-spectrum, see Appendix~\ref{app:omega-irreducibility}.

Once the Kato AGPs $A_\mu^\K$ from Eq.~\eqref{eq:Kato_objs} are known, one can define Wilson line operators
\begin{equation}
\label{eq:W_mu}
   W_\mu(\thetavec) = \mathcal{T}_s\mathrm{e}^{-i\int_0^{2\pi}\mathrm{d}s\,A_\mu^\K(\thetavec+s\hat e_\mu)}
\end{equation}
around the $\mu$-th noncontractible loop of the torus $\T^D$ with tangent vector $\hat e_\mu$. If the holonomy group of the space $\Pi_n$ is abelian, then $[W_\mu(\thetavec),W_\nu(\thetavec)]=0$ on $\im\Pi_n[\thetavec]$ for all $\thetavec$; a nonvanishing commutator therefore certifies a nonabelian holonomy. We use this as a diagnostic for geometric reducibility.

Within $\mathrm{im}\Pi_n$, the Fubini-Study metric $G_n$ is a scalar-valued tensor in the abelian case ($r_n=1$) and a matrix-valued tensor in the nonabelian case ($r_n>1$). To probe further geometric physical properties, we represent the metric $G_n$ in a gauge-invariant way by defining the so-called Finsler metrics~\cite{mitscherling2025gauge}
\begin{equation}
\label{eq:g1g2mitscherling}
\begin{aligned}
    g^{(1)}_{n,\mu\nu}&=\text{tr}\,G_{n,\mu\nu},\\
    g^{(2)}_{n,\mu\nu\rho\sigma}&=\text{tr}\left[G_{n,(\mu\nu}G_{n,\rho\sigma)}\right],\\
    \Delta g^{(2)}_{n,\mu\nu\rho\sigma}&=2\left[g^{(1)}_{n,(\mu\nu}g^{(1)}_{n,\rho\sigma)} -g^{(2)}_{n,\mu\nu\rho\sigma}\right],
\end{aligned}
\end{equation}
where the parentheses $(\cdots)$ denote normalized symmetrization over indices $\mu\nu\rho\sigma$. The first Finsler metric is the familiar Fubini-Study metric, which measures the geodesic quantum distance between two states~\cite{mitscherling2025gauge}; the second Finsler metric measures the distance between degenerate states. The third quantity $\Delta g^{(2)}_{n,\mu\nu\rho\sigma}{=}0$ for abelian invariant spaces. To visualize the local quantum geometry on the torus $\mathbb{T^2}$, we take $\vec v(\phi)=(\cos\phi,\sin\phi)^\mathsf{T}$ for $\phi\in\left[0,2\pi\right)$, and define the directional contractions
\begin{equation}
\label{eq:metric-indicatrices}
\begin{aligned}
    r_n^{(1)}(\phi)=g^{(1)}_{n,\mu\nu}v_\mu v_\nu,\qquad r_n^{(2)}(\phi)=\Delta g^{(2)}_{n,\mu\nu\rho\sigma}v_\mu v_\nu v_\rho v_\sigma\,.
\end{aligned}
\end{equation}
For a grid of points on the torus, we display the $\phi$-dependent petal-shaped polar plots of $[r_n^{(1)}(\phi)]^{1/2}$ and $[r_n^{(2)}(\phi)]^{1/4}$ in the spirit of Tissot indicatrices~\cite{tissot1881memoire}, and color-code their area according to the strength of
\begin{equation}
\label{eq:metric-indicatrices-area}
\begin{aligned}
    g_n^{(1)}=g^{(1)}_{n,\mu\mu},\qquad \Delta g_n^{(2)}=\sqrt{\Delta g^{(2)}_{n,\mu\mu\nu\nu}}\,.
\end{aligned}
\end{equation}
These quantities are derived from the Finsler metrics in Eq.~\eqref{eq:g1g2mitscherling}; they visualize the local abelian ($[r_n^{(1)}(\phi)]^{1/2}$, $g_n^{(1)}$) and nonabelian ($[r_n^{(2)}(\phi)]^{1/4}$, $\Delta g_n^{(2)}$) quantum geometry at each torus point.

\textbf{Topology of invariant spaces.---}%
In turn, the fiber bundle's topological properties are captured by the corresponding Chern classes. While the concept is general and applies to any torus of dimension $D$, we illustrate it for $D=2$ using the first Chern number:
\begin{eqnarray}
\label{eq:Chern_n}
    C_n = \frac{1}{4\pi}\int_{\T^2} \text{tr}( F_{n,\mu\nu} )\, \mathrm d\theta^\mu\wedge\mathrm d\theta^\nu, \qquad C_n\in\mathbb{Z} \, ,
\end{eqnarray}
where $F_{n,\mu\nu}$ is the Berry curvature tensor and $\mathrm{tr}$ is the Hilbert-space trace, since $F_{n,\mu\nu}$ is a scalar tensor in abelian cases, and a matrix-valued one in nonabelian cases. A nonzero Chern number describes a global twist of the fibers. Below, we use the term `(un)twisted' to refer to trivial/nontrivial bundle topology, to set the discussion apart from various other notions of topology in physics. For $D{>}2$, other invariants of the operator components of $F_n$ can be used to define higher Chern numbers. Topologically trivial (untwisted) bundles require all Chern numbers to vanish \cite{qi2008topological}.

We are now fully equipped to provide a classification for quasiperiodic drives based on their dynamical geometric and topological properties. We consider exclusively quasiperiodic drives with incommensurate frequencies $\omegavec$ which cover the torus ergodically, since commensurate drives are described by conventional Floquet theory.

\textbf{Model.---}%
To illustrate our classification with concrete examples, we use CD driving~\cite{kolodrubetz2017geometry} to design a toy model which exhibits the requisite dynamical geometric and topological properties~\cite{crowley2019topological}. To inscribe independently tunable dynamical geometric and topological features, we start with two qubits $A,B$, subject to a $D{=}2$-tone quasiperiodic drive $\thetavec_t {=} \omegavec t {+} \thetavec_0{\in}\T^2$, described by the Hamiltonians

\begin{eqnarray}
\label{eq:HA_HB}
    H^\alpha_{A}(\thetavec) &=& (\cos\theta_1+\cos\theta_2)\sigma_A^z +\epsilon\sigma_{A}^{a}, \nonumber\\
    H^\tau_{B}(\thetavec) &=& \left(\vec{d}(\thetavec)+\vec{a}_{\mathrm{CD}}(\thetavec)\right)\cdot \vec{\sigma}_B\, ,
\end{eqnarray}
where $a {=} z,x$ parametrizes the direction of the perturbation for qubit $A$ of strength $\epsilon$; when $\epsilon{\ll}1$, qubit $A$ has abelian invariant spaces for $a{=}z$, and a nonabelian invariant space for $a{=}x$. We denote this abelianness (A) or nonabelianness (\=A) of the Hamiltonian $H^\alpha_{A}$ via $\alpha{=}\mathrm{A},\mathrm{\bar A}$. Qubit $B$ is a counterdiabatically driven Qi-Wu-Zhang (QWZ) model \cite{qi2008topological,crowley2019topological} with $\vec d{=}(\sin\theta_1,\sin\theta_2,m{+}\cos\theta_1{+}\cos\theta_2)^\mathsf{T}$, where the AGP $\vec{a}_{\mathrm{CD}} = \vec{d}\times\omega^\mu\partial_\mu{\vec{d}}/(2|\vec{d}|^2)$ suppresses any nonadiabatic transitions. The QWZ model is topological for $0<|m|{<}2$, where $\vec d/|\vec{d}|$ wraps exactly once around the unit sphere, and trivial when $|m|{>}2$. We denote this twistedness (T) or untwistedness (\=T) in the Hamiltonian $H^\tau_{B}$ via $\tau{=}\mathrm{T},\mathrm{\bar T}$.

To keep the model generic, i.e., not exactly solvable, we move away from the fine-tuned reverse-engineered point and explicitly perturb the Hamiltonian $H^\tau_{B} \to \centernot{H}\!\!\!^{\,\tau}_{B}$ by keeping only Fourier modes with $|\kvec|\leq 1$, $\kvec\in\mathbb{Z}^2$; although it destroys the CD property of $\centernot{H}\!\!^{\,\tau}_{B}$, this mode truncation dresses but does not remove its topological features, as we show below. Finally, we bring in nonabelian geometric features by entangling the qubits using a two-qubit change-of-frame rotation $R(\thetavec)$; it leaves the ranks, dynamical irreducibility, and Chern numbers of the invariant spaces unchanged, but alters their Berry connection by the projected pure-gauge term $i(\partial_\mu R)R^\dagger$, turning the abelian holonomy of the decoupled rank-two spaces $\mathbb{C}^2_A\otimes\mathrm{im}(\Pi_B)$ into a nonabelian one. The full quasiperiodically driven Hamiltonian then reads
\begin{eqnarray}
\label{eq:model_H}
    H^{\alpha\tau}(\thetavec) &=& R(\thetavec)\left(H^\alpha_{A}(\thetavec) +  \centernot{H}\!\!^{\,\tau}_{B}(\thetavec)\right) R(\thetavec)^\dagger+i\omega^\mu(\partial_\mu R)R^\dagger, \nonumber\\
    R(\thetavec) &=& \exp\left(i\frac{\theta_1}{2}\sigma_{A}^x\sigma_{B}^z\right) \exp\left(i\frac{\theta_2}{2}\sigma_{A}^y\sigma_{B}^z\right)\, ,
\end{eqnarray}
and contains a total of $19$ Fourier modes,\footnote{Note that, whereas $H^{\alpha\tau}(\thetavec)$ has a finite number of modes, the average energy operator $\mAE^{\alpha\tau}[\thetavec]$ and its AGP $A^\mAE(\thetavec)$ have infinitely many modes.} cf.~Appendix~\ref{app:sec:model}. We fix $\epsilon{=}10^{-2}$, $|\omegavec|{=}0.1$, and use the golden ratio $\omega_2/\omega_1{=}(1{+}\sqrt{5})/2$ throughout; this leaves $m{\in}\{1,3\}$ and $a{\in}\{x,z\}$ as the only free parameters to control the topological and dynamical geometric properties of the corresponding effective average energy Hamiltonian $\mAE^{\alpha\tau}$, respectively. We note that simpler toy models exist to separately describe each of the classes discussed below; here, we opted for the Hamiltonian $H^{\alpha\tau}(\thetavec)$ because (i) it is a single model to exhibit all four classes, and (ii) as a driven four-level system it can be realized experimentally. Due to the mode truncation, there is no known closed-form expression for the average energy or Kato decompositions~\eqref{eq:qpGFT_decomp}; we therefore use a variational principle to find $A^\mAE$ and $\mAE^{\alpha\tau}$ numerically, see Sec.~\ref{subsec:varl}.

\begin{figure}[t!]
	\centering
    \includegraphics[width=0.50\textwidth]{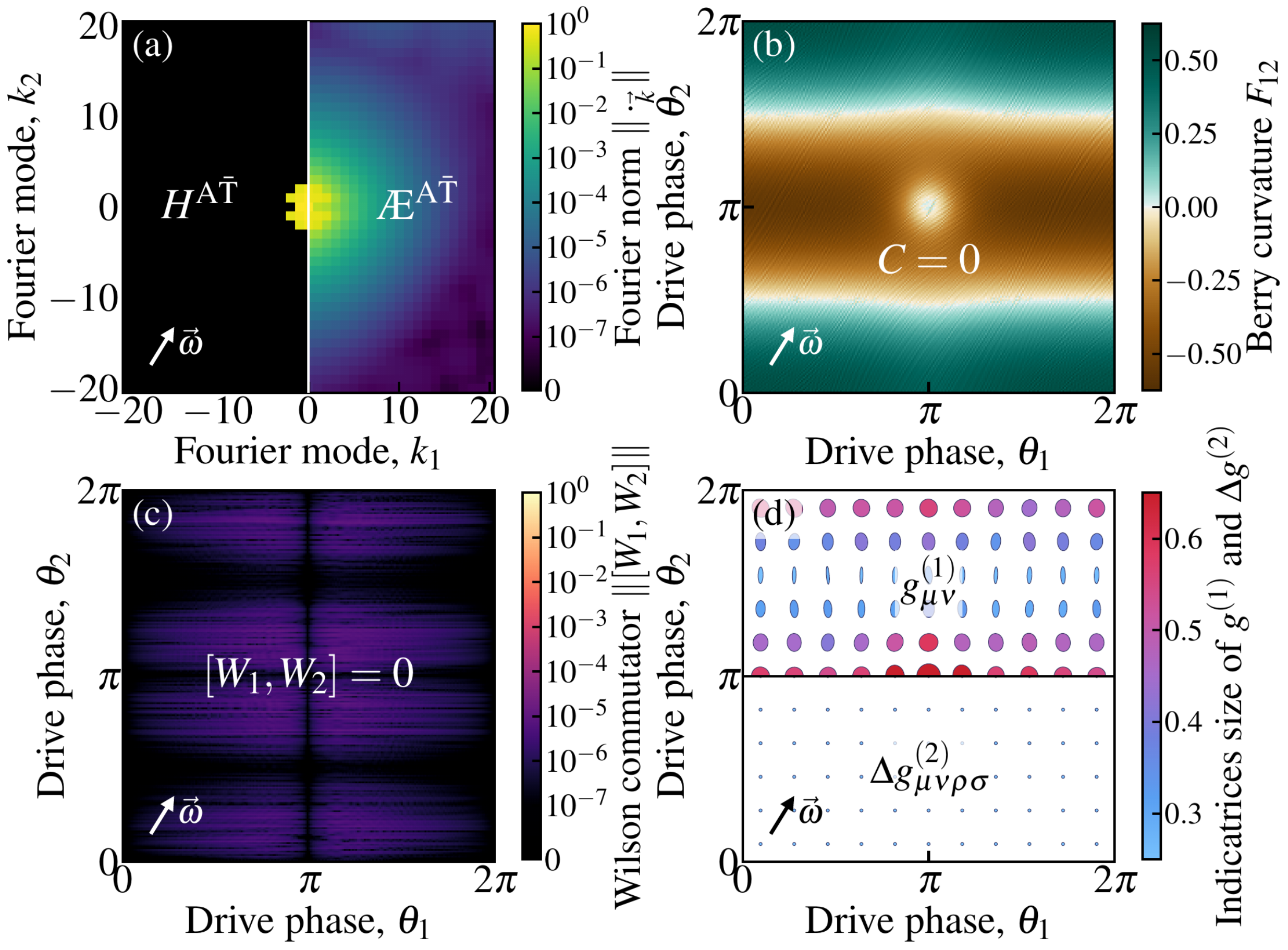}
	\caption{ \textbf{Abelian untwisted (A\=T) drives: an example.} Quantum geometry and topology of the average-energy ground state manifold, associated with the quasiperiodic drive $H^{\mathrm{ A}\mathrm{\bar T}}(\thetavec_t)$ from Eq.~\eqref{eq:model_H}; the remaining eigenspaces behave similarly. (a) Matrix norm of the Fourier-mode decomposition of the drive $H^{\mathrm{ A}\mathrm{\bar T}}(\thetavec)$ and the average energy $\mAE^{\mathrm{ A}\mathrm{\bar T}}[\thetavec]$. Whereas the drive Hamiltonian is truncated to a finite number of modes, the average-energy operator has infinitely many modes; (b) Berry curvature $F_{12}(\thetavec)$ of the $\mAE^{\mathrm{ A}\mathrm{\bar T}}[\thetavec]$ ground state, which integrates to a Chern number $C=0\pm 10^{-6}$, cf.~Eq.~\eqref{eq:Chern_n}, justifying that the eigenspace is untwisted; (c) Wilson line commutator~\eqref{eq:W_mu} vanishes $[W_1,W_2]=0 \pm 10^{-5}$, indicating the abelian geometric holonomy. (d) Tissot indicatrices~\eqref{eq:metric-indicatrices} $[r_n^{(1)}(\phi)]^{1/2}$ (top), and $[r_n^{(2)}(\phi)]^{1/4}$ (bottom) of the Finsler metrics $g^{(1)}_{\mu\nu}$ and $\Delta g^{(2)}_{\mu\nu\rho\sigma}$ defined in Eq.~\eqref{eq:g1g2mitscherling}, indicating the abelian quantum geometry of the ground state manifold. The joint colorbar denotes the corresponding indicatrix sizes, $g_n^{(1)}$, $\Delta g_n^{(2)}$, see Eq.~\eqref{eq:metric-indicatrices-area}. Data obtained using the variational principle from Sec.~\ref{subsec:varl} for perturbation strength $\epsilon=10^{-2}$, drive parameters $|\omegavec|=0.1, \omega_2/\omega_1=(1+\sqrt{5})/2$, and Fourier mode cutoff $K=110$. The arrow shows the direction of the quasiperiodic drive velocity $\omegavec$. We dropped the ground-state label. }
	\label{fig:abelian_nontwisting}
\end{figure}

\textbf{Classification of quasiperiodic drives.---}%
Our classification of quasiperiodic drives, laid out in Secs.~\ref{subsec:AN}-\ref{subsec:NT} below, and in Table~\ref{table:Floquet4classes}, transfers concepts from static systems parametrized on the torus $\T^D$ to quasiperiodically driven systems. Similar to Floquet systems whose properties are determined by the effective quasienergy and its eigenstates, we investigate the dynamically invariant stationary-state eigenprojectors $\Pi_n$ of the average energy $\mAE$, which provides an effective description of the dynamics for quasiperiodically driven systems, cf.~Eq.~\eqref{eq:U_qp}. We refer to a quasiperiodic drive as abelian/nonabelian (A/\=A) and untwisted/twisted (\=T/T) depending on the corresponding properties of the eigenbundles of its invariant spaces $\im(\Pi_n[\thetavec])$; the existence of a single invariant projector $\Pi_n[\thetavec]$ whose bundle is nonabelian or twisted renders the generating quasiperiodic drive nonabelian or twisted, respectively, and for \=AT drives we require at least one space which is both nonabelian and twisted. Finally, we emphasize that the term `nonabelian' used throughout this paper refers exclusively to dynamically irreducible invariant spaces (for otherwise we can reduce these into smaller invariant spaces).

\subsection{\label{subsec:AN}Abelian untwisted (A\=T) drives}

The simplest class is that of \textit{abelian untwisted (A\=T) drives}. The abelian property implies that all dynamically irreducible invariant spaces are one-dimensional; hence, the drive possesses $N$ orthogonal dynamically invariant spaces spanned by the rank-one invariant projectors $\Pi_n$, i.e., $r_n=1$ for all $n$. The term `untwisted' here refers to the drive failing to induce a nontrivial topology in any of the fiber bundles associated with these effective invariant projectors. Further, because the bundles are not twisted (i.e., they are all topologically trivial), we can pick a globally smooth gauge for the overall phase of the wavefunction and define smooth individual ket states $\ket{n^\F[\thetavec]}$ everywhere on the torus $\T^D$, such that $\Pi_n[\thetavec]=\ket{n^\F[\thetavec]}\bra{n^\F[\thetavec]}$ holds.

An example of an A\=T drive is given by the Hamiltonian~\eqref{eq:model_H}, with $a{=}z$ and $m{=}3$. Figure~\ref{fig:abelian_nontwisting}(a) shows the Fourier-mode decompositions of $H^{\mathrm{A}\mathrm{\bar T}}_{\kvec}$ and the numerically computed average energy operator $\mAE^{\mathrm{A}\mathrm{\bar T}}_{\kvec}$ with the colorbar showing a logarithmic scale: although the lab-frame Hamiltonian $H^{\mathrm{A}\mathrm{\bar T}}$ is truncated to a finite number of modes, the average energy $\mAE^{\mathrm{A}\mathrm{\bar T}}$ has infinitely many modes. Panel (b) shows the trace of the Berry curvature of the lowest-$\mae$ ground state over the torus; its Chern number vanishes, indicating the corresponding eigenstate is topologically trivial, or untwisted. Panel (c) tests the abelian property of the ground state manifold by computing the commutator of the two great-circle Wilson lines: we find $[W_1(\thetavec),W_2(\thetavec)]\equiv 0$ up to numerical precision and hence the invariant space associated with the effective projector $\Pi_{\rm GS}$ is abelian. Finally, panel (d) shows its Fubini-Study metric $g_{\mu\nu}^{(1)}$ and connected second Finsler metric $\Delta g_{\mu\nu\rho\sigma}^{(2)}$: we find $\Delta g_{\mu\nu\rho\sigma}^{(2)}=0$ up to numerical precision, corroborating the evidence that $\Pi_{\rm GS}$ is abelian. The same features are shared by the other three invariant spaces $\Pi_{n\neq \rm GS}$ [not shown].

We now demonstrate that Floquet's theorem for periodically driven systems transfers over directly to abelian untwisted quasiperiodic drives. Moreover, we will define globally smooth micromotion $P(\thetavec;\thetavec_0)$ and quasienergy $H^\F[\thetavec]$ operators on the torus $\T^D$, which constitute the integral Floquet decomposition
\begin{equation}
\label{eq:qp-Floquet_decomp}
    U(t,0) = P(\thetavec_t;\thetavec_0)\, \mathrm e^{-i t H^\F[\thetavec_0]}\, .
\end{equation}
The U$(1)$ gauge freedom of the eigenkets $|n^{\F}[\thetavec_t]\rangle$ in A\=T quasiperiodically driven systems is partially fixed to $\mathbb{Z}^D$ by a globally smooth gauge choice, guaranteed to exist by the trivial topology; in this gauge, the Floquet eigenstates $|n^{\F}[\thetavec_t]\rangle$ accumulate a linear phase in time, in analogy to Geometric Floquet theory~\cite{schindler2025geometric}. The leftover gauge freedom
\begin{equation}
    \ket{n^\F[\thetavec]} \longrightarrow \mathrm{e}^{i\kvec_n\cdot\thetavec}\ket{n^\F[\thetavec]}, \qquad \kvec_n\in\mathbb{Z}^D
\end{equation}
can be parametrized by a vector of integers $\kvec_n\in\mathbb{Z}^D$. To avoid gauge ambiguities, in the following, we start from the completely fixed average energy gauge, and work our way back to define the quasienergies via the analogue of the relation $\epsilon_n^\F =\mae_n - \gamma_n/T$; we then relax the complete gauge fixing, switching from the average-energy to the quasiperiodic Floquet gauge, $\epsilon_n^\F \longrightarrow\epsilon_n^\F+\kvec_n\cdot\omegavec$.

Given the eigenprojectors $\Pi_n[\thetavec]$, the average energies $\mae_n$ associated with the lab-frame $H(\thetavec)$ are given directly by
\begin{equation}
\label{eq:qp-ae}
    \mae_n = \int_{\T^D} \frac{ \mathrm{d}^D\theta }{(2\pi)^D} \,\mathrm{tr}\left(H(\thetavec)\Pi_n[\thetavec]\right)\, .
\end{equation}

\begin{figure}[t!]
\centering
\includegraphics[width=\columnwidth]{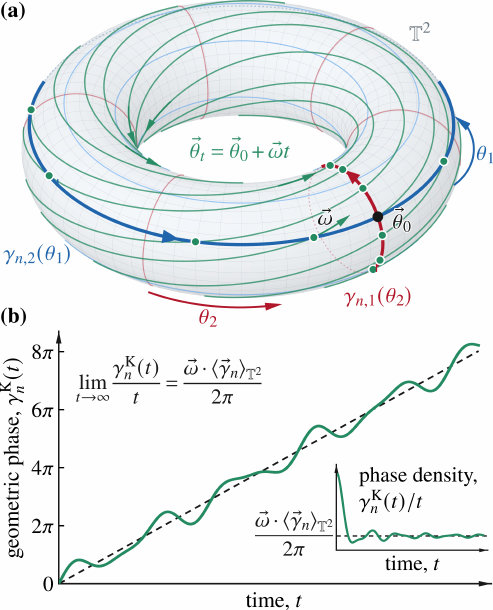}
\caption{ \textbf{Geometric phase for abelian untwisted quasiperiodic drives without a closed orbit.} (a) Noncontractible cycles (red and blue), displaced cycles (thin), and a quasiperiodic trajectory (green) on $\T^2$; green dots mark crossings with the cycles. The phases defined in Eq.~\eqref{eq:qp-geom_phase} depend only on the transverse coordinate: $\gamma_{n,1}(\theta_2)$ and $\gamma_{n,2}(\theta_1)$. (b) Schematic unwrapped Kato-gauge geometric phase $\gamma_n^\K(t)=\arg\langle n^\F[\thetavec_t]|W^\K(t;0)|n^\F[\thetavec_0]\rangle$ (green) and its mean linear growth (dashed). The inset shows the same phase divided by time, $\gamma_n^\K(t)/t$; the dashed line marks $\omegavec\cdot\langle\vec\gamma_n\rangle_{\T^2}/(2\pi)$. The accumulated phase grows without bound, while its long-time density remains finite and gives the geometric contribution to the quasienergy in Eq.~\eqref{eq:qp-epsF_ae_gamma}. }
\label{fig:qpGeometricPhase}
\end{figure}

Defining the geometric phases $\gamma_n$ requires more care, as the quasiperiodic trajectory does not close back onto itself in a finite time. However, since the states $\ket{n^\F[\thetavec]}$ can be defined globally on the torus, and because incommensurate quasiperiodic trajectories cover the entire torus ergodically, a well-defined notion of a geometric phase along the $\mu$-th noncontractible loop of the torus exists for A\=T drives, cf.~Appendix~\ref{app:sec:untwisted_abelian},
\begin{equation}
\label{eq:qp-geom_phase}
    \gamma_{n,\mu}(\thetavec) = \arg\text{tr}\left(\Pi_n[\thetavec]W_\mu(\thetavec)\right)\, ,
\end{equation}
where the Wilson line along the $\mu$-th noncontractible loop of the torus $\T^D$ is defined in Eq.~\eqref{eq:W_mu}. Notice that, for $D>1$, $\gamma_{n,\mu}(\thetavec)$ is independent of $\theta_\mu$: shifting the starting point along the $\mu$-th noncontractible loop leaves the phase unchanged. Weighting all directions $\mu$ by the winding rate $\omega^\mu/(2\pi)$, averaging over the torus we can assign each eigenket $\ket{n^\F}$ a finite scalar geometric phase density $\omega^\mu\langle\gamma_{n,\mu}\rangle_{\T^D}/2\pi$ with units of energy or inverse time, where $\braket{\cdot}_{\mathbb{T}^D}$ denotes the torus average\footnote{The torus average of a smooth scalar function $f(\thetavec)$ is $\braket{f}_{\mathbb{T}^D}=(2\pi)^{-D}\int_{\mathbb{T}^D}\D^D \theta f(\thetavec)$. When $\omegavec$ is incommensurate, this also equals the long-time average $\lim_{T\to\infty}T^{-1}\int_0^T f(\thetavec_t)\D t$ by ergodicity~\cite{puig2002reducibility}.} of $D$ scalar functions of the geometric phase $\gamma_{n,\mu}(\thetavec)$ smoothly defined on the torus. This suggests that quasiperiodic drives in the A\=T class can be intuitively thought of as periodic drives with infinite period when the quasiperiodic trajectory closes as $t\to\infty$; thus, although the corresponding geometric phase accumulated over infinite time formally diverges, its density remains finite, see Fig.~\ref{fig:qpGeometricPhase}.

Since all quantities entering the right-hand sides of Eqs.~\eqref{eq:qp-ae} and~\eqref{eq:qp-geom_phase} are ultimately defined in terms of the invariant projectors $\Pi_n$ and the known drive $H(\thetavec)$, we can define generalized quasienergies as
\begin{equation}
\label{eq:qp-epsF_ae_gamma}
    \epsilon_n^\F = \mae_n - \frac{1}{2\pi} \langle\vec\gamma_{n}\rangle_{\T^D}\cdot\omegavec +\kvec_n\cdot\omegavec \qquad \kvec_n\in\mathbb{Z}^D\, ,
\end{equation}
where $\kvec_n$ encodes the residual $\mathbb{Z}^D$ gauge freedom.

Having determined the quasienergies, the quasienergy operator is defined by
\begin{equation}
    H^\F[\thetavec] = \sum_n \epsilon_n^\F\, \Pi_n[\thetavec]\, .
\end{equation}
As expected, $\epsilon_n^\F$ does not depend on the starting point on the torus $\thetavec_0$, and hence $H^\F[\thetavec]$ is a globally smooth dynamical invariant. In turn, the corresponding Floquet AGP is found using the analogue of the Floquet decomposition~\eqref{eq:HF_decomp} $A^\F(\thetavec) = H(\thetavec) - H^\F[\thetavec]$; from it, we can define the micromotion operator as
\begin{equation}
    P(\thetavec_t;\thetavec_0) = \mathcal{T}_s\exp\left(-i \int^t_0\mathrm ds\, A^\F(\thetavec_s) \right).
\end{equation}
The existence of globally smooth kets provides smooth transport between any two points of $\T^D$. Partially fixing their gauge, $\U(1)\to\mathbb{Z}^D$, absorbs the time-dependence of their accumulated phase into this micromotion transport, leaving only a linearly growing phase $\mathrm e^{-it\epsilon_n^\F}$, for almost all frequency vectors $\omegavec$, see Appendix~\ref{app:sec:untwisted_abelian} for the explicit construction.  Thus, $P(\thetavec;\thetavec_0)$ is globally smooth on the torus; the above expression evaluates it along the physical trajectory $\thetavec=\thetavec_t$.

The formal similarity of the above equations to periodically driven systems should be taken with a grain of salt when it comes to using numerical algorithms. In practice, we find that a frequency-lattice Hilbert space calculation of the quasienergies fails to select a stable quasienergy representative, since the copies $\epsilon_n^\F+\kvec\cdot\omegavec$ are dense on the real line $\mathbb{R}$ for quasiperiodic drives; every finite quasienergy interval therefore contains infinitely many equivalent copies. Fixing a finite cutoff on the number of Fourier harmonics kept in the quasienergy matrix acting on the frequency-lattice Hilbert space, an eigensolver typically selects a cutoff-dependent representative and mixes the eigenvectors whenever frequency-lattice resonances occur. For these reasons, the numerically stable physical object is the globally smooth invariant projector $\Pi_n$, rather than the frequency-lattice eigenvector ket $|n\rangle\rangle$ defined in Eq.~\eqref{eq:Sambe}.

Notice how the quasienergy folding problem is circumvented by the average energies $\mae_n$: they do not suffer from such a quasi-continuous folding ambiguity since the gauge freedom is fixed completely. Moreover, the average-energy decomposition holds unconditionally, i.e., for all frequency vectors.

However, finding variationally the corresponding average energy operator $\mAE$ and hence its dynamically invariant eigenprojectors requires working in frequency-lattice \textit{operator} space, cf.~Sec.~\ref{subsec:varl}.

An immediate consequence of Eq.~\eqref{eq:qp-Floquet_decomp} is that the abelian untwisted class of quasiperiodic drives contains all Floquet drives as a proper subset. We emphasize that this A\=T class includes \textit{incommensurate} drives. One can readily check that for $D=1$ the above expressions reduce to those from Sec.~\ref{sec:GFT-floquet}. Indeed, any U$(N)$ bundle over the circle $S^1=\T$ is necessarily reducible\footnote{The notion of dynamical reducibility coincides with geometric reducibility on the circle $\T$, and follows trivially from the single Wilson line commuting with itself.} and topologically trivial, which are the mathematical structures required for Floquet's theorem.

\subsection{\label{subsec:AT}Abelian twisted (AT) drives}

Next, we turn our attention to abelian twisted (AT) drives. Similar to the A\=T class, all dynamically irreducible invariant spaces are exclusively rank-one, and the corresponding fiber bundles are abelian. However, here (at least some of) the bundles are twisted (i.e., topologically nontrivial), and carry nonzero Chern numbers. We refer to such quasiperiodic drives as twisted. Figure~\ref{fig:abelian_twisting} shows an example of an abelian twisted drive; further examples of AT drives are known from the literature \cite{crowley2019topological,crowley2020half}.

\begin{figure}[t!]
	\centering
    \includegraphics[width=0.50\textwidth]{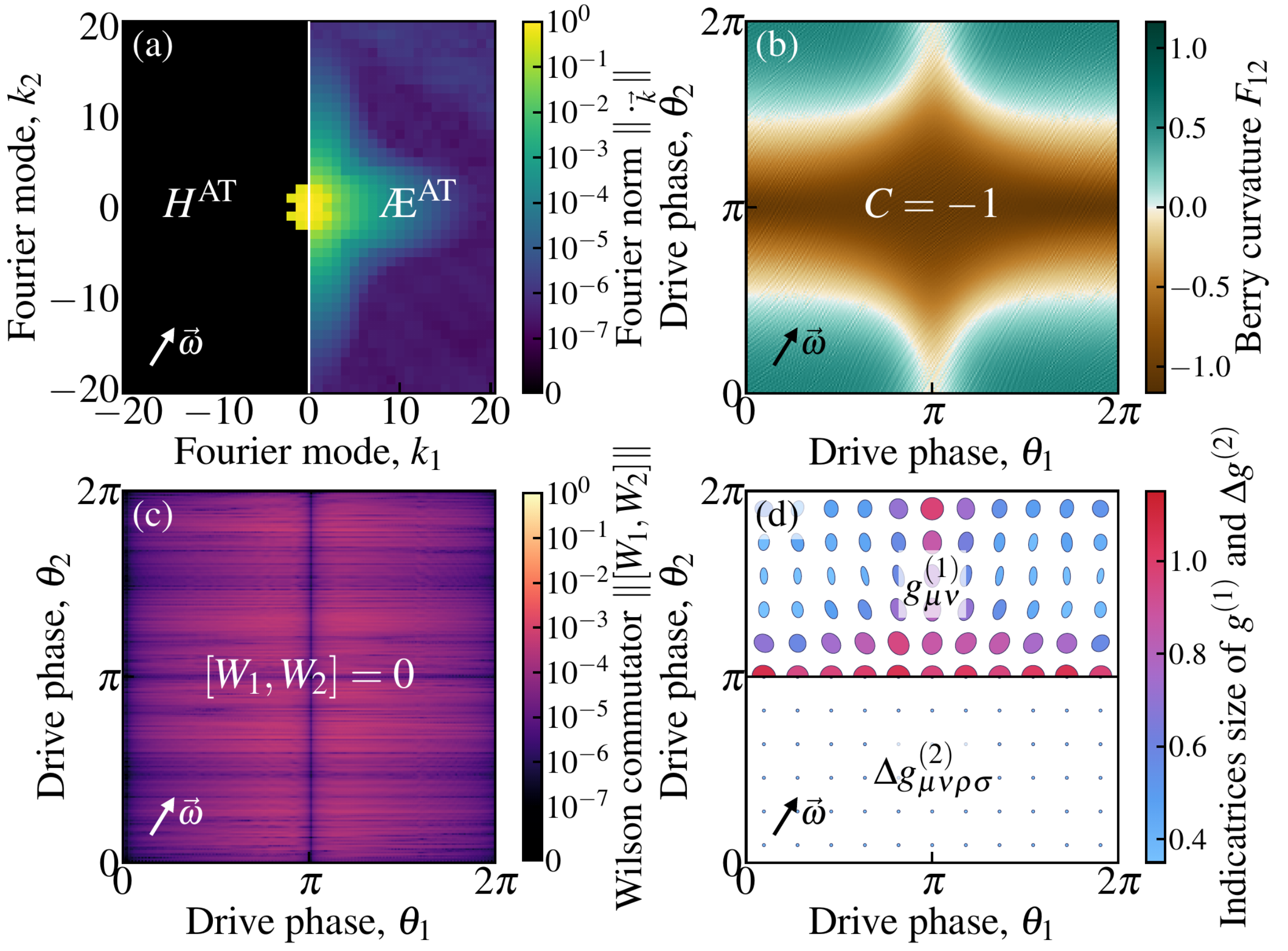}
	\caption{\textbf{Abelian twisted (AT) drives: an example.} Same quantities and numerical parameters as in Fig.~\ref{fig:abelian_nontwisting}, for $H^{\mathrm A\mathrm T}$. The average-energy operator in (a) remains globally smooth despite the nonzero Chern number $C=-1\pm10^{-5}$ in (b). (c) The Wilson-line commutator vanishes up to $10^{-4}$. (d) The vanishing connected second Finsler metric confirms the abelian quantum geometry.}
	\label{fig:abelian_twisting}
\end{figure}

As a consequence of the nontrivial topological twist, there exists no globally smooth gauge choice for the overall phase of the stationary state wavefunctions; therefore, abelian twisted drives possess no globally smooth stationary wavefunctions. For the same reason, one cannot define a smooth micromotion operator on the entire torus $\T^D$ either. A related problem occurs with defining quasienergies: intuitively, since ``$\epsilon^\F = \mae - \gamma/T$'' and because the geometric phase $\gamma$ should be given by the integral of the Berry connection, the absence of a smooth gauge for the connection precludes the existence of $\thetavec_0$-independent quasienergies; hence, the quasienergy operator can no longer be the global quasiperiodic dynamical invariant whose constant eigenvalues generate the full phase evolution. We therefore witness the breakdown of Floquet's theory for AT quasiperiodic drives due to a topological obstruction in the invariant spaces spanned by the invariant projectors $\Pi_n$. For this reason, AT drives do not admit an exact integral Floquet decomposition~\eqref{eq:qp-Floquet_decomp} for their unitary evolution.

In contrast to the stationary states, the dynamically invariant projectors $\Pi_n[\thetavec]$ do not depend on the eigenket's evolution phase, and remain globally smooth functions on the entire torus $\T^D$; hence, so is the corresponding AGP $A^\mAE(\thetavec)$. Therefore, the average energy operator $\mAE[\thetavec]$ is a globally defined dynamical invariant. As a result, the geometric decomposition~\eqref{eq:U_qp} continues to hold for AT drives despite the topological obstruction. This is a direct manifestation of the superior generality of the adiabatic theorem (here in the form of counterdiabatic driving) compared to Floquet's theorem. Thus, abelian twisted drives provide minimal ingredients to go strictly beyond Floquet physics.

The lack of a globally smooth gauge becomes manifest on a singular $D{-}2$ dimensional submanifold of the torus $\T^D$, which is a set of measure zero. Therefore, almost all trajectories never meet the singular set\footnote{Yet, the almost-everywhere defined micromotion winds rapidly when the trajectory passes close to it.}. Starting from a simple product between the same projector at two points $\Pi_n[\thetavec]\Pi_n[\thetavec_0]$, one can define an almost-globally smooth micromotion operator between $\thetavec$ and $\thetavec_0$ and a phase-accumulation rate fluctuating around a constant quasienergy to retain as much of the Floquet structure as possible. The $D{-}2$ dimensional submanifold of singularities is exactly the $\thetavec$ points where the product $\Pi_n[\thetavec]\Pi_n[\thetavec_0]$ vanishes. We invite the interested reader to check out Appendix~\ref{app:sec:twisted_abelian} where we provide a detailed discussion.

\subsection{\label{subsec:NN}Nonabelian untwisted (\=A\=T) drives}

Another class of quasiperiodic drives that goes strictly beyond Floquet physics is given by nonabelian untwisted (\=A\=T) drives. Similar to their abelian counterparts, here the stationary state projectors $\Pi_n$ give rise to topologically trivial (untwisted) invariant spaces. However, the dimension of at least one dynamically invariant space is larger than one, i.e., $r_n=\text{rank}(\Pi_n)>1$ for some $n$; moreover, dynamical irreducibility precludes these nonabelian invariant spaces from being reduced any further. We call such quasiperiodic drives nonabelian untwisted.

For example, consider the Hamiltonian $H^\mathrm{\bar A}_A$ for qubit $A$, cf.~Eq.~\eqref{eq:HA_HB}. At $\epsilon=0$ the two eigenstates of $\sigma^z_A$ are rank-one invariant spaces with Kato energies $\pm(\cos\theta_1+\cos\theta_2)$, whose torus averages coincide: $\mae_\uparrow=\mae_\downarrow=0$. The transverse term $\epsilon\sigma^x_A$ is therefore a resonant perturbation, and as long as $|\epsilon|$ remains small, it can merge the two degenerate invariant spaces into a single one. Dynamically, this happens at the crossings of the unperturbed levels, i.e., along the curves $\cos\theta_1+\cos\theta_2=0$ on the torus, where the instantaneous gap is $2|\epsilon|$: for $|\epsilon|\ll|\vec\omega|\ll 1$ (since only near the adiabatic limit can the constant $\epsilon \sigma^x$ be considered a perturbation), as in our numerics, the crossings are traversed diabatically, and each passage induces a Landau-Zener transition between the two states, with the relative phase accumulated irregularly between successive crossings. These quasiperiodically repeated Landau--Zener transitions prevent a smooth rank-one invariant splitting on the torus and leave a single dynamically irreducible, nonabelian invariant space (see Appendix~\ref{sec:LandscapeQP} for further KAM analysis).

Figure~\ref{fig:nonabelian_nontwisting} provides an example of a nonabelian untwisted drive. A nonconstructive example of a nonabelian untwisted drive known from the literature is a spin-$1/2$ particle in a quasiperiodic magnetic field with no globally smooth polarization axis preserved by its three-dimensional precession~\cite{eliasson2002ergodic}.

\begin{figure}[t!]
	\centering
    \includegraphics[width=0.50\textwidth]{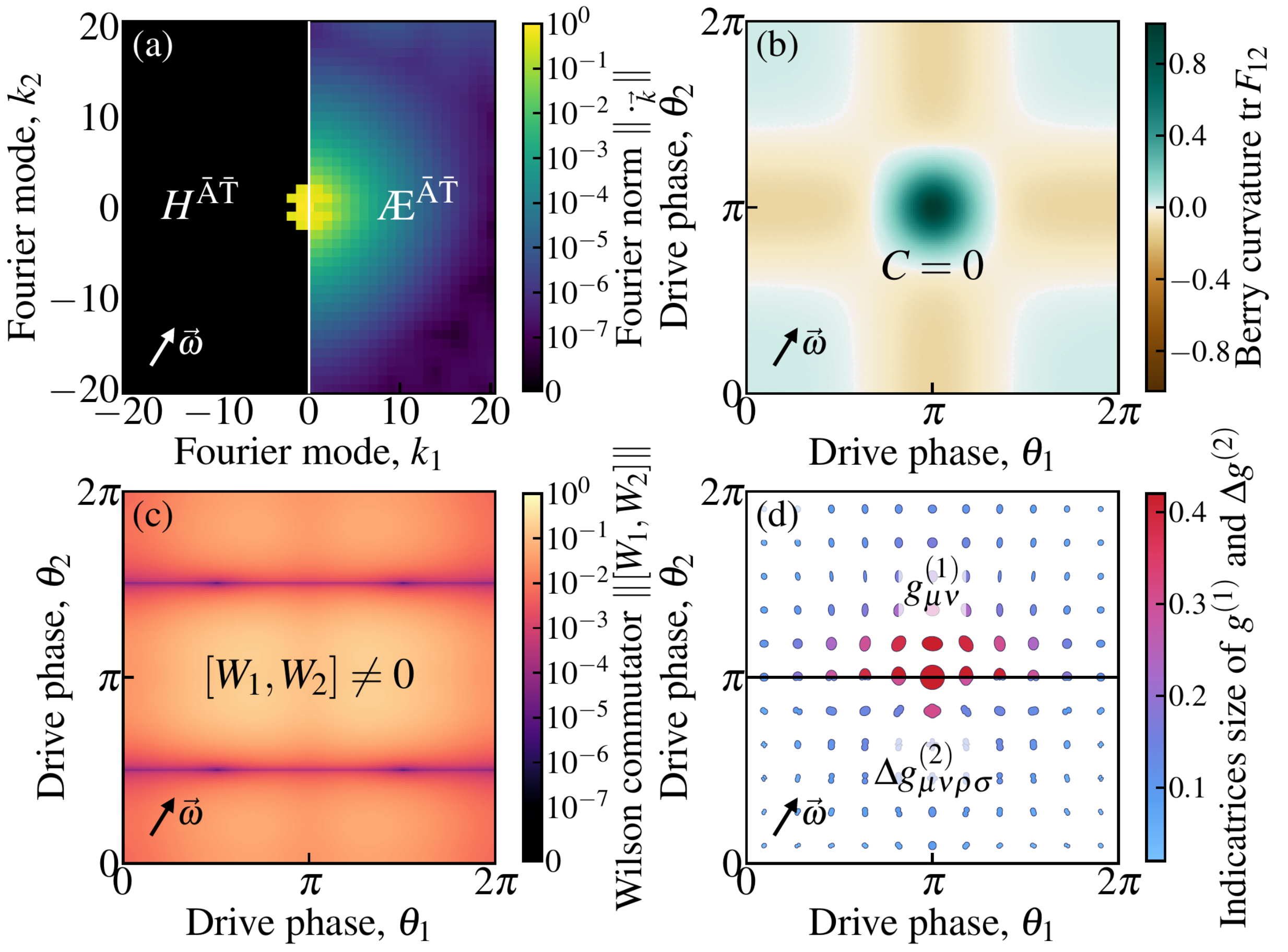}
	\caption{\textbf{Nonabelian untwisted (\=A\=T) drives: an example.} Same quantities and numerical parameters as in Fig.~\ref{fig:abelian_nontwisting}, for $H^{\mathrm{\bar A}\mathrm{\bar T}}$ at $K=100$. (a) The average-energy Fourier components remain localized. (b) The trace of the ground-space curvature, $\mathrm{tr}\,F_{12}$, gives $C=0\pm10^{-11}$. (c) The nonzero Wilson-line commutator and (d) the finite connected second Finsler metric reveal the nonabelian geometry of the rank-two ground state manifold.}
	\label{fig:nonabelian_nontwisting}
\end{figure}

Due to dynamical irreducibility, the invariant space $\Pi_n$ cannot be decomposed into rank-one subspaces that obey the stationary-state condition (i.e., they only accumulate a phase under the exact quasiperiodic drive). Therefore, the notion of individual stationary-state eigenkets is not defined within the nonabelian dynamically irreducible invariant spaces of \=A\=T drives. Nevertheless, the triviality of the bundle enables them to have a globally smooth basis, although their individual kets are not stationary-state eigenkets anymore. As a consequence, the micromotion is smoothly defined on the torus for invariant spaces of rank $r_n>1$, but the Wilson line operators can be intrinsically nonabelian: transport generated by the corresponding AGP can mix states within the invariant space $\im(\Pi_n)$, even though population never leaks in or out of it under an \=A\=T quasiperiodic drive.

For the same reason, the quasienergy operator cannot fully capture the internal dynamics within a nonabelian invariant space. In addition to the scalar $\epsilon_n^\F$, the internal dynamics generator has a non-scalar irreducible-tensor component; hence, this induces an additional time-ordering w.r.t.~the Floquet gauge in the remaining dynamics reduced by the smooth micromotion in the evolution decomposition, see Appendix~\ref{app:sec:untwisted_nonabelian}. Intuitively, this can be understood from the defining relation ``$\epsilon^\F = \braket{E^\K(\thetavec)-\omega^\mu\gamma_\mu(\thetavec)/(2\pi)}_{\thetavec\in\mathbb{T}^D}$'': nonabelianness implies that the Kato energy $E^\K_n$ should be promoted to a matrix, $\Pi_n H \Pi_n$ describing the internal dynamics within the invariant space $\Pi_n$.
Likewise, the geometric phase $\gamma_{n,\mu}$ is promoted to a Wilczek-Zee holonomy matrix $W_{n,\mu}=W_\mu \Pi_n$~\cite{wilczek1984appearance} acting within the invariant space $\im(\Pi_n)$. Since the infinite-time average of the time-ordered evolution of matrices is no longer equivalent to their torus-average, we find that, for \=A\=T drives, Floquet theory fails due to a nonabelian obstruction.%
\footnote{ A decomposition of the form~\eqref{eq:qp-Floquet_decomp} with constant $H^{\F}$ and smooth $P$ would make the eigenprojectors of $H^{\F}$ transported by $P$ smooth dynamically invariant subprojectors of $\Pi_n$ (and if $H^{\rm F}\propto\mathds{1}_{r_n\times r_n}$ on $\im\Pi_n$, every subspace would be invariant), contradicting dynamical irreducibility. } Therefore, the unitary evolution operator of nonabelian untwisted drives does not admit an exact Floquet decomposition of the form in Eq.~\eqref{eq:qp-Floquet_decomp}.

However, following a similar procedure as in Sec.~\ref{subsec:AN}, it is possible to derive a generalized nonabelian version of Floquet theory by separating the $\U(1)$ phase of the quasienergy $\epsilon^\F_n$ and the noncommuting remainder of the internal dynamics; we invite the interested reader to Appendix~\ref{app:sec:untwisted_nonabelian} for a discussion of a generalized Floquet decomposition of the nonabelian untwisted class.

By contrast, the infinitesimal average energy decomposition~\eqref{eq:qpGFT_decomp} avoids this nonabelian issue by construction: the average energy gauge separates cleanly the constant scalar part of the dynamics -- the average energy $\mae_n$ -- from all the other irreducible-tensor components, see Eq.~\eqref{eq:AE_objs}. Thus, one can think of the average energy operator $\mAE[\thetavec]$ as the torus-averaged scalar component of the irreducible tensor decomposition of the Kato Hamiltonian $H^\K(\thetavec)$; the irreducible traceless dynamical component is absorbed by the gauge potential $A^\mAE(\thetavec)$. Therefore, the average-energy decomposition of the evolution~\eqref{eq:U_qp} uses a simple exponential $\mathrm{e}^{-it\mAE[\thetavec_0]}$ to describe the dynamics in the $W^\mAE(t;0)$-transported basis. By contrast, a generalized nonabelian Floquet decomposition requires the time-ordering of an $r_n\times r_n$ matrix to describe the internal dynamics within the smooth basis transported by the micromotion, see Appendix~\ref{app:sec:untwisted_nonabelian}.

\subsection{\label{subsec:NT}Nonabelian twisted (\=AT) drives}

The most general class comprises all nonabelian-irreducible topologically nontrivial (i.e., twisted) quasiperiodic drives. Their hallmark is the existence of at least one invariant space that is both nonabelian and twisted. Similar to the AT class, the topological obstruction precludes the existence of globally defined stationary-state kets and globally smooth micromotion on the torus $\T^D$; however, like for the AT drives, an almost everywhere smooth version of the micromotion and a generalized quasienergy operator can be globally defined on the torus except for singularities forming a submanifold of dimension at most $D-2$. Moreover, similar to the \=A\=T class, the nonabelian geometry and dynamics of nonabelian twisted drives induce an irreducible component that cannot be long-time averaged to a scalar quasienergy: the reduced internal dynamics must carry a time ordering.

However, the invariant projectors remain globally smooth despite their higher rank $r_n>1$. Consequently, the average-energy decomposition remains valid with Eq.~\eqref{eq:U_qp} unchanged also for nonabelian twisted drives. The average energy $\mAE$ is a globally well-defined dynamical invariant, and the corresponding gauge potential $A^\mAE$ induces a nonabelian holonomy captured by the Wilson line operator $W^\mAE$. Figure~\ref{fig:nonabelian_twisting} shows an example of a nonabelian twisted drive.

\begin{figure}[t!]
	\centering
    \includegraphics[width=0.50\textwidth]{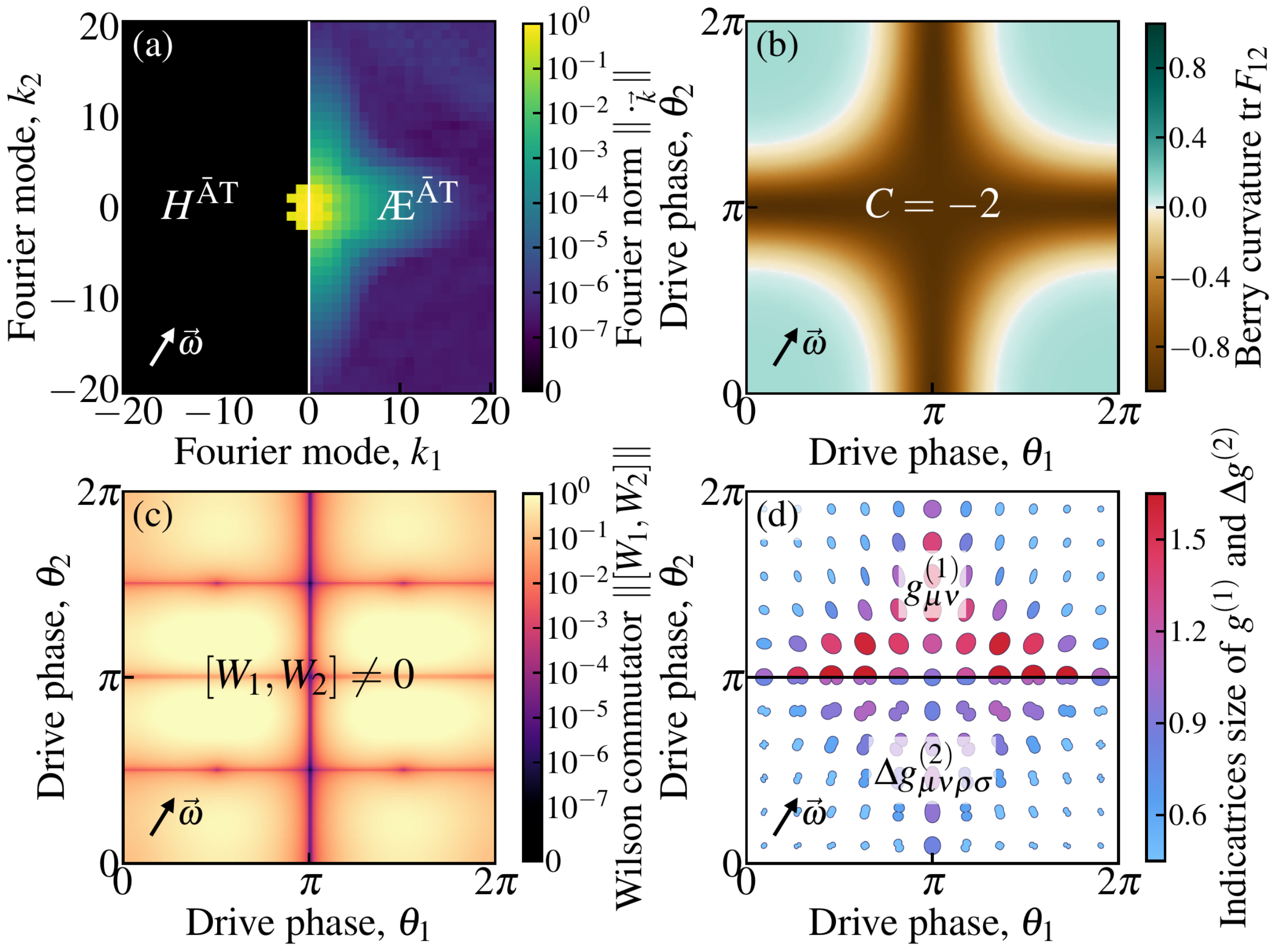}
	\caption{\textbf{Nonabelian twisted (\=AT) drives: an example.} Same quantities and numerical parameters as in Fig.~\ref{fig:nonabelian_nontwisting}, for $H^{\mathrm{\bar A}\mathrm T}$. (a) The average-energy operator remains globally smooth. (b) The ground-space curvature gives $C=-2\pm10^{-9}$. (c) The nonzero Wilson-line commutator and (d) the finite connected second Finsler metric show that the twisted ground state manifold is also nonabelian.}
	\label{fig:nonabelian_twisting}
\end{figure}

One can always build a nonabelian twisted drive by combining two systems under nonabelian untwisted and abelian twisted drives, respectively, as we did in Eq.~\eqref{eq:HA_HB}. Another example of a studied model in the nonabelian twisted class is the four-tone counterdiabatic construction of the Qi–Hughes–Zhang model~\cite{qi2008topological,long2022coupled}. Its rank-two invariant spaces carry vanishing first Chern numbers but an odd second Chern number. Topologically, such a bundle does not admit any smooth rank-one subbundle on $\mathbb{T}^4$~\cite{SadovekSchutte2026}; thus, its nonabelian dynamically invariant spaces are also dynamically irreducible%
\footnote{ This specific example of nonabelian spaces is dynamically irreducible for a different reason than the nonabelian drives $H^\mathrm{\bar A\bar T}$ and $H^\mathrm{\bar A T}$ in Figs.~\ref{fig:nonabelian_nontwisting} and~\ref{fig:nonabelian_twisting}: $H^\mathrm{\bar A\bar T}$ and $H^\mathrm{\bar A T}$ have no smooth subprojector $\Phi_n<\Pi_n$ that solves the transport equation Eq.~\eqref{eq:AaeTransportEquation} (the corresponding subspaces are mixed by the dynamics); the QHZ model has no smooth rank-one subprojectors whatsoever, invariant or not, due to the odd second Chern number: irreducibility is forced by topology, not by dynamics. }.

\begin{figure*}[t!]
    \centering
    \includegraphics[width=\textwidth]{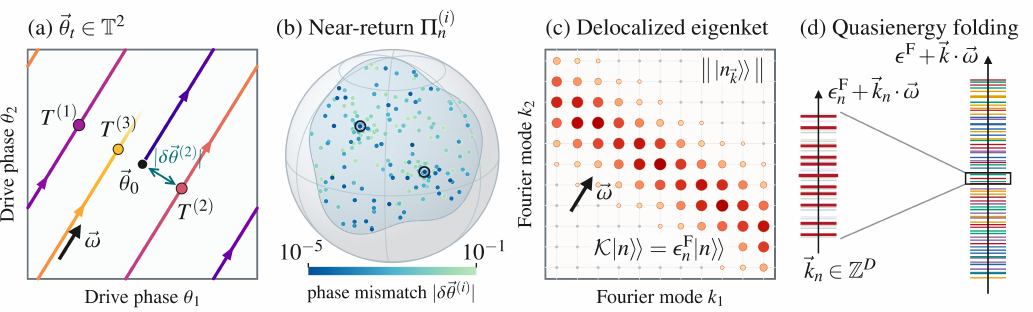}
    \caption{ \textbf{Illustration of near return physics, frequency-lattice eigenstate delocalization, and dense quasienergy folding.} (a) Torus trajectory $\thetavec_t$, colored from purple to yellow with increasing time $t$; dots mark $\thetavec_0$ (black) and three near-return times $T^{(i)}$. (b) Return projectors $\Pi_n^{(i)}$ are eigenprojectors of the evolution operator at return times $U(T^{(i)},0)$, see Eq.~\eqref{eq:DiophProjectors}. They form a shaded region in the projective Hilbert space, colored by the size of the return phase mismatch $|\delta\thetavec^{(i)}|$; circles mark the limit of two return-time subsequences. (c) Schematic Fourier components of a delocalized eigenket of the frequency-lattice Schr\"odinger operator $\mathcal K$ defined in Eq.~\eqref{eq:Sambe}; dot sizes and colors depict the norm $\|\ket{n_{\kvec}}\rangle\|$. (d) Quasienergies $\epsilon^\F_n+\kvec\cdot\omegavec$ are densely folded for incommensurate frequency vectors $\omegavec$ (right); left axis shows a zoom of the black-boxed interval, with all but the red family of $\epsilon^\F_n$ faded. The quasienergy replicas remain dense within the boxed interval; hence, restricting quasienergy to a finite window does not select a single representative of each physical state for $D{\geq}2$. Schematic data used for illustration. }
    \label{fig:schematicPreviousMethods}
\end{figure*}

In addition to the average-energy decomposition universal for the four classes, we provide a generalized Floquet decomposition of dynamically irreducible invariant spaces in the \=AT-class, respecting nonabelian and topological constraints, in Appendix~\ref{app:sec:twisted_nonabelian}. Its micromotion transports the invariant space between the points $\thetavec_0$ and $\thetavec_t$, without accumulating any preceding transport history (unlike the corresponding Wilson lines); the remaining internal dynamics is described by an $r_n\times r_n$ time-ordered rotation.

\section{\label{sec:num_algos}Computational techniques for the average energy \texorpdfstring{$\mAE$}{AE} and its gauge potential \texorpdfstring{$A^\mAE$}{A(AE)}}

We now turn to computing the average energy and its dynamically invariant eigenprojectors. We first explain why replacing quasiperiodic dynamics by near returns or periodic approximants need not recover these invariant spaces. We then extend the variational construction for periodic drives~\cite{mondal2026floquet} to frequency-lattice \textit{operator} space. For simplicity, we restrict the discussion to two-tone ($D{=}2$) drives, although the formalism is generic.

\subsection{\label{subsec:dioph}Failure of Diophantine approximations}

Quasiperiodic drives do not have a return period; instead, the corresponding trajectory explores the torus $\T^D$ ergodically. Nevertheless, the trajectory returns arbitrarily close to its starting point $\thetavec_0$ infinitely often. We call $T^{(i)}$ a sequence of near-return times if there exist integer vectors $\kvec^{(i)}\in\mathbb{Z}^D$ such that the return phase mismatch obeys [see Fig.~\ref{fig:schematicPreviousMethods}(a)]
\begin{equation}\label{eq:diophantine-return}
    \delta\thetavec^{(i)} = \thetavec_{T^{(i)}} - \thetavec_0 -2\pi\kvec^{(i)} =\omegavec T^{(i)}-2\pi\kvec^{(i)} \stackrel{i\to\infty}{\to} \vec 0\, .
\end{equation}
The mismatch $|\delta\thetavec^{(i)}|$ need not decrease monotonically, and different sequences with the same limit exist.

The Diophantine construction~\cite{crowley2019topological} treats such near returns as approximate periodic cycles: for $D=2$, the continued-fraction convergents $p^{(i)}/q^{(i)}$ of the irrational ratio $\omega_2/\omega_1$ define one such sequence, $T^{(i)}=2\pi q^{(i)}/\omega_1$ and $\kvec^{(i)}=(q^{(i)},p^{(i)})$, with $|\delta\thetavec^{(i)}|=2\pi|q^{(i)}\,\omega_2/\omega_1-p^{(i)}|$.

To identify the dynamically invariant spaces, one typically diagonalizes the unitary evolution $U(T^{(i)},0)$ of the exact original quasiperiodic drive and follows its rank-one spectral eigenspaces $\Pi_n^{(i)}$:
\begin{equation}\label{eq:DiophProjectors}
U(T^{(i)},0)=\sum_{n=1}^N u_n^{(i)}\Pi_n^{(i)},
\end{equation}
where $u_n^{(i)}$ are the eigenvalues of $U(T^{(i)},0)$. Eventually, these eigenprojectors (or, in the nonabelian case, groups of $r_m$ of them summed together) form convergent sequences whose limit is ${\Pi}^{(\infty)}_m=\lim_{i\to\infty}\sum_{s=1}^{r_m} \Pi^{(i)}_{m,s}$, where $\Pi^{(i)}_{m,s}$ are the corresponding $r_m$ out of $N$ rank-1 eigenprojectors $\Pi_n^{(i)}$ at time $T^{(i)}$ forming the converging group; this is known as the return-time method~\cite{folta2014stationkeeping,maggia2017advances}. Alternatively, one can first replace the exact incommensurate-frequency vector $\omegavec$ by its Diophantine approximation $2\pi\kvec^{(i)}/T^{(i)}$, and treat the approximate time-periodic system within Floquet theory; these Floquet eigenprojectors, or their sum, also form convergent sequences. We refer to the latter as the Floquet-approximant method~\cite{ho1983semiclassical,crowley2019topological}.

Although the sequence-dependent construction ensures, by definition, the dynamical invariance of the limit ${\Pi}^{(\infty)}_m$ within this particular sequence at a fixed $\thetavec_0$:
\begin{equation}
\begin{aligned}
    &U(T^{(i)},0){\Pi}^{(\infty)}_m[\thetavec_0] U(T^{(i)},0)^\dagger \\
    &\quad=\sum_{s=1}^{r_m} \Pi^{(i)}_{m,s}[\thetavec_0] + \mathcal O\left(\left\|{\Pi}^{(\infty)}_m[\thetavec_0] - \sum_{s=1}^{r_m} \Pi^{(i)}_{m,s}[\thetavec_0] \right\|\right) \\
    &\quad\stackrel{i\to\infty}{\to}{\Pi}^{(\infty)}_m[\thetavec_0] \, ,
\end{aligned}
\end{equation}
it does not automatically guarantee the dynamical invariance for all Diophantine-approximated sequences defined in Eq.~\eqref{eq:diophantine-return} or all $\thetavec_0$ on the torus. To see this, we recall that the smooth dynamical invariance gives $[U(T^{(i)},0),\Pi_m[\thetavec_0]]=\mathcal O(|\delta\thetavec^{(i)}|)$ following Eq.~\eqref{eq:U_qp}. However, notice that the eigenvalues $u_{m,s}^{(i)}$ between different grouped spaces $\sum_{s=1}^{r_m}\Pi^{(i)}_{m,s}[\thetavec_0]$ from Eq.~\eqref{eq:DiophProjectors} wind around the unit circle independently of this error of dynamical invariance in $U(T^{(i)},0)$. Hence, along the subsequences for which two eigenvalues approach each other faster than $O(|\delta\thetavec^{(i)}|)$ decreases, degenerate perturbation theory can mix the corresponding eigenprojectors $\Pi^{(i)}_{m,s}$ of $U(T^{(i)},0)$ by an amount of order one~\cite{davis1970rotation}; this can result in a wrong candidate $\lim_{i\to\infty}\sum^{r_m}_{s=1} \Pi^{(i)}_{m,s}$ that converges in some Diophantine-approximated sequences but diverges in others. Therefore, the limit of the sequence-selected eigenprojectors depends on the subsequence, as schematically shown in Fig.~\ref{fig:schematicPreviousMethods}(b).

Appendix~\ref{app:diophantine-counterexamples} provides two explicit counterexamples in the abelian untwisted class, one for each of the two approximate Diophantine methods. The corresponding sequence-selected eigenspaces converge to the wrong candidate space. We therefore seek a method which imposes dynamical invariance over the torus inherently, not relying on a rational-approximation sequence.

\subsection{\label{subsec:varl}Variational principle}

The average energy retains the part of the lab-frame Hamiltonian that is invariant under the drive. Following the periodic construction of Geometric Floquet theory~\cite{mondal2026floquet}, it is the projection of the lab-frame Hamiltonian $H$ onto the kernel of the Liouvillian, $\ker\LL$:
\begin{eqnarray}
\label{eq:AE_varl}
    \mAE[\thetavec] &=& \lim_{t\to\infty}\frac{1}{t}\int_0^t \mathrm d s \, U(s,0)^\dagger H(\omegavec s+\thetavec)U(s,0) \nonumber\\
    &=& (1-\LL^+\LL)H(\thetavec),
\end{eqnarray}
where $\LL^+$ is the Moore-Penrose pseudoinverse of the Liouvillian $\LL(\cdot)=[H(\thetavec),\cdot]-i\omega^\mu\partial_{\mu}(\cdot)$. The dependence on $\omegavec$ is essential: changing the drive direction changes which spaces remain invariant.

The Liouvillian $\LL^\dagger=\LL$ is a Hermitian superoperator on the frequency-lattice operator space over the torus $\T^D$, with respect to the inner product
\begin{equation}
\label{eq:inner_product}
    \langle X,Y\rangle_{\T^D} =\int_{\T^D}\frac{\mathrm{d}^D\theta}{(2\pi)^D} \text{tr}(X^\dagger Y)\, .
\end{equation}
Its zero modes give the dynamical invariants of $H(\thetavec_t)$.

Since the $\mAE$ operator is a dynamical invariant, it satisfies $\LL(\mAE)=0$. By the linearity of the Liouvillian $\LL$, we have
\begin{equation}
    \LL(A^\mAE)=\LL(H-\mAE)=\LL(H)\,.
\end{equation}
However, $X=A^\mAE$ is not the only solution to the problem $\LL(X)=\LL(H)$; it leaves the freedom to add any dynamical invariant to the solution $X$, shifting a dynamical invariant between the gauge potential and the average energy. Minimizing $\|X\|^2=\langle X,X\rangle_{\T^D}$ removes this freedom, see Appendix~\ref{app:minimum-norm-refinement}. The minimum-norm (Moore-Penrose pseudoinverse) solution is therefore orthogonal to every dynamical invariant and gives
\begin{equation}
\label{eq:varl_AE-AGP}
\begin{aligned}
    A^\mAE &=\underset{X: \LL(X)=\LL(H)}{\operatorname{argmin}}\|X\|^2 =\LL^+\LL H\, .
\end{aligned}
\end{equation}
Once the approximate AGP has been determined, we can readily compute the average energy using $\mAE^{\mathrm{varl}}=H-A^\mAE$, see Eq.~\eqref{eq:qpAE_decomp}. By Eq.~\eqref{eq:AE_varl} above, we find $\LL(\mAE^{\mathrm{varl}})\approx 0$ within the error of the approximation.

Physically, all dynamically invariant components of $H$ remain inside $\mAE$; none is absorbed into $A^\mAE$, which is an element of $(\ker\LL)^\perp$ [see Appendix~\ref{app:subsec:spetral_types}]. Therefore, orthogonality to each invariant projector fixes uniquely the AGP $A^\mAE$ and, by Eq.~\eqref{eq:varl_AE-AGP}, also the average energy:
\begin{equation}\label{eq:aeEnergy}
    \langle\Pi_n,A^\mAE\rangle_{\T^D}=0, \qquad \mae_n=\frac{1}{r_n}\langle\Pi_n,H\rangle_{\T^D}.
\end{equation}
Therefore, generically, the spectral projectors of $\mAE$ resolve the dynamically irreducible invariant spaces, see Appendix~\ref{app:omega-irreducibility}.

We implement this projection in frequency-lattice operator space. Attaching an auxiliary state $\ket{\kvec}$ to each Fourier harmonic represents an operator $O(\thetavec)=\sum_{\kvec}O_{\kvec}\mathrm{e}^{i\kvec\cdot\thetavec}$ in the basis $E_{mn}\otimes\ket{\kvec}$, where $E_{mn}=\ket{m}\bra{n}\hat{=} \ket{m}\otimes\ket{n}$ are matrix-valued canonical basis vectors. With the same Fourier convention for the known lab-frame drive Hamiltonian $H$, Eq.~\eqref{eq:Lqp} gives
\begin{equation}
\label{eq:L-frequency}
\begin{aligned}
    (\LL O)^{\kvec}_{mn} &=\LL^{\kvec,\vec{l}}_{mn,pq}O^{\vec{l}}_{pq},\\
    \LL^{\kvec,\vec{l}}_{mn,pq} &=H^{\kvec-\vec{l}}_{mp}\delta_{nq} -\delta_{mp}H^{\kvec-\vec{l}}_{qn} + (\kvec\cdot\omegavec)\delta_{\kvec,\vec{l}} \delta_{mp}\delta_{nq}.
\end{aligned}
\end{equation}
Repeated indices (also vector-valued) are summed over. The two convolution terms depending on the difference $\vec k - \vec l$ describe the action of $H$ on the ket and bra indices; the third diagonal term accounts for the evolution of the drive phases.

\begin{figure}[t!]
    \centering\includegraphics[width=0.48\textwidth]{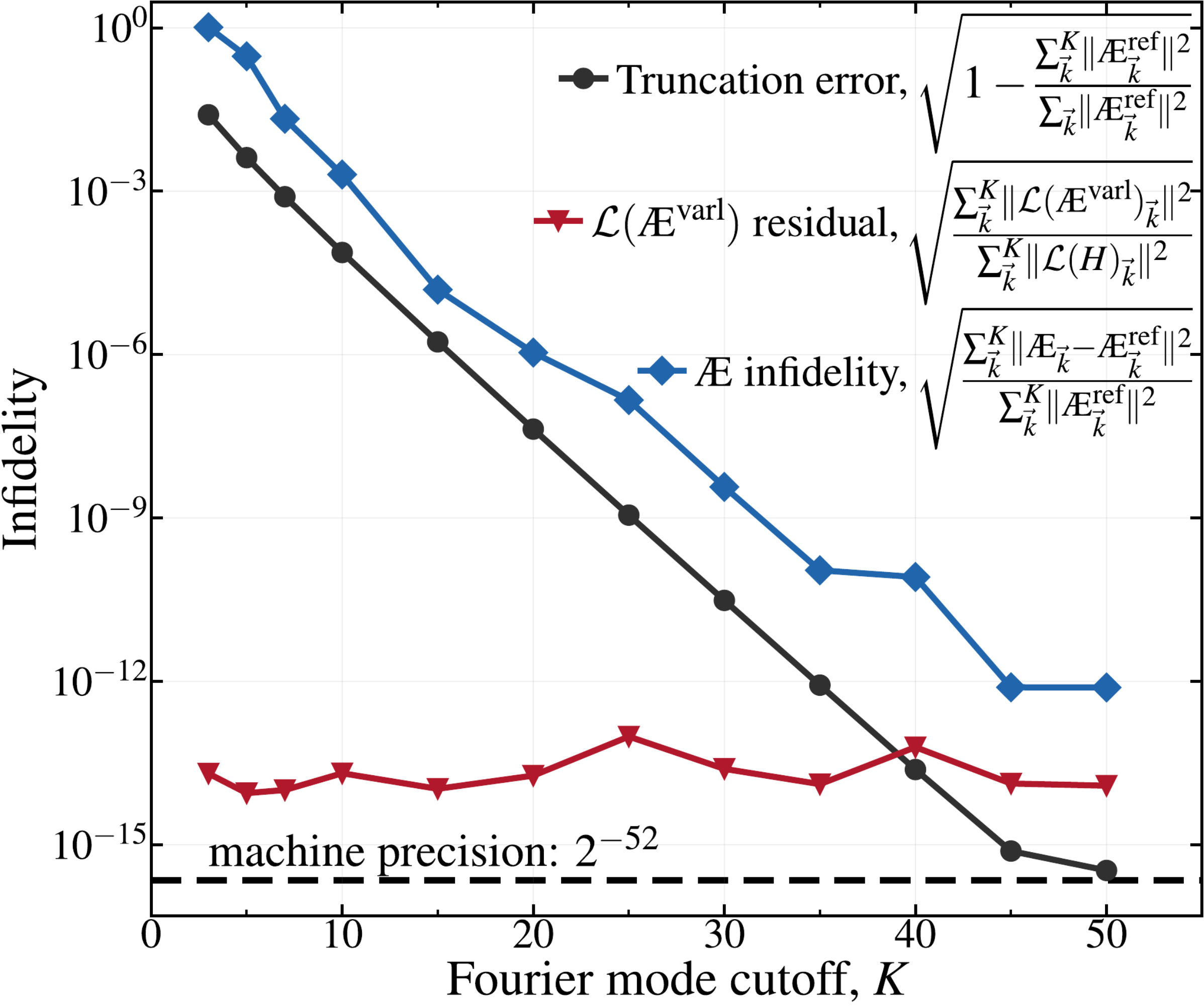}
    \caption{\textbf{Convergence of the variational average-energy operator} for the analytically solvable nonabelian twisted drive $H^{\mathrm{\bar A}\mathrm T,\text{rev-eng}}$ from Eq.~\eqref{eq:AnalyH}. Black circles show the relative Fourier-mode norm of the analytical reference $\mAE^{\rm ref}$ outside $|k_1|,|k_2|\leq K$, separating the omitted Fourier weight from the variational error. Red triangles show the normalized variational residual $\|\LL(\mAE^{\mathrm{varl}})\|/\|\LL(H)\|$, measuring the violation of dynamical invariance; its value reaches $10^{-14}$. Blue diamonds show the relative infidelity of the reconstructed $\mAE$ against $\mAE^{\mathrm{ref}}$, which falls below $10^{-12}$ at a Fourier cutoff $K=45$. Data obtained for perturbation strength $\epsilon=10^{-2}$ and drive parameters $|\omegavec|=0.1$, $\omega_2/\omega_1=(1+\sqrt5)/2$.}
    \label{fig:varlConvergence}
\end{figure}

We use Eq.~\eqref{eq:L-frequency} to solve the variational principle of the minimum-norm problem~\eqref{eq:varl_AE-AGP}. We make an ansatz for the approximate AGP expanded in the fixed basis $E_{mn}\otimes\ket{\kvec}$, placing a Fourier cutoff $K$ on the number of modes retained
\begin{equation}
    |k_1|,|k_2|\leq K.
\end{equation}
The minimization can be performed using, e.g., regularized eigenprojection or an iterative solver, such as MinResQLP \cite{liu2025pseudoinverse}. This produces an approximate average energy operator $\mAE^{\mathrm{varl}}=H-A^\mAE$ for the finite Fourier cutoff $K$. Implementation details are given in Appendix~\ref{app:varl-accuracy}.

We assess convergence by (i) comparing the reconstructed $\mAE^\text{varl}$ with an analytically solvable reference model $\mAE^\text{ref}$, and (ii) testing the dynamical invariance of the variational solution $\mAE^\text{varl}$ independently. In Fig.~\ref{fig:varlConvergence}, the relative $\mAE$-infidelity falls below $10^{-12}$ at a cutoff of $K=45$ [blue diamonds], whereas the residual measuring dynamical invariance is of order $10^{-14}$ [red triangles]. For comparison, the truncation error of the reference $\mAE^\text{ref}$ shows the omitted Fourier weight separately [black circles]. In Appendix~\ref{app:varl-accuracy} we extend the comparison to larger cutoff values $K$, and benchmark different numerical algorithms to solve the variational equation~\eqref{eq:varl_AE-AGP}. Finally, we note that convergence with increasing the cutoff $K$ can be proved formally for any incommensurate $\omegavec$ and finite-dimensional Hilbert space, with the arithmetic of $\omegavec$ affecting only the convergence rate; this makes the variational principle a controlled approximation, see Appendix~\ref{app:varlConvergenceProof}.

Once we have obtained the approximation $\mAE^{\mathrm{varl}}[\thetavec]$, we diagonalize it for each fixed $\thetavec\in\T^D$, and examine its eigenspaces as the cutoff parameter $K$ increases.
We find that a finite cutoff can result in an artificial splitting of a dynamically irreducible nonabelian invariant space into nearly degenerate rank-one subspaces [Fig.~\ref{fig:ae_splitting}]. When this happens, these subspaces are completely delocalized over the frequency lattice, see Fig.~\ref{fig:mergeLocalized}(a,b). However, this delocalization is unphysical because the joint space (i.e., the proper ground-state manifold) recovers the correct nonabelian rank-two invariant space smoothly, see Fig.~\ref{fig:mergeLocalized}(c). %
In practice, we use two criteria to decide whether nearly degenerate rank-one eigenspaces of $\mAE^{\rm varl}$ form a single irreducible space: (i) their splitting must decrease with the cutoff $K$ at least as fast as the omitted Fourier weight of $\text{\AE}^{\rm varl}$, see Fig.~\ref{fig:ae_splitting}, (ii) the rank-one projectors must be delocalized on the frequency lattice while their sum is localized, as shown in Fig.~\ref{fig:mergeLocalized}. Both criteria are satisfied by all nonabelian spaces reported in Sec.~\ref{sec:classification}. Once a converged dynamically invariant space has been correctly identified, Eq.~\eqref{eq:aeEnergy} fixes the corresponding average energy $\mae_n$, giving the reconstructed $\mAE=\sum_n\mae_n\Pi_n$. The properties of these dynamically irreducible spectral projectors then determine the geometry and topology used in Sec.~\ref{sec:classification}.

\textit{\bf Comparison with quasienergy diagonalization.---}%
Our variational principle for the average energy $\mAE$ involves the Liouvillian superoperator $\LL$ and is therefore naturally formulated on the frequency lattice \textit{operator} space. By contrast, previously, quasiperiodic drives have been studied by analyzing the frequency-lattice Schr\"odinger operator $\mathcal K=H-i\omega^\mu\partial_\mu$ on the frequency lattice state space, spanned by the basis $\{\ket{n}\otimes\ket{\vec k}\}$~\cite{ho1983semiclassical,crowley2019topological}; this approach seeks to find the quasiperiodic eigenkets $\ket{n}\rangle$
\begin{equation}\label{eq:Sambe}
\begin{aligned}
    \mathcal K\ket{n}\rangle&=\epsilon^\F_n\ket{n}\rangle,\quad \mathcal K^{\kvec,\vec l}_{m,n} =H^{\kvec-\vec l}_{m,n} +(\kvec\cdot\omegavec)\delta_{\kvec,\vec l}\,\delta_{m,n}.
\end{aligned}
\end{equation}
As we discussed in Sec.~\ref{subsec:AT}, a globally smooth eigenket $\ket{n}\rangle$ can only define an untwisted rank-one invariant projector. Such a ket-state cannot represent a twisted rank-one projector due to a global topological obstruction to defining its smooth overall phase~\cite{crowley2019topological}. Likewise, it cannot resolve an irreducible nonabelian invariant space either, where the evolution mixes the internal states. Using a frequency lattice state-space ansatz, the obtained eigenket appears delocalized over the frequency lattice in the presence of these obstructions, as illustrated in Fig.~\ref{fig:schematicPreviousMethods}(c). Another noteworthy obstacle when diagonalizing the frequency-lattice Schr\"odinger operator $\mathcal K$ is preserving the ordering of the quasienergies $\epsilon^\F_n$ as the Fourier cutoff $K$ is increased: already for drives in the abelian untwisted class, gauge-transformed copies of the same state produce dense quasienergies [see Fig.~\ref{fig:schematicPreviousMethods}(d) and Eq.~\eqref{eq:qp-epsF_ae_gamma}], which breaks the level continuity with $K$ in practice. Because the gauge-equivalent copies are dense, no spectral unfolding procedure is well defined; by contrast, the average-energy gauge is completely fixed and unique.

\begin{figure}[t!]
    \centering
    \includegraphics[width=0.48\textwidth]{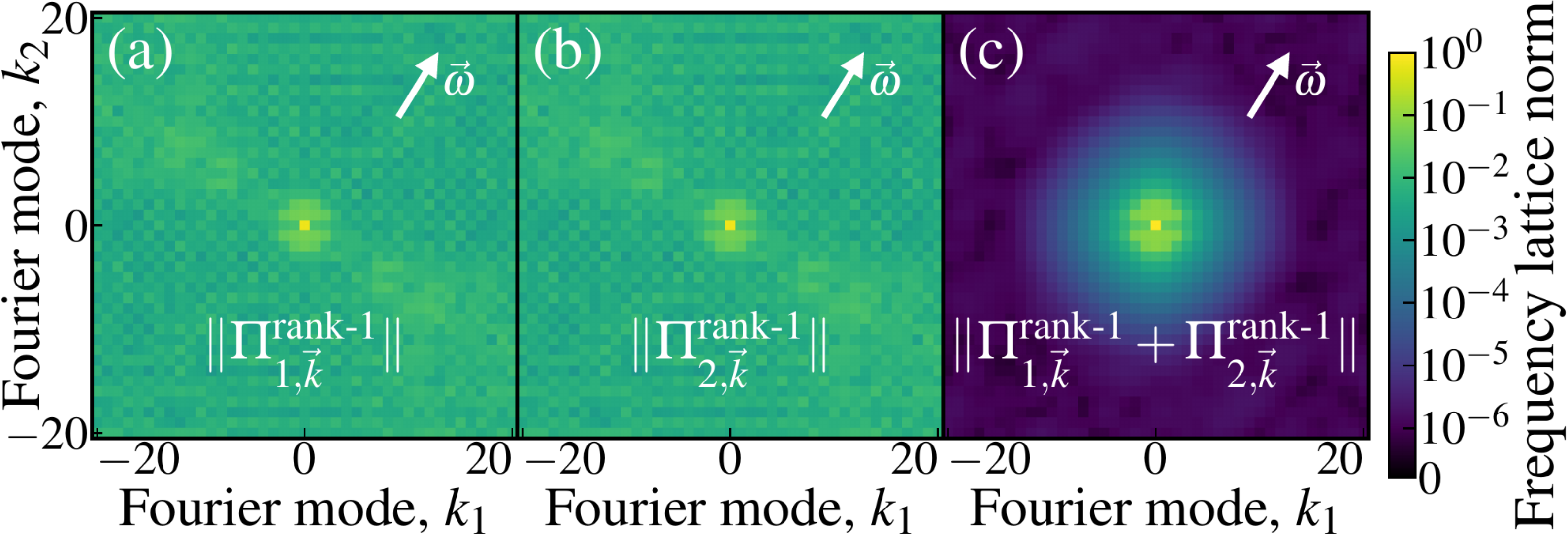}
    \caption{ \textbf{Forcing nonabelian dynamically irreducible invariant spaces into rank-one eigenprojectors of the variational $\mAE^{\rm varl}$ leads to unphysical delocalization; the nonabelian rank-two invariant projector is globally smooth and local.} (a,b) Fourier-mode norms of rank-one spectral projectors $\Pi_1^{\mathrm{rank\text{-}1}}$ and $\Pi_2^{\mathrm{rank\text{-}1}}$ obtained by diagonalizing $\mAE^{\mathrm{varl}}$ exhibit delocalization over the frequency lattice. (c) The rank-two sum $\Pi_{\mathrm{GS}}=\Pi_1^{\mathrm{rank\text{-}1}}+\Pi_2^{\mathrm{rank\text{-}1}}$ defining the physical ground-state manifold is smooth. Data for the nonabelian untwisted model $H^\mathrm{\bar A\bar T}$, see Eq.~\eqref{eq:model_H} and Fig.~\ref{fig:nonabelian_nontwisting}, with parameters $\epsilon=10^{-2},|\omegavec|=0.1, \omega_2/\omega_1={(1+\sqrt{5})}/{2}$, and a Fourier cutoff $K=45$. }
    \label{fig:mergeLocalized}
\end{figure}

In Fig.~\ref{fig:SambeBad} we illustrate the failure of frequency lattice state space to produce physical states for an abelian twisted drive: the eigenket of the frequency-lattice Schr\"odinger operator $\mathcal{K}$ appears as delocalized noise [panel (a)]; by contrast, both the corresponding zero-mode of $\LL$ [panel (b)] and the dynamically invariant eigenprojector of $\mAE$ [panel (c)] remain localized on the frequency lattice and therefore physical.

We note in passing that computing the approximate Liouvillian eigenmodes alone does not enforce their projection property: the operator $X_n$ reconstructed from them in Fig.~\ref{fig:SambeBad}(b) has an idempotency error of order $10^{-5}$. By contrast, diagonalizing $\mAE^{\mathrm{varl}}$ instead gives eigenprojectors satisfying the property $\Pi_n\Pi_m=\delta_{nm}\Pi_n$ pointwise on the torus; the Fourier mode decomposition of $\Pi_n$ has a background noise of order $10^{-5}$, as shown in Fig.~\ref{fig:SambeBad}(c).

\textbf{Computing Kato \& generalized Floquet objects.---}%
Once the projectors converge, Eq.~\eqref{eq:Kato_objs} gives the Kato decomposition, which separates cleanly the instantaneous geometric from the dynamical properties of the evolution. The Kato AGPs $A^\K_\mu(\thetavec)$ define the quantum geometric tensor, Finsler metrics, Berry curvature, and Wilson lines in Eqs.~\eqref{eq:QGT}-\eqref{eq:g1g2mitscherling}.

The generalized Floquet decomposition we derive in Appendix~\ref{app:sec:4classes} writes each invariant space's contribution $U_n(t,0)=U(t,0)\Pi_n[\thetavec_0]$ to the full evolution $U=\sum_n U_n$ as a micromotion together with internal dynamics. The micromotion is defined between two points $\thetavec, \thetavec_0\in\T^D$ on the torus, and so is the remaining internal-dynamics generator. For nonabelian drives, the internal dynamics needs an $r_n\times r_n$ time ordering. For twisted drives, the definition of the micromotion and the internal-dynamics generator features vortex-like singularities on a set of measure zero representing topological charges that sum up to give the corresponding Chern number. Kato and average-energy decompositions also admit formulations within each invariant space, see Appendix~\ref{app:SummaryDecomposition}; however, they require time-ordering of the corresponding $N\times N$ AGPs to construct a Wilson line for each invariant space. The Wilson lines themselves depend on the trajectory and are, hence, not defined globally on the torus; as a consequence, topological charges do not appear naturally in the Kato or average-energy decompositions.

In summary, defining a variational principle on the frequency-lattice operator space allows us to naturally avoid dynamical geometric and topological obstructions associated with the smoothness of quasienergy states. The procedure consists of computing a controlled approximation to the average energy operator $\mAE[\thetavec]$, and then determining its globally smooth eigenprojectors $\Pi_n[\thetavec]$. Our variational principle does not depend on any Diophantine sequences; hence, it defines a new method for calculating dynamically irreducible invariant spaces for quasiperiodically driven systems, and potentially more general structured nonequilibrium drives~\cite{wu2026geometric}.

\begin{figure}[t!]
    \centering
    \includegraphics[width=0.48\textwidth]{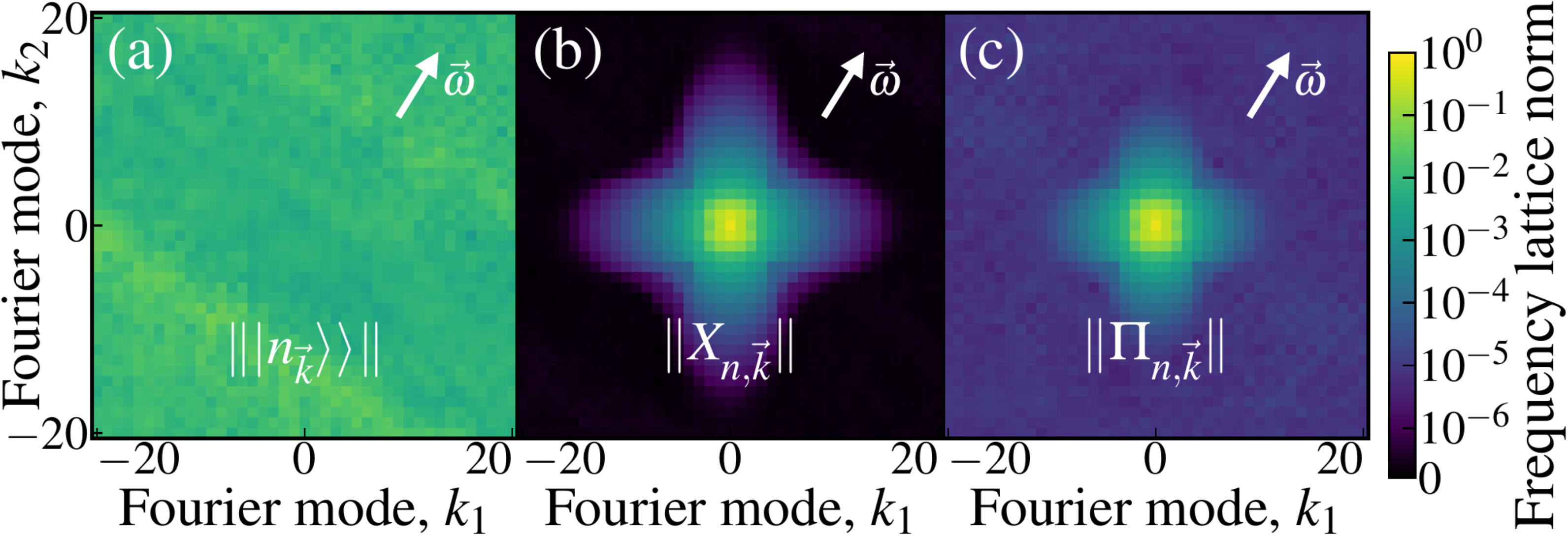}
    \caption{ \textbf{Delocalized frequency-lattice state-space eigenket and its localized frequency-lattice Liouvillian zero-mode and invariant projector.} (a) Fourier-mode norm of the eigenket $\ket n\rangle$ of the frequency-lattice Schr\"odinger operator $\mathcal K$ obtained on the frequency-lattice state-space. The state appears delocalized due to a topological obstruction~\cite{crowley2019topological}. (b) The corresponding approximate zero-mode $X_n$ of the Liouvillian $\LL$, obtained by diagonalization followed by optimizing the projector property $X_n^2=X_n$, is local since projectors are insensitive to the topological singularity in the overall wavefunction phase. (c) The corresponding dynamically invariant eigenprojector $\Pi_n$ of the variationally computed average energy operator $\mAE$ comes out naturally local. Data for the abelian twisted model $H^\mathrm{AT}$, see Eq.~\eqref{eq:model_H} and Fig.~\ref{fig:abelian_twisting}, with parameters $\epsilon=10^{-2}, |\omegavec|=0.1, \omega_2/\omega_1=(1+\sqrt{5})/2$, and a Fourier cutoff $K=45$. }
    \label{fig:SambeBad}
\end{figure}

\section{\label{sec:ergodicity}Nonabelian Ergodicity}

When analyzed carefully, the geometric theory of quasiperiodically driven systems implies a new form of quantum ergodicity, inherently different from the notions of Hilbert space ergodicity~\cite{pilatowsky2024hilbert} and ergodicity based on level-spacing statistics~\cite{dalessio2016quantum}.

Putting topology aside, the rank of the average-energy eigenprojectors $r_n=\text{rank}(\Pi_n[\thetavec])$ allows us to classify each quasiperiodic drive as abelian or nonabelian, cf.~Sec.~\ref{sec:classification}. Importantly, the nonabelian drives in question are dynamically irreducible, i.e., the gauge group $\U(r_n)$ cannot be reduced any further to a block-diagonal subgroup that still reproduces the evolution. One may therefore ask what the physical meaning of these degenerate invariant spaces is.

As before, we assume that the drive is nondegenerate, in the sense that the kernel of the Liouvillian, $\ker\LL=\mathrm{span}\{\Pi_n\}$, is spanned by the average-energy eigenprojectors $\Pi_n$.%
\footnote{ This excludes the accidental degeneracies discussed in Appendix~\ref{app:omega-irreducibility}. For degenerate drives, see Ref.~\cite{zenodo}. } Consider a family of quasiperiodic drives labeled by a parameter $\Delta$ that controls the largest rank $\max_n(r_n)$ among all dynamically irreducible invariant projectors $\Pi_n$, and assume that tuning $\Delta$ causes invariant spaces to coalesce. The system undergoes a series of transitions that reorganize the internal structure of the space of dynamical invariants, $\ker\LL$. For a nondegenerate drive, $\dim\ker\LL$ equals the number of dynamically irreducible invariant spaces [see Fig.~\ref{fig:liouvillianKernelSpectrum}] and is therefore at most $N$. The space of dynamical invariants shrinks by one each time two invariant eigenspaces merge, because dynamical irreducibility ensures that no new internal invariants emerge to replace them. The coalescence process ends when a single invariant projector remains, spanning the entire Hilbert space, $\Pi=\mathds{1}_{N\times N}$, with $\dim\ker\LL=1$.

\begin{figure}[t!]
    \centering
    \includegraphics[width=\columnwidth]{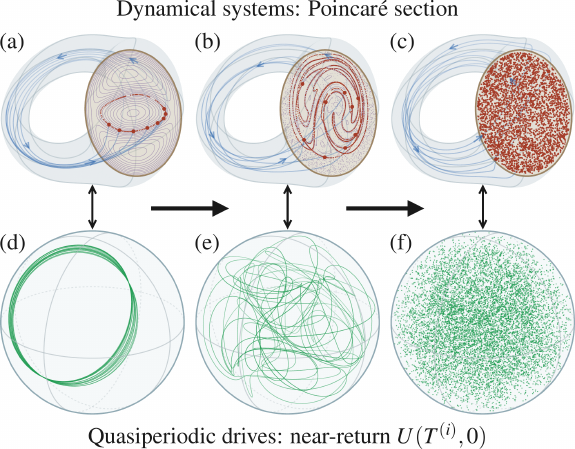}
    \caption{ \textbf{Nonabelian ergodicity: a classical analogy.} Closed classical systems: (a) Poincar\'e sections of a Hamiltonian flow (blue curve with arrows) in phase space; brown points trace the KAM invariant tori; (b) Under perturbation, the KAM tori break into finer and finer invariant tori, until showing a fragmented pattern of cantori in the section; (c) When the perturbation is strong enough, a chaotic sea emerges and occupies the entire region. In quantum systems, (d) near-return unitary evolution $U(T^{(i)},0)$ winds densely on the return group $G$ (green curve) defined in Eq.~\eqref{eq:return_holonomy}; for abelian $G=U(1)\times U(1)\times\cdots$, the projection of this winding is a Lissajous pattern. (e) The abelian return dynamics gradually distorts as invariant spaces coalesce into larger and larger spaces to preserve invariance; (f) In the end, the only invariant space left is $\Pi=\mathds{1}_{N\times N}$ and the return group $G$ generically becomes $\U(N)$ in the absence of further constraints: the system exhibits complete nonabelian ergodicity. }
    \label{fig:schematicNonabelianErgodicity}
\end{figure}

This procedure is reminiscent of how ergodicity emerges in the phase space of closed classical systems, as illustrated in Fig.~\ref{fig:schematicNonabelianErgodicity}. A mixed phase space is typically divided into regular regions defined by stable invariant tori (labeled by action-angle variables -- the classical counterpart of the invariant projectors $\Pi_n$), separated by ergodic regions called the chaotic sea. Notably, the tori originate from conservation laws and constrain the dynamics to lower-dimensional manifolds. As a model parameter is tuned so that a conservation law is lifted, these restrictive tori break down into Cantor sets (cantori)~\cite{mackay1984transport}. Before fully merging into the global chaotic sea, escaping trajectories are temporarily confined to larger, intermediate regions of phase space bounded by the cantori. Thus, this process gradually turns islands of stability into ergodic chaotic-sea regions. When the last invariant torus breaks, the chaotic sea becomes connected across the ergodic phase space.

\textbf{Definition of nonabelian ergodicity.---}%
Back to quantum quasiperiodically driven systems, we can now define the analogous notion of \textit{nonabelian ergodicity}. For simplicity, suppose an arbitrary initial state $\ket{n_0}$ lies in the dynamically irreducible invariant space defined by the projector $\Pi_n[\thetavec_0]$.
As its trajectory returns to points $\thetavec_{T^{(i)}}$ defined in Eq.~\eqref{eq:diophantine-return} arbitrarily close to the initial point $\thetavec_0$ with $|\delta\thetavec^{(i)}|\to0$, its return evolution $U(T^{(i)},0)$ evolves $\ket{n_0}$ to $U(T^{(i)},0)\ket{n_0}\in\im\Pi_n[\thetavec_{T^{(i)}}]$; this gives a sequence of states approximately still in $\im\Pi_n[\thetavec_0]$, because $\Pi_n$ is a dynamical invariant:%
\footnote{ Unlike Diophantine-approximated methods where one uses the return times to find dynamical invariant spaces, here we have already obtained the invariant spaces from the variational principle; instead, we use the return times as a tool to probe nonabelian ergodicity. }
\begin{eqnarray}
    U({T^{(i)}},0)\Pi_n[\thetavec_0]U({T^{(i)}},0)^\dagger &=&\Pi_n[\thetavec_{T^{(i)}}]\\
    &=&\Pi_n[\thetavec_{0}] +\mathcal O(|\delta\thetavec^{(i)}|)\,. \nonumber
\end{eqnarray}
The accumulated near-return evolution $U({T^{(i)}},0)$ restricted to $\im\Pi_n[\thetavec_0]$ defines the return group $G_n$.
Because the latter is dynamically irreducible, there are no remaining conservation laws within the space $\Pi_n$ that would split it further; hence, in general, given sufficient time, the return dynamics $U(T^{(i)},0)\ket{n_0}$ eventually comes arbitrarily close to every quantum state in the invariant space $\im\Pi_n$, and thus $G_n=\U(r_n)$.%
\footnote{ When $G_n=\U(r_n)$, the return dynamics explores the space $\im\Pi_n$ from every initial state $\ket{n_0}\in\im\Pi_n[\thetavec_0]$.}

\begin{figure*}[t!]
    \centering
    \includegraphics[width=\textwidth]{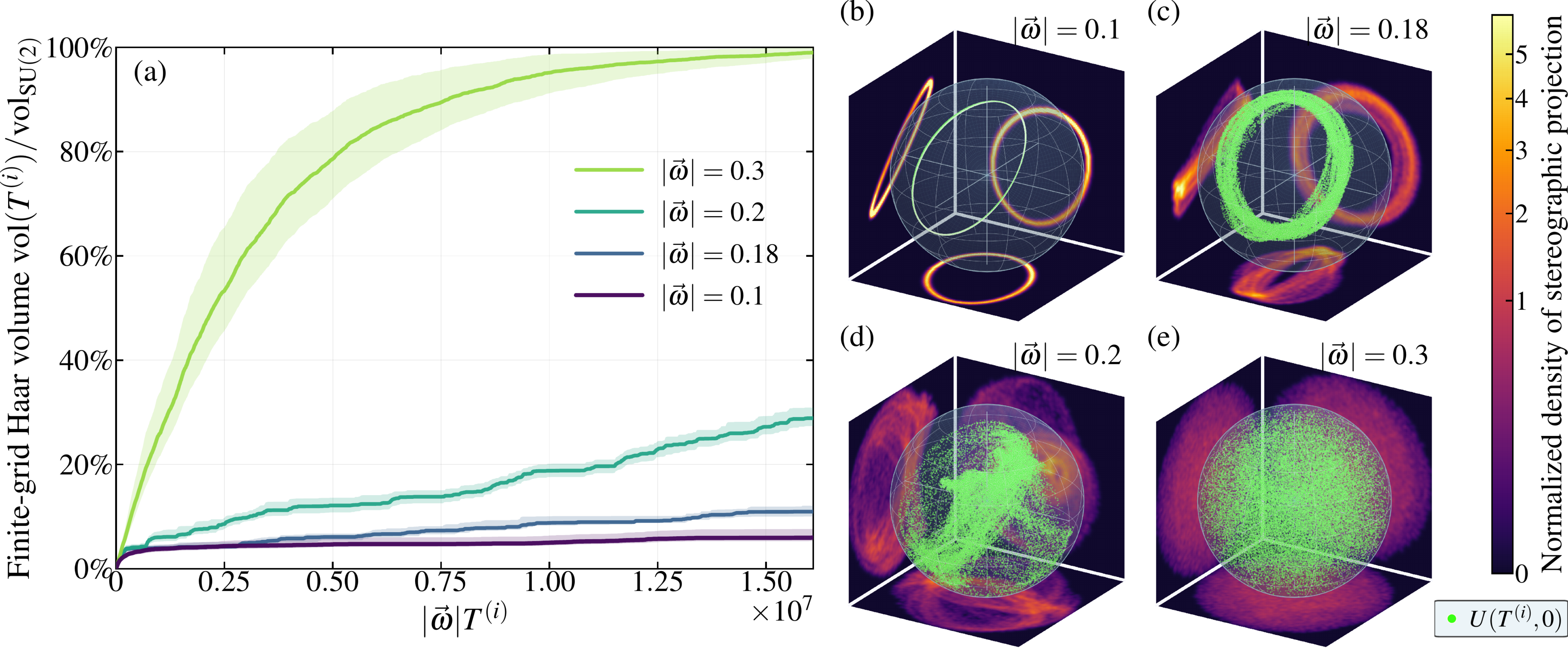}
    \caption{ \textbf{Nonabelian ergodicity in the return dynamics of the quasiperiodically driven truncated QWZ qubit at different drive frequency strengths.} (a) Solid curves and shaded bands around them show the pointwise medians and 10\%-90\% percentiles of the volume ratios $\text{vol}(T^{(i)})/\text{vol}_{\mathrm{SU}(2)}$ over equal-Haar-volume grids containing $\text{vol}_{\rm SU(2)}=4000$ to 13500 cells; (b)-(e) Green markers show the return matrices $U(T^{(i)},0)\in\mathrm{SU}(2)\simeq S^3$ at $|\omegavec|=0.1,0.18,0.2,0.3$, respectively, stereographically projected from a fixed pole in $\mathrm{SU}(2)$ and radially compactified to the open unit ball in $\mathbb R^3$ for visualization. The three rear faces show normalized planar marginal densities on $128\times128$ grids. Here $\text{vol}(T^{(i)})$ is the total Haar volume of cells visited by the near-return evolution up to $T^{(i)}$ with phase mismatch $|\delta\thetavec^{(i)}|<\delta$: the volume ratio measures the fraction of $\mathrm{SU}(2)$ explored by the drive. At $|\omegavec|=0.1$ the return matrices remain on an abelian cycle; at $|\omegavec|=0.3$ they fill $\mathrm{SU}(2)$, indicating nonabelian ergodicity; at $|\omegavec|=0.18$ and 0.2, the rear faces show Lissajous-like curves, representing an abelian return dynamics that becomes more complex. The Hamiltonian is the truncated $\centernot{H}\!\!^{\,\mathrm{T}}_{B}$, with $\omega_2/\omega_1=(1+\sqrt5)/2$, and return tolerance $\delta=0.05$.}
    \label{fig:geometricErgodicitySU2}
\end{figure*}

In the return limit, by definition, each dynamically invariant space comes back to itself. An arbitrary initial state $\ket{\psi_0}\in\mathbb H$ can therefore rotate within these invariant spaces only. The set of limiting return evolutions on the full Hilbert space hence forms a group $G$:
\begin{equation}\label{eq:return_holonomy}
    G=\left\{\lim_{i\to\infty}U(T^{(i)},0): |\delta\thetavec^{(i)}| \stackrel{i\to\infty}{\to} 0\right\} .
\end{equation}
Here the limit runs over all convergent sequences of return unitaries with $T^{(i)}\to\infty$.
Because the spaces defined by $\Pi_n$ are dynamically invariant, in general $G=\U(r_1)\times\U(r_2)\times\cdots\times\U(r_J)$,%
\footnote{ Whereas dynamical irreducibility forbids the gauge group of $\im\Pi_n$ to reduce to a block-diagonal subgroup $\U(r')\times \U(r_n-r')$ that reproduces the evolution, it does not prevent $G$ from reducing to a proper subgroup $G\subsetneq \U(r_1)\times\U(r_2)\times\cdots\times\U(r_J)$. This can happen when accidental degeneracies or symmetry constraints correlate different blocks or restrict evolution within a block. The traceless-qubit example below has $G=\U(1)\subsetneq\U(1)\times\U(1)$ in the abelian case and $G=\mathrm{SU}(2)\subsetneq\U(2)$ in the nonabelian case. } where $J$ is the number of invariant spaces and $\sum_{n=1}^{J}r_n=N$. The unitary factors $\U(r_n)$ are the quantum analogue of the invariant tori in closed classical systems. Related evolution closures are known as \textit{minimality} in the cocycle literature~\cite{nerurkar1997construction,tkachenko2000reducibility,eliasson2002ergodic,karaliolios2018continuous}.

Tuning the parameter $\Delta$ causes the dynamically irreducible invariant spaces to merge, thus enlarging the dynamically accessible part of the return-dynamics space; in other words, initial quantum states can now ergodically mix across a larger invariant space under the quasiperiodic drive. If the dynamically irreducible invariant spaces keep on merging with changing $\Delta$, ultimately, when a single invariant projector $\Pi=\mathds{1}_{N\times N}$ is left, $G=\U(N)$ in general, and the return dynamics explores all the unitary automorphisms of the Hilbert space. Algebraically, there are then no globally smooth non-constant quasiperiodic dynamical invariants left, in analogy to the fully ergodic classical phase space.

In this limit, the average-energy decomposition~\eqref{eq:qpGFT_decomp} carries no information beyond an overall phase: $\mAE=\mae\,\mathds{1}_{N\times N}$ with $\mae=\langle\mathrm{tr}H\rangle_{\T^D}/N$; no ground state can be singled out, and the entire evolution is generated by the gauge potential $A^{\mAE}=H-\mae\mathds{1}_{N\times N}$ through the Wilson line $W^{\mAE}$. In the language of Ref.~\cite{schindler2025geometric}, the drive is purely geometric. The associated bundle is the trivial bundle $\T^D\times\mathbb{H}$ with structure group $\U(N)$, constituting the natural setting for complete nonabelian ergodicity.

\textbf{Example.---}%
To test nonabelian ergodicity numerically, we investigate the return evolution group [Eq.~\eqref{eq:return_holonomy}] generated by $U(T^{(i)},0)$ for the quasiperiodically driven truncated QWZ model with Hamiltonian $\centernot{H}\!\!^{\,\mathrm{T}}_{B}$, see Eq.~\eqref{eq:HA_HB} and surrounding text. We deliberately choose this single-qubit toy model to visualize the emergence of nonabelian ergodicity, cf.~Fig.~\ref{fig:geometricErgodicitySU2}. Moreover, for $m{=}1$, the non-truncated QWZ Hamiltonian $H^{\mathrm{T}}_{B}$ always has two rank-one twisted dynamically invariant spaces. In the abelian case, $G=\U(1)\times\U(1)$ for a single qubit. However, because the truncated Hamiltonian $\centernot{H}\!\!^{\,\mathrm{T}}_{B}$ is in addition traceless, we have $\det U(t,0){=}1$ and the two wavefunction phases encoded by $\U(1)\times\U(1)$ are locked to be the opposite of one another; this leaves a single independent cycle $\U(1)\simeq S^1$ where the return dynamics $U(T^{(i)},0)$ is ergodic. However, although the dynamically irreducible invariant spaces $\im\Pi_n[\thetavec]$ sweep the entire Bloch sphere as $\thetavec$ varies (because they are twisted), the return dynamics is not ergodic in the space of unitaries $\U(2)$; i.e., the drive does not exhibit complete nonabelian ergodicity.

The truncation to the lowest non-constant Fourier modes, $H^{\mathrm{T}}_{B} \to \centernot{H}\!\!^{\,\mathrm{T}}_{B}$, breaks the counterdiabatic construction in the definition of $H^{\mathrm{T}}_{B}$, see Eq.~\eqref{eq:HA_HB}; hence, there is no guarantee that the invariant spaces of $\centernot{H}\!\!^{\,\mathrm{T}}_{B}$ remain abelian. We find that, by fixing the quasiperiodic drive direction $\omega_2/\omega_1{=}(1{+}\sqrt5)/2$ and increasing $\Delta{=}|\omegavec|$ away from the adiabatic regime, the dynamically irreducible invariant spaces eventually merge. Na\"ively, one could expect $G{=}\U(2)$; however, because the qubit Hamiltonian $\centernot{H}\!\!^{\,\mathrm{T}}_{B}$ is also traceless, we have to factor out the corresponding constraint $\det U(t,0)=1$. The return dynamics $U(T^{(i)},0)$ is then ergodic inside $G=\mathrm{SU}(2)\simeq S^3$.

To investigate this prediction, we divide $G=\mathrm{SU}(2)$ into equal-volume cells; the number of cells in the finite-resolution grid gives its volume $\text{vol}_{\rm SU(2)}$. We then solve numerically for the quasiperiodic dynamics of $\centernot{H}\!\!^{\,\mathrm{T}}_{B}(\thetavec_t)$ with $\thetavec_t{=}\omegavec t{+}\thetavec_0$, up to a maximal observation time $T^{(i)}_\text{max}{\sim}10^8$, and retain the closest approach point on the torus in every connected interval satisfying $|\delta\thetavec^{(i)}|<\delta$, with $\delta{=}0.05$ a fixed constant.
Finally, we estimate the volume visited by the quasiperiodic drive by counting whether each equal-volume cell of $\mathrm{SU}(2)$ has been visited at least%
\footnote{ Ergodicity does not imply uniformity of the return-dynamics distribution.} once up until $T^{(i)}_\text{max}$. This defines the Haar volume $\text{vol}(T^{(i)}_\text{max})$ of the return dynamics. The Haar volume ratio $\text{vol}(T^{(i)}_\text{max})/\text{vol}_{\rm SU(2)}$ measures how ergodically the return dynamics behaves within the finite resolution $\delta$.

Fig.~\ref{fig:geometricErgodicitySU2} exhibits the resulting crossover from abelian to nonabelian ergodicity upon increasing $\Delta=|\omegavec|$. For $|\omegavec|{=}0.1$ close to the adiabatic limit, the return dynamics remains confined to a $U(1)$ cycle, see Fig.~\ref{fig:geometricErgodicitySU2} (b), and the return group $G=\U(1)$ is abelian. At $|\omegavec|{=}0.18$, the Haar volume ratio still saturates to a near-zero fraction; the return dynamics remains a single closed $\T$ cycle, so $G=\U(1)$, but the data start displaying a Lissajous-curve-like behavior [panel (c)]. At $|\omegavec|{=}0.2$, the Haar volume ratio does not saturate up until $T^{(i)}_\text{max}$; the Lissajous-like return curve becomes distorted, but $G{=}\U(1)$ is still abelian  [panel (d)]. The observed thickness is a finite-tolerance effect: the sampled set $\{U(T^{(i)},0):|\delta\thetavec^{(i)}|<\delta\}$ is a two-parameter thickening of the $U(1)$ cycle, whose width is set by the transport of the invariant projectors over the return mismatch $|\delta\thetavec^{(i)}|<\delta$ [cf.~Fig.~\ref{fig:schematicNonabelianErgodicity}(e)]; it grows with $|\omegavec|$ because the projectors vary more rapidly over the torus as the merging point is approached. Finally, for $|\omegavec|{=}0.3$ the return dynamics covers almost all equal-volume cells of the ball $\rm SU(2)$, as visible by the green cloud  [panel (e)]; the return-dynamics group $G=\mathrm{SU}(2)$ is now nonabelian, indicating the two invariant eigenprojectors $\Pi_1, \Pi_2$ have merged. The progressively stronger distortion of the sampled return trajectories in Fig.~\ref{fig:geometricErgodicitySU2} is a numerical observation at finite time and resolution. The abelian-to-nonabelian transition itself remains discrete: we find numerically that the two rank-one invariant spaces merge into a single rank-two invariant space at $|\omegavec|{=}0.3$, and the return group changes from $\U(1)$ to $\mathrm{SU}(2)$.

\textbf{Discussion.---}%
We use the full evolution $U$ to probe this dynamical mixing because both the geometric and the dynamical phases contribute to the return motion. Notice that the definition of the Kato transport implemented by the Wilson line $W^\K$ depends sensitively on the number of invariant spaces; when they merge, the notion of geometric transport it measures changes abruptly. Similarly, the average-energy transport implemented by $W^\mAE$ not only exhibits the same discontinuity, but also removes the phases $\mathrm{e}^{-it\mae_n}$. Even a rank-one invariant space with a phase growing linearly in time can have trivial $W^\K$ and $W^\mAE$, while its full evolution winds around a $\U(1)$ phase cycle. The Wilson lines alone would miss the dynamical phases and therefore they are insufficient to capture nonabelian ergodicity on their own.

Physically, nonabelian ergodicity means that the drive mixes the quantum states in a way that makes them absolutely inseparable: i.e., there is no basis inside their nonabelian-irreducible eigenspace whose states are stationary w.r.t.~the dynamics. This implies that periodic drives, which are always irreducible-abelian for any finite Hilbert space dimension by Floquet's theorem, are not fully ergodic. In the thermodynamic limit, Floquet's theorem can break down, leading to the occurrence of arbitrarily large irreducible-nonabelian invariant spaces in which states are inseparably mixed by the drive; the Floquet many-body system is then in its ergodic phase~\cite{mondal2026floquet}. Finite-size data obtained from exact diagonalization show clear evidence of the corresponding average-energy spectral implosion~\cite{schindler2025geometric, mondal2026floquet} that ultimately leads to a completely degenerate $\mae$-spectrum. Interestingly, unlike Floquet systems, the nonabelian dynamics of many quasiperiodic drives enables them to exhibit this mechanism already in finite Hilbert spaces. Last, we note that nonabelian ergodicity relies only on the recurrence of the driving protocol, i.e., on the drive trajectory sampling the torus ergodically; hence, the definition applies beyond the family of quasiperiodic drives~\cite{wu2026geometric}.

\textbf{Comparison with Hilbert-space ergodicity.---}%
Before we close, we emphasize that nonabelian ergodicity concerns the group of return unitaries rather than the orbit of one state in projective Hilbert space $\mathbb{CP}^{N-1}$, and is in this sense stronger than Hilbert-space ergodicity~\cite{pilatowsky2024hilbert}. The latter depends on an initial state $\ket{\psi_0}\in\mathbb{CP}^{N-1}$, and postulates that a set of unitaries $\{ V_i\}$ generated, e.g., by some time evolution, renders $\{V_i\ket{\psi_0}\}$ dense in $\mathbb{CP}^{N-1}$, a space of $2N-2$ real dimensions. By contrast, complete nonabelian ergodicity requires that the return-time limit of the sequence of unitaries $U(T^{(i)},0)$ be dense in $\U(\mathbb{H})=\U(N)$, a space of $N^2$ real dimensions.  Nonabelian ergodicity is even stronger than unitary ergodicity~\cite{pilatowsky2024hilbert}: the latter considers the evolution $U(t,0)$ at all times $t\geq 0$, whereas nonabelian ergodicity probes the rotations accessible at near returns. The same truncated QWZ drive at $|\omegavec|=0.1$ illustrates this distinction. Close to the adiabatic limit, its two twisted rank-one invariant spaces remain distinct and move over the entire Bloch sphere. Together with the evolution phase, this motion lets the full evolution $U(t,0)$ explore $\mathrm{SU}(2)$. With a constant offset in the Hamiltonian, the evolution would have unitary ergodicity since $U(t,0)$ is expected to be dense in $\U(2)$. However, at near-return times $T^{(i)}$, we found in Fig.~\ref{fig:geometricErgodicitySU2}(b) that $U(T^{(i)},0)$ is not nonabelian-ergodic, because the spaces return to themselves and only the abelian $\U(1)$-phase cycle remains.

\FloatBarrier
\section{\label{sec:outro}Discussion \& Outlook}

We have developed a geometric description of quasiperiodically driven quantum dynamics by identifying their dynamically invariant spaces -- the generalization of Floquet states. Analyzing their dynamical geometric and topological properties, we propose a generic four-fold classification of quasiperiodic drives as (un)twisted and dynamically irreducible (non)abelian. To classify a given drive, we design a variational principle in frequency-lattice operator space to approximate its average-energy operator $\mAE[\thetavec]$, whose eigendecomposition defines the invariant eigenspaces $\Pi_n[\thetavec]$, sorted by the corresponding average-energy eigenvalues $\mae_n$; this allows us to avoid problems of previous approaches stemming from dense quasienergy folding or from the absence of globally smooth eigenkets. We show that neither twisted nor nonabelian drives admit globally smooth stationary kets, whereas the invariant projectors remain globally smooth and therefore physical. We expect the twisted classes to be stable to small perturbations of model parameters as a result of topological protection, as long as the invariant space does not merge with another one; in contrast, the (non)abelian property is not as robust, and the corresponding parameter space is expected to exhibit intervening Arnold-tongue resonances~\cite{puig2002reducibility}.

Our framework provides different exact decompositions of the unitary evolution for quasiperiodic drives. The average-energy decomposition~\eqref{eq:U_qp} is universal to all classes, and contains no time-ordering of the dynamical invariant $\mAE[\thetavec_t]$; the Wilson line operator $W^\mAE(t;0)$ is defined in terms of the time-ordered AGP $A^\mAE(\thetavec_t)$, and is computed along the drive trajectory $\thetavec_t$ as it is not globally defined on the torus $\T^D$. We generalize Floquet theory to incommensurate quasiperiodic drives in the abelian untwisted class, and construct the corresponding micromotion and quasienergy operators, cf.~Eq.~\eqref{eq:qp-Floquet_decomp}. Tracing the reason for the failure of conventional Floquet theory back to topological or dynamical geometric obstructions, we derive a generalized (nonabelian) Floquet theory for the remaining classes [Appendix~\ref{app:sec:4classes}]; it uses a micromotion and an internal dynamics $\U(r_n)$ generator acting within $\im\Pi_n$, both defined almost everywhere on the torus to decompose the evolution within each invariant space. Depending on the class, this internal dynamics can reduce to a scalar phase or generalize to a time-ordered nonabelian rotation. Unlike the average energy decomposition, the Floquet decompositions require Diophantine frequency vectors.

Investigating the physical meaning of the nonabelian-irreducible dynamically invariant spaces, we propose the concept of nonabelian ergodicity. It characterizes the dense exploration of the unitary return group accessible to a given evolution, and bears a formal analogy with ergodicity in classical phase space.

Our four-fold classification is based on the properties of the dynamically invariant eigenprojectors of the average energy, which we compute using the variational principle from Sec.~\ref{sec:num_algos}. A natural question is whether one can formulate necessary and/or sufficient criteria for a certain quasiperiodically driven system $H(\thetavec)$ to belong to a given class without computing the invariant projectors. In addition, it remains to sort out the relative `size' of each class depending on the number $D$ of incommensurate tones in the drive: intuitively, for strong, generic drives, we expect non-finetuned drives to fall into the nonabelian untwisted class, with the average rank of the effective eigenprojectors growing with $D$.

This work focuses on the general framework, including drive-induced geometry and topology, but without investigating the effect of additional physical parameters such as quasimomentum or external magnetic fluxes~\cite{gavensky2025streda}. We anticipate future work to develop the methodology we propose further: for instance, the framework defines the objects needed to study topological insulators in effective average-energy bands beyond the paradigm of Floquet systems. Under a change of frame, the invariant spaces transform covariantly; their ranks, dynamical irreducibility and Chern numbers are frame-independent, whereas their quantum geometry and the average energies, like any energy away from equilibrium, are properties of the lab frame~\cite{schindler2025geometric}. For AT drives, the geometric phase along a noncontractible loop of the torus winds by $2\pi C_n$ as a function of the transverse phase, which is a direct signature of twistedness; the Chern number $C_n$ quantizes the average rate of energy transfer between the drive tones for states in $\Pi_n$~\cite{martin2017topological,crowley2019topological,long2021nonadiabatic}. \=A\=T drives, on the other hand, can exhibit a nonabelian Wilczek-Zee connection whose holonomy along noncontractible loops of the torus or in the space of additional parameters could implement geometric braiding. Combining the two, it is open whether \=AT drives can implement topologically protected braiding via such a nonabelian holonomy. These and other new phenomena exhibited by quasiperiodic drives can now be investigated in a clean and well-defined setting. We therefore envision the emergence of a paradigm for quantum control beyond Floquet engineering~\cite{eckardt2017colloquium}.

Our four-fold classification has distinct signatures readily accessible in current two-tone experiments on qubits and NV centers~\cite{boyers2020exploring,malz2021topological}: (i) The average energy $\mae_n$ of a stationary state is its time-averaged energy, $\mae_n=\lim_{t\to\infty}t^{-1}\int_0^t\mathrm{d}s\,\langle\psi(s)|H(\thetavec_s)|\psi(s)\rangle$ for $|\psi(0)\rangle\in\im\Pi_n[\thetavec_0]$; the merging of invariant spaces appears as a measurable degeneracy of these averages. (ii) The return fidelity $F^{(i)}=|\langle\psi(0)|U(T^{(i)},0)|\psi(0)\rangle|^2$ at near-return times distinguishes the abelian and nonabelian classes: for a state in a rank-one invariant space, $F^{(i)}\to1$ as $|\delta\thetavec^{(i)}|\to0$; in turn, within a nonabelian space of rank $r_n$ the return fidelities of a generic state are expected to be distributed over the interval $[0,1]$, with a distribution set by the properties of the return group $G_n$ (Sec.~\ref{sec:ergodicity}). The transition from the first to the second behavior as a drive parameter is varied can give a direct probe of the merging of dynamically invariant spaces. (iii) Twistedness can be detected by the average rate of energy transfer between the drive tones, which is quantized to $C_n\,\omega_1\omega_2/2\pi$ for states in a rank-one space~\cite{martin2017topological,crowley2019topological,long2021nonadiabatic} and, for a rank-$r_n$ space, equals $(C_n/r_n)\,\omega_1\omega_2/2\pi$ when averaged over states in $\im\Pi_n$. A systematic study of the above three protocols, including their robustness to the finite return tolerance and to decoherence, is left for future work.

Another exciting direction is the study of many-body systems exposed to quasiperiodic drives. This will likely require more efficient algorithms to compute the average energy (or directly its ground state). Recently, it was shown that Floquet states can exhibit ground-state order in large interacting quantum systems~\cite{mondal2026floquet}; we expect this to extend to certain drives in the abelian untwisted class. More generally, the intriguing possibility of identifying new many-body phases of matter without a Floquet counterpart~\cite{friedman2022topological} in the twisted and nonabelian classes is an exciting future direction. Their stability requires a careful analysis of heating and energy absorption~\cite{else2020long}. In that respect, the interplay between various types of many-body localization and the rank of the dynamically invariant spaces may give rise to new mechanisms to stabilize quasiperiodically driven systems. This is inextricably related to the onset of ergodicity and its facilitation by quasiperiodic drives; in Sec.~\ref{sec:ergodicity} we connected the latter to the coalescence of dynamically invariant spaces, as a model parameter is tuned. The framework we propose gives access to spectral objects defined without a Floquet operator, and enlarges the toolbox for the investigation of quantum chaos beyond periodically driven systems.

Finally, we remark that the framework we developed is not restricted to quasiperiodic drives. Indeed, the notion of the average-energy operator $\mAE$ as a dynamical invariant and its AGP $A^\mAE$ as well as the variational principle to compute them generalize directly to other geometric drives on compact manifolds~\cite{wu2026geometric}, and presumably other kinds of aperiodic drives~\cite{zhao2021random,moon2025experimental,liu2026prethermalization}.

\textit{Acknowledgments.---}%
We thank Dan S.~Borgnia, Anushya Chandran, Wen-Wei Ho, David M.~Long, Yang Liu, Anatoli Polkovnikov, Paul M.~Schindler, Qiyun Yang, and Xin-Chi Zhou for insightful discussions.

The authors used GPT Astra 6, GPT Sol 6.1, and Claude Fable 5.1 during the preparation of this manuscript to improve readability, generate octopuses of the schematic figures, and assist with code optimization and plotting. The authors assume full responsibility for the scientific content, accuracy, and integrity of this paper.

Funded by the European Union (ERC, QuSimCtrl, 101113633). Views and opinions expressed are, however, those of the authors only and do not necessarily reflect those of the European Union or the European Research Council Executive Agency. Neither the European Union nor the granting authority can be held responsible for them. MB was supported in part by grant NSF PHY-2309135 to the Kavli Institute for Theoretical Physics (KITP).

\textit{Data availability statement.---}%
The data and code associated with this manuscript are available in the Zenodo repository~\cite{zenodo}, DOI: \href{https://doi.org/10.5281/zenodo.24107408}{10.5281/zenodo.24107408}. Automatic access is restricted; access credentials will be provided to human researchers upon reasonable request via Zenodo for academic and verification purposes.

\else %
\fi %

\ifsuppmode

\renewcommand{\thefigure}{A\arabic{figure}}
\renewcommand{\theHfigure}{supp.\arabic{figure}}
\renewcommand{\theHtable}{supp.\arabic{table}}
\setcounter{figure}{0}
\setcounter{equation}{0}
\renewcommand{\thesection}{S\arabic{section}}
\setcounter{section}{0}

\twocolumngrid

\ifcombinedmode
\else
\maketitle
\fi %

\tableofcontents

\begin{appendix}
\renewcommand{\theHequation}{appendix.\Alph{section}.\arabic{equation}}

\section{Irreducibility of Invariant Spaces}
\label{app:omega-irreducibility}

In this appendix, we discuss two notions of irreducibility for invariant spaces; we then compare and contrast them using concrete examples.

A dynamically invariant space $\Pi_n[\thetavec]$ is called \textit{dynamically irreducible} if it cannot be split into smaller, smooth dynamically invariant proper subspaces. Conversely yet equivalently, $\Pi_n[\thetavec]$ is dynamically reducible if it has a smooth subprojector $\Phi[\thetavec]$ on $\T^D$ with $\im \Phi[\thetavec] \subsetneq \im \Pi_n[\thetavec]$ (in short, $\Phi<\Pi_n$) that satisfies
\begin{equation}\label{eq:omega-reduction}
    i\omega^\mu\partial_\mu\Phi[\thetavec]=[A^\mAE(\thetavec),\Phi[\thetavec]], \qquad 0<\operatorname{rank}\Phi<r_n,
\end{equation}
where $r_n=\tr\,\Pi_n[\thetavec]$ is the rank of $\Pi_n[\thetavec]$ that sets the dimension of its space, $r_n=\dim\im\Pi_n[\thetavec]$.

The smoothness condition on the entire torus $\T^D$ is essential: transporting an arbitrary ket vector along a single fixed trajectory does not ensure that its projector approaches its initial value when the drive phases return arbitrarily close [Fig.~\ref{fig:schematicPreviousMethods}(a)]. Otherwise, despite being a dynamical invariant in time $t\in\mathbb{R}$, $\Phi[t]$ is not smoothly defined on the entire torus $\mathbb{T}^D$.

Within an average-energy eigenspace, $\Pi_n\mAE\Pi_n=\mae_n\Pi_n$, every subprojector $\Phi<\Pi_n$ commutes with $\mAE$. Thus the same dynamical reduction can be equivalently formulated using $H$ instead of $A^\mAE=H-\mAE$ since
\begin{equation}\label{eq:omega-reduction-generators}
    [A^\mAE,\Phi] =[H,\Phi]=[\omega^\mu A^\K_\mu+\Pi_nH\Pi_n,\Phi].
\end{equation}
Note that the subprojector $\Phi$ does not commute with the Kato Hamiltonian in general, $[\Pi_nH\Pi_n,\Phi]=[H^\K,\Phi]\neq 0$, as $H^\K$ may contain an irreducible traceless part, cf.~Eq.~\eqref{eq:tensor_decomp}. Hence, dynamical reducibility is defined w.r.t.~the average-energy AGP $A^\mAE(\thetavec)$ and fixed to the drive trajectory, defined by $\omegavec$ for quasiperiodic drives.

On the other hand, \textit{geometric reducibility} is a global concept. It asks whether a splitting survives transport in every direction $\mu$ on the torus:
\begin{equation}
\label{eq:global-reduction}
    i\partial_\mu\Phi=[A_\mu^\K,\Phi], \quad 0<\operatorname{rank}\Phi<r_n,\quad\forall\mu
\end{equation}
Geometric irreducibility is defined w.r.t.~the Kato AGP $A^\K_\mu$; by contrast, dynamical irreducibility concerns the physical evolution along the drive as captured by the average energy AGP $A^\mAE$, including both its geometry and its internal dynamics.

Therefore, neither notion of irreducibility implies the other; the following examples make this clear.

\textbf{Dynamically irreducible but geometrically reducible.} A counterexample is the qubit model of $H=H_A^{\rm\bar A}$ defined in Eq.~\eqref{eq:HA_HB}. The small transverse perturbation $\epsilon \sigma^x$ in the traceless Hamiltonian $H_A^{\rm\bar A}$ makes it a dynamically irreducible nonabelian (untwisted) drive in a two-dimensional total Hilbert space, as discussed in detail in Appendix~\ref{sec:LandscapeQP}. Thus it only has a single invariant projector, the identity $\mathds{1}$. Hence, all parallel-transport AGPs $A_\mu^\K$ vanish trivially, and the AGP $A^{\mAE}$ equals the entire traceless Hamiltonian $H_A^{\rm\bar A}$ which contains the small transverse perturbation $\epsilon \sigma^x$:
\begin{equation}
    \begin{gathered}
    \Pi=\mathds{1}_{2\times 2},\quad \mAE=0,\quad A^\mAE=H,\\
    A^\K_\mu=0=A^\K,\quad H^\K=H.
    \end{gathered}
\end{equation}
However, every rank-one subprojector $\Phi<\Pi$ that is constant on the torus defines a geometric reduction globally, since Eq.~\eqref{eq:global-reduction} is also satisfied. Thus, this is an example of a dynamically irreducible but geometrically reducible invariant space.

\textbf{Geometrically irreducible but dynamically reducible.} We now show that the converse can also be true. Consider a three-level system with two incommensurate frequencies $\omegavec=(1,\sqrt{2})^\mathsf{T}$ which define the quasiperiodic drive $\thetavec_t = \omegavec t+\thetavec_0$ with $\thetavec_0=\vec{0}$. Define two orthogonal invariant spaces via
\begin{equation}
\begin{aligned}
    V(\thetavec)&=\operatorname{diag}(\mathrm e^{-i\theta_1},\mathrm e^{-i\theta_2},1),\\
    \ket{3[\thetavec]}&=\frac{1}{\sqrt{3}}V(\thetavec)(1,1,1)^\mathsf{T},\\
    \Pi_3[\thetavec]&=\ket{3[\thetavec]}\bra{3[\thetavec]},\qquad \Pi_{1+2}[\thetavec]=\mathds{1}_{3\times 3}-\Pi_3[\thetavec].
\end{aligned}
\end{equation}
The Kato AGPs of these two invariant spaces and a Hamiltonian transporting them are given by
\begin{equation}\label{eq:geometric-irreducible-example}
\begin{aligned}
    A^\K_\mu(\thetavec)&=\frac{i}{2}\bigl([\partial_\mu\Pi_{1+2}[\thetavec],\Pi_{1+2}[\thetavec]] +[\partial_\mu\Pi_3[\thetavec],\Pi_3[\thetavec]]\bigr),\\
    H(\thetavec)&=H^\K(\thetavec) + \omega^\mu A^\K_\mu(\thetavec),\qquad H^\K(\thetavec)=\Pi_3[\thetavec]=\mAE[\thetavec].
\end{aligned}
\end{equation}
The corresponding average energy $\mAE[\thetavec]$ has two eigenprojectors: a rank-two $\Pi_{1+2}[\thetavec]$ with a doubly degenerate $\mae_{1+2}=0$, and a rank-one $\Pi_3[\thetavec]$ with $\mae_3=1$ representing the dynamical invariant spaces.

However, in the co-moving frame defined by $V(\thetavec)$, the frame-transformed Hamiltonian is actually constant:
\begin{eqnarray}
    && V(\thetavec)^\dagger H(\thetavec) V(\thetavec) -iV(\thetavec)^\dagger \omega^\mu\partial_\mu V(\thetavec) \\
    &=& \frac{2-\sqrt{2}}{3}\Pi_3[\vec{0}] -\Pi_{1+2}[\vec{0}]\operatorname{diag}(1,\sqrt{2},0)\Pi_{1+2}[\vec{0}]. \nonumber
\end{eqnarray}
Writing the two orthonormal basis states of $\im \Pi_{1+2}[\thetavec]$ as
\begin{equation}
\begin{aligned}
    \ket{1[\thetavec]}&=\frac{V(\thetavec)}{\sqrt{6}}(2,-1,-1)^{\mathsf T}, &\ket{2[\thetavec]}&=\frac{V(\thetavec)}{\sqrt{2}}(0,1,-1)^{\mathsf T},
\end{aligned}
\end{equation}
one can see that the Hamiltonian in the transformed frame has three distinct constant rank-one eigenprojectors, among which two are subprojectors of $\Pi_{1+2}[\vec0]$. Transforming the two rank-one eigenprojectors within $\im\Pi_{1+2}[\vec0]$ back by $V(\thetavec)$ gives smooth dynamical invariants $\Pi_1[\thetavec]$ and $\Pi_2[\thetavec]$. Therefore the invariant space $\Pi_{1+2}[\thetavec]$ is dynamically reducible. This is expected, because in the construction of $H$ above we intentionally configured the average energies $\mae_1=\mae_2=0$ to be accidentally degenerate.

On the other hand, because the invariant space $\Pi_{1+2}[\thetavec]$ is two-dimensional and hence all its subprojectors are necessarily rank-one, geometric reducibility according to Eq.~\eqref{eq:global-reduction} would require the holonomy group to be abelian; this requires the two Wilson lines along the noncontractible loops within $\Pi_{1+2}$ to commute: $[W_1,W_2]=0$. However, a straightforward calculation shows that the Wilson line operators
\begin{equation*}
\begin{aligned}
    W_1(\thetavec)&=\mathrm e^{4\pi i/3}\ket{1[\thetavec]}\bra{1[\thetavec]}+\ket{2[\thetavec]}\bra{2[\thetavec]} \\&\quad+\mathrm e^{2\pi i/3}\ket{3[\thetavec]}\bra{3[\thetavec]},\\
    W_2(\thetavec)&=\mathrm e^{4\pi i/3}\frac{-\ket{1[\thetavec]}+\sqrt{3}\ket{2[\thetavec]}}{2}\frac{-\bra{1[\thetavec]}+\sqrt{3}\bra{2[\thetavec]}}{2}\\&\quad+\frac{\sqrt{3}\ket{1[\thetavec]}+\ket{2[\thetavec]}}{2}\frac{\sqrt{3}\bra{1[\thetavec]}+\bra{2[\thetavec]}}{2} \\&\quad+\mathrm e^{2\pi i/3}\ket{3[\thetavec]}\bra{3[\thetavec]},
\end{aligned}
\end{equation*}
do not commute. Therefore, we find that $\Pi_{1+2}[\thetavec]$ is geometrically irreducible.

We thus see that the invariant space $\Pi_{1+2}$ of the Hamiltonian $H$ serves as an example of both a dynamically reducible invariant space when viewed as a dynamical invariant $\Pi_{1+2}[\thetavec_t]$ of the drive $H(\thetavec_t)$, and -- at the same time -- a geometrically irreducible bundle $\Pi_{1+2}[\thetavec]$ over the torus $\T^2$.

In general, the spectral projectors of the average-energy operator $\mAE=\sum_n\mae_n\Pi_n$ directly separate the dynamically irreducible invariant spaces, because $\mAE=H-A^\mAE$ tends to split any invariant spaces into finer ones as a result of the minimum-norm property of the average-energy AGP $A^\mAE$ (see Appendix \ref{app:minimum-norm-refinement}). It is intuitively clear that the eigenprojectors of $\mAE$ are dynamically irreducible, since $\mAE\in\ker\LL$ and the Liouvillian knows about the trajectory via $\omegavec$. Only accidental degeneracy between average energies can hide a finer splitting.

For example, qubit $A$ in Eq.~\eqref{eq:HA_HB} at $\epsilon=0$ has
\begin{equation*}
    H(\thetavec)=(\cos\theta_1+\cos\theta_2)\sigma^z\, .
\end{equation*}
Its evolution is a pure $z$-rotation
\begin{equation*}
    U(t,0)=\exp[-if(\thetavec_t)\sigma^z],
\end{equation*}
where
\begin{equation*}
    f(\thetavec_t)=\frac{1}{\omega_1}\left(\sin \theta_{t,1} -\sin \theta_{0,1}\right)+\frac{1}{\omega_2}\left(\sin\theta_{t,2}-\sin\theta_{0,2}\right).
\end{equation*}
Choosing an arbitrary rank-one projector $\Pi[\thetavec_0]$, we define a smooth projector on the torus $\Pi[\thetavec]=\exp(-if(\thetavec)\sigma^z)\Pi[\thetavec_0]\exp(if(\thetavec)\sigma^z)$ that is rank-one and dynamically invariant because
\begin{equation*}
    i \omega^\mu\partial_\mu \Pi[\thetavec]=\omega^\mu\partial_\mu f(\thetavec)[\sigma^z,\Pi[\thetavec]]=[H(\thetavec),\Pi[\thetavec]]\, .
\end{equation*}
However, its average energy $\mae=\braket{\text{tr}(H(\thetavec)\Pi[\thetavec])}_{\mathbb{T}^D}\propto\braket{\cos\theta_1+\cos\theta_2}_{\mathbb{T}^D}=0$ vanishes accidentally. Thus, the average-energy operator $\mAE=0$ also vanishes, and its single spectral eigenspace is the rank-two full Hilbert space of the qubit; however, it is dynamically reducible by Eqs.~\eqref{eq:omega-reduction} and~\eqref{eq:omega-reduction-generators}.

A more interesting example, for $D{=}1$, of degenerate (and hence nonabelian) average-energy eigenspaces that are dynamically reducible is provided by discrete time crystals~\cite{schindler2025geometric,mondal2026floquet}: there the average-energy gap closes in the thermodynamic limit, as a result of symmetry breaking and the AGP $A^\mAE$ for the ground-state manifold is therefore nonabelian; yet, the eigenspace corresponding to the ground-state manifold is dynamically reducible, since the corresponding quasienergies differ by $\pi/T$ and single out distinct rank-one invariant subspaces. There exists a deeper mathematical reason for this phenomenon: all $\U(2)$ fiber bundles over the circle $\T$ are reducible; additionally, they are also topologically trivial. This places the family of Floquet systems in the abelian untwisted class.

The distinction between geometric irreducibility and dynamical irreducibility matters also for reverse engineering models using counterdiabatic driving \cite{kolodrubetz2017geometry}. For the reasons explained above, prescribing a static rank-$r_n$ projector on the torus and adding its counterdiabatic drive guarantees transport of that invariant space, but does not establish its dynamical irreducibility. For the invariant space to be dynamically irreducible, the exact CD time evolution must possess no strictly smaller smooth invariant subspaces. This is a \textit{dynamical} requirement that cannot be easily reverse-engineered using counterdiabatic driving, as it requires finding a quasiperiodic $r_n\times r_n$ Hamiltonian for which Eq.~\eqref{eq:omega-reduction} admits no smooth solution. Precisely for this reason, the abelian/nonabelian classification of quasiperiodic drives in Sec.~\ref{sec:classification} is dynamical, i.e., different from the corresponding classification of static systems. Notice also the subtle difference between reverse-engineering dynamical irreducibility and twistedness: to control twistedness, one only needs to add the counterdiabatic drive to a topological Hamiltonian, such as the QWZ model; but in order to control abelianness in the dynamically irreducible sense used throughout this work, one has to first identify an inherently dynamically irreducible nonabelian drive $H^\alpha_A$, and then embed it in a Hamiltonian $H^{\alpha\tau}$.

In summary, geometric and dynamical irreducibility are logically independent. Both types of reducibility require special structure. For a proper nonabelian average-energy eigenspace $\Pi_n$, geometric reducibility requires (at least) two orthogonal subprojectors preserved by parallel transport in every torus direction; in fact, such eigenprojectors constitute a zero-measure set in the space of smooth rank-$r_n$ projectors~\cite{zanardi1999holonomic,pachos2000nonabelian}. By contrast, dynamical reducibility manifests in accidentally degenerate average energies of (at least) two independent dynamically invariant projectors. Thus, in the absence of symmetries enforcing such degeneracy, we expect dynamically reducible eigenspaces to also constitute measure-zero exceptions in generic families of drives. The two forms of irreducibility are therefore generically expected to coexist, without either implying the other. In the periodic Floquet case ($D{=}1$), however, any degenerate space is at the same time dynamically and geometrically reducible to rank-one subspaces that can be described by single wavefunction-kets. Unless stated otherwise, all nonabelian invariant spaces discussed throughout this work are assumed dynamically irreducible (see Appendix~\ref{sec:LandscapeQP}).

\section{Evolution Decomposition: Basis Transport and Internal Dynamics}
\label{sec:SuppDynamicsReduction}

\subsection{Periodically Driven Systems}

\subsubsection{Floquet Decomposition}

Floquet theory resolves the full $\mathrm{U}(N)$ dynamics in an $N$-dimensional Hilbert space into $N$ rank-one invariant spaces. In each space, a smooth periodic micromotion transports the corresponding wavefunction, while a $\mathrm{U}(1)$ phase accumulating at a constant rate describes the dynamics within it, which we call internal dynamics. We want to generalize this dynamics-induced reduction, $\mathrm{U}(N)\to\mathrm{U}(1)\times\cdots\times\mathrm{U}(1)$, to quasiperiodic drives. As in the main text, to keep the notation close to the quasiperiodic case, we use the angle $\theta\in\T$ rather than time itself; let $\theta_t=\omega t+\theta_0$, and denote the period by $T=2\pi/\omega$. The unitary evolution obeys
\begin{equation}\label{eq:SchrodingerEvolution}
    i\frac{\mathrm{d}}{\mathrm{d}t}U(t,0)=H(\theta_t)U(t,0), \qquad U(0,0)=\mathds{1}_{N\times N},
\end{equation}
and its composition law reads as
\begin{equation}\label{eq:EvolutionProperty}
    U(t_2,0)=U(t_2,t_1)U(t_1,0).
\end{equation}
The initial phase $\theta_0$ fixes the trajectory, so we write $U(t,0)$ without an additional phase label.

Floquet's theorem provides a rotating frame $P(\theta;\theta_0)$ that is periodic in both drive-phase arguments, and in which the generator is constant in time (i.e., on the circle $\T$):
\begin{subequations}
\label{eq:FloquetGoal}
\begin{align}
    H^\F[\theta_0] &=P(\theta;\theta_0)^\dagger H(\theta)P(\theta;\theta_0) -i\omega P(\theta;\theta_0)^\dagger\partial_\theta P(\theta;\theta_0),\\
    U(t,0)&=P(\theta_t;\theta_0)\, \mathrm{e}^{-itH^\F[\theta_0]}.
\end{align}
\end{subequations}
We choose the Floquet gauge $P(\theta_0;\theta_0)=\mathds{1}_{N\times N}$~\cite{Bukov_2015general_HFE}. The two factors on the right-hand side of the second line are the micromotion and the macromotion. In this periodic setting, the micromotion operator $P$ is globally smooth on $\T$, and hence periodic, while the macromotion is generated by a time-independent Hamiltonian $H^\F$ (also known as the quasienergy operator)\footnote{To define the decomposition, we have to choose a branch of the logarithm in $H^\F[\theta_0]=i\log U(T,0)/T$, following which Eq.~\eqref{eq:FloquetGoal} defines the micromotion operator $P$. Different branches generate the familiar Floquet-zone freedom~\cite{Bukov_2015general_HFE}.}:
\begin{equation}\label{eq:HfloquePeriodic}
    H^\F[\theta_0]=iT^{-1}\log U(T,0)\,.
\end{equation}

The spectral decomposition of the quasienergy operator is
\begin{equation}\label{eq:SpectralHF}
    H^\F[\theta_0]=\sum_n\epsilon_n^\F\, \Pi_n[\theta_0]\,,
\end{equation}
and the $\epsilon^\F_n$ are called the quasienergies. By the spectral theorem for normal operators, the Hermitian quasienergy operator $H^\F[\theta_0]$ can always be diagonalized. In the case of degenerate quasienergies, every degenerate eigenspace can be resolved into rank-one eigenprojectors $\Pi_n[\theta_0]$ by choosing a partition of rank-one subspaces within that eigenspace.

Changing the starting phase transports the eigenprojectors, and hence the Floquet Hamiltonian, via the micromotion:
\begin{subequations}\label{eq:constantEF}
\begin{align}
\Pi_n[\theta]&=P(\theta;\theta_0)\Pi_n[\theta_0] P(\theta;\theta_0)^\dagger, \\
    H^\F[\theta]&=P(\theta;\theta_0)H^\F[\theta_0] P(\theta;\theta_0)^\dagger.
\end{align}
\end{subequations}
There is no topological obstruction to following these rank-one invariant spaces smoothly around the drive circle $\T$. A globally continuous Floquet basis can therefore be chosen, with
\begin{equation}\label{eq:Floquet state}
    \ket{n^\F[\theta]}=P(\theta;\theta_0)\ket{n^\F[\theta_0]}.
\end{equation}
Following its eigenvectors around the drive therefore resolves the motion into $N$ independent rank-one invariant spaces.

For each dynamically invariant space, we define the physically relevant evolution within that space
\begin{equation}
    U_n(t,0):=U(t,0)\Pi_n[\theta_0].
\end{equation}
Its Floquet form is
\begin{subequations}\label{eq:FullFloquetTheory}
\begin{align}
    U_n(t,0)&=P_n(\theta_t;\theta_0)\, \mathrm{e}^{-it\epsilon_n^\F},\\
    P_n(\theta;\theta_0)&=\ket{n^\F[\theta]}\bra{n^\F[\theta_0]}.
\end{align}
\end{subequations}
Here $P_n$ maps the initial rank-one invariant space to its position at time $t$; it does not describe the internal accumulated phase. The second factor $\mathrm{e}^{-it\epsilon_n^\F}\in\mathrm{U}(1)$ is the complete internal dynamics of that space's $U_n(t,0)$. Thus Floquet theory supplies both parts separately: smooth transport of invariant spaces and scalar internal evolution.

\subsubsection{Geometric Decompositions and their Physical Meaning}

We first compare three equivalent descriptions (gauge choices) of the evolution induced by periodic drives, for which every irreducible invariant space is rank-one. The projectors in Eq.~\eqref{eq:SpectralHF} obey $\LL(\Pi_n)=0$, with $\LL$ the Liouvillian defined in Eq.~\eqref{eq:L}. They therefore follow the invariant spaces transported by the exact dynamics without transitions between them.

\textbf{Floquet gauge.---}%
Floquet eigenstates possess the discrete gauge freedom~\cite{schindler2025geometric}
\begin{subequations}
\label{eq:ZgaugeTransf}
\begin{align}
    \epsilon_n^\F&\longrightarrow\epsilon_n^\F+k_n\omega, \qquad k_n\in\mathbb{Z},\\
    \ket{n^\F[\theta]}&\longrightarrow \mathrm{e}^{ik_n\theta}\ket{n^\F[\theta]}.
\end{align}
\end{subequations}

\textbf{Kato gauge.---}%
The Kato decomposition expresses the same evolution directly in terms of these projectors~\cite{schindler2025geometric},
\begin{subequations}
\begin{align}
    H(\theta)&=H^\K(\theta)+A^\K(\theta),\\
    H^\K(\theta)&=\sum_n\Pi_n[\theta]H(\theta)\Pi_n[\theta],\\
    A^\K(\theta)&=\omega A^\K_\theta(\theta),\quad A^\K_\theta(\theta)=-\frac{1}{2}\sum_n[\Pi_n[\theta],i\partial_\theta\Pi_n[\theta]].
\end{align}
\end{subequations}
Thus, we find $H^\K(\theta)=\sum_nE_n^\K(\theta)\Pi_n[\theta]$, where $E_n^\K(\theta)=\text{tr}(H(\theta)\Pi_n[\theta])$ for Floquet systems.

In the Kato gauge, we define the parallel transporter (a.k.a.~Wilson line operator) as the time-ordered exponential of the Kato AGP $A^\K$
\begin{equation}
    W^\K(t;0)=\mathcal{T}_s\mathrm{e}^{-i\int_{0}^{t}\mathrm{d}s\,A^\K(\theta_s)}.
\end{equation}
Its restriction to any fixed invariant space $\Pi_n$, together with the Kato energy, gives the evolution in each rank-one space,
\begin{equation}\label{eq:WhyKatoNotGoodAsAE}
    U_n(t,0)=W^\K(t;0)\Pi_n[\theta_0]\mathrm{e}^{-i\int_{0}^{t}\mathrm{d}s\,E_n^\K(\theta_s)}.
\end{equation}
The restriction $W^\K\Pi_n[\theta_0]$ isometrically transports a basis of the initial invariant space to the space at $\theta_t$, as does $P_n$. However, unlike micromotion transport, parallel transport also accumulates the geometric phase, so successive visits to the same drive phase generally give different values, for example, $W^\K(T;0)\neq W^\K(2T;0)$. Thus, $W^\K(t;0)$ is defined from time $0$ to $t$ along the drive trajectory, but it is not globally defined on the circle $\T$ since it is not periodic.

\begin{table*}[t!]
\label{table:smallTable}
\begin{center}
\renewcommand{\arraystretch}{1.08}
\begin{tabular}{p{3.5cm}|p{1.5cm}|p{6.8cm}|p{4.0cm}}
\hline
\multicolumn{1}{c|}{$U_n(t,0)$ \textbf{Decomposition}} & \multicolumn{1}{c|}{\textbf{Equation}} & \multicolumn{1}{c|}{\textbf{Ket/wavefunction transport in invariant space $\Pi_n$}} & \multicolumn{1}{c}{\textbf{Internal dynamics}} \\
\hline
\textit{\textbf{Floquet}} (F) & Eq.~\eqref{eq:FullFloquetTheory} & periodic (globally smooth) micromotion $P_n(\theta_t;\theta_0)$ & $\exp[-it\epsilon_n^\F]$ \\
\textit{\textbf{Kato}} (K) & Eq.~\eqref{eq:WhyKatoNotGoodAsAE} & parallel-transport Wilson line $W^\K(t;0)\Pi_n[\theta_0]$ & $\exp[-i\int_{0}^{t}\!\mathrm{d}s\,E_n^\K(\theta_s)]$ \\
\textit{\textbf{Average energy}} (\AE) & Eq.~\eqref{eq:AEdefs} & Wilson line $W^\mAE(t;0)\Pi_n[\theta_0]$ & $\exp[-it\mae_n]$ \\
\hline
\end{tabular}
\caption{ \textbf{Periodic drives: evolution $U_n(t,0)$ within an invariant space $\Pi_n[\theta]$, separated into ket transport and internal dynamics.} The Floquet ($\F$) decomposition naturally reduces the full Hilbert-space evolution $U$ into the evolution $U_n$ within each invariant space via (i) a micromotion $P_n$ defined as a smooth periodic function on the circle, and (ii) a constant quasienergy $\epsilon_n^\F$. The Kato ($\K$) decomposition separates the geometric phase from the dynamical phase of the evolution. The average energy ($\mAE$) decomposition absorbs part of the dynamical evolution into its ket transport, but leaves a constant average energy $\mae_n$ over the circle $\T$ in the internal dynamics; it defines a gauge-free dynamical invariant operator $\mAE[\theta]$. }
\end{center}
\end{table*}

\textbf{Average-energy gauge.---}%
The average-energy gauge keeps the projector construction but replaces the phase-dependent Kato energies by their circle-averages, introducing the average-energy eigenvalues $\mae_n=(2\pi)^{-1}\int_0^{2\pi}\mathrm{d}\theta\,E_n^\K(\theta)$:
\begin{subequations}
\label{eq:AEdefs}
\begin{align}
    \mAE[\theta]&=\sum_n\mae_n\Pi_n[\theta]=H(\theta)-A^\mAE(\theta),\\
    \label{eq:A_AE-Floquet}
    A^\mAE(\theta)&=A^\K(\theta)+\sum_n\bigl(E_n^\K(\theta)-\mae_n\bigr)\Pi_n[\theta],\\
    U(t,0)&=\mathcal{T}_s\mathrm{e}^{-i\int_{0}^{t}\mathrm{d}s\,A^\mAE(\theta_s)}\mathrm{e}^{-it\mAE[\theta_0]},\\
    U_n(t,0)&=\mathcal{T}_s\mathrm{e}^{-i\int_{0}^{t}\mathrm{d}s\,A^\mAE(\theta_s)}\Pi_n[\theta_0]\mathrm{e}^{-it\mae_n}.
\end{align}
\end{subequations}
We can also define the average-energy-gauge Wilson line operator $W^\mAE(t;0)=\mathcal{T}_s\exp[-i\int_{0}^{t}\mathrm{d}s\,A^\mAE(\theta_s)]$. It coincides with the Wilson line in the Kato gauge every complete cycle, $W^\K(T;0) = W^\mAE(T;0)$, but the two operators generally differ in between stroboscopic times.

Similar to $A^\K$, $A^\mAE$ is a smooth periodic generator on the circle $\T$. Notice that both their Wilson lines $W^\K$ and $W^\mAE$ are generally not periodic functions of the drive phase; they are defined as functions of time but not on the circle $\T$.

Let us now compare and contrast the three gauge choices/decompositions (Floquet, Kato, and \AE) in Table~\ref{table:smallTable}. The same evolution $U_n(t,0)$ within space $\Pi_n$ admits these three distinct decompositions, each having its own distinct advantages:
\begin{enumerate}
    \item[($\F$):] The Floquet decomposition uses a periodic micromotion $P_n(\theta;\theta_0)$ defined on the base manifold $\mathbb{T}$ and a linearly growing phase generated by constant quasienergy $\epsilon_n^\F$. Despite having a residual Floquet gauge freedom, cf.~Eq.~\eqref{eq:ZgaugeTransf}, one can fix the gauge using the gauge-free average energy $\mae_n$ and geometric phase $\gamma_n$ by setting $\epsilon_n^\F=\mae_n-\omega\gamma_n/(2\pi)$;
    \item[($\K$):] The Kato decomposition is useful for separating the geometric effects (the physical quantities that only depend on $\Pi_n$ and its derivatives $\partial_\theta\Pi_n$, e.g., the geometric phase $\gamma_n$) from the internal dynamical effects (the quantities only depending on $\Pi_n$ and $H$, e.g., the Kato energy $E^\K$ as the integrand of the dynamical phase). Both the geometric and dynamical quantities are embedded separately in the two parts of the Kato decomposition: the history-dependent transport $W^\K(t;0)$ and Kato energy $E^\K(\theta)$;
    \item[($\mAE$):] The average energy decomposition is a trade-off between the two other decompositions: it absorbs part of the dynamical effect from the Kato decomposition into the history-dependent transport $W^\mAE$, and only leaves the constant average energy $\mae_n$ in the internal dynamics. Thus, it defines a gauge-free dynamical invariant $\mAE$. We shall see below that the average-energy decomposition and its variational principle~\cite{mondal2026floquet} are the key to computing dynamically irreducible invariant spaces $\Pi_n$ for quasiperiodic drives, without which one cannot perform the Kato or the generalized Floquet decomposition.
\end{enumerate}

\subsection{Dynamical Invariants for Quasiperiodic Systems}
\label{app:subsec:dyn_invs}

For a periodic drive, an orthonormal eigenbasis of the one-cycle Floquet unitary $U(T,0)$ provides a complete set of rank-one invariant spaces, including when its eigenvalues are degenerate. Transporting these spaces around the drive circle $\T$ produces the projectors used above.

Unlike Floquet drives, a quasiperiodic trajectory has no exact return time; hence, there is no finite-time dynamical operation that identifies the corresponding dynamically invariant spaces. However, one may still obtain them as the globally smooth projector-valued solutions $\Pi_n[\thetavec]$ of the equation
\begin{equation}
    \LL(\Pi_n)\stackrel{!}{=}0,\qquad \LL(\cdot) := [H(\thetavec), \cdot]  - i \omega^\mu\partial_\mu(\cdot)\, .
\end{equation}
Their existence, continuity, and convergence under the variational principle are discussed in Sec.~\ref{sec:Diophantine}; here we recap the basic idea of obtaining $\mAE$ and $\Pi_n$ from the variational principle for completeness.

Considering the exact objects (i.e., without placing any finite Fourier cutoff), $H(\thetavec)$ is orthogonally split into two components, one belonging to the kernel of the quasiperiodic Liouvillian $\LL$ and the other to its orthogonal complement,
\begin{equation}
    \mAE[\thetavec]=(1-\LL^{+}\LL)H(\thetavec), \qquad A^\mAE(\thetavec)=\LL^{+}\LL H(\thetavec).
\end{equation}
Equivalently,
\begin{equation}\label{eq:pseudoinverse}
\begin{gathered}
    A^\mAE(\thetavec)=\underset{X(\thetavec): \LL(X)=\LL(H)}{\operatorname{argmin}} \int_{\T^D}\frac{\mathrm{d}^D\varphi}{(2\pi)^D}\text{tr}\bigl(X(\vec\varphi)^\dagger X(\vec\varphi)\bigr).
\end{gathered}
\end{equation}
This is the minimum-norm problem used in our variational principle, and how we define $\mAE$ in Sec.~\ref{subsec:varl} of the main text.

For generic quasiperiodic drives, there is no Floquet theorem to demand that all invariant projectors $\Pi_n$ must be rank-one; we therefore derive a nonabelian generalization in Sec.~\ref{app:sec:4classes}. As we discuss in Appendix~\ref{app:omega-irreducibility}, apart from accidental degeneracies between distinct invariant spaces, the spectral projector of $\mAE$ at eigenvalue $\mae_n$ directly identifies a dynamically irreducible invariant space of rank $r_n=\text{tr}\,\Pi_n$ defined by the eigenprojector $\Pi_n$.

\subsubsection{Universal Average-Energy Decomposition}
\label{app:AEqp}

Previous studies have shown that, in general, there does not exist a Floquet theory that reconstructs the unitary evolution for quasiperiodically driven systems \cite{verdeny2016quasi}. In the following subsection \ref{sec:2obstructions}, we will see why physically such a Floquet decomposition cannot exist for all quasiperiodic systems. In contrast, once the average-energy projectors $\Pi_n$ are known, the average-energy decomposition can be readily used to reconstruct the evolution for any quasiperiodic system, without a reference to Floquet theory. Intuitively, this is because the Kato and average-energy gauges are based on the geometric concepts of adiabaticity, which are more general.

Setting $r_n=\text{rank}(\Pi_n)=\tr\Pi_n$, we define
\begin{subequations}
\begin{align}
    \mae_n&=\int_{\T^D}\frac{\mathrm{d}^D\theta}{(2\pi)^D} \frac{1}{r_n}\text{tr}\left(\Pi_n[\thetavec]H(\thetavec)\right),\\
    \mAE[\thetavec]&=\sum_n\mae_n\Pi_n[\thetavec]=H(\thetavec)-A^\mAE(\thetavec),\\
    \label{eq:A_AE}
    A^\mAE(\thetavec)&=A^\K(\thetavec)+\sum_n\left(\Pi_n[\thetavec]H(\thetavec)\Pi_n[\thetavec]-\mae_n\Pi_n[\thetavec]\right)\, .
\end{align}
\end{subequations}
Notice the subtle but important difference between the generic Eq.~\eqref{eq:A_AE} for quasiperiodically driven systems and its Floquet counterpart~\eqref{eq:A_AE-Floquet}: the equality $\Pi_n[\thetavec]H(\thetavec)\Pi_n[\thetavec]= E^\K_n(\thetavec)\Pi_n[\thetavec]$ holds automatically for abelian dynamically irreducible spaces; for nonabelian dynamically irreducible spaces its left-hand side can also contain a traceless tensor component, see Eq.~\eqref{eq:tensor_decomp}.

The full evolution, its restriction to each invariant space, and the Wilson line are
\begin{subequations}
\label{eq:AEgeneralRank}
\begin{align}
    \label{eq:U_AE_appendix}
    U(t,0)&=W^\mAE(t;0)\mathrm{e}^{-it\mAE[\thetavec_0]},\\
    \label{eq:Un_AE}
    U_n(t,0)&=W^\mAE(t;0)\Pi_n[\thetavec_0]\mathrm{e}^{-it\mae_n}, \\
    \label{eq:W_AE}
    W^\mAE(t;0)&=\mathcal{T}_s\exp[-i\int_{0}^{t}\mathrm{d}s\,A^\mAE(\thetavec_s)] .
\end{align}
\end{subequations}
The first factor in the average-energy decomposition~\eqref{eq:U_AE_appendix} is the Wilson line operator $W^\mAE$ defined from time $0$ to $t$, using a time-ordering on the entire Hilbert space. It contains the transport of all invariant spaces and all internal evolution left after extracting their average-energy phases, including geometric phases and noncommuting rotations. The remaining dynamics in Eq.~\eqref{eq:U_AE_appendix} retains a simple exponential of the dynamical invariant $\mAE$, assigning the phase $\exp[-it\mae_n]$ to space $n$, see Eq.~\eqref{eq:Un_AE}.

Note that, despite having a simple exponential form for the reduced dynamics, the basis transport $W^\mAE(t;0)$ of the average-energy decomposition is still history-dependent.

\begin{table*}[t!]
\begin{center}
\renewcommand{\arraystretch}{1.08}
\begin{tabular}{p{3.5cm}|p{1.5cm}|p{6.8cm}|p{4.0cm}}
\hline
\multicolumn{1}{c|}{\shortstack[c]{$U_n(t,0)$ \textbf{Decomposition}}} & \multicolumn{1}{c|}{\textbf{Equation}} & \multicolumn{1}{c|}{\textbf{Basis transport in invariant space $\Pi_n$}} & \multicolumn{1}{c}{\textbf{Internal dynamics}} \\
\hline
\textit{\textbf{Kato}} (K) & Eq.~\eqref{eq:KatoGeneralRank} & parallel-transport Wilson line $W^\K(t;0)\Pi_n[\thetavec_0]$ & $\mathcal{T}_s\exp[-i\int_{0}^{t}\!\mathrm{d}s\,M_n^\K(s,0)]$ \\
\textit{\textbf{Average energy}} (\AE) & Eq.~\eqref{eq:AEgeneralRank} & Wilson line $W^\mAE(t;0)\Pi_n[\thetavec_0]$ & $\exp[-it\mae_n]$ \\
\hline
\end{tabular}
\caption{\textbf{Quasiperiodic drives: evolution $U_n(t,0)$ within an invariant space $\Pi_n[\thetavec]$, separated into basis transport and internal dynamics.} The Kato ($\K$) decomposition separates geometric transport from the dynamical phase and rotations generated by $M_n^\K$, the projected Hamiltonian in the parallel-transported basis. The average energy ($\mAE$) decomposition absorbs part of the dynamical evolution into its basis transport, but leaves a constant average energy $\mae_n$ in the internal dynamics; it defines a gauge-free dynamical invariant operator $\mAE[\thetavec]$. Both decompositions apply to all four classes in Table~\ref{table:table}. }
\label{table:KatoAE}
\end{center}
\end{table*}

\subsubsection{Higher-Rank Invariant Spaces and Kato Decomposition}\label{app:KatoQP}

The Kato AGP along the torus direction $\mu$ parallel-transports the invariant spaces:
\begin{subequations}\label{eq:KatoConnection}
\begin{align}
    A^\K_\mu(\thetavec)&=-\frac{1}{2}\sum_n[\Pi_n[\thetavec],i\partial_\mu\Pi_n[\thetavec]],\\
    A^\K(\thetavec)&=\omega^\mu A^\K_\mu(\thetavec).
\end{align}
\end{subequations}
The Wilson line along one prescribed $\mu$-th canonical cycle of the torus $\T^D$ and its restriction to invariant space $n$ are
\begin{subequations}\label{eq:KatoWilsonLoop}
\begin{align}
    W_\mu(\thetavec)&=\mathcal{T}_s\mathrm{e}^{-i\int_0^{2\pi}\mathrm{d}s\,A^\K_\mu(\thetavec+s\hat e_\mu)},\\
    W_{n,\mu}(\thetavec)&=W_\mu(\thetavec)\Pi_n[\thetavec].
\end{align}
\end{subequations}
Here $\hat e_\mu$ is the unit vector along $\theta_\mu$, and $\mathcal{T}_s$ is $s$-ordering along the cycle. Note that this prescribed cycle is distinct from the physical quasiperiodic trajectory $\thetavec_t=\omegavec t+\thetavec_0$. For $r_n=1$, the restricted Wilson line has a single eigenphase -- the Berry phase along the $\mu$-th canonical cycle,
\begin{equation}\label{eq:BerryFromWilson}
    \gamma_{n,\mu}(\thetavec)=\arg\text{tr}\left(\Pi_n[\thetavec]W_\mu(\thetavec)\right) \ \operatorname{mod}2\pi,
\end{equation}
whereas for $r_n>1$ the restricted Wilson line is a $\mathrm{U}(r_n)$ rotation that defines the holonomy group. For $D>1$, we define $\gamma_{n,\mu}(\thetavec)$ on $\T^D$, although it is independent of $\theta_\mu$, since shifting the starting point along the same $\mu$-cycle leaves the phase unchanged.

The Hamiltonian projected onto a higher-rank invariant space has the irreducible tensor decomposition
\begin{equation}
\label{eq:tensor_decomp}
    \Pi_nH\Pi_n=\frac{\text{tr}(\Pi_nH)}{r_n}\Pi_n+\widetilde H_n, \qquad \text{tr}(\widetilde H_n)=0.
\end{equation}
In a basis parallel-transported by $W^\K$ (i.e., fixing the parallel transport gauge), these are the scalar and traceless parts of the internal Hamiltonian $\Pi_nH\Pi_n$ inside the space $\Pi_n$: they generate a common $\mathrm{U}(1)$ dynamical phase and an $\mathrm{SU}(r_n)$ dynamical rotation, respectively. The scalar part is also named the Kato energy $E_n^{\K}=\text{tr}(\Pi_nH)/r_n$. Since the decomposition is irreducible, the traceless component cannot be represented by a single scalar quasienergy. The evolution within each invariant space can be expressed as \begin{subequations}
\label{eq:KatoGeneralRank}
\begin{align}
    \label{eq:Un_K}
    U_n(t,0)&=W^\K(t;0)\Pi_n[\thetavec_0]\mathcal{T}_s\exp[-i\int_{0}^{t}\!\mathrm{d}s\,M_n^\K(s,0)], \\
    \label{eq:W_K}
    W^\K(t;0)&=\mathcal{T}_s\exp[-i\int_{0}^{t}\mathrm{d}s\,A^\K(\thetavec_s)], \\
    \label{eq:Mn_K}
    M_n^\K(t,0)&=W^\K(t;0)^\dagger\Pi_n[\thetavec_t]H(\thetavec_t)\Pi_n[\thetavec_t]W^\K(t;0),
\end{align}
\end{subequations}
where $M_n^\K(t,0)$ is the rank-$r_n$ dynamical generator in the parallel-transport basis, and $W^\K(t;0)$ is the parallel-transport Wilson line obtained from a time-ordering over the entire Hilbert space, just like $W^\mAE$. Note that they are all operators defined as functions of time but not on the torus, meaning that in general $\|W^\K(t+T^{(i)};0)-W^\K(t;0)\|\gg O(|\delta\thetavec^{(i)}|)$ at near-return times $T^{(i)}$ with the return phase mismatch $\delta\thetavec^{(i)}$ defined in Eq.~\eqref{eq:diophantine-return}.

\subsection{Obstructions to Extending Floquet Theory beyond Periodic Drives}
\label{sec:2obstructions}

\begin{table*}[t]
\centering
\setlength{\tabcolsep}{6pt}
\renewcommand{\arraystretch}{1.45}
\begin{tabular}{>{\centering\arraybackslash}p{3.3cm}|p{6.3cm}|p{6.3cm}}
\hline
\shortstack[c]{\vspace{0.01cm}\\
\textit{\textbf{Generalized}} \textit{\textbf{Floquet}} ($\mathrm{F}$)\\
\textbf{Decomposition}} & \multicolumn{1}{c|}{\shortstack[c]{\textbf{untwisted}\\(topologically trivial)}} & \multicolumn{1}{c}{\shortstack[c]{\textbf{twisted}\\(topological)}} \\
\hline
\vspace*{2.1cm} \textbf{abelian} $r_n=1$
&
\[
U_n(t,0)=\ket{n^\F[\thetavec_t]}\mathrm{e}^{-it\epsilon_n^\F}\bra{n^\F[\thetavec_0]}\quad \eqref{eq:AbelUntFloquet}
\]
\begin{itemize}
    \item globally smooth Floquet state $\ket{n^\F[\thetavec]}$ exists on the torus $\T^D$; it defines micromotion $P_n(\thetavec;\thetavec_0)=\ket{n^\F[\thetavec]}\bra{n^\F[\thetavec_0]}$;
    \item quasienergy $\epsilon_n^\F$ is constant on $\T^D$; it equals the average energy $\mae_n$ minus the mean geometric phase accumulated per unit time $\omega^\mu/(2\pi) \left\langle\gamma_{n,\mu}\right\rangle_{\T^D}$, see Eq.~\eqref{eq:AbelUntFloquetE}.
\end{itemize}
This is the most straightforward generalization of Floquet theory to quasiperiodic drives. &
\[
\begin{aligned}
U_n(t,0)={}&P_n(\thetavec_t;\thetavec_0)\,\mathrm{e}^{-it\epsilon_n^\F(\thetavec_0)}\quad \eqref{eq:AbelTwiFloquet} \\
&\times\mathrm{e}^{-i\int_{0}^t\mathrm{d}s\,[\varepsilon_n(\thetavec_s;\thetavec_0)-\epsilon_n^\F(\thetavec_0)]}
\end{aligned}
\]
\begin{itemize}
    \item no globally smooth Floquet ket exists;
    \item along almost every trajectory, the generalized micromotion $P_n(\thetavec_t;\thetavec_0)$ is smooth except for isolated codimension-two singularities where $\Pi_n[\thetavec_t]$ is orthogonal to $\Pi_n[\thetavec_0]$;
    \item quasienergy $\epsilon_n^\F(\thetavec_0)$ depends smoothly on $\thetavec_0$; the total phase rate fluctuates around it.
\end{itemize}
\\ \hline
\vspace*{2.3cm} \textbf{nonabelian} $r_n>1$ \textit{and} \textbf{dynamically irreducible}
&
\[
\begin{aligned}
U_n(t,0)={}&V_n^\F(\thetavec_t) \,\mathrm{e}^{-it\epsilon_n^\F}\quad \eqref{eq:U(1)SU(rank)} \\
&\times\mathcal{T}_s\mathrm{e}^{-i\int_{0}^{t}\mathrm{d}s\, M_n^\F(\thetavec_s)}\,V_n^{\F}(\thetavec_0)^\dagger
\end{aligned}
\]
\begin{itemize}
    \item a globally smooth basis $V^\F_n(\thetavec)$ exists, as does the micromotion $P_n(\thetavec;\thetavec_0)=V_n^\F(\thetavec)V_n^\F(\thetavec_0)^\dagger$;
    \item the internal dynamics consists of a $\mathrm{U}(1)$ scalar quasienergy phase generated by $\epsilon^\F_n$, and a nonabelian internal $\mathrm{SU}(r_n)$ rotation;
    \item the time-ordered $r_n\times r_n$ generator of the $\mathrm{SU}(r_n)$ internal dynamics $M_n^\F(\thetavec_s)$ is irreducible, Hermitian, traceless and cannot be replaced by any constant operator.
\end{itemize}

&
\[
\begin{aligned}
U_n(t,0)={}&P_n(\thetavec_t;\thetavec_0)\,\mathrm{e}^{-it\epsilon^\F_n(\thetavec_0)}\quad \eqref{eq:THEmostGeneralCase} \\
&\times\mathcal{T}_s\mathrm{e}^{-i\int_{0}^{t}\mathrm{d}s\,M^\F_n(\thetavec_s;\thetavec_0)}
\end{aligned}
\]
\begin{itemize}
    \item no globally smooth basis exists on $\T^D$;
    \item the internal dynamics is a $\U(r_n)$ time-ordered rotation, generated by $M^\F_n(\thetavec_s;\thetavec_0)$ which is almost globally smooth away from the singularities of pair ($\thetavec_s,\thetavec_0$) where $\Pi_n$ becomes orthogonal.
    \item along almost every trajectory the generalized micromotion $P_n(\thetavec_t;\thetavec_0)$ is smooth, except for codimension-two singularities.
\end{itemize}
\\ \hline
\end{tabular}

\caption{ \textbf{Classification of dynamically invariant spaces: generalized Floquet decomposition of the evolution operator $U_n$ within each invariant space.} The full evolution decomposes as $U(t,0)= \bigoplus_n U_n(t,0)$; the evolution $U_n$ within a dynamically irreducible invariant space $\Pi_n$ is separated into micromotion and internal dynamics. Untwisted drives: the micromotion $P_n$ transports the basis of $\Pi_n$ smoothly on the torus $\T^D$. The remaining internal dynamics in the abelian untwisted case is only a linearly growing $\U(1)$-phase generated by the quasienergy $\epsilon^\F_n$. Twisted drives: the micromotion $P_n$ transports the basis of $\Pi_n$ smoothly almost everywhere on the torus $\T^D$. The remaining internal dynamics in the abelian twisted case retains a leftover zero-average scalar phase-rate fluctuation $\varepsilon_n(\thetavec_s;\thetavec_0)-\epsilon_n^\F(\thetavec_0)$. The nonabelian drives, in addition to the quasienergy phase $\epsilon^\F_n$, have a Hermitian rank-$r_n$ generator $M^\F_n$, both defined everywhere on the torus $\T^D$ up to a set of co-dimension-2 for twisted drives.}
\label{table:table}
\end{table*}

Floquet theory for periodically driven systems is not only a statement about the full evolution operator $U(t,0)$: it also factorizes the unitary evolution $U_n(t,0) = U(t,0)\Pi_n[\theta_0]$ within space $\Pi_n$ into the transport of a rank-one invariant space and a scalar phase within it. The effective generalized Floquet theory of quasiperiodic dynamics retains this physical form of a two-factor factorization,  where the first factor is still a micromotion that transports the basis of invariant spaces and the second factor is the remaining internal dynamics generator, but they are no longer always a globally smooth micromotion or just a scalar $\U(1)$-phase generator.

\textbf{Dynamical Geometric Obstruction.---}%
First, consider a dynamically irreducible invariant space $\Pi_n[\thetavec]$ with rank $r_n>1$. If its evolution could be written using a globally smooth basis $V_n(\thetavec)$ and a constant Hermitian $r_n\times r_n$ matrix $h^\F$, then every rank-one eigenprojector $\pi_a$ of $h^\F$ would define a globally smooth invariant projector $V_n(\thetavec)\pi_aV_n(\thetavec)^\dagger$ inside $\Pi_n[\thetavec]$. This contradicts dynamical irreducibility (see Appendix \ref{app:omega-irreducibility}). The internal dynamics of such an invariant space must therefore remain an explicitly time-ordered $\mathrm{U}(r_n)$ rotation.

\textbf{Topological Obstruction.---}%
Second, a twisted (i.e., topologically nontrivial) invariant space has no globally smooth basis over the drive torus $\T^D$. A reference-based map can still be constructed by projecting a basis at $\thetavec_0$ into the invariant space at $\thetavec$. In terms of local orthonormal bases $V_n$, this map becomes singular when the overlap matrix loses rank, i.e.,
\begin{equation}
    \det\!\left[V_n(\thetavec)^\dagger V_n(\thetavec_0)\right]=0.
\end{equation}
For a two-tone drive ($D{=}2$), this complex condition generically defines a set of codimension two, hence isolated points on $\T^2$. A one-dimensional quasiperiodic trajectory starting from almost every $\thetavec_0$ never hits these points exactly, although its dense motion approaches them arbitrarily closely. The reference-based map is consequently smooth along almost every trajectory, with increasingly rapid variation near these points, but it cannot be made globally smooth on the torus.

Hence, there are two independent physical yes/no questions for the evolution $U_n$ within every fixed invariant space $n$: (i) is its internal evolution scalar or matrix-valued, and (ii) does its moving invariant space admit a globally smooth basis? The answers to these questions yield the four-fold classification (abelian/nonabelian, untwisted/twisted) of a dynamically irreducible invariant space for any quasiperiodic drive. The four classes in Table~\ref{table:table} and Sec.~\ref{app:sec:4classes} below display their corresponding basis transport and explicit $\mathrm{U}(r_n)$ internal dynamics, which combine to form the generalized Floquet ($\F$) decomposition of evolution $U_n$ within each invariant space.

We remark that the Kato ($\K$) and the average energy ($\mAE$) decompositions are instead universal for all four classes (see Table~\ref{table:KatoAE}). Therefore, the same quasiperiodically driven evolution $U_n$ within each invariant space still admits three different decompositions, as in the case of periodic drives; they have their own distinct advantages, see Appendix \ref{app:SummaryDecomposition}.

\subsection{Generalized Floquet Decomposition}
\label{app:sec:4classes}

Below, we focus on the evolution of a single invariant space $\Pi_n$ of an arbitrary quasiperiodic drive. The total evolution operator can then be reconstructed from the relation $U(t,0)= \bigoplus_n U_n(t,0)$. The dynamical geometric-topological classification also applies to each space $\Pi_n$, and further divides them into four classes, as it classifies the drives in Sec.~\ref{sec:classification}.

\subsubsection{Abelian Untwisted (\texorpdfstring{$\rm A\bar T$}{A T-bar}) Spaces}
\label{app:sec:untwisted_abelian}

This class is the direct quasiperiodic counterpart of Floquet theory: the invariant projector $\Pi_n$ is rank-one, a globally smooth ket $\ket{n^\F[\thetavec]}$ exists with $\Pi_n[\thetavec] = \ket{n^\F[\thetavec]}\bra{n^\F[\thetavec]}$, and the evolution within space $\Pi_n$ factorizes into a smooth quasiperiodic transport map and a $\mathrm{U}(1)$ phase accumulating at a constant rate given by the quasienergy, $\epsilon^\F_n$.

However, the quasienergy $\epsilon^\F_n$ and its corresponding ket $\ket{n^\F[\thetavec]}$ have a $\mathbb{Z}^D$ gauge freedom, which generalizes the $\mathbb{Z}$ gauge freedom of Eq.~\eqref{eq:ZgaugeTransf} in the periodic case:
\begin{subequations}
\label{eq:ZDgauge}
\begin{align}
    \epsilon_n^\F&\longrightarrow\epsilon_n^\F+\kvec_n\cdot\omegavec\,, \qquad \kvec_n\in\mathbb{Z}^D,\\
    \ket{n^\F[\thetavec]}&\longrightarrow \mathrm{e}^{i\kvec_n\cdot\thetavec}\ket{n^\F[\thetavec]}\,.
\end{align}
\end{subequations}
This dense $\mathbb{Z}^D$ unfolding freedom explains why a direct diagonalization of the frequency-lattice Schr\"odinger operator from Eq.~\eqref{eq:Sambe} does not select a stable quasienergy representative even in this simplest quasiperiodic class. Incommensurability of the drive translates into the copies $\epsilon_n^\F+\kvec\cdot\omegavec$ being dense on the real axis; hence, restricting the spectrum to a finite interval does not select a unique representative. At a finite Fourier cutoff, an eigensolver selects cutoff-dependent representatives and may mix modes through arbitrarily close resonances on the frequency lattice. Moreover, the quasienergy levels shift with increasing Fourier cutoff, and no continuity of the levels can be established due to the dense gauge freedom.

\begin{figure}[t!]
	\centering
    \includegraphics[width=0.48\textwidth]{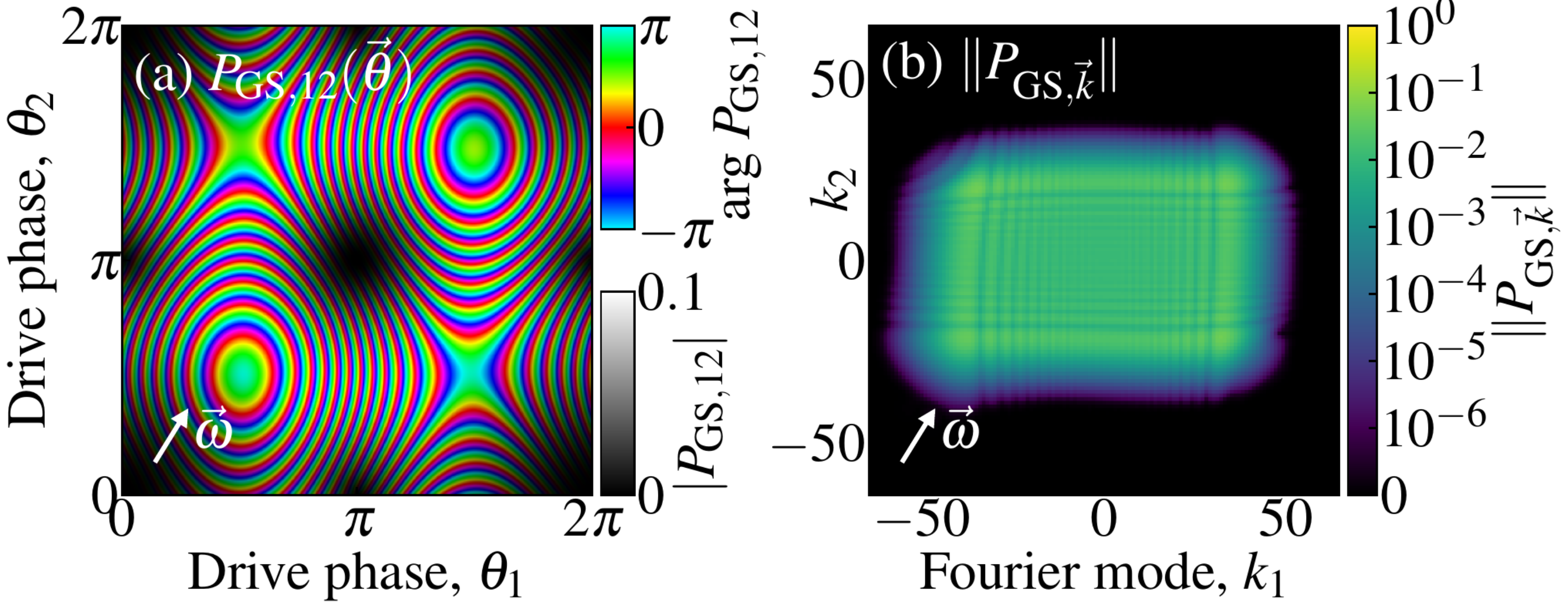}
	\caption{\textbf{Micromotion of an A\=T average-energy ground state.} (a) Matrix element $P_{\mathrm{GS},12}(\thetavec)$, with hue and brightness showing its phase and magnitude. (b) Fourier-mode contributions $\|P_{\mathrm{GS},\kvec}\|$. The ground-state micromotion $P_{\rm GS}$ is globally smooth on the torus, and localized on the frequency lattice. We use the abelian untwisted Hamiltonian $H^{\mathrm{A}\mathrm{\bar{T}}}$ from Eq.~\eqref{eq:model_H} with $\thetavec_0=(0.5,0.2)$, and $|\omegavec|=0.1$, $\omega_2/\omega_1=(1+\sqrt5)/2$ and $\epsilon=10^{-2}$. }
	\label{fig:Floquet-AbeUnt}
\end{figure}

For these reasons, we have to fix the generalized $\mathbb{Z}^D$ Floquet gauge using gauge-free objects: the average energy and the geometric phase. We now explicitly construct a gauge-fixed quasienergy $\epsilon^\F_n$ and its corresponding stationary ket $\ket{n^\F[\thetavec]}$ for the abelian untwisted spaces.

The geometric phase along the $\mu$-th canonical cycle of the drive torus is
\begin{subequations}
\begin{align}
    \gamma_{n,\mu}(\thetavec) &=\oint\mathrm{d}\theta_\mu\, \bra{n^\F[\thetavec]}i\partial_\mu\ket{n^\F[\thetavec]}\\
    &=\arg\text{tr}\left(\Pi_n[\thetavec]W_\mu(\thetavec)\right) \ \operatorname{mod}\, 2\pi.
\end{align}
\end{subequations}
The phase $\gamma_{n,\mu}(\thetavec)$ is independent of the starting point along the same $\theta_\mu$-cycle, although it can depend smoothly on the other drive phases. As a transverse coordinate $\theta_\nu$ winds once around the torus, $\gamma_{n,\mu}$ winds by $-2\pi C_{n,\mu\nu}$, with $C_{n,\mu\nu}$ the Chern number of the $(\mu,\nu)$ subtorus. For untwisted spaces, all these windings vanish, so $\gamma_{n,\mu}$ admits a continuous branch on the entire torus $\mathbb{T}^D$, unique up to a global shift by $2\pi\mathbb{Z}$; this is what allows us to average it over the torus below.

By continuously unfolding these phases, the quasienergy associated with the rank-one invariant space is completely gauge-fixed, see Eq.~\eqref{eq:ZDgauge}, to
\begin{equation}
\label{eq:AbelUntFloquetE}
    \epsilon_n^\F =\mae_n-\frac{\omega^\mu}{2\pi} \left\langle\gamma_{n,\mu}\right\rangle_{\T^D}.
\end{equation}
The torus average over the geometric phase appears because the physical quasiperiodic trajectory (assumed to have incommensurate frequencies) samples the entire torus uniformly. Its long-time geometric phase per unit time is therefore the sum of the cycle phases, weighted by the corresponding frequencies and averaged over the torus.

To prove the validity of Eq.~\eqref{eq:AbelUntFloquetE} and to connect it to the physical evolution, we compare a state with itself when the drive phases nearly return to their initial values [Fig.~\ref{fig:schematicPreviousMethods}(a)]. As in Eq.~\eqref{eq:diophantine-return}, the Diophantine approximation provides times $T^{(i)}\to\infty$ and integer vectors $\kvec^{(i)}$ such that
\begin{equation}
    \delta\thetavec^{(i)}=\omegavec T^{(i)}-2\pi\kvec^{(i)}, \qquad |\delta\thetavec^{(i)}|\to0.
\end{equation}
Since the known rank-one invariant projector $\Pi_n[\thetavec]$ is smooth, its evolved space then approaches the initial one. The return evolution therefore acts approximately as a scalar phase within $\im(\Pi_n[\thetavec_0])$, which we extract from its normalized overlap:
\begin{equation}\label{eq:returnPhase}
    \mathrm{e}^{i\varphi_n^{(i)}}= \frac{\text{tr}\left(\Pi_n[\thetavec_0] U(T^{(i)},0)\right)} {\left|\text{tr}\left(\Pi_n[\thetavec_0] U(T^{(i)},0)\right)\right|}.
\end{equation}
The denominator is required because at finite $i$ the final space lies at $\thetavec_0+\delta\thetavec^{(i)}$ and is not exactly the initial one; the normalization extracts only the phase of their overlap. Projecting the Kato evolution in Eq.~\eqref{eq:WhyKatoNotGoodAsAE} onto the smooth ket gives its continuous unfolding,
\begin{equation}
\begin{split}
    \varphi_n^{(i)}{=}&{-}\!\! \int_{0}^{T^{(i)}}\mathrm{d}s\, E_n^\K(\thetavec_s) +\int_{0}^{T^{(i)}}\mathrm{d}s\, \bra{n^\F[\thetavec_s]}i\omega^\mu\partial_\mu \ket{n^\F[\thetavec_s]}\\
    &+O(|\delta\thetavec^{(i)}|).
\end{split}
\end{equation}
The final error term on the right-hand side comes from identifying the rank-one space at $\thetavec_0+\delta\thetavec^{(i)}$ with the initial space. Smoothness makes this mismatch at most linear with $|\delta\thetavec^{(i)}|$. Dividing by $T^{(i)}$ and replacing the long-time average by the drive-phase average (justified for globally smooth functions and ergodic drives) gives
\begin{equation}\label{eq:scalarGenerator}
    \frac{\varphi_n^{(i)}}{T^{(i)}}\longrightarrow -\mae_n+\frac{\omega^\mu}{2\pi} \left\langle\gamma_{n,\mu}\right\rangle_{\T^D}.
\end{equation}
Matching $\mathrm{e}^{i\varphi_n^{(i)}}$ to $\mathrm{e}^{-i\epsilon_n^\F T^{(i)}}$ establishes Eq.~\eqref{eq:AbelUntFloquetE}.

Starting from a globally smooth ket $\ket{n[\thetavec]}$, we absorb the fluctuations of its internal phase into $\ket{n^\F[\thetavec]}=\mathrm{e}^{i\chi_n(\thetavec)}\ket{n[\thetavec]}$ by choosing the gauge
\begin{equation}\label{eq:untwistedU1phase}
\begin{split}
    \omega^\mu\partial_\mu\chi_n(\thetavec)&=\epsilon_n^\F-\varepsilon_n(\thetavec),\\
    \varepsilon_n(\thetavec)&=\bra{n[\thetavec]}(H-i\omega^\mu\partial_\mu)\ket{n[\thetavec]}.
\end{split}
\end{equation}
Since by definition of average energy and averaged geometric phases, $\epsilon_n^\F=\langle\varepsilon_n\rangle_{\T^D}$, the right-hand side has zero torus average. Whenever $\varepsilon_n(\thetavec)$ is smooth on the torus, the Fourier solution $\chi_{n,\kvec}=i\varepsilon_{n,\kvec}/(\kvec\cdot\omegavec)$ for $\kvec\neq0$ is also smooth for almost all $\omegavec$~\cite{puig2002reducibility}. With this definition of a globally smooth $\ket{n^\F[\thetavec]}$, we obtain the quasiperiodic Floquet decomposition
\begin{subequations}\label{eq:AbelUntFloquet}
\begin{align}
    U_n(t,0)&=P_n(\thetavec_t;\thetavec_0)\, \mathrm{e}^{-it\epsilon_n^\F},\\
    P_n(\thetavec;\thetavec_0) &=\ket{n^\F[\thetavec]}\bra{n^\F[\thetavec_0]}.
\end{align}
\end{subequations}
Figure~\ref{fig:Floquet-AbeUnt} shows that the globally smooth micromotion has rapidly decaying Fourier components.

Equation~\eqref{eq:AbelUntFloquet} provides an alternative factorization of $U_n$ into the globally smooth micromotion $P_n$ restricted to space $\Pi_n$ and the scalar internal quasienergy phase $\epsilon_n^\F$.

\begin{figure}[t!]
	\centering
    \includegraphics[width=0.48\textwidth]{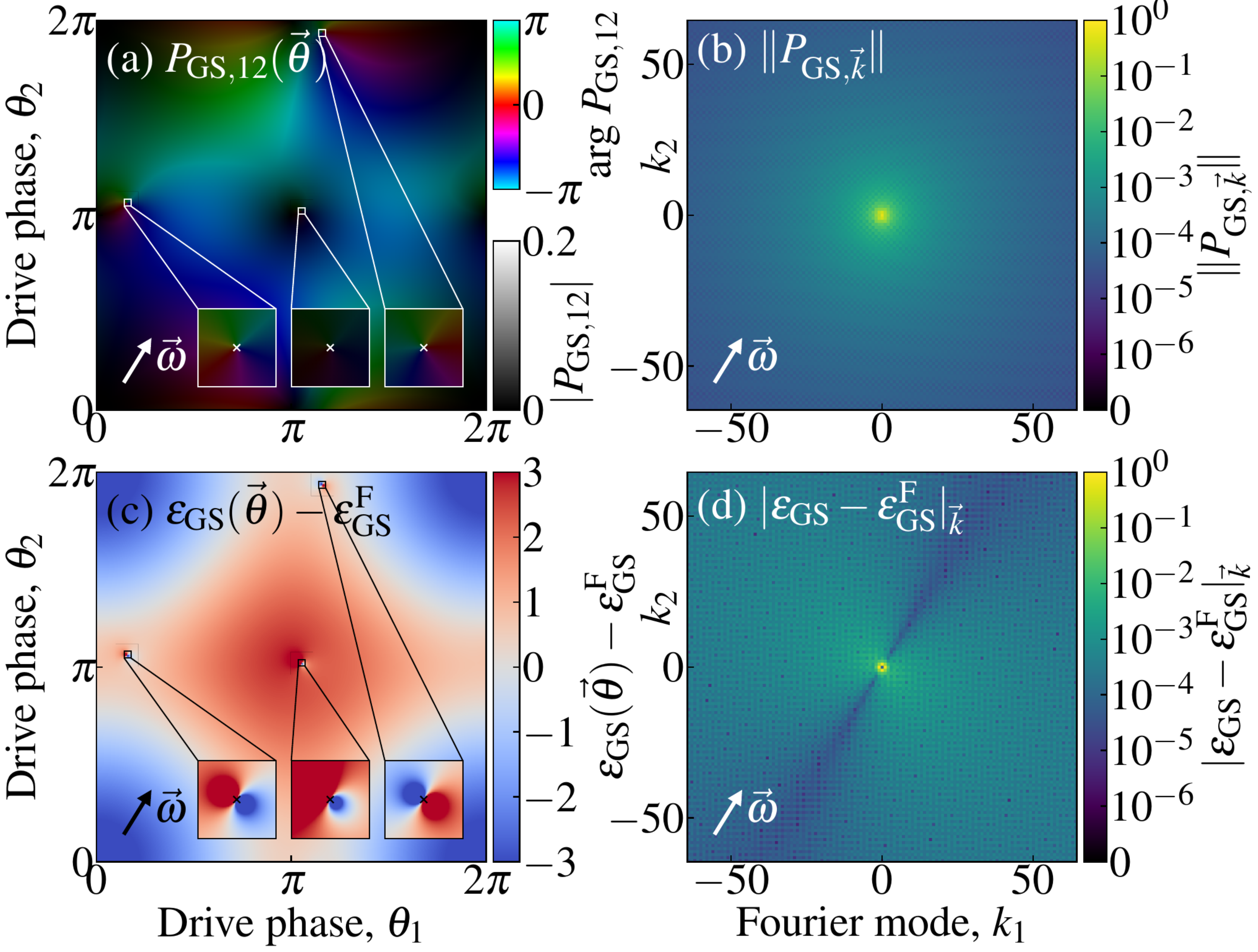}
	\caption{\textbf{Micromotion and scalar phase-rate fluctuation of an AT average-energy ground state.} (a,b) Micromotion and its Fourier decomposition norm, with the color conventions of Fig.~\ref{fig:Floquet-AbeUnt}. Insets enlarge the boxed singularities; crosses mark their locations. (c) Scalar phase-rate fluctuation $\varepsilon_{\mathrm{GS}}(\thetavec)-\epsilon_{\mathrm{GS}}^\F$ on a signed linear color scale; (d) norms of its Fourier coefficients. Both quantities have singularities on the torus and algebraically decaying Fourier components. Model: abelian twisted Hamiltonian $H^{\mathrm A\mathrm T}$, cf. Eq.~\eqref{eq:model_H}; parameters as in Fig.~\ref{fig:Floquet-AbeUnt}.}
    \label{fig:Floquet-AbeTwi}
\end{figure}

\subsubsection{Abelian Twisted (\texorpdfstring{$\rm AT$}{AT}) Spaces}
\label{app:sec:twisted_abelian}

In this class, the invariant projector $\Pi_n[\thetavec]$ is globally smooth and has rank one, so the internal dynamics is again $\mathrm{U}(1)$. However, a nonzero Chern number prevents any ket spanning the invariant space from being smooth over the entire torus $\T^D$. Below, we save what we can from conventional Floquet theory and devise a generalized Floquet description for quasiperiodic dynamics within this class; it combines a reference-based map, smooth away from a singular set of codimension two, with a scalar internal phase.

A description relative to the initial space $\Pi_n[\thetavec_0]$ can be defined for all points $\thetavec\in\T^D$ of the torus where the space is not orthogonal to $\Pi_n[\thetavec_0]$. The normalized reference-based map
\begin{equation}
    P_n(\thetavec;\thetavec_0) =\frac{\Pi_n[\thetavec]\Pi_n[\thetavec_0]} {\sqrt{\text{tr}\left(\Pi_n[\thetavec] \Pi_n[\thetavec_0]\right)}}
\end{equation}
maps the state in the initial rank-one space at $\thetavec_0$ to the corresponding state in the space $\Pi_n[\thetavec]$ at $\thetavec$ without changing its norm. Its connection is
\begin{subequations}
\label{eq:referenceConnection}
\begin{align}
    a_{\mu,n}(\thetavec;\thetavec_0) &:=i\,\text{tr}\left(\Pi_n[\thetavec_0] P_n^\dagger\partial_\mu P_n\right)\\
    &=-\frac{\text{tr}\left( [i\partial_\mu\Pi_n[\thetavec],\Pi_n[\thetavec]] \Pi_n[\thetavec_0]\right)} {2\,\text{tr}\left(\Pi_n[\thetavec]\Pi_n[\thetavec_0]\right)}.
    \label{eq:referenceConnection_den}
\end{align}
\end{subequations}
In any local ket representation, the connection can be rewritten as
\begin{equation}\label{eq:twistedGeomIntegran}
    a_{\mu,n}(\thetavec;\thetavec_0) =i\braket{n[\thetavec]|\partial_\mu n[\thetavec]} +\partial_\mu\arg\braket{n[\thetavec_0]|n[\thetavec]}.
\end{equation}
The first term on the right-hand side is the Berry connection. The second term compensates for the local rephasing of the ket. Thus, the connection $a_{\mu,n}$ is independent of the $U(1)$-gauge choice for a local ket basis wherever the overlap is nonzero. However, it does depend on the initial space via $\thetavec_0$, and becomes singular when the two rank-one spaces are orthogonal, see the denominator in Eq.~\eqref{eq:referenceConnection_den}.

The exact scalar generator of the $\mathrm{U}(1)$ phase in this description is therefore $\thetavec_0$-dependent:
\begin{equation}\label{eq:rankOneReferenceGenerator}
    \varepsilon_n(\thetavec;\thetavec_0) =E_n^\K(\thetavec)-\omega^\mu a_{\mu,n}(\thetavec;\thetavec_0),
\end{equation}
which fixes the $\mathbb{Z}^D$ quasienergy-shift freedom. The evolution within space $\Pi_n$ along any nonsingular part of the trajectory is
\begin{equation}\label{eq:rankOneReferenceEvolution}
    U_n(t,0)=P_n(\thetavec_t;\thetavec_0)\, \mathrm{e}^{-i\int_{0}^{t}\mathrm{d}s\, \varepsilon_n(\thetavec_s;\thetavec_0)}.
\end{equation}
This formula is exact and requires no choice of ket-states. As in Floquet theory, using the drive-phase average, we then define the quasienergy associated with the chosen invariant space
\begin{equation}\label{eq:referenceQuasienergy}
    \epsilon_n^\F(\thetavec_0) :=\mae_n-\omega^\mu \left\langle a_{\mu,n}(\thetavec;\thetavec_0)\right\rangle_{\T^D}.
\end{equation}
The value $\epsilon_n^\F(\thetavec_0)$ is defined for the chosen reference point $\thetavec_0$ and unfolding gauge, without the usual Floquet-gauge ambiguity that shifts it by $\kvec\cdot\omegavec$. Its dependence on $\thetavec_0$ reflects the global twist of the fiber bundle, i.e., the nontrivial topology of invariant space $\Pi_n$. An alternative construction proposed in Ref.~\cite{crowley2019topological} unwraps the geometric phases onto $\mathbb{R}^D$; its resulting quasienergy is then generally multivalued on $\T^D$, remains $\thetavec_0$-dependent, and does not provide a globally smooth transport of basis.

We retain $P_n$ as the micromotion, which is smooth away from the reference singularities, and keep the scalar phase fluctuations explicitly in the internal evolution:
\begin{equation}
\label{eq:AbelTwiFloquet}
\begin{split}
    U_n(t,0)={}&P_n(\thetavec_t;\thetavec_0)\, \mathrm{e}^{-it\epsilon_n^\F(\thetavec_0)}\\
    &\times\mathrm{e}^{-i\int_{0}^{t}\mathrm{d}s\, [\varepsilon_n(\thetavec_s;\thetavec_0)-\epsilon_n^\F(\thetavec_0)]}.
\end{split}
\end{equation}
This separates the average phase accumulation per unit time from its zero-average but unbounded phase-rate fluctuations $\varepsilon_n-\epsilon^\F_n$ due to singularities induced by the twisted topology:
\begin{equation}
\begin{split}
\underbrace{\int_{0}^{T}\mathrm{d}s\,\varepsilon_n(\thetavec_s;\thetavec_0)}_{\text{long-time accumulation}} -&\underbrace{T\,\epsilon_n^\F(\thetavec_0)}_{\text{torus accumulation}}\\=&\underbrace{\int_{0}^{T}\mathrm{d}s\, [\varepsilon_n(\thetavec_s;\thetavec_0)-\epsilon_n^\F(\thetavec_0)]}_{o(T)\text{ when $T$ large, but not bounded}}.
\end{split}
\end{equation}

The reference micromotion $P_n$ is plotted in Fig.~\ref{fig:Floquet-AbeTwi}. Almost every quasiperiodic trajectory avoids the codimension-two singular set exactly, but approaches it arbitrarily closely; the rapid variation of $P_n$ near a reference singularity is compensated by the zero-average, unbounded phase-rate fluctuation $\varepsilon_n-\epsilon^\F_n$ plotted in Fig.~\ref{fig:Floquet-AbeTwi}. The Chern number fixes the total winding, but additional singularities with opposite windings can occur in pairs on the torus; they originate from the vanishing denominator in Eq.~\eqref{eq:referenceConnection_den}; in the frequency lattice, the vortex-like singularities render the Fourier modes of micromotion $P$ decaying only algebraically, instead of a localized blob in the centre.

For an untwisted rank-one space, the globally smooth Floquet ket defines $P_n$ by Eq.~\eqref{eq:AbelUntFloquet}, without dividing by its overlap with a reference ket. A zero of that overlap therefore does not make this micromotion singular. For a twisted space, an everywhere nonsingular reference map would generate a globally smooth ket, which the topology forbids.

\begin{figure}[t!]
	\centering
    \includegraphics[width=0.48\textwidth]{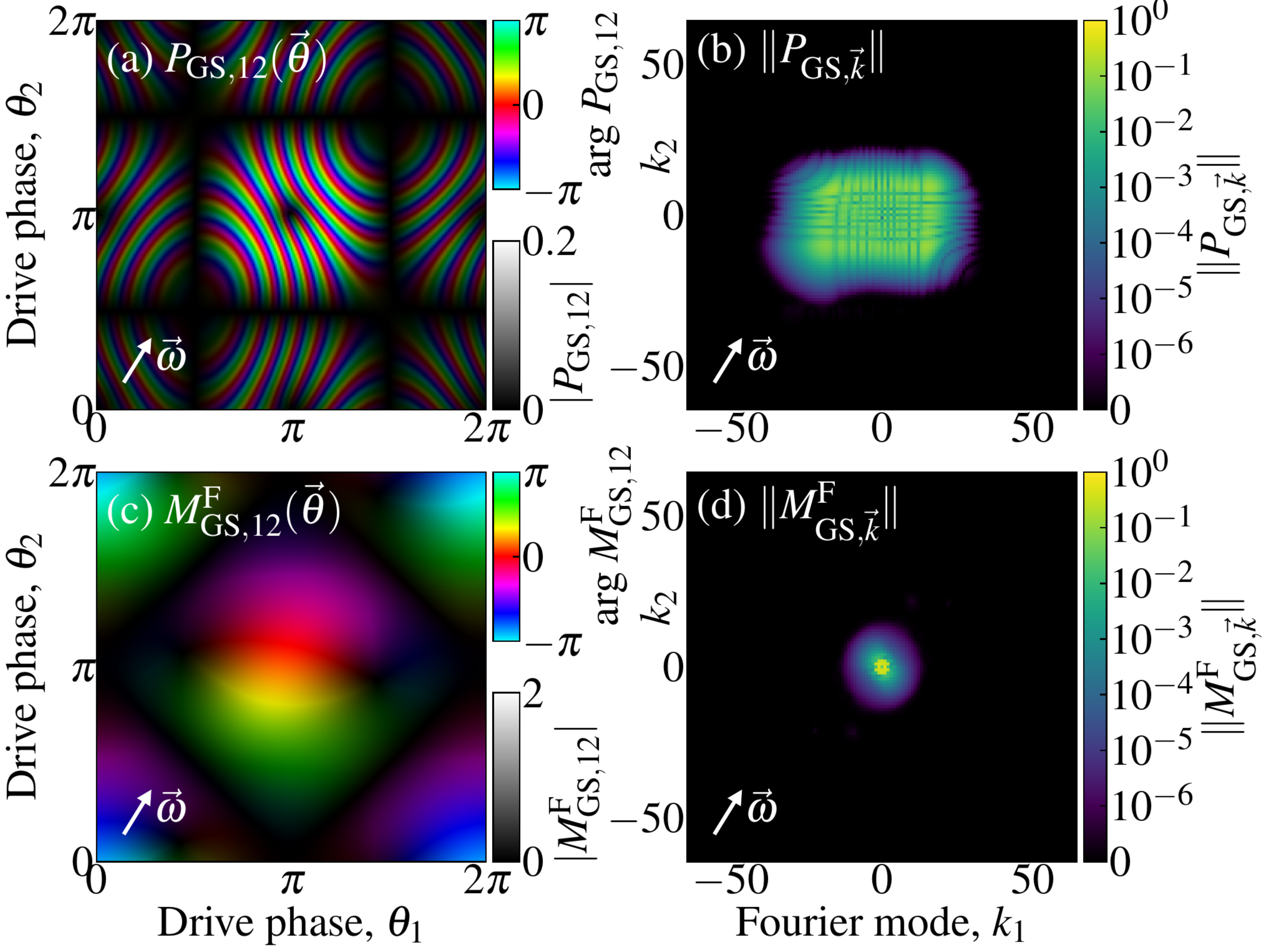}
	\caption{\textbf{Micromotion and internal dynamics of a \=A\=T average-energy ground state manifold.} (a,c) Matrix elements $P_{\mathrm{GS},12}(\thetavec)$ and $M^\F_{\mathrm{GS},12}(\thetavec)$; (b,d) their Fourier-mode contributions, with the color conventions of Fig.~\ref{fig:Floquet-AbeUnt}. Both $P_{\rm GS}$ and $M^\F_{\rm GS}$ are globally smooth on the torus and localized on the frequency lattice. Model: nonabelian untwisted Hamiltonian $H^{\mathrm{\bar A}\mathrm{\bar T}}$ from Eq.~\eqref{eq:model_H}; parameters as in Fig.~\ref{fig:Floquet-AbeUnt}.}
	\label{fig:Floquet-NonUnt}
\end{figure}

\subsubsection{Nonabelian Untwisted (\texorpdfstring{$\rm \bar A\bar T$}{A-bar T-bar}) Spaces}
\label{app:sec:untwisted_nonabelian}

We now consider a nonabelian invariant space $\Pi_n$ which is dynamically irreducible. For such drives, $r_n>1$, so the internal dynamics remains a time-ordered $\mathrm{U}(r_n)$ rotation, while the untwisted (i.e., topologically trivial) character permits a globally smooth orthonormal basis. This basis gives a smooth map between the moving invariant spaces and exposes the remaining $r_n$-dimensional dynamics explicitly.

Let $V_n(\thetavec)$ be the $N\times r_n$ matrix whose columns are these basis states, so that $V_n^\dagger V_n=\mathds{1}_{r_n\times r_n}$ and $V_nV_n^\dagger=\Pi_n[\thetavec]$. The evolution within space $\Pi_n$ is
\begin{subequations}\label{eq:U(rank)}
\begin{align}
U_n(t,0)&=V_n(\thetavec_t)G_n(t,0)V_n(\thetavec_0)^\dagger,\\
    G_n(t,0)&=\mathcal{T}_s\mathrm{e}^{-i\int_{0}^t \mathrm{d}s\,M_n(\thetavec_s)},\\
    M_n(\thetavec)&=V_n(\thetavec)^\dagger H(\thetavec)V_n(\thetavec) -iV_n(\thetavec)^\dagger\omega^\mu\partial_\mu V_n(\thetavec).
\end{align}
\end{subequations}
The map $V_n(\thetavec)V_n(\thetavec_0)^\dagger$ transports the basis from $\thetavec_0$ to $\thetavec$; it is the generalization of a globally smooth micromotion operator. The matrix $G_n(t,0)$ evolves its $r_n\times r_n$ coefficients from $0$ to $t$. Under a different globally smooth basis defined by the change-of-frame rotation $R_n(\thetavec)\in\mathrm{U}(r_n)$,
\begin{subequations}\label{eq:UgGauge}
\begin{align}
    V_n(\thetavec)&\longrightarrow V_n(\thetavec)R_n(\thetavec),\\
    M_n(\thetavec)&\longrightarrow R_n(\thetavec)^\dagger M_n(\thetavec) R_n(\thetavec) -iR_n(\thetavec)^\dagger \omega^\mu\partial_\mu R_n(\thetavec).
\end{align}
\end{subequations}
Accordingly, the internal dynamics propagator $G_n(t,0)$ transforms as $G_n(t,0){\to} R_n(\thetavec_t)^\dagger G_n(t,0)R_n(\thetavec_0)$, leaving $U_n(t,0)$ unchanged. No globally smooth choice of basis makes $M_n$ constant over $\T^D$ without producing lower-rank invariant projectors, as shown in Sec.~\ref{sec:2obstructions}; the internal time ordering therefore remains a fundamental hallmark of nonabelian drives.

However, we can still separate the scalar $\mathrm{U}(1)$ phase from the intrinsic $\mathrm{SU}(r_n)$ rotation.  Define
\begin{equation}\label{eq:eFnonabelUntwi}
    \epsilon_n^\F=\frac{1}{r_n} \left\langle\text{tr}\, M_n(\thetavec) \right\rangle_{\T^D}.
\end{equation}
This separation of the overall phase depends on the chosen globally smooth basis, while the physical evolution $U_n$ is unchanged by basis rotations. Let us choose the common phase $V_n^\F=\mathrm{e}^{i\chi_n}V_n$ from the zero-average equation
\begin{equation}
    \omega^\mu\partial_\mu\chi_n(\thetavec) =\epsilon_n^\F-\frac{1}{r_n}\text{tr}\, M_n(\thetavec).
\end{equation}

The transformed internal-dynamics generator matrix is
\begin{equation}
    \begin{aligned}
    M_n^\F&=M_n+\omega^\mu\partial_\mu\chi_n\,\mathds{1}_{r_n\times r_n} -\epsilon_n^\F\mathds{1}_{r_n\times r_n}\\
    &=M_n-\frac{1}{r_n}\text{tr}\, M_n(\thetavec)\,\mathds{1}_{r_n\times r_n},
    \end{aligned}
\end{equation}
so $M_n^\F$ is traceless. This is the standard irreducible tensor decomposition of a Hermitian $r_n\times r_n$ matrix into its scalar and traceless components: the former produces the common $\mathrm{U}(1)$ phase, while the latter produces the intrinsic $\mathrm{SU}(r_n)$ rotation.

Equation~\eqref{eq:U(rank)} then becomes
\begin{equation}\label{eq:U(1)SU(rank)}
\begin{split}
    U_n(t,0)=V_n^\F(\thetavec_t) \mathrm{e}^{-it\epsilon_n^\F}\mathcal{T}_s\mathrm{e}^{-i\int_{0}^t \mathrm{d}sM_n^\F(\thetavec_s)} V_n^\F(\thetavec_0)^\dagger.
\end{split}
\end{equation}
The micromotion \begin{equation} P_n=V_n^\F(\thetavec)V_n^\F(\thetavec_0)^\dagger
\end{equation} and $M_n^\F$ are plotted in Fig.~\ref{fig:Floquet-NonUnt}. They both show global smoothness on the torus and localization in the frequency lattice.
Here nonabelianness requires time ordering over $M^\F_n$ in order to obtain the internal dynamics. We remark that this time-ordered factor cannot be removed from a dynamically irreducible nonabelian space.

\subsubsection{Nonabelian Twisted (\texorpdfstring{$\rm \bar AT$}{A-bar T}) Spaces}
\label{app:sec:twisted_nonabelian}

\begin{figure}[t!]
	\centering
    \includegraphics[width=0.48\textwidth]{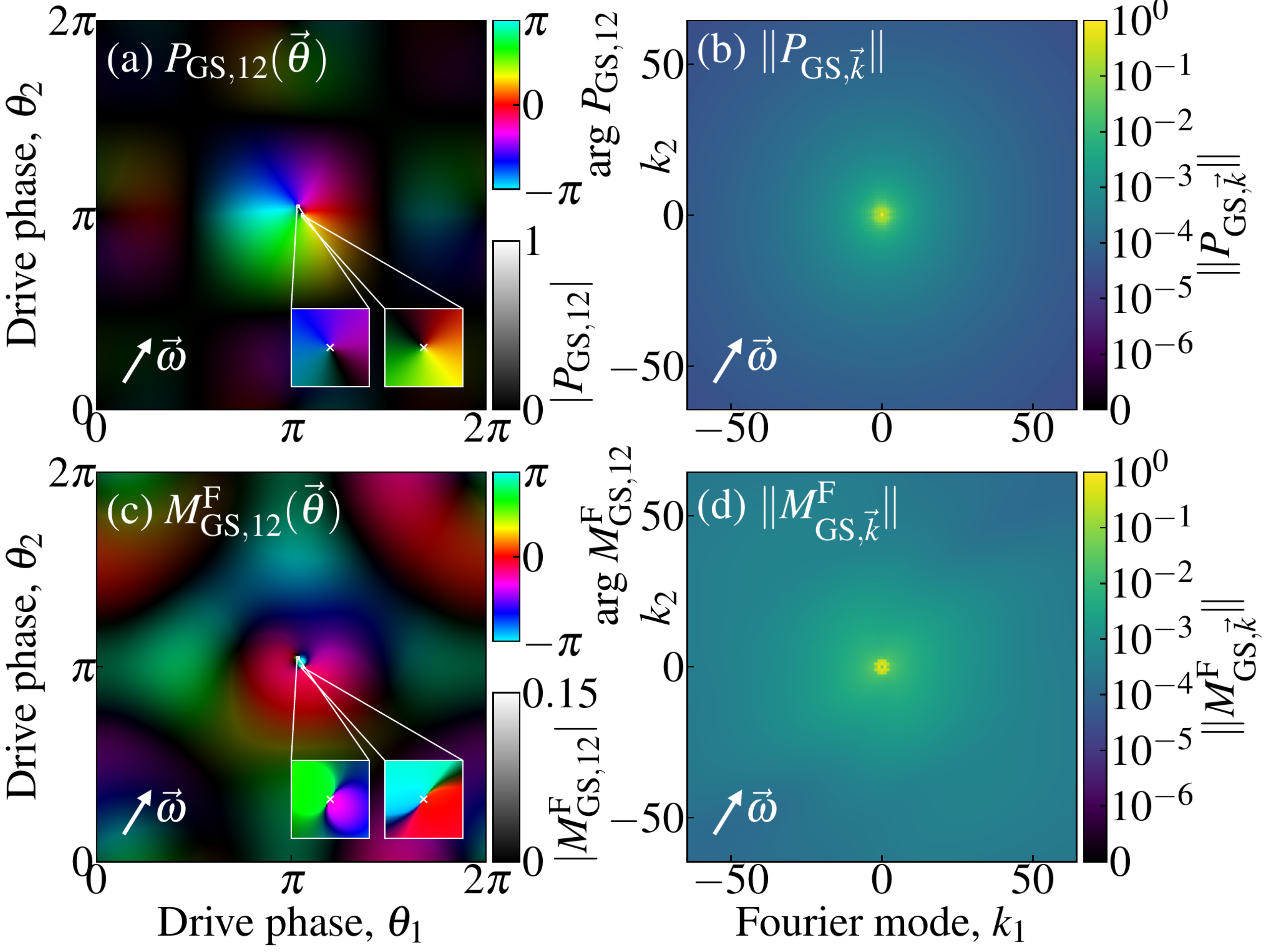}
	\caption{\textbf{Micromotion and internal dynamics of a nonabelian twisted (\=AT) average-energy ground state manifold.} (a-d) Same quantities and color conventions as in Fig.~\ref{fig:Floquet-NonUnt}. Insets enlarge the boxed singularities; crosses mark their locations. Both $P_{\rm GS}$ and $M^\F_{\rm GS}$ have singularities at the same drive phases and algebraically decaying Fourier components. Model: nonabelian twisted Hamiltonian $H^{\mathrm{\bar A}\mathrm T}$, cf.~Eq.~\eqref{eq:model_H}; parameters as in Fig.~\ref{fig:Floquet-AbeUnt}.}
	\label{fig:Floquet-NonTwi}
\end{figure}

This is the most general class: no globally smooth basis exists due to the topological obstruction, and the internal evolution remains a time-ordered $\mathrm{U}(r_n)$ rotation. We therefore construct the transport map directly from the smooth projectors and express the internal dynamics in the fixed initial invariant space.

Algorithm~\ref{alg:qp-reconstruction} presents the complete procedure of the geometric theory of quasiperiodically driven systems: from the variational principle, to the classification of the invariant spaces, to the average-energy, Kato and generalized Floquet decompositions of $U_n(t,0)$ within invariant space $\Pi_n$.

Let us fix the initial drive phase $\thetavec_0$ and define, within the dynamically irreducible invariant space $\Pi_n$,
\begin{equation}\label{eq:referenceOverlapRoot}
    S_n(\thetavec;\thetavec_0) :=\left(\Pi_n[\thetavec_0]\Pi_n[\thetavec]\Pi_n[\thetavec_0]\right)^{1/2}.
\end{equation}
Where this operator is invertible, let $S_n^{-1}$ denote its inverse within the initial invariant space, extended as zero on the orthogonal complement within the total Hilbert space. Since the joint projection $\Pi_n[\thetavec]\Pi_n[\thetavec_0]$ changes norms and overlaps, the factor $S_n^{-1}$ acts in the initial space to compensate for this change, giving the normalized transport map
\begin{equation}\label{eq:referenceTransport}
    P_n(\thetavec;\thetavec_0) :=\Pi_n[\thetavec] S_n^{-1}(\thetavec;\thetavec_0) \Pi_n[\thetavec_0].
\end{equation}
The resulting $P_n$ is not transitive\footnote{Transport between two drive phases generally differs from transport through an intermediate phase: $P_n(\thetavec_2;\thetavec_1)P_n(\thetavec_1;\thetavec_0)\neq P_n(\thetavec_2;\thetavec_0)$.} unlike in the untwisted case, but it is still an isometry that preserves inner products:
\begin{equation}
    P_m^\dagger P_n=\delta_{mn}\Pi_n[\thetavec_0], \qquad P_mP_n^\dagger=\delta_{mn}\Pi_n[\thetavec].
\end{equation}

Using the transport map $P_n$, the dynamics within the moving space can be represented inside the fixed initial space. The corresponding $\mathrm{U}(r_n)$ generator is
\begin{align}
    M_n(\thetavec;\thetavec_0)={}& P_n(\thetavec;\thetavec_0)^\dagger H(\thetavec)P_n(\thetavec;\thetavec_0)\nonumber\\
    &-iP_n(\thetavec;\thetavec_0)^\dagger \omega^\mu\partial_\mu P_n(\thetavec;\thetavec_0).
    \label{eq:twistedMatrixGenerator}
\end{align}
By construction, it is Hermitian and acts only in the initial invariant space $\Pi_n[\thetavec_0]$. In any fixed orthonormal basis of that space, Eq.~\eqref{eq:twistedMatrixGenerator} is an $r_n\times r_n$ matrix. The exact evolution within space $\Pi_n$ reads as
\begin{equation}\label{eq:mostGeneralCase}
    U_n(t,0)=P_n(\thetavec_t;\thetavec_0) \mathcal{T}_s\mathrm{e}^{-i\int_{0}^t \mathrm{d}sM_n(\thetavec_s;\thetavec_0)}.
\end{equation}
Equation~\eqref{eq:mostGeneralCase} separates basis transport $P_n(\thetavec_t;\thetavec_0)$ from the remaining nonabelian $\mathrm{U}(r_n)$ internal dynamics in the fixed initial space. Substituting $U_n=P_nG_n$ into the Schr\"odinger equation and multiplying by $P_n^\dagger$ gives $i\partial_t G_n=M_nG_n$, with $M_n$ from Eq.~\eqref{eq:twistedMatrixGenerator}. Once $\Pi_n$ is known, the micromotion $P_n$ requires no trajectory-dependent integral. Both the micromotion and the internal-dynamics generator $M_n$ are defined on the torus.

As in the abelian twisted case, we extract the average scalar phase accumulation while retaining its zero-average but unbounded fluctuations in the internal generator. The quasienergy remains the drive-phase average of the normalized trace of $M_n$.
\begin{subequations}\label{eq:ThreeMusketeers}
\begin{equation}
    \epsilon_n^\F(\thetavec_0)=\frac{1}{r_n} \left\langle\text{tr}\,M_n(\thetavec;\thetavec_0) \right\rangle_{\thetavec\in\T^D},
\end{equation}
\begin{equation}
    M^\F_n(\thetavec;\thetavec_0)=M_n(\thetavec;\thetavec_0) -\epsilon_n^\F(\thetavec_0)\Pi_n[\thetavec_0].
\end{equation}
\end{subequations}

The almost everywhere smooth micromotion $P_n$, the Floquet quasienergy $\epsilon^\F_n$, and the $\mathrm{U}(r_n)$ generator of internal dynamics $M^\F_n$ from Eqs.~\eqref{eq:ThreeMusketeers} define the most general form of a gauge-free generalized Floquet decomposition for evolution within each invariant space, valid for any quasiperiodically driven closed quantum system:
\begin{equation}\label{eq:THEmostGeneralCase}
    U_n(t,0)=P_n(\thetavec_t;\thetavec_0) \,\mathrm{e}^{-it\epsilon^\F_n(\thetavec_0)}\,\mathcal{T}_s\mathrm{e}^{-i\int_{0}^t \mathrm{d}s M^\F_n(\thetavec_s;\thetavec_0)}~.
\end{equation}

Figure~\ref{fig:Floquet-NonTwi} shows singularities at the same drive phases in $P_n$ and $M_n^\F$, together with their algebraically decaying Fourier tails. The singularities still correspond to drive phases $\thetavec$ with a non-invertible $S_n(\thetavec;\thetavec_0)$ in Eq.~\eqref{eq:referenceOverlapRoot} at fixed $\thetavec_0$, and they sum to the twisted topology of $\Pi_n$. When $\thetavec_t$ passes by the singularities of internal rotation generator $M^\F_n$, its unbounded components compensate the rapid change of the micromotion $P_n$ near these points.

\begin{algorithm}[t!]
\begin{algorithmic}[1]
\Require Fourier harmonics $H_{\kvec}$, $\omegavec$%
\State \textbf{Construct} $\LL$ from $ H_{\kvec}$ and $\omegavec$ \Comment{Eq.~\eqref{eq:L-frequency}}
\State $\mAE^{\rm varl}\gets H-{\operatorname{argmin}}_{\LL X=\LL H}\|X\|^2$ \Comment{Eqs.~\eqref{eq:AE_objs},\eqref{eq:varl_AE-AGP}}
\State  \textbf{Diagonalize} $\mAE^{\mathrm{varl}}$ and \textbf{merge} degen.~spaces to find $\Pi_n$
\State $\mae_n\gets\langle\Pi_n,H\rangle_{\T^D}/\mathrm{tr}\,\Pi_n$, $A^\K_\mu\gets \sum_n (i\partial_\mu\Pi_n)\,\Pi_n$ \Comment{Eqs.~\eqref{eq:aeEnergy},\eqref{eq:Kato_objs}}
\State \textbf{Classify} space $\Pi_n$ via abelianness and topology
\State $\mAE$ decomp.: $W^\mAE, A^\mAE$ from $ \mAE\gets\sum_n\mae_n\Pi_n$ \Comment{Eq.~\eqref{eq:U_qp}}
\State Kato decomp.: $W^\K,H^\K$ from $A^\K\gets \omega^\mu A^\K_\mu$ \Comment{Eq.~\eqref{eq:KatoGeneralRank}}
\State Generalized Floquet decomp.:
\Statex $S_n^{-1}\gets(\Pi_n[\thetavec_0]\Pi_n[\thetavec]\Pi_n[\thetavec_0])^{-1/2}$ \Comment{Eq.~\eqref{eq:referenceOverlapRoot}}
\Statex \quad \textit{abelian:} $S_n^{-1}\gets\bigl[\mathrm{tr}(\Pi_n[\thetavec]\Pi_n[\thetavec_0])\bigr]^{-1/2}\Pi_n[\thetavec_0]$
\State $P_n\gets\Pi_n[\thetavec]S_n^{-1}\Pi_n[\thetavec_0]$ \Comment{Eq.~\eqref{eq:referenceTransport}}
\Statex \quad \textit{untwisted:} \textbf{choose} a smooth basis $V_n$, $P_n\gets V_n(\thetavec)V_n(\thetavec_0)^\dagger$
\State $M_n\gets P_n^\dagger H P_n-iP_n^\dagger\omega^\mu\partial_\mu P_n$
\Statex $\epsilon_n^\F\gets \langle\mathrm{tr}M_n\rangle_{\T^D}/r_n$, $M_n^\F\gets M_n-\epsilon_n^\F\Pi_n[\thetavec_0]$ \Comment{Eq.~\eqref{eq:ThreeMusketeers}}
\Statex \quad \textit{untwisted:} \textbf{solve} $\omega^\mu\partial_\mu\chi_n=\epsilon_n^\F-\mathrm{tr}M_n/r_n$ with $\chi_n(\thetavec_0)=0$;
\Statex \quad $P_n\gets\mathrm e^{i\chi_n} P_n$, $M_n^\F\gets M_n-\Pi_n[\thetavec_0]\mathrm{tr}M_n/r_n$ \Comment{Eq.~\eqref{eq:U(1)SU(rank)}}
\State $U_n(t,0)\gets P_n(\thetavec_t;\thetavec_0)\mathrm e^{-i\epsilon_n^\F t}\,\mathcal T_s\mathrm e^{-i\int_{0}^t M_n^\F(\thetavec_s;\thetavec_0)\D s}$ \Comment{Eq.~\eqref{eq:THEmostGeneralCase}}
\Statex \quad \textit{abelian:} \textbf{omit} $\mathcal T_s$ as $(\varepsilon_n-\epsilon_n^\F)\Pi_n[\thetavec_0]\gets M_n^\F$ \Comment{Eq.~\eqref{eq:AbelTwiFloquet}}
\end{algorithmic}
\par\kern2pt\hrule
\caption{ \textbf{Effective operators for quasiperiodic drives.} The variational principle determines the invariant spaces, from which we reconstruct the evolution in the three gauges. Indented formulas give the \textit{abelian} and \textit{untwisted} simplifications in place of the corresponding general expressions. }
\label{alg:qp-reconstruction}
\end{algorithm}

In the untwisted construction of Eq.~\eqref{eq:U(1)SU(rank)}, a globally smooth basis is used to absorb the scalar phase fluctuations of Eq.~\eqref{eq:THEmostGeneralCase} into the micromotion, leaving a traceless internal generator. In turn, for a twisted but abelian invariant space, $M^\F_n=[\varepsilon_n-\epsilon_n^\F]\Pi_n[\thetavec_0]$, and Eq.~\eqref{eq:THEmostGeneralCase} reduces to Eq.~\eqref{eq:AbelTwiFloquet}. Hence, the generalized Floquet decomposition in Eq.~\eqref{eq:THEmostGeneralCase} of the nonabelian twisted class is universally applicable in the other three classes.

Equations~\eqref{eq:mostGeneralCase} and \eqref{eq:THEmostGeneralCase} become singular when $S_n$ has a zero eigenvalue, meaning that a nonzero initial state has zero projection onto the current invariant space. As in Sec.~\ref{sec:2obstructions}, the singular set has codimension two, or dimension $D-2$, and almost every one-dimensional quasiperiodic trajectory avoids these singularities.

Given that the nonabelian twisted class is the most general case for invariant spaces of any quasiperiodic drive, the total micromotion operator $P(\thetavec;\thetavec_0)=\sum_nP_n(\thetavec;\thetavec_0)$ that represents the basis transport, and the quasienergy operator $H^\F[\thetavec;\thetavec_0]=\sum_n \epsilon_n^\F(\thetavec_0)\Pi_n[\thetavec]$ are always well-defined. The $\thetavec_0$ dependency embeds the singularity induced by twistedness, and $H^\F[\thetavec_t;\thetavec_0]$ is a dynamical invariant at any fixed $\thetavec_0$.

The Floquet AGP $A^\F(\thetavec;\thetavec_0)=i\omega^\mu(\partial_\mu P(\thetavec;\thetavec_0))P(\thetavec;\thetavec_0)^\dagger$ gives the corresponding decomposition of the laboratory-frame Hamiltonian:
\begin{equation}\label{eq:generalized-Floquet-infinitesimal}
\begin{split}
H(\thetavec)={}&H^\F[\thetavec;\thetavec_0]+A^\F(\thetavec;\thetavec_0)\\
&+P(\thetavec;\thetavec_0)\left[\sum_n M_n^\F(\thetavec;\thetavec_0)\right]P(\thetavec;\thetavec_0)^\dagger\,.
\end{split}
\end{equation}

These operators define a gauge-free, geometric, generalized Floquet theory for any closed quantum evolution $U(t,0)$ under a quasiperiodic drive, valid for almost all initial phases $\thetavec_0$:
\begin{equation}\label{eq:GeneralQP-GFT}
\begin{split}
    U(t,0)={}&P(\thetavec_t;\thetavec_0)\, \mathrm{e}^{-itH^\F[\thetavec_0;\thetavec_0]} \\&\quad\times\sum_n\Pi_n[\thetavec_0]\mathcal{T}_s\mathrm{e}^{-i\int_{0}^t \mathrm{d}s\,M^\F_n(\thetavec_s;\thetavec_0)}.
\end{split}
\end{equation}

\subsection{Comparison of Evolution Decompositions within Each Invariant Space}
\label{app:SummaryDecomposition}

Given a dynamically irreducible invariant space of any quasiperiodic drive, depending on whether its rank is $r_n=1$ and its topology is trivial, it belongs to one of the four abelian/nonabelian and untwisted/twisted classes. The most natural way to factorize the evolution $U_n(t,0)$ within space $\Pi_n$ in a generalized Floquet decomposition depends on the classification: abelian classes can be viewed as a special case of nonabelian classes; untwisted classes are a special case of twisted ones, see Fig.~\ref{fig:octopusQPCD}(f).

The four classes are summarized in Table~\ref{table:table}, which displays their corresponding basis transport and explicit $\mathrm{U}(r_n)$ internal dynamics in the generalized Floquet ($\F$) decomposition of the evolution operator $U_n(t,0)$ restricted to the $n$-th invariant space. The numerical infidelity of the generalized Floquet decomposition is shown in Fig.~\ref{fig:reconstructedEvolutionError}. Overall, the error of reconstructed ground-space evolution $U_{\rm GS}(t,0)$ grows with the evolution time $t$. As the Fourier cutoff $K$ for the variational principle increases from 10 to 50, the infidelity of the reconstructed $\mae$-ground-state evolution $U_\mathrm{GS}(t,0)$ for the abelian untwisted case drops as the invariant projector $\Pi_n$ gets more accurate. For sufficiently high $K=45,50$, all four classes show an infidelity of about $10^{-10}$ for the reconstructed evolution $U_\mathrm{GS}(t,0)$ at time $t=10^5$, showing the approximation is controlled.

\begin{figure}[t!]
    \centering
    \includegraphics[width=0.48\textwidth]{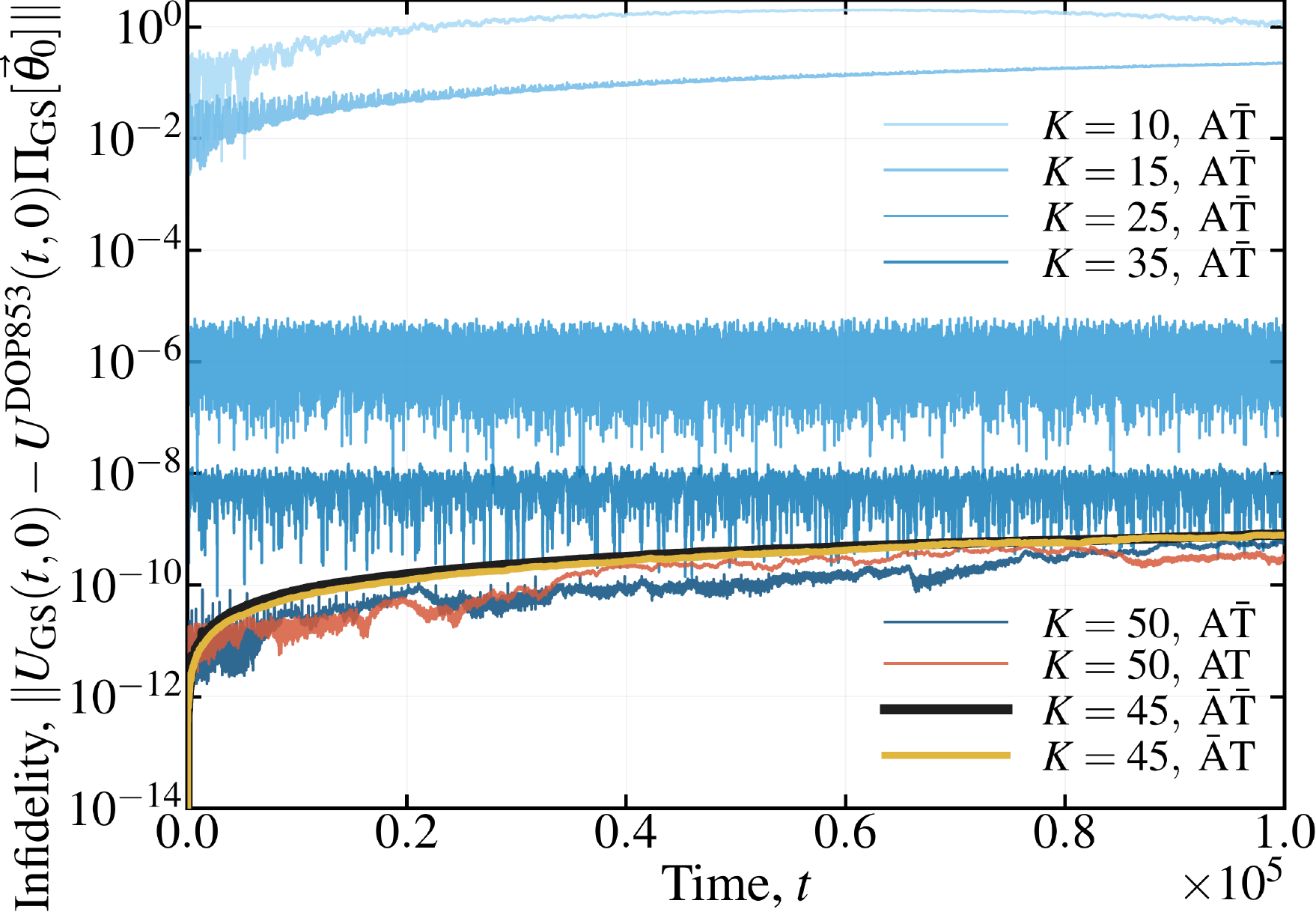}
    \caption{\textbf{Infidelity of the generalized Floquet reconstruction.} The norm $\|U_{\mathrm{GS}}(t,0)-U^{\mathrm{DOP853}}(t,0)\Pi_{\mathrm{GS}}[\thetavec_0]\|$ compares the reconstructed ground-space evolution using generalized Floquet decompositions from Table~\ref{table:table} with the one using direct integration. The error grows with time but decreases controllably as the Fourier cutoff $K$ increases. At $K=45,50$, all four classes reach an infidelity of order $10^{-10}$ at $t=10^5$. Colors distinguish the four drive classes, while blue shades show the convergence with $K$ in the abelian untwisted case. Data obtained for the Hamiltonians $H^{\alpha\tau,\text{rev-eng}}$ in Eq.~\eqref{eq:AnalyH}, $\thetavec_0=(0.5,0.2)^\mathsf T$, and $|\omegavec|=0.1$, $\omega_2/\omega_1=(1+\sqrt5)/2$ and $\epsilon=10^{-2}$. The absolute and relative tolerances of the eighth-order Dormand--Prince integrator (DOP853) are set to $10^{-14}$.}
    \label{fig:reconstructedEvolutionError}
\end{figure}

All three decompositions start from the invariant projectors $\Pi_n$, obtained by solving Eq.~\eqref{eq:Lqp}, which requires information about the full operator Hilbert space. Once the invariant spaces $\Pi_n$ are known, the internal dynamics of $U_n(t,0)$ can be described by $r_n\times r_n$ blocks in any of the three gauges.\footnote{Fixing the Kato or average-energy gauges requires history-dependent Wilson lines along the drive; this requires information about how a given space varies on the torus within the full Hilbert space.} The difference is which motion is absorbed into the transported basis. Therefore, the same invariant space of quasiperiodically driven evolution $U_n(t,0)$ admits three equivalent but distinct decompositions as in the Floquet case, each with its own distinguished advantages:
\begin{enumerate}
    \item[($\K$):] the Kato decomposition in Appendix \ref{app:KatoQP} is useful for separating the geometric effects (the physical quantities that only depend on $\Pi_n$ and its derivatives $\partial_\mu\Pi_n$), e.g., the geometric phase $\gamma_{n,\mu}$, the quantum geometric tensor $Q_{n,\mu\nu}=\Pi_n A^\K_\mu A^\K_\nu\Pi_n$ defined from the directional AGPs -- from the dynamical effects (quantities only depending on $\Pi_n$ and $H$, e.g., the Kato energy $E^\K$ and the traceless rank-$r_n$ dynamical generator $\widetilde{H}_n$). Both the geometric and the dynamical quantities are separately embedded in the two parts of the Kato decomposition.
    \item[($\mAE$):] the average-energy gauge instead extracts the constant dynamical phase $\exp(-it\mae_n)$ in each invariant space. Its Wilson line $W^\mAE$ carries the geometric transport together with the remaining dynamical fluctuations and nonabelian rotations, and therefore also retains the trajectory history, similar to $W^\K$ in the Kato decomposition. This decomposition has the same form in all four classes, while $\mAE[\thetavec]$ remains globally smooth and dynamically invariant. Its variational construction provides the invariant projectors needed by all three descriptions, up to accidental average-energy degeneracies [Appendix~\ref{app:omega-irreducibility}].
    \item[($\F$):] in the generalized Floquet decomposition, $P_n(\thetavec;\thetavec_0)$ instead fixes the basis transport using the globally defined invariant space $\Pi_n$, without integrating the time-dependent trajectory. The internal generator is defined on the same torus, exposing the quasienergy phase, the scalar fluctuations in the case of twisted spaces, and the $\U(r_n)$ rotations in the case of nonabelian spaces. The micromotion obeys $P_n(\thetavec;\thetavec_0)=P_n(\thetavec_0;\thetavec)^\dagger$ and preserves quasiperiodicity when transforming observables, providing a rotating frame for quasiperiodic Floquet engineering. In the untwisted classes, its smoothness gives rapidly decaying Fourier components, while in the twisted classes, it retains the topological singularities and thus is almost globally smooth on the torus and has algebraically decaying Fourier components [Appendix~\ref{app:sec:4classes}].
\end{enumerate}

The average-energy decomposition requires no arithmetic condition on $\omegavec$ (Appendix~\ref{app:subsec:spetral_types}). By contrast, the almost-everywhere smooth generalized Floquet decomposition additionally requires Diophantine $\omegavec$, i.e., it holds for almost all frequency vectors $\omegavec$.

\section{Model and Numerical Validation}
\label{app:sec:model}

\begin{figure}[t!]
    \centering
    \includegraphics[width=0.48\textwidth]{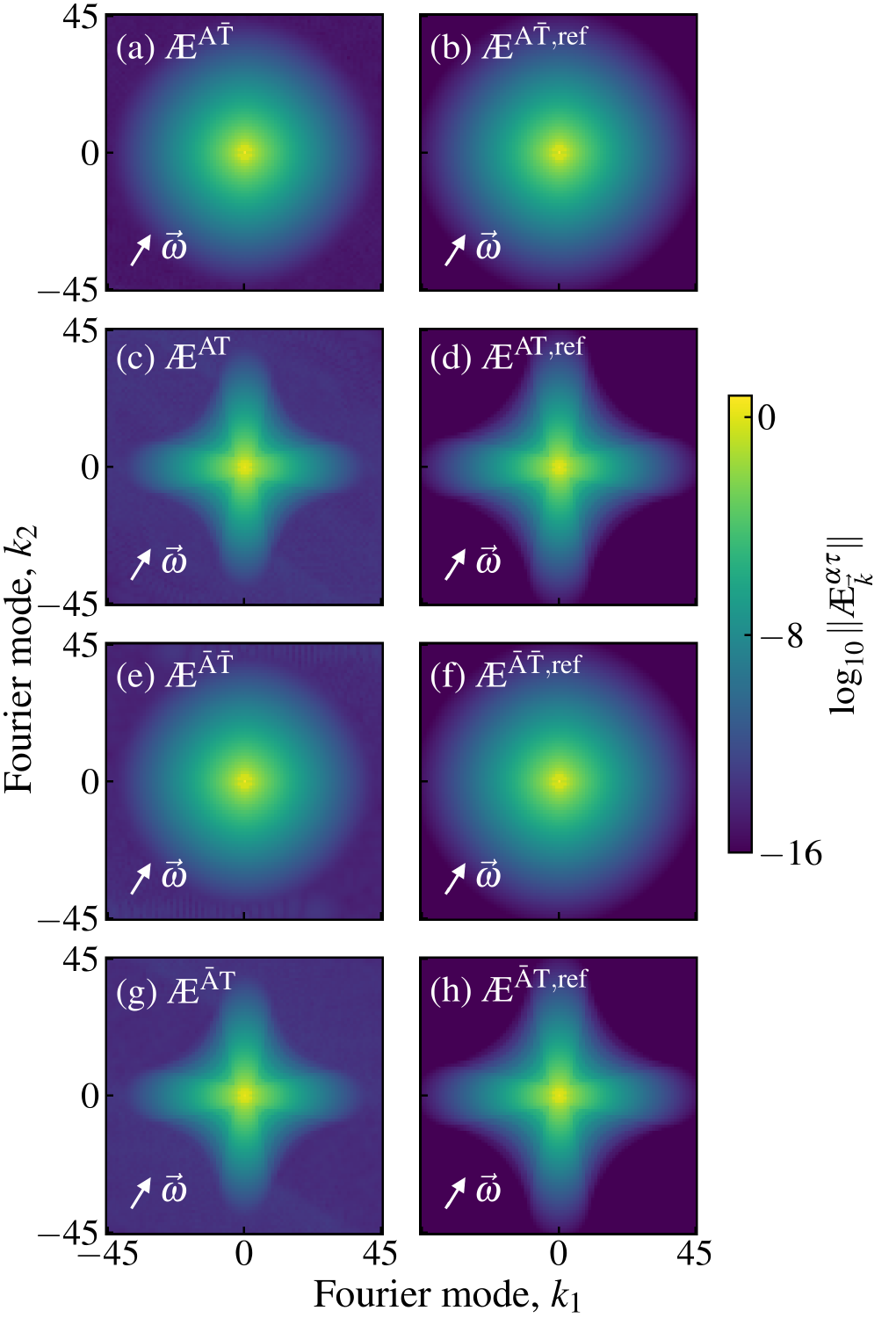}
    \caption{\textbf{Average-energy reconstruction in the four drive classes.} Matrix norms of the Fourier components of the reconstructed average energy $\mAE^{\alpha\tau}$ (left) and the analytical reference $\mAE^{\alpha\tau,\mathrm{ref}}$ (right), for the exactly solvable reverse-engineered drives in Eq.~\eqref{eq:AnalyH}. Rows show abelian untwisted, abelian twisted, nonabelian untwisted, and nonabelian twisted drives, from top to bottom. The common color scale shows $\log_{10}\|\mAE^{\alpha\tau}_{\kvec}\|$; white arrows indicate $\omegavec$. The variational solution reproduces the analytical Fourier structure in all four classes. Data obtained using the variational principle of Sec.~\ref{subsec:varl}, at $K=45$, with parameters $|\omegavec|=0.1$, $\omega_2/\omega_1=(1+\sqrt5)/2$ and $\epsilon=10^{-2}$.}
    \label{fig:analyAEnumVSref}
\end{figure}

In this section, we first use the non-truncated version of the model defined in Eq.~\eqref{eq:model_H} and its analytical solution to benchmark the variational principle [Sec.~\ref{subsec:varl}]; then we shift our focus to the truncated version illustrated in Figs.~\ref{fig:abelian_nontwisting}, \ref{fig:abelian_twisting}, \ref{fig:nonabelian_nontwisting}, and \ref{fig:nonabelian_twisting}, and provide complementary numerical results.

\subsection{Analytically Solvable Four-Class Model}

For qubit $B$ defined in Eq.~\eqref{eq:HA_HB}, the exact counterdiabatic field reads explicitly as
\begin{equation}
\begin{split}
    &\vec{a}_{\mathrm{CD}}(\thetavec)=\frac{1}{2d^2(\thetavec)}\\
    &\qquad\times\begin{bmatrix} -\omega_1\sin\theta_1\sin\theta_2 -\omega_2[1+(m+\cos\theta_1)\cos\theta_2]\\
    \omega_2\sin\theta_1\sin\theta_2 +\omega_1[1+(m+\cos\theta_2)\cos\theta_1]\\
    -\omega_1\cos\theta_1\sin\theta_2 +\omega_2\cos\theta_2\sin\theta_1
    \end{bmatrix},
\end{split}
\end{equation}
where $d(\thetavec)=|\vec d(\thetavec)|$.

Using $R=\exp\left(i\theta_1\sigma_{A}^x\sigma_{B}^z/2\right) \exp\left(i\theta_2\sigma_{A}^y\sigma_{B}^z/2\right)$ from Eq.~\eqref{eq:model_H}, we define the quasiperiodic drive with nontruncated Fourier modes as a reverse-engineered model, to benchmark the variational principle:

\begin{equation}\label{eq:AnalyH}
    H^{\alpha\tau,\text{rev-eng}}=R (H^\alpha_{A}+H^\tau_{B}) R^\dagger +i\omega^\mu(\partial_\mu R)R^\dagger~,
\end{equation}
where
\begin{equation}
\begin{split}
    i\omega^\mu(\partial_\mu R)R^\dagger ={}&-\frac{\omega_1}{2}\sigma_{A}^x\sigma_{B}^z\\
    &-\frac{\omega_2}{2}\sigma_{A}^y\sigma_{B}^z\cos\theta_1\\
    &+\frac{\omega_2}{2}\sigma_{A}^z\sin\theta_1.
\end{split}
\end{equation}
Its exact abelian average-energy operator is:
\begin{equation}
\begin{split}
    \mAE^{\mathrm{A}\tau,\mathrm{ref}}[\thetavec] =R(\thetavec)\left[ \epsilon\sigma_{A}^z+ \frac{\langle d(\thetavec)\rangle_{\T^2}}{d(\thetavec)} \vec{d}(\thetavec)\cdot\vec{\sigma}_{B} \right]R(\thetavec)^\dagger.
\end{split}
\end{equation}
For the nonabelian models, the corresponding small-$\epsilon$, small-$|\omegavec|$ average-energy operator is
\begin{equation}
    \mAE^{\mathrm{\bar A}\tau,\mathrm{ref}}[\thetavec] =R(\thetavec)\left[ \frac{\langle d(\thetavec)\rangle_{\T^2}}{d(\thetavec)} \vec{d}(\thetavec)\cdot\vec{\sigma}_{B} \right]R(\thetavec)^\dagger.
\end{equation}

For all four analytically solvable models, the reconstructed result at Fourier mode cutoff $K=45$ has infidelity below $10^{-12}$ relative to $\|\mAE^{\mathrm{ref}}\|$; the mode-resolved comparison is shown in Fig.~\ref{fig:analyAEnumVSref}.

The truncated models in Eq.~\eqref{eq:model_H} retain only the five lowest Fourier modes $(k_1,k_2)=(0,0),(0,1),(1,0),(0,-1),(-1,0)$ of qubit $B$ before applying the change-of-frame transformation $R$. These models are no longer exactly solvable. We define the truncation operator as
\begin{equation}
\label{eq:T1trunc}
   \mathsf{T}_1[O](\thetavec) =\sum_{|k_1|+|k_2|\leq1} O_{\kvec}\mathrm{e}^{i\kvec\cdot\thetavec}\,,
\end{equation}
which gives
\begin{equation}\label{eq:QWZtrunc}
    \centernot{H}\!\!^{\,\tau}_{B}=\mathsf{T}_1[H^\tau_{B}]\,.
\end{equation}
We also round the retained coefficients to absolute precision $10^{-7}$; their values are listed in Table~\ref{tab:H-modes-four-classes}. Additional results from solving the variational principle are illustrated in Appendix~\ref{sec:MinResQLPresults} below.

\subsection{Accuracy and Efficiency of the Variational Principle}
\label{app:varl-accuracy}

\begin{figure}[t!]
\centering\includegraphics[width=0.48\textwidth]{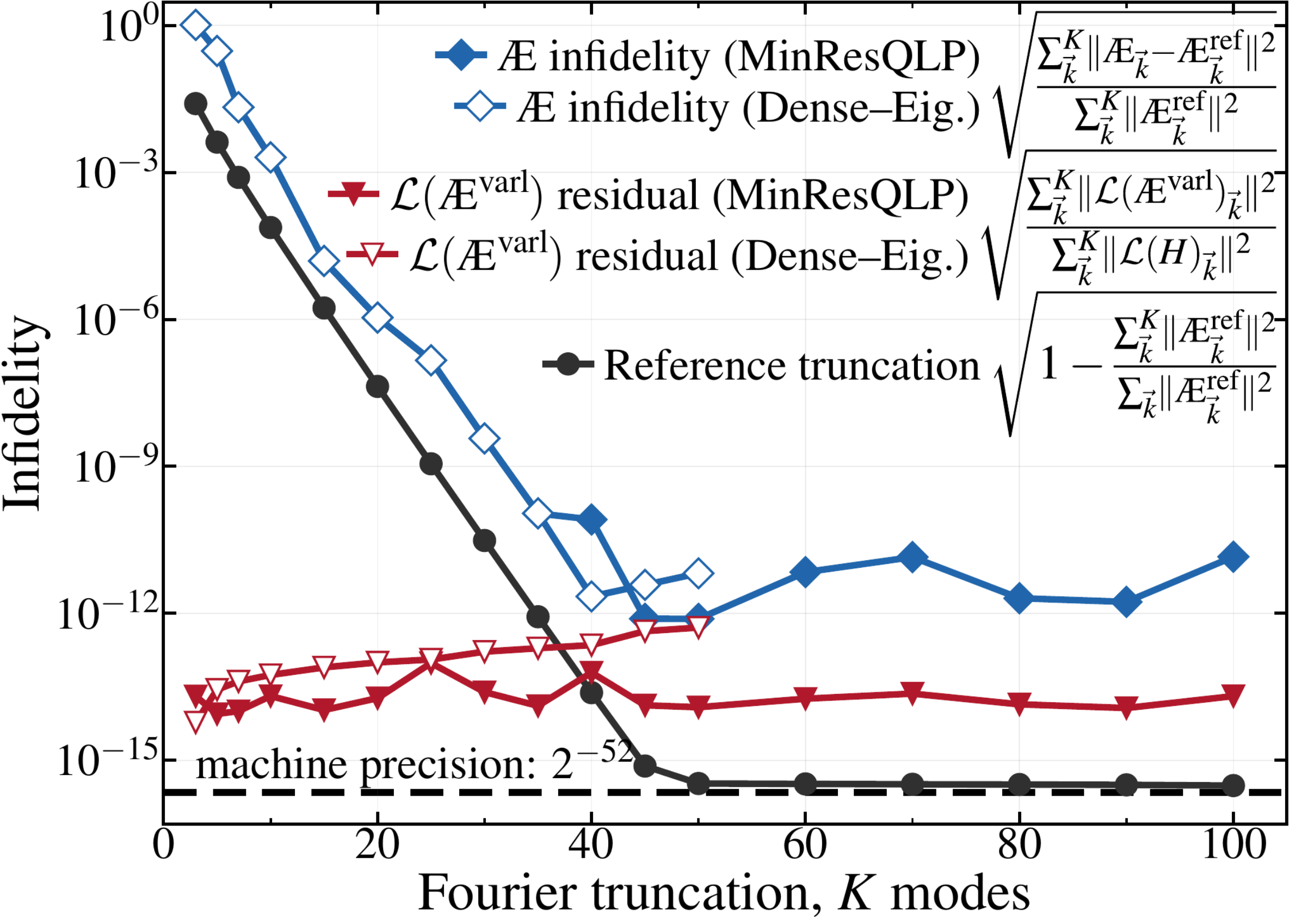}
	\caption{\textbf{Convergence of the variational average energy: comparison of numerical methods.} Same quantities, model, and drive parameters as in Fig.~\ref{fig:varlConvergence}, extended to larger Fourier cutoffs $K$. Filled blue diamonds and red triangles show MinResQLP results up to $K=100$; open markers show dense-eigendecomposition (Dense-Eig) results up to $K=50$. The variational infidelity saturates near $10^{-12}$, while the reference-truncation error (black circles) continues down to machine precision, marked by the dashed line at $2^{-52}$. Data obtained from the variational principle using the nontruncated version of the nonabelian twisted $H^{\rm \bar A T,\text{rev-eng}}$ from Eq.~\eqref{eq:AnalyH} with parameters as in Fig.~\ref{fig:analyAEnumVSref}.}
    \label{fig:FidelityComparison}
\end{figure}

We truncate the frequency lattice to $|k_1|,|k_2|\leq K$ and compute a variational average-energy operator in two numerically independent ways. The first uses MinResQLP~\cite{liu2025pseudoinverse}, an iterative Krylov-based algorithm tailored for solving our minimum-norm linear problem $\LL(A^{\mAE})=\LL(H)$ in Eq.~\eqref{eq:pseudoinverse}, followed by $\mAE^{\mathrm{varl}}=H-A^{\mAE}$.  The second performs a dense eigendecomposition (Dense-Eig) of the Liouvillian superoperator $\LL$ and projects $H$ onto modes whose eigenvalues have magnitude below the fixed threshold of $10^{-14}$, in order to obtain $\mAE^{\mathrm{varl}}$.

At cutoff $K$, the Liouvillian matrix has dimension $N^2(2K+1)^D$. Dense eigendecomposition stores the full matrix, requiring memory quadratic in this dimension. MinResQLP keeps a fixed number of Krylov vectors and applies the frequency-space convolution without forming the matrix, so its memory grows linearly. This allows the larger Fourier cutoffs used in Fig.~\ref{fig:FidelityComparison}. Furthermore, because the latter only looks for the pseudoinverse solution, it is also inherently faster than the former since the dense eigendecomposition has a complexity that scales cubically with $N^2(2K+1)^D$.

\begin{figure}[t!]
    \centering
    \includegraphics[width=0.99\columnwidth]{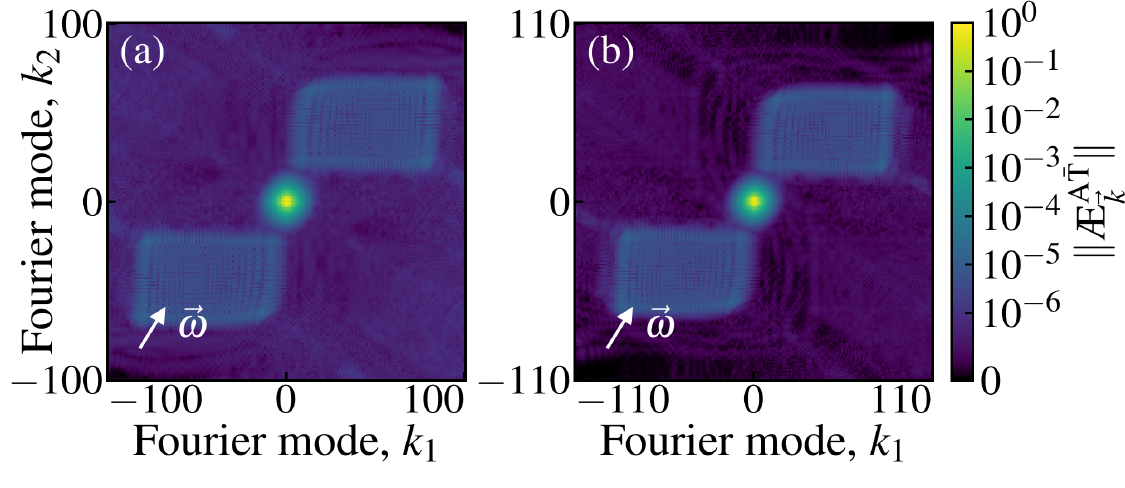}
    \caption{\textbf{Average-energy Fourier components of an abelian untwisted drive.} Matrix norms $\|\mAE^{\mathrm A\mathrm{\bar T}}_{\kvec}\|$ for the Hamiltonian in Eq.~\eqref{eq:model_H}, at Fourier cutoffs (a) $K=100$ and (b) $K=110$. Both panels use the same logarithmic color scale; arrows show the drive velocity $\omegavec$. The rectangular patches persist as the weaker background decreases with $K$. Data obtained using the variational principle of Sec.~\ref{subsec:varl}, with the non-reverse-engineered abelian untwisted model $H^{\rm A \bar T}$ and drive parameters as in Fig.~\ref{fig:abelian_nontwisting}.}
	\label{fig:AE_AT}
\end{figure}

\begin{figure}[t!]
    \centering
        \includegraphics[width=0.99\columnwidth]{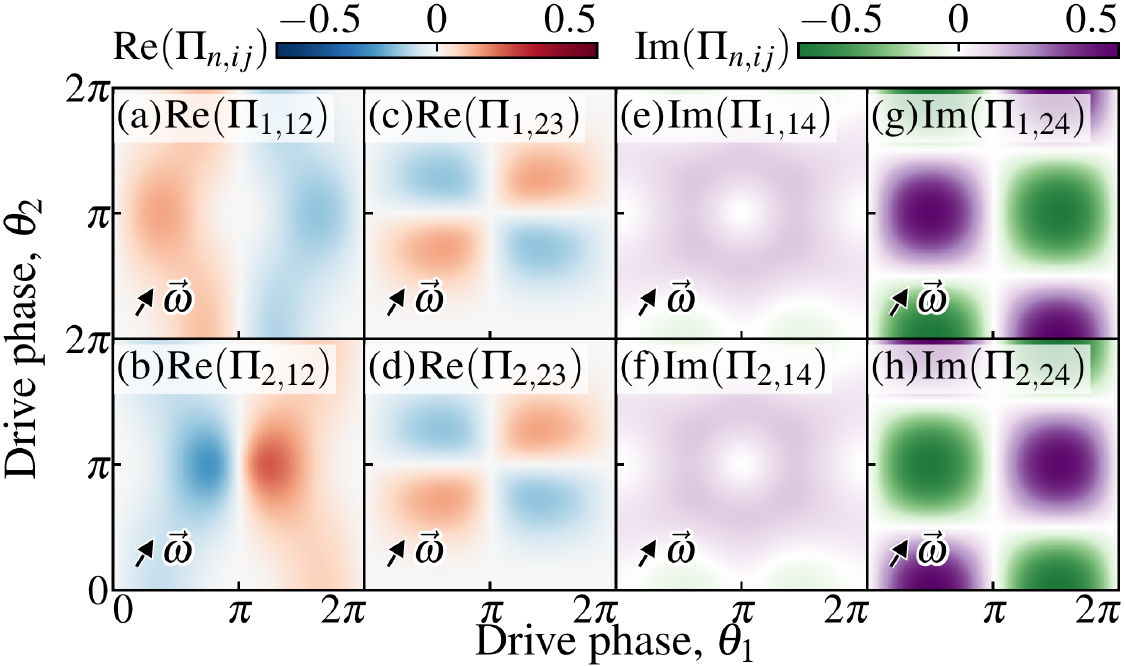}
	\caption{\textbf{Invariant projectors of an abelian untwisted drive.} Matrix components of $\Pi_n[\thetavec]$, with columns showing $\mathrm{Re}(\Pi_{n,12})$ and $\mathrm{Re}(\Pi_{n,23})$ (blue--red), followed by $\mathrm{Im}(\Pi_{n,14})$ and $\mathrm{Im}(\Pi_{n,24})$ (green--purple). The top and bottom rows show $n=1$ and $n=2$, respectively; arrows indicate $\omegavec$. The displayed invariant projectors are smooth on the torus. Data obtained by diagonalizing the variational average-energy operator at $K=110$, for the abelian untwisted model $H^{\rm A \bar T}$ and drive parameters as in Fig.~\ref{fig:abelian_nontwisting}.}
	\label{fig:projector_AT}
\end{figure}

Figure~\ref{fig:FidelityComparison} shows the convergence with $K$ of the nonabelian twisted model $H^{\mathrm{\bar A}\mathrm T,\text{rev-eng}}$ from Eq.~\eqref{eq:AnalyH}. The reference-truncation error indicates the error of the finite-$K$ truncated analytical reference $\mAE^{\rm \bar A T, ref}$ due to omission of Fourier modes: it reaches machine precision after $K=50$. Independently, the normalized Liouvillian residual $\|\LL(\mAE^{\rm \bar A T, varl})\|/\|\LL(H)\|$ measures how accurately the variational $\mAE^{\rm \bar A T, varl}$ satisfies the dynamical-invariant equation~\eqref{eq:Lqp}; using MinResQLP this residual is kept at around $10^{-14}$. Finally, the infidelity of the numerical $\mAE^{\rm \bar A T}$ relative to the analytical reference $\mAE^{\rm \bar A T, ref}$ also saturates at $10^{-12}$ once $K\geq 45$.

\subsection{Complementary results on non-solvable models}\label{sec:MinResQLPresults}

\begin{figure}[t!]
    \centering
    \includegraphics[width=0.99\columnwidth]{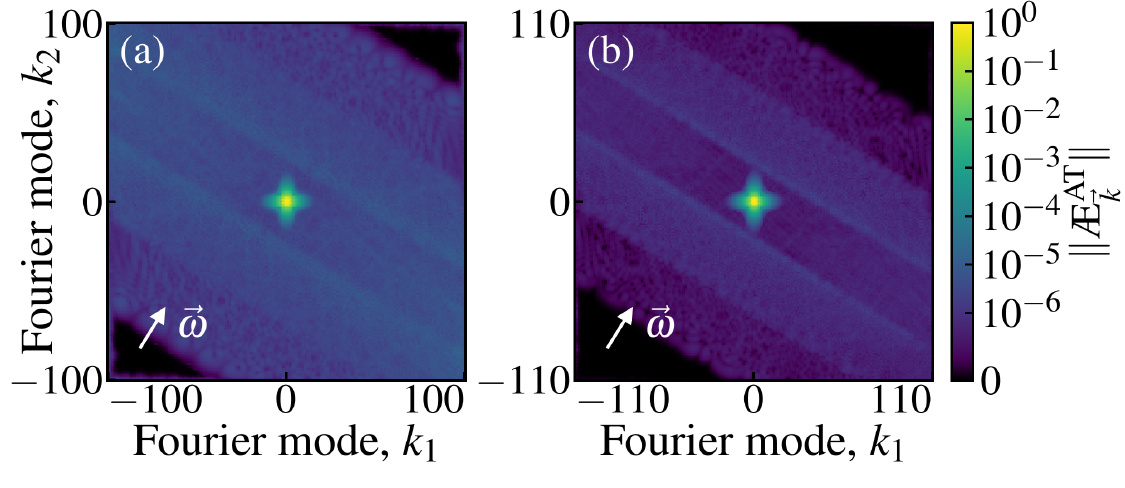}
    \caption{\textbf{Average-energy Fourier components of an abelian twisted drive.} Same quantities and color conventions as in Fig.~\ref{fig:AE_AT}, for $H^{\mathrm A\mathrm T}$ at (a) $K=100$ and (b) $K=110$. The weak strips perpendicular to $\omegavec$ remain visible at $K=110$. Method and drive parameters as in Fig.~\ref{fig:abelian_twisting}.}
	\label{fig:AE_AnT}
\end{figure}

\begin{figure}[t!]
	\centering
    \includegraphics[width=0.99\columnwidth]{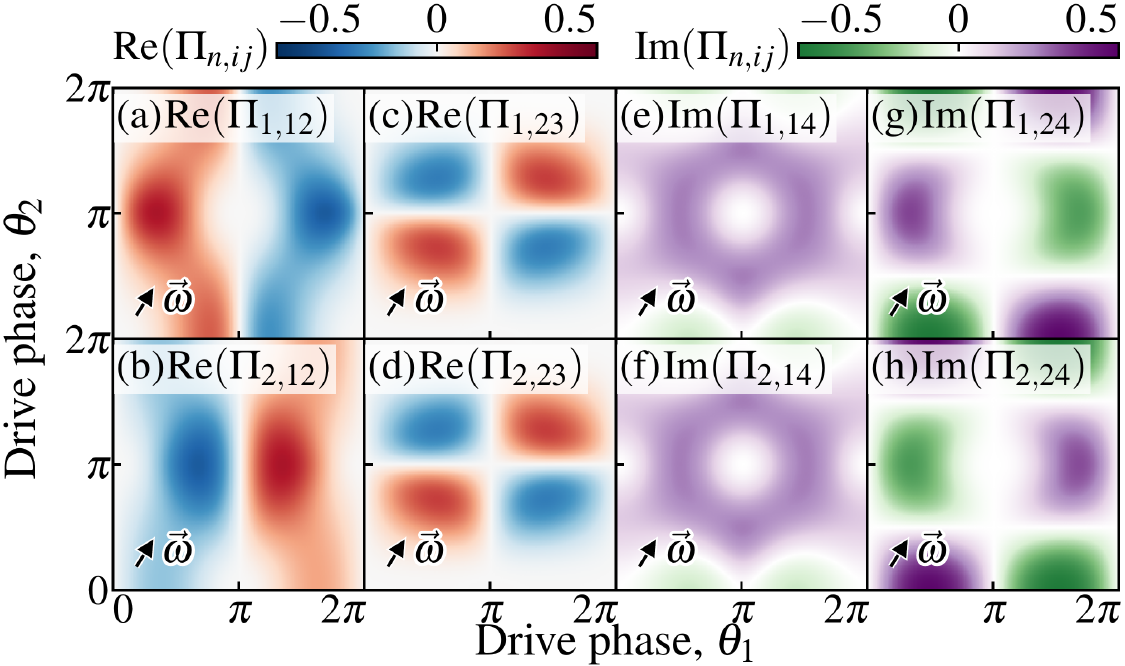}
	\caption{\textbf{Invariant projectors of an abelian twisted drive.} Same matrix components and panel arrangement as in Fig.~\ref{fig:projector_AT}, for $H^{\mathrm A\mathrm T}$ at $K=110$. The projectors remain globally smooth despite their nontrivial topology. Model and drive parameters as in Fig.~\ref{fig:abelian_twisting}.}
	\label{fig:projector_AnT}
\end{figure}

We now fix the truncated five-mode Hamiltonian from Eq.~\eqref{eq:QWZtrunc}, and increase the variational Fourier cutoff $K$. This allows us to distinguish the physical deformation caused by the discarded CD harmonics from the noisy numerical background caused by the finite variational window.

\subsubsection{Abelian untwisted (A\=T) case}

The two approximately rectangular patches in Fig.~\ref{fig:AE_AT} persist as $K$ increases. They reflect the deformation of the invariant spaces caused by discarding higher harmonics of the CD field, and therefore belong to the five-mode drive itself. By contrast, the much weaker oblique strips, perpendicular to $\omegavec$, decrease with $K$: the finite variational window has difficulty resolving these Fourier modes with small $\kvec\cdot\omegavec$, see Appendix~\ref{app:varlConvergenceProof}, but their contribution is negligible. The rank-one projectors remain smooth over the torus [Fig.~\ref{fig:projector_AT}], consistent with the abelian untwisted character of the drive.

\subsubsection{Abelian twisted (AT) case}

\begin{figure}[t!]
    \centering
    \includegraphics[width=0.99\columnwidth]{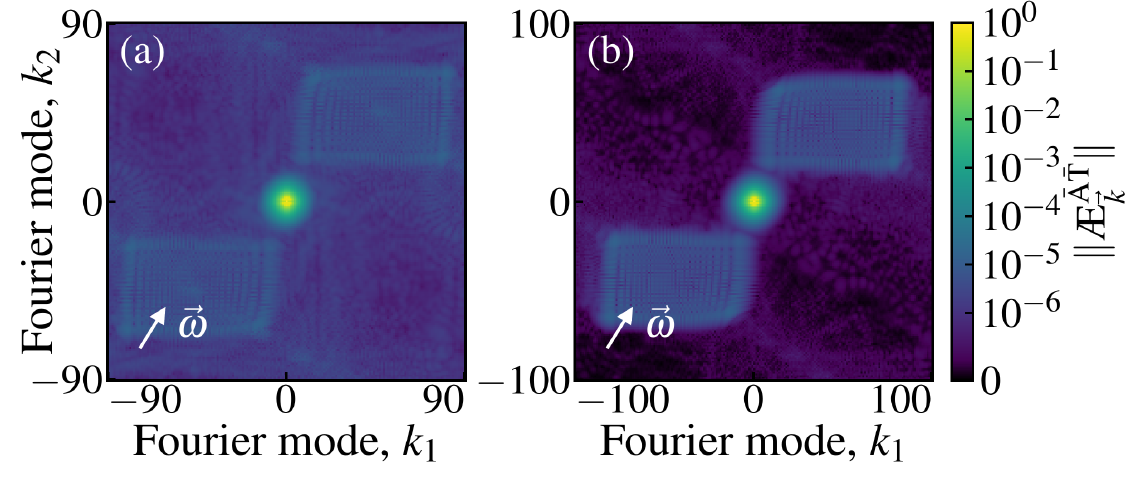}
    \caption{\textbf{Average-energy Fourier components of a nonabelian untwisted drive.} Same quantities and color conventions as in Fig.~\ref{fig:AE_AT}, for $H^{\mathrm{\bar A}\mathrm{\bar T}}$ at (a) $K=90$ and (b) $K=100$. Method and parameters as in Fig.~\ref{fig:nonabelian_nontwisting}.}
	\label{fig:AE_nAT}
\end{figure}

\begin{figure}[t!]
	\centering
    \includegraphics[width=0.99\columnwidth]{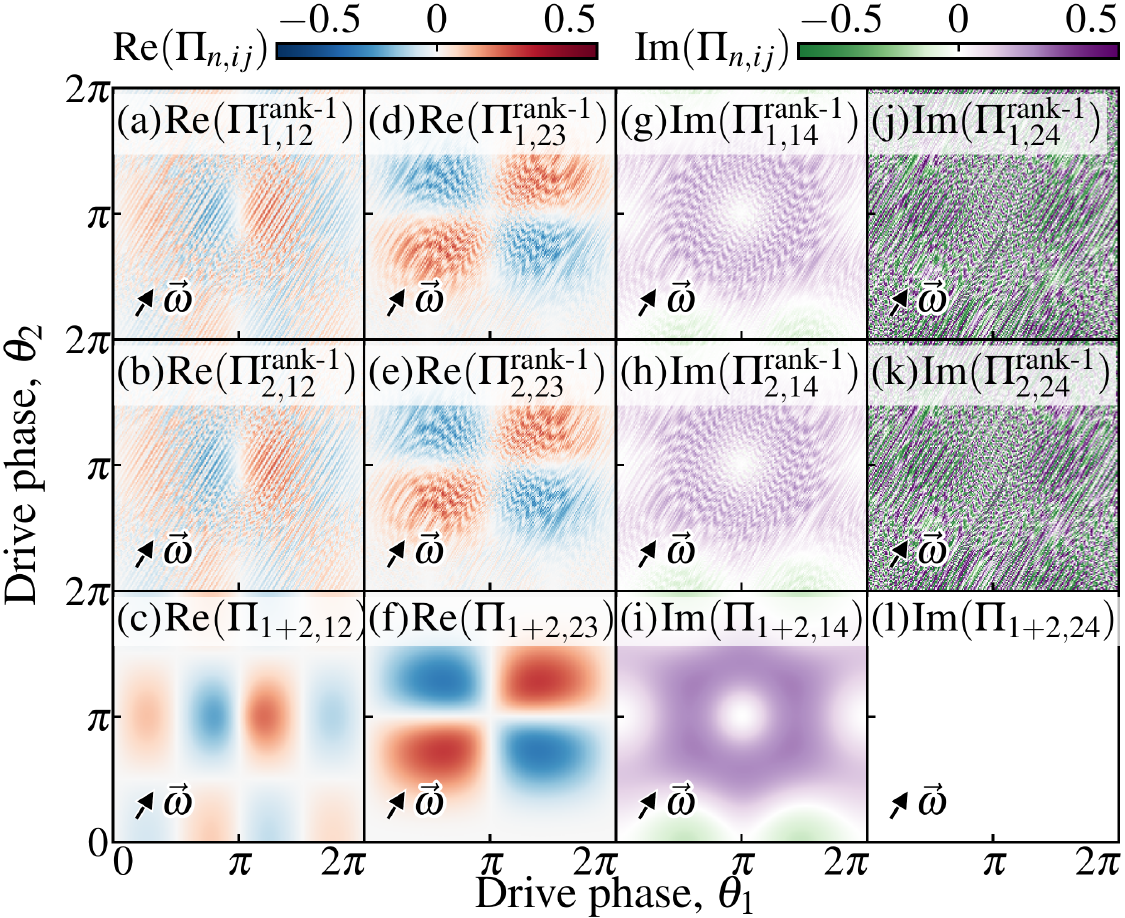}
	\caption{\textbf{Rank-one eigenprojectors and merged invariant spaces of a nonabelian untwisted drive.} The upper two rows show matrix components of the rank-one eigenprojectors $\Pi^{\mathrm{rank\text{-}1}}_1$ and $\Pi^{\mathrm{rank\text{-}1}}_2$ of $\mAE^{\mathrm{varl}}$. The bottom row shows the paired projector $\Pi_{1+2}$; matrix components and column order are as in Fig.~\ref{fig:projector_AT}. Arrows indicate $\omegavec$. The paired projectors are smooth despite the rapidly varying rank-one components. Method and drive parameters as in Fig.~\ref{fig:nonabelian_nontwisting}, at $K=100$.}
	\label{fig:projector_nAT}
\end{figure}

At $m=1$, the stronger CD field makes its five-mode truncation a larger perturbation than at $m=3$. The resulting KAM deformation of $\mAE$, due to the quasiperiodic resonances discussed in Appendix~\ref{sec:LandscapeQP}, spreads over more Fourier modes and is not yet fully resolved at $K=110$. Away from the central localized region, Fig.~\ref{fig:AE_AnT} shows a weak strip perpendicular to $\omegavec$ whose intensity still decreases with $K$. This remaining finite-window background obscures part of the deformation. The torus components in Fig.~\ref{fig:projector_AnT} show smooth rank-one invariant projectors: their nontrivial topology does not obstruct a smooth projector description.

\subsubsection{Nonabelian untwisted (\=A\=T) case}

\begin{figure}[t!]
    \centering
        \includegraphics[width=0.99\columnwidth]{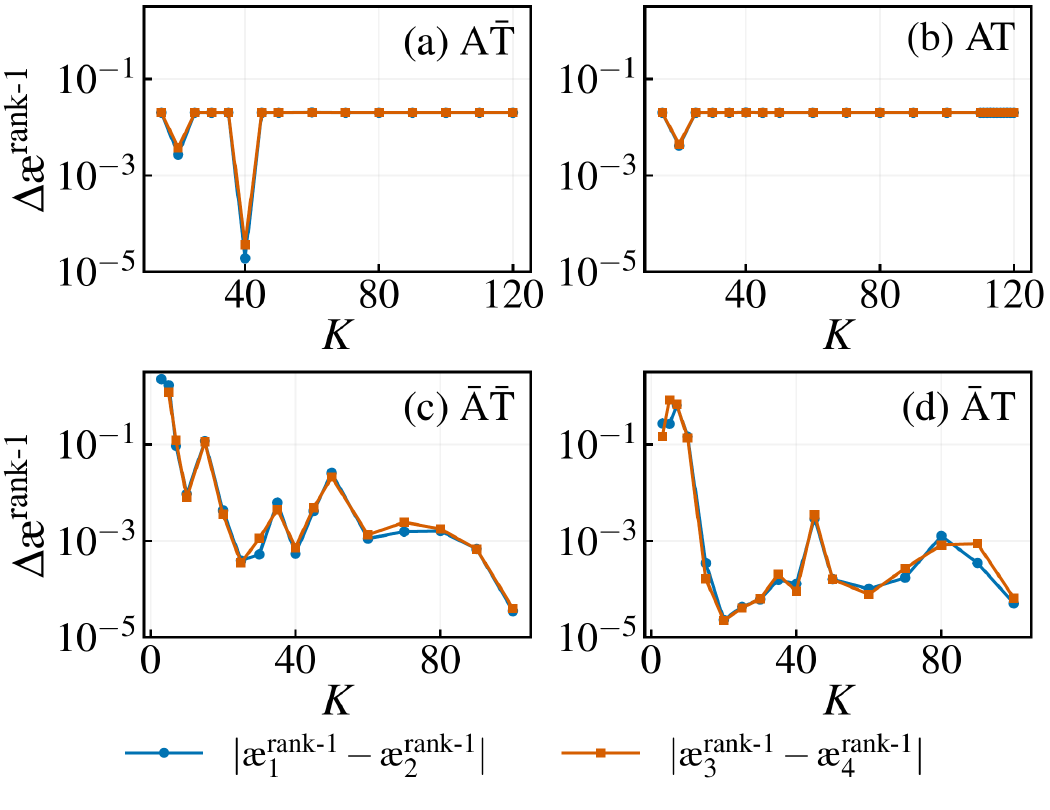}
	\caption{\textbf{Average-energy pair splittings in the four drive classes.} Splittings $|\mae^{\text{rank-1}}_1-\mae^{\text{rank-1}}_2|$ (blue) and $|\mae^{\text{rank-1}}_3-\mae^{\text{rank-1}}_4|$ (orange) between the eigenvalues of the finite-cutoff $\mAE^{\mathrm{varl}}$, as functions of $K$. Panels (a,b) show abelian untwisted and twisted truncated drives; (c,d) show nonabelian untwisted and twisted truncated drives. The nonabelian pairs approach degeneracy as $K$ increases, while the abelian splittings remain of order $\epsilon$. Data obtained using the variational principle of Sec.~\ref{subsec:varl}, with drive parameters as in Fig.~\ref{fig:nonabelian_nontwisting}.}
	\label{fig:ae_splitting}
\end{figure}
As in the abelian untwisted case, Fig.~\ref{fig:AE_nAT} shows persistent rectangular patches caused by truncating the drive, together with a shrinking background from the finite variational window. This similarity with the abelian case is due to the fact that the truncation does not act on the qubit $A$ which controls the (dynamically irreducible) abelianness.

However, for the nonabelian drive, increasing $K$ must also recover the degeneracy within each rank-two invariant space. At finite $K$, diagonalizing $\mAE^{\mathrm{varl}}=\sum_n\mae_n^\text{rank-1}\Pi_n^\text{rank-1}$ gives four rank-one eigenprojectors with small pair splittings [Fig.~\ref{fig:ae_splitting}(c)]. These splittings fluctuate with $K$ but become much smaller than the perturbation scale $\epsilon$, whereas the abelian splittings remain of order $\epsilon$.

The rapid variation of the individual projectors cancels when each pair is added, giving the two smooth rank-two projectors $\Pi_{1+2}$ and $\Pi_{3+4}$ [one pair is shown in Fig.~\ref{fig:projector_nAT}]. Thus, the rank-one splitting of $|\mae^\text{rank-1}_n-\mae^\text{rank-1}_m|\ll 1$ is a numerical artifact, and does not describe four separate invariant spaces.

\subsubsection{Nonabelian twisted (\=AT) case}

\begin{figure}[t!]
    \centering
    \includegraphics[width=0.99\columnwidth]{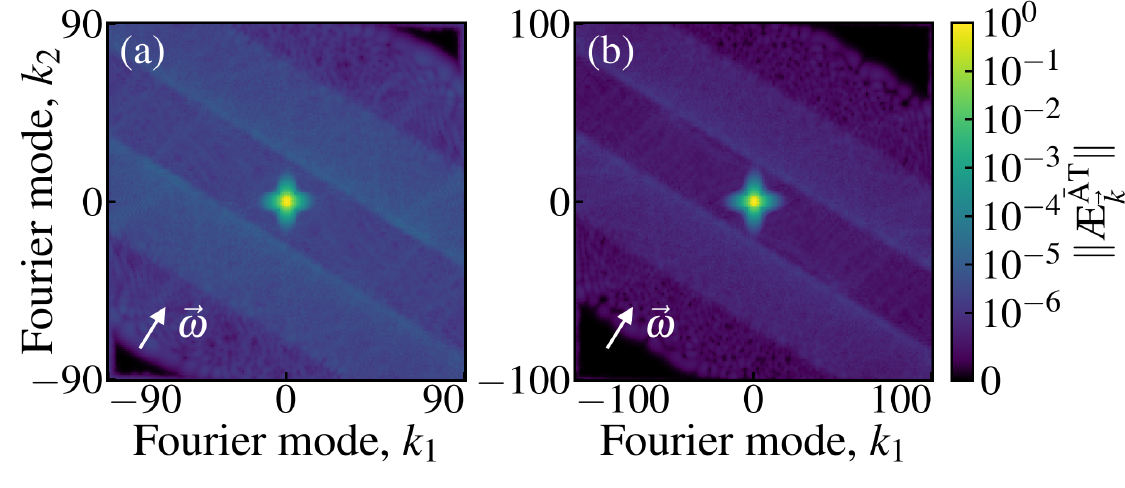}
    \caption{\textbf{Average-energy Fourier components of a nonabelian twisted drive.} Same quantities and color conventions as in Fig.~\ref{fig:AE_AT}, for $H^{\mathrm{\bar A}\mathrm T}$ at (a) $K=90$ and (b) $K=100$. Method and drive parameters as in Fig.~\ref{fig:nonabelian_twisting}.}
	\label{fig:AE_nAnT}
\end{figure}

\begin{figure}[t!]
	\centering
    \includegraphics[width=0.99\columnwidth]{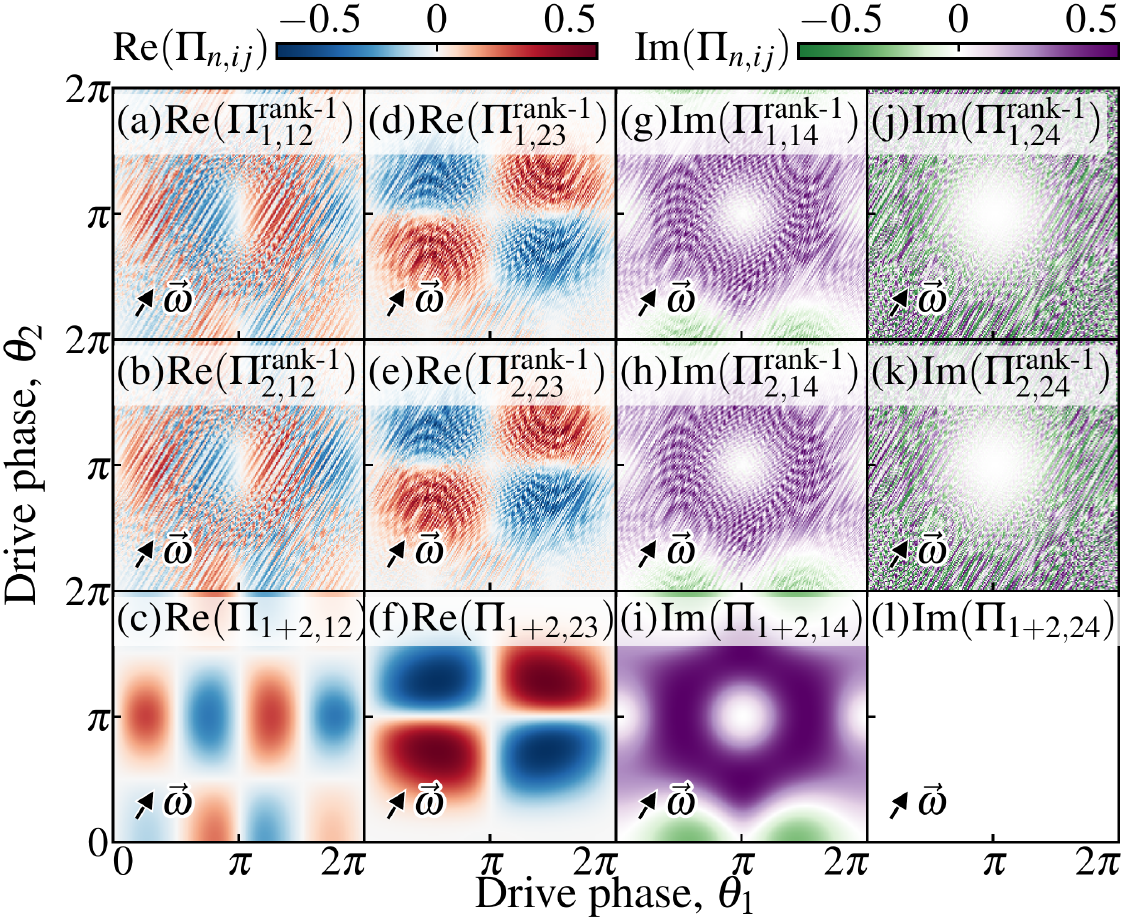}
	\caption{\textbf{Rank-one eigenprojectors and merged invariant spaces of a nonabelian twisted drive.} Same matrix components and panel arrangement as in Fig.~\ref{fig:projector_nAT}, for $H^{\mathrm{\bar A}\mathrm T}$ at $K=100$. The rank-two projectors remain smooth also in the twisted class. Model and drive parameters as in Fig.~\ref{fig:nonabelian_twisting}.}
	\label{fig:projector_nAnT}
\end{figure}

\begin{figure}[t!]
    \centering
    \includegraphics[width=0.99\columnwidth]{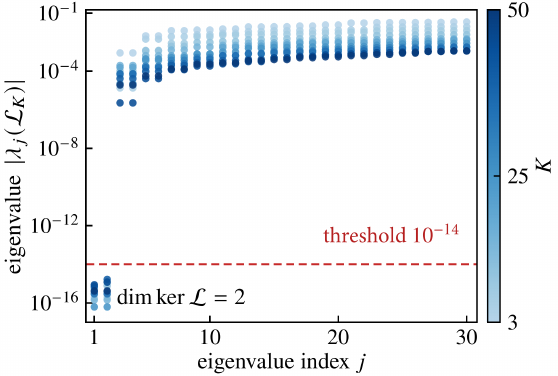}
    \caption{\textbf{Liouvillian zero modes of a nonabelian twisted drive.} The 30 smallest eigenvalues $|\lambda_j(\LL_K)|$ by absolute value for $H^{\mathrm{\bar A}\mathrm T}$, with Fourier cutoffs $K=3$--$50$ indicated by color. The red dashed line marks the kernel threshold. The colorbar shows the truncation cutoff $K$. The two eigenvalues at machine precision give $\dim\ker\LL=2$. Model and drive parameters as in Fig.~\ref{fig:nonabelian_twisting}.}
    \label{fig:liouvillianKernelSpectrum}
\end{figure}

The twisted case gives the same pairwise reconstruction: the small rank-one splittings in Fig.~\ref{fig:ae_splitting}(d) decrease overall with $K$, while $\Pi_{1+2}$ and $\Pi_{3+4}$ remain smooth on the torus [shown for $\Pi_{1+2}$ in Fig.~\ref{fig:projector_nAnT}].

The Liouvillian spectrum provides another independent numerical check. In Fig.~\ref{fig:liouvillianKernelSpectrum}, two modes remain at machine precision while the remaining displayed modes lie orders of magnitude higher, above the kernel threshold. The calculation therefore gives $\dim\ker\LL=2$, matching two rank-two invariant spaces rather than four independent rank-one spaces.

Their nontrivial topology is compatible with smooth projectors; the Fourier background continues to decrease as the variational window is enlarged [Fig.~\ref{fig:AE_nAnT}] just like in Fig.~\ref{fig:AE_AnT} of the abelian twisted case; once again, this is because the truncation does not act on qubit $A$ that controls abelianness.

\section{Landscape of Nonabelian Drives}
\label{sec:LandscapeQP}

Generic quasiperiodic drives are likely to be nonabelian, since the set of nonabelian drives contains a countable intersection of open dense subsets of the space of all quasiperiodic drives~\cite{nerurkar1997construction,nerurkar2007construction,eliasson2002ergodic}.\footnote{However, when the time-varying part of a quasiperiodic drive is sufficiently weak compared with its nondegenerate constant part, nonabelian regions can be rare in measure along a generic one-parameter scan~\cite{her2008full,krikorian1999presque}; moreover, high-frequency ($|\omegavec|\gg1$) drives are also predominantly abelian~\cite{verdeny2016quasi}.}

In the language of dynamical systems, the nonabelian drives appear where rank-one invariant spaces are mixed and resonant. These resonances form nested Arnold-tongue-like regions~\cite{puig2002reducibility} with a fractal-like boundary. Consequently, an arbitrarily small parameter change can encounter a high-order resonance even when the drive remains abelian for most nearby parameter values.

These resonances can be studied through a microscopic mechanism called the small-divisor structure of quasiperiodic perturbation theory~\cite{kolmogorov1954conservation,eliasson2002ergodic}. Dressing perturbatively two rank-one dynamically invariant spaces introduces denominators
\begin{equation}\label{eq:smallDivisorF}
    \left|\epsilon^\F_n-\epsilon^\F_m-\kvec\cdot\omegavec\right| , \qquad \kvec\in\mathbb{Z}^D.
\end{equation}
Resonances between the levels $\epsilon^\F_n$ and $\epsilon^\F_m$ are related by a ``frequency-lattice driving photon'' of energy $\kvec\cdot\omegavec$, where the small denominator in Eq.~\eqref{eq:smallDivisorF} decays more rapidly than a certain power of the wavelength of these ``photons'' $|\kvec|^{1-D}$. Away from these resonances, globally smooth dynamically invariant spaces remain rank-one and thus abelian~\cite{eliasson2002ergodic,her2008full,xue2020reducibility}. Near a quasiperiodic resonance, where $|\epsilon^\F_n-\epsilon^\F_m-\kvec\cdot\omegavec|\approx 0$ but does not exactly vanish, these spaces can remain abelian while becoming strongly delocalized on the frequency lattice; this signals the proximity to points where the rank-one description breaks down, and already motivates combining the resonant spaces. On resonance, $|\epsilon^\F_n-\epsilon^\F_m-\kvec\cdot\omegavec|= 0$, when there is no globally smooth rank-one solution of Eq.~\eqref{eq:L}, the originally rank-one spaces instead form an irreducible nonabelian dynamically invariant space. From a practical point of view, for finite Fourier-mode truncation, the latter two cases are both effectively nonabelian, as shown in Sec.~\ref{sec:ergodicity}.

An exactly degenerate abelian drive with $\mae_n=\epsilon^\F_n=0$ (hence no geometric phases) is a natural starting point for studying the onset of nonabelian dynamics because its rank-one solution is not protected by an $\mae$-gap, and average energies are analytically easier to manipulate than quasienergies. Starting from $H_0(\thetavec)=(\cos\theta_1+\cos\theta_2)\sigma^z$, a longitudinal perturbation $\epsilon\sigma^z$ preserves the commuting rank-one spaces, whereas a transverse perturbation $\epsilon\sigma^x$ couples them through dense crossings of the two energy levels of $H_0$.

Along the quasiperiodic trajectory, the drive $h(t)=\cos\theta_1(t)+\cos\theta_2(t)$ repeatedly changes sign. The transverse perturbation opens a gap $2|\epsilon|$ where the two unperturbed energy levels cross. Near a crossing at $t_c$, the Hamiltonian has the Landau-Zener form $H(t)\simeq (\D h(t_c)/\D t)(t-t_c)\sigma^z+\epsilon\sigma^x$, allowing transitions between its instantaneous eigenstates. The relative phase accumulated between crossings affects the subsequent transitions. For two incommensurate frequencies, this sequence does not repeat periodically. When $\epsilon$ gets smaller and smaller, so does the energy gap of these Landau-Zener crossings, which densely sample the level-crossing lines $\cos\theta_1+\cos\theta_2=0$ on $\T^2$ because $\omega_2/\omega_1$ is irrational. Physically, this means that when the perturbation $\epsilon \sigma^x$ is weak enough, the accumulated quasiperiodic crossings would obstruct the existence of a smooth rank-one projector solution $\Pi_1$ to the equation for dynamical invariants $\LL(X)=0$.

To express this mixing on the frequency lattice, we change to the reference frame $R(\thetavec)=\mathrm{e}^{-if(\thetavec)\sigma^z}$, with $f=\sin\theta_1/\omega_1+\sin\theta_2/\omega_2$, which removes the evolution generated by $H_0$. The remaining Hamiltonian is
\begin{equation}\label{eq:qubitA-Bessel}
\begin{aligned}
    H_{\mathrm{rot}} &=R^\dagger HR-iR^\dagger\omega^\mu\partial_\mu R\\
    &=\epsilon\bigl[\cos(2f)\sigma^x-\sin(2f)\sigma^y\bigr],\\
    \langle H_{\mathrm{rot}}\rangle_{\T^2}&=\Delta_0\sigma^x,\\
    \Delta_0&=\epsilon J_0(2/\omega_1)J_0(2/\omega_2),
\end{aligned}
\end{equation}
where $J_0$ is the Bessel function of the first kind. The time-averaged Hamiltonian plays the role of a leading-order approximated Floquet Hamiltonian $H^\F$ \cite{verdeny2016quasi}, and it has constant eigenvalues $\pm\Delta_0$. Its two eigenspaces are coupled by the remaining Fourier components, oscillating at $\kvec\cdot\omegavec$. To lowest order, the KAM expansion corrects their projectors by dividing these couplings by $\kvec\cdot\omegavec\pm2\Delta_0$~\cite{eliasson2002ergodic,karaliolios2018continuous}. Near the adiabatic limit where $|\omegavec|\to0$, $J_0(2/\omega_1)$ and $J_0(2/\omega_2)$ oscillate; for a Diophantine frequency vector with $|\omegavec|\approx 0$, the weak Fourier amplitude of mode $\kvec$ can therefore strongly mix the two spaces near the small KAM divisors $|\epsilon^\F_1-\epsilon^\F_2-\kvec\cdot\omegavec|$ as
\begin{equation}\label{eq:qubitA-resonance}
    |\kvec\cdot\omegavec \pm2\Delta_0| \simeq0\,.
\end{equation}
If such mixing at successively higher Fourier modes prevents the individual projectors from converging to smooth functions on the torus, the rank-two invariant space cannot be separated into rank-one dynamically invariant subspaces~\cite{eliasson2002ergodic,karaliolios2018continuous}.

\section{Mathematical Details}\label{sec:Diophantine}

We first establish the existence of continuous invariant spaces in Sec.~\ref{app:invariant-space-existence}. Sections~\ref{app:varlConvergenceProof} and~\ref{app:minimum-norm-refinement} prove convergence of the variational construction and explain why minimum norm favors refinement into dynamically irreducible spaces. Section~\ref{app:diophantine-counterexamples} gives counterexamples to the Diophantine-approximated methods, and Sec.~\ref{app:avila-comparison} compares our construction with previous work, including Avila's global theory.

\subsection{Existence and Continuity of Invariant Spaces}\label{app:invariant-space-existence}

We assume a finite-dimensional Hermitian Hamiltonian $H(\thetavec)$ with $C^1(\T^D)$ dependence on the drive phases and an incommensurate frequency vector $\omegavec$.  A continuous Hermitian operator $O[\thetavec]$ defined on the torus $\T^D$ is a dynamical invariant when, for every $\thetavec_0$,
\begin{equation}\label{eq:invariantCovarianceProof}
    O[\thetavec_t]=U(t,0)O[\thetavec_0]U(t,0)^\dagger, \qquad \thetavec_t=\omegavec t+\thetavec_0,
\end{equation}
or, equivalently, $\LL(O)=0$. The covariant formulation then implies that the identity and the complement of any invariant projector are invariant.

Consider the nonzero continuous invariant projectors $0<\Pi\leq\mathds{1}_{N\times N}$. Their ranks are constant over the connected drive-phase space because $\text{tr}\,\Pi[\thetavec]$ is continuous and integer-valued.  Choose a projector of the smallest available rank.  It is dynamically irreducible: a proper nonzero invariant subprojector would have smaller rank.  Removing this projector and repeating the argument in its orthogonal complement gives
\begin{equation}\label{eq:irreducibleResolutionProof}
    \mathds{1}_{N\times N}=\sum_{n=1}^{J}\Pi_n[\thetavec], \qquad \LL(\Pi_n)=0, \qquad J\leq N,
\end{equation}
after at most $N$ steps.  This proves the existence of a maximally refined decomposition into continuous dynamically invariant spaces.  The construction also includes the fully irreducible case $J=1$ and $\Pi_1=\mathds{1}_{N\times N}$.  The decomposition need not be unique when there is accidental degeneracy.

Incommensurability makes the decomposition recoverable once the full space of continuous zero modes is known.  Indeed, an invariant operator that vanishes at one phase point vanishes along its dense orbit and hence everywhere.  Thus evaluation at one point embeds the real vector space of continuous Hermitian zero modes into the $N^2$-dimensional real vector space of Hermitian matrices. Denote this zero-mode space by $\mathcal{V}_{\mathrm{dyn}}$.  If a basis $\{O_a\}$ of it has been obtained, an invariant projector $\Pi$ can be tested using
\begin{equation}
    \mathcal{V}_{\Pi}=\{\Pi O\Pi:O\in\mathcal{V}_{\mathrm{dyn}}\}.
\end{equation}
Then $\Pi$ is dynamically irreducible if and only if $\mathcal{V}_{\Pi}=\mathbb{R}\Pi$.  If this condition fails, a nonscalar Hermitian $B\in\mathcal{V}_{\Pi}$ has phase-independent eigenvalues because it is transported unitarily.  Its spectral projectors are continuous invariant proper subprojectors of $\Pi$. Repeating this finite matrix diagonalization terminates after finitely many rank reductions and produces Eq.~\eqref{eq:irreducibleResolutionProof}.

\subsection{Convergence of the Finite-cutoff Variational Principle}
\label{app:varlConvergenceProof}

Fourier-mode truncation slightly spoils dynamical invariance. We show that restoring the omitted modes while gradually tightening the invariance condition recovers the dynamically invariant part of $H$, without requiring a gap around the Liouvillian zero modes.

We use $\|O\|^2=\langle O,O\rangle_{\T^D}$ with the inner product in Eq.~\eqref{eq:inner_product}. The subscript $K$ denotes the retained Fourier modes $|k_\mu|\leq K$, and $\LL_K O=(\LL O)_K$ for $O=O_K$: the original Liouvillian acts on the retained variable and its output is truncated to the same modes. For the smooth finite-dimensional Hermitian drive considered here, $\LL$ is self-adjoint on the space of operators with a square-integrable derivative along $\omegavec$. Fourier truncation generally shifts the nontrivial invariant modes away from zero, so at finite $K$ we allow a small invariance error through the regularized variational problem:
\begin{equation}
\label{eq:regularizedAEConvergence}
\begin{aligned}
    \mAE_{K,\eta} &=\underset{O=O_K=O^\dagger}{\operatorname{argmin}} \left\{\|\LL_K O\|^2+\eta^2\|O-H_K\|^2\right\}\\
    &=\frac{\eta^2}{\LL_K^2+\eta^2}H_K.
\end{aligned}
\end{equation}
The first term enforces dynamical invariance; the second keeps as much of the Hamiltonian as possible in the invariant part. The resulting filter retains modes close to zero and suppresses the others. Narrowing its width $\eta$ as more Fourier modes are restored should therefore leave precisely $\mAE$.

Formally, to see how slowly the width must shrink, we truncate the exact invariant $\mAE$. Fourier truncation commutes with the derivative along $\omegavec$, so
\begin{equation}
\label{eq:physicalModeRecoveryProof}
    \LL_K\bigl((\mAE)_K\bigr) =-\bigl[H,\mAE-(\mAE)_K\bigr]_K \longrightarrow0.
\end{equation}
Thus its apparent failure of invariance comes entirely from the omitted Fourier tail. Since $H$ is bounded, this error vanishes as the tail is restored. We choose $\eta_K\to0$ slowly enough that
\begin{equation}\label{eq:shrinkingWindowProof}
    \frac{\|\LL_K((\mAE)_K)\|}{\eta_K}\longrightarrow0.
\end{equation}
This prevents the filter from discarding the dynamically invariant component while its zero mode is still displaced by truncation; no gap around zero is required.

The truncated exact invariant is an allowed trial operator in Eq.~\eqref{eq:regularizedAEConvergence}. Minimality therefore gives
\begin{equation}
\begin{aligned}
    &\|\mAE_{K,\eta_K}-H_K\|^2\\
    &\quad\leq\|(\mAE)_K-H_K\|^2 +\frac{\|\LL_K((\mAE)_K)\|^2}{\eta_K^2},\\
    &\|\LL_K(\mAE_{K,\eta_K})\|^2\\
    &\quad\leq\|\LL_K((\mAE)_K)\|^2 +\eta_K^2\|(\mAE)_K-H_K\|^2.
\end{aligned}
\end{equation}
The second bound eliminates the invariance error. Testing against any fixed finite Fourier-mode operator then shows that every weak limit obeys the original equation $\LL(O)=0$. The first bound says that this invariant is no farther from $H$ than $\mAE$ itself. Since $\mAE$ is the unique closest invariant, every such limit is $\mAE$; the same distance bound upgrades weak convergence to convergence in norm:
\begin{equation}\label{eq:QKConvergenceProof}
    \|\mAE_{K,\eta_K}-\mAE\|\longrightarrow0.
\end{equation}
The calculation therefore recovers the full dynamically invariant contribution of $H$. This proves convergence of $\mAE$ obtained through the $K$-truncated variational principle in the torus-averaged norm.

Intuitively, the finite Fourier-mode cutoff $K$ degrades the zero-mode property of the approximate truncated invariant, $\|\LL_K((\mAE)_K)\|=\delta_K>0$. Since truncation commutes with the derivative along $\omegavec$ but not with the commutator, $\delta_K$ is controlled by the Fourier weight of $\mAE$ outside the cutoff, $\delta_K\le 2\max_{\thetavec\in\T^D}\|H(\thetavec)\|\,\|\mAE-(\mAE)_K\|$. When the exact $\mAE$ is analytic on $\T^D$ and hence exponentially localized in Fourier space, the error is exponentially suppressed, $\delta_K\sim\exp(-K/\xi)$, for some model-dependent localization length $\xi$; when $\mAE$ is only $C^s$, the decay is algebraic, $\delta_K\sim K^{-s}$. In either case, for sufficiently large cutoff value $K$ the violation of the dynamical invariant property can be controllably pushed below machine precision.

Notice that the filter width $\eta_K$ is constrained from both sides. (i) From below, condition~\eqref{eq:shrinkingWindowProof} requires $\delta_K\ll\eta_K$, so that the truncation-displaced invariants are retained by the filter; this is the only condition used in the convergence proof. (ii) From above, the filter should not admit any spurious near-zero modes of $\LL_K$. The closest ones are the shifted invariants $E_{\vec l}\,\Pi_n$, with $E_{\vec l}=e^{i\vec l\cdot\vec\theta}$, which are exact eigenmodes of $\LL$ with eigenvalue $\vec l\cdot\omegavec$, and are present in the truncated problem for $|\vec l|\lesssim 2K$. For a Diophantine frequency vector, $|\vec l\cdot\omegavec|\ge\gamma|\vec l|^{-\tau}$ for all $\vec l\in\mathbb Z^D\setminus\{0\}$, their eigenvalues are bounded away from zero by $\gamma K^{-\tau}$ at cutoff $K$.\footnote{For instance, for $D=2$ and the golden ratio, we have $\tau=1$. And almost all $\omegavec\in\mathbb{R}^D$ are Diophantine frequency vectors.} Hence, the filter excludes them provided $\eta_K\ll\gamma K^{-\tau}$.

A controlled approximation therefore requires a window
\begin{equation}
\label{eq:varl_regimes}
  e^{-K/\xi}\;\ll\;\eta_K\;\ll\;\gamma K^{-\tau} \quad \text{or} \quad K^{-s}\;\ll\;\eta_K\;\ll\;\gamma K^{-\tau}
\end{equation}
which exists for $K\gg\xi\,\tau\log K$ in the analytic case; for algebraic decay, $\delta_K\sim K^{-s}$, the window exists only if $s>\tau$. The width of the window thus reflects the competition between the Fourier localization of the invariants and the arithmetic of $\omegavec$: for Liouville frequencies\footnote{Liouville frequencies permit resonances closer than any inverse power of the Fourier order. They form a set of zero Lebesgue measure; almost every frequency vector is Diophantine~\cite{puig2002reducibility}.} the lower Diophantine bound $\gamma K^{-\tau}$ fails along a subsequence of $K$ and the window closes intermittently. Note that the convergence of $\mAE_{K,\eta_K}$ to $\mAE$ as $K\to\infty$ holds under the lower constraint in Eq.~\eqref{eq:varl_regimes} alone; the upper constraint controls the convergence rate.

In practice, the dense-eigendecomposition kernel threshold and the MinResQLP stopping tolerance play the role of a fixed $\eta$, which is consistent with the above once $K$ exceeds a few localization lengths, i.e., once $\delta_K$ has dropped below that threshold.

\subsection{Minimum Norm and Dynamical Irreducibility}
\label{app:minimum-norm-refinement}

We now show why the spectral decomposition of the converged $\mAE$ in general directly separates dynamically irreducible invariant spaces. Splitting a reducible invariant space into finer dynamically invariant subspaces lowers the AGP norm $\|A^\mAE\|$. To show this, consider a smooth invariant rank-two space containing two orthogonal smooth rank-one invariant subspaces,
\begin{equation}
    \Pi_{1+2}=\Pi_1+\Pi_2,\qquad \Pi_1\Pi_2=0, \qquad \LL(\Pi_1)=\LL(\Pi_2)=0.
\end{equation}
Equation~\eqref{eq:aeEnergy} fixes their average energies and that of the combined space:
\begin{equation}
\begin{aligned}
    \mae_1&=\langle\Pi_1,H\rangle_{\T^D},\qquad \mae_2=\langle\Pi_2,H\rangle_{\T^D},\\
    \mae_{1+2}&=\frac{\mae_1+\mae_2}{2}.
\end{aligned}
\end{equation}
Subtracting either $\mae_{1+2}\Pi_{1+2}$ or $\mae_1\Pi_1+\mae_2\Pi_2$ from $H$ gives two different AGPs $A^\mAE$ satisfying $\LL(A^\mAE)=\LL(H)$. Keep every other invariant space explicitly in both AGPs:
\begin{equation}
\begin{aligned}
    A^\mAE_{\mathrm{merged}} &=H-\mae_{1+2}\Pi_{1+2} -\sum_{m: \Pi_m\Pi_{1+2}=0}\mae_m\Pi_m,\\
    A^\mAE_{\mathrm{split}} &=H-\mae_1\Pi_1-\mae_2\Pi_2 -\sum_{m: \Pi_m\Pi_{1+2}=0}\mae_m\Pi_m.
\end{aligned}
\end{equation}
The sums run over all remaining invariant spaces, with arbitrary ranks $r_m$, so $\Pi_{1+2}+\sum_{m: \Pi_m\Pi_{1+2}=0}\Pi_m=\mathds{1}_{N\times N}$. Both operators act on the full $N$-dimensional Hilbert space and satisfy $\LL(A^\mAE)=\LL(H)$. Their Kato transport generators also involve the complete projector resolutions:
\begin{equation}
\begin{aligned}
    A^\K_{\mathrm{merged}} &=-\frac{i\omega^\mu}{2}\Bigl([\Pi_{1+2},\partial_\mu\Pi_{1+2}]+\sum_{m:\Pi_m\Pi_{1+2}=0}[\Pi_m,\partial_\mu\Pi_m]\Bigr),\\
    A^\K_{\mathrm{split}} &=-\frac{i\omega^\mu}{2}\Bigl(\sum_{j=1}^{2}[\Pi_j,\partial_\mu\Pi_j]+\sum_{m:\Pi_m\Pi_{1+2}=0}[\Pi_m,\partial_\mu\Pi_m]\Bigr).
\end{aligned}
\end{equation}

For each internal block, we use $\Pi_mH\Pi_m=E_m^\K\Pi_m+\widetilde H_m$ from Eq.~\eqref{eq:tensor_decomp}, where $E_m^\K=\mathrm{tr}(\Pi_mH)/r_m$ and $\mathrm{tr}\,\widetilde H_m=0$. Equation~\eqref{eq:A_AE} then separates the average-energy AGP into Kato transport, a scalar $\mathrm{U}(1)$ term $(E_m^\K-\mae_m)\Pi_m$, and the traceless internal rotation $\widetilde H_m$. Their cross terms vanish pointwise: Kato transport has no diagonal block in its own projector resolution, and the scalar term is orthogonal to the traceless one. With the torus averages and norms used above, the two full squared norms are therefore
\begin{equation}\label{eq:sqNormAs}
\begin{aligned}
    \|A^\mAE_{\mathrm{merged}}\|^2 &=\|A^\K_{\mathrm{merged}}\|^2+\Bigl\langle2(E_{1+2}^\K-\mae_{1+2})^2 +\mathrm{tr}(\widetilde H_{1+2}^2)\Bigr\rangle_{\T^D}\\
    &\quad+\sum_{m: \Pi_m\Pi_{1+2}=0}\Bigl\langle r_m(E_m^\K-\mae_m)^2 +\mathrm{tr}(\widetilde H_m^2)\Bigr\rangle_{\T^D},\\
    \|A^\mAE_{\mathrm{split}}\|^2 &=\|A^\K_{\mathrm{split}}\|^2+\Bigl\langle\sum_{j=1}^{2}(E_j^\K-\mae_j)^2 \Bigr\rangle_{\T^D}\\
    &\quad+\sum_{m: \Pi_m\Pi_{1+2}=0}\Bigl\langle r_m(E_m^\K-\mae_m)^2 +\mathrm{tr}(\widetilde H_m^2)\Bigr\rangle_{\T^D}.
\end{aligned}
\end{equation}
Here the rank-one blocks have no traceless part, while every higher-rank complementary block retains its full $\mathrm{SU}(r_m)$ contribution. These complementary terms are the same in the two expressions in Eq.~\eqref{eq:sqNormAs}. The Kato transport terms, however, differ. Using $i\omega^\mu\partial_\mu\Pi_j=[H,\Pi_j]$, this change is
\begin{equation}
\begin{aligned}
    A^\K_{\mathrm{split}}-A^\K_{\mathrm{merged}} &=\frac12\Bigl([\Pi_1,i\omega^\mu\partial_\mu\Pi_2]+[\Pi_2,i\omega^\mu\partial_\mu\Pi_1]\Bigr)\\
    &=\Pi_1H\Pi_2+\Pi_2H\Pi_1.
\end{aligned}
\end{equation}
It acts inside $\Pi_{1+2}$, where $A^\K_{\mathrm{merged}}$ has zero diagonal block. Its cross term with the original transport consequently vanishes, giving pointwise
\begin{equation}    \mathrm{tr}\bigl[(A^\K_{\mathrm{split}})^2\bigr]
     -\mathrm{tr}\bigl[(A^\K_{\mathrm{merged}})^2\bigr]=2\mathrm{tr}(\Pi_1H\Pi_2H).
\end{equation}
The same rotation was already present inside the merged block. Indeed, $E_{1+2}^\K=(E_1^\K+E_2^\K)/2$ and
\begin{equation}
\begin{aligned}
    \widetilde H_{1+2} &=\frac{E_1^\K-E_2^\K}{2}(\Pi_1-\Pi_2) +\Pi_1H\Pi_2+\Pi_2H\Pi_1,\\
    \mathrm{tr}(\widetilde H_{1+2}^2) &=\frac{(E_1^\K-E_2^\K)^2}{2} +2\mathrm{tr}(\Pi_1H\Pi_2H).
\end{aligned}
\end{equation}
Thus the extra Kato transport exactly replaces the off-diagonal part of the merged internal rotation. Splitting does not remove this physical motion; it transfers the same contribution between the two terms of the decomposition.

Once these transport and internal-rotation terms cancel, the full norm difference reduces to
\begin{equation}\label{eq:minimum-norm-splitting}
\begin{aligned}
    &\|A^\mAE_{\mathrm{merged}}\|^2 -\|A^\mAE_{\mathrm{split}}\|^2\\
    &=\Bigl\langle\sum_{j=1}^{2}\bigl[ (E_j^\K-\mae_{1+2})^2-(E_j^\K-\mae_j)^2 \bigr]\Bigr\rangle_{\T^D}\\
    &=\sum_{j=1}^{2}\Bigl[ 2(\mae_j-\mae_{1+2})\langle E_j^\K\rangle_{\T^D} +\mae_{1+2}^2-\mae_j^2\Bigr]\\
    &=\mae_1^2+\mae_2^2-2\mae_{1+2}^2 =\frac{(\mae_1-\mae_2)^2}{2}\geq0.
\end{aligned}
\end{equation}
The last step uses $\langle E_j^\K\rangle_{\T^D}=\mae_j$. The decrease is therefore a statement about the torus-averaged norm. It comes entirely from keeping the distinct conserved average energies in $\mAE$.

For two subspaces of arbitrary ranks $r_1,r_2$, we have $r_{1+2}=r_1+r_2$ and $r_{1+2}E_{1+2}^\K=r_1E_1^\K+r_2E_2^\K$. The split expression also retains $\widetilde H_1$ and $\widetilde H_2$. Their contributions are present in the merged block through the identity
\begin{equation}
\begin{aligned}
    \mathrm{tr}(\widetilde H_{1+2}^2) &=\sum_{j=1}^{2}\Bigl[\mathrm{tr}(\widetilde H_j^2) +r_j(E_j^\K-E_{1+2}^\K)^2\Bigr]\\
    &\quad+2\mathrm{tr}(\Pi_1H\Pi_2H).
\end{aligned}
\end{equation}
The same cancellation, with $\mae_{1+2}=(r_1\mae_1+r_2\mae_2)/(r_1+r_2)$, gives
\begin{equation}
    \|A^\mAE_{\mathrm{merged}}\|^2 -\|A^\mAE_{\mathrm{split}}\|^2 =\frac{r_1r_2}{r_1+r_2}(\mae_1-\mae_2)^2.
\end{equation}
Further splitting lowers the norm whenever the independent dynamically irreducible subspaces have different average energies. Apart from the rare exceptions of accidental degeneracy, the spectral eigenspaces of $\mAE$ are identified with the dynamically irreducible invariant spaces.

Finally, consider a smooth rank-one resolution of a dynamically irreducible rank-two invariant space $\Pi_{1+2}$. No rank-one subprojectors $\Pi_1,\Pi_2$, such that $0<\Pi_1<\Pi_{1+2}$ and $\Pi_2=\Pi_{1+2}-\Pi_1$, can be dynamically invariant, by definition. This implies $\LL(\Pi_1)=-\LL(\Pi_2)\neq0$. Consequently,
\begin{equation}
\begin{aligned}
    A^\K_{\mathrm{split}}-A^\K_{\mathrm{merged}} &=\Pi_1H\Pi_2+\Pi_2H\Pi_1\\
    &\quad+\frac12[\Pi_1-\Pi_2,\LL(\Pi_1)].
\end{aligned}
\end{equation}
Then the defining equation $\LL(A^\mAE)=\LL(H)$ would exclude this splitting, because
\begin{equation}
\begin{aligned}
    \LL(A^\mAE_{\mathrm{split}})-\LL(H) &=-(\mae_1-\mae_2)\LL(\Pi_1),\\
    \bigl\|\LL(A^\mAE_{\mathrm{split}})-\LL(H)\bigr\|^2&=\frac{(\mae_1-\mae_2)^2}{2} \sum_{j=1}^{2}\|\LL(\Pi_j)\|^2\geq0.
\end{aligned}
\end{equation}
This is the positive invariance residual penalized in Sec.~\ref{app:varlConvergenceProof}. Assigning different average energies to these subspaces violates the AGP equation; assigning equal energies leaves $A^\mAE_{\mathrm{split}}=A^\mAE_{\mathrm{merged}}$ and produces no spectral splitting. Thus, apart from accidental degeneracies, the variational principle provides an $\mAE$ whose eigenprojectors directly correspond to dynamically irreducible invariant spaces.

\subsection{Counterexamples for Diophantine-approximation Methods}
\label{app:diophantine-counterexamples}

We discuss two methods trying to reduce the dynamics of quasiperiodic drives~\cite{folta2014stationkeeping,maggia2017advances,ho1983semiclassical,crowley2019topological}, by using an integer vector sequence that approximates the direction of the irrational $\omegavec$. Both methods tend to transform the original quasiperiodic problem to a sequence of periodic problems whose limit is the original one. These methods, dependent on some chosen integer vector sequences, are called Diophantine-approximated.

Both constructions start from Diophantine approximations to the drive frequency vector $\omegavec$. The return-time method diagonalizes the evolution of the original drive at near return times $T^{(i)}$ depending on a sequence of integer vectors $\kvec^{(i)}\in\mathbb{Z}^D$ such that $\omegavec T^{(i)}-2\pi\kvec^{(i)}\to0$; the Floquet-approximant method changes the frequencies to approximate ones $\omegavec^{(i)}=2\pi\kvec^{(i)}/T^{(i)}$ that close the trajectory exactly, and diagonalizes the resulting periodic evolution. In both cases, one follows the eigenspaces along a chosen sequence of improving approximations.

For the \textit{return-time method}~\cite{folta2014stationkeeping,maggia2017advances}, we integrate the Schr\"odinger equation with $H(\omegavec t+\thetavec_0)$ up to each $T^{(i)}$, and diagonalize $U(T^{(i)},0)$. We form its spectral projectors $\Pi_n^{(i)}$ at the initial phase and match them between successive steps by their overlap $\mathrm{tr}(\Pi_n^{(i)}\Pi_m^{(i-1)})$. Eventually, these rank-one projectors, or their sum, form convergent sequences; the corresponding limit is expected to give the dynamically invariant spaces $\Pi_n[\thetavec_0]$.

For the \textit{Floquet-approximant method}~\cite{ho1983semiclassical,crowley2019topological}, we replace the original $H(\thetavec=\omegavec t+\thetavec_0)$ by periodic ones $H^{(i)}(t)=H(\omegavec^{(i)}t+\thetavec_0)$. We then integrate this periodic Hamiltonian for one period $T^{(i)}$. We denote its evolution by $U^{(i)}(t,0)$, keeping $U(t,0)$ for the original irrational drive. We diagonalize $U^{(i)}(T^{(i)},0)$ and follow its Floquet projector sequence to obtain the dynamically invariant spaces $\Pi_n[\thetavec_0]$ from their converging limit or the converging limit of their sum.

As mentioned, both methods need to first select a sequence of Diophantine-approximated integer vectors. However, the convergence of approximated eigenspaces within this particular  Diophantine-approximated sequence does not imply the convergence in all  Diophantine-approximated sequences. The examples below show how this inherent dependency on the Diophantine-approximated sequence could miss the true dynamically invariant space in both methods. In the following, we choose $D=2$ and use
\begin{equation}
    \omegavec=(1,\varrho), \qquad \varrho=\frac{1+\sqrt5}{2}.
\end{equation}

\subsubsection{Eigenspaces at Selected Return Times}

We construct a smooth abelian, untwisted qubit drive and compare two families of near returns. In one family the accumulated relative phase nearly vanishes, so a small residual rotation of the states determines the return eigenspaces; the other resolves the original invariant splitting. Their different limits show why diagonalizing a selected return sequence need not recover the invariant projectors.

\begin{figure}[t!]
    \centering
    \includegraphics[width=0.48\textwidth]{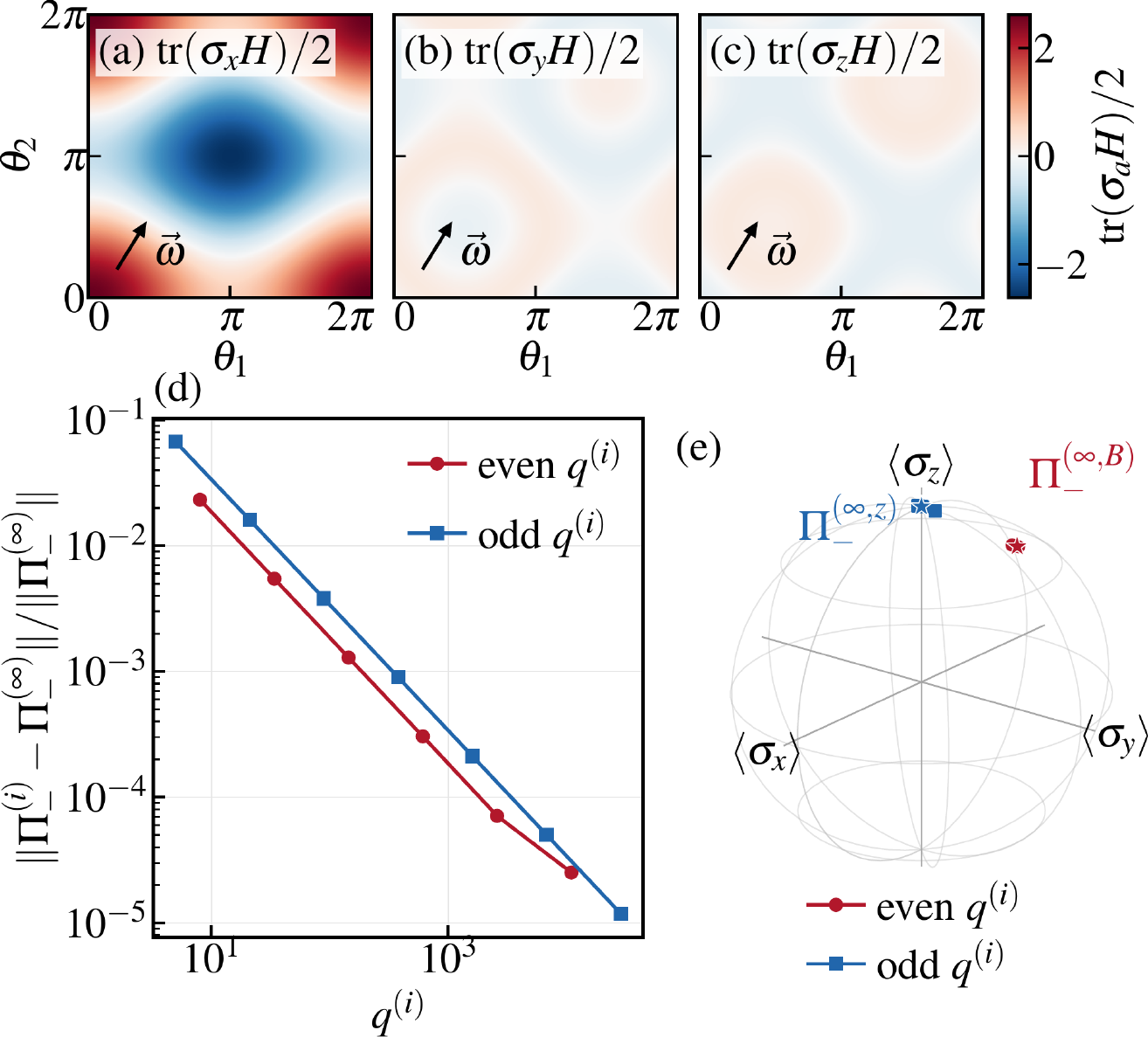}
    \caption{\textbf{Nonconvergence of return eigenspaces under different Diophantine-approximated return-time sequences.} (a--c) Hamiltonian from Eq.~\eqref{eq:returnCounterexampleModel} defined on the torus; (d) infidelity of the return eigenprojectors along two Diophantine sequences $\|\Pi_-^{(i)}-\Pi_-^{(\infty)}\|/\|\Pi_-^{(\infty)}\|$ versus Fibonacci denominator $q^{(i)}$, where $\Pi_-^{(\infty)}$ is the limit of each sequence. Red circles and blue squares denote even- and odd-denominator returns; (e) the corresponding Bloch vectors; red and blue stars mark $\Pi_-^{(\infty,B)}$ and $\Pi_-^{(\infty,z)}$. $\lambda=1$, $\omegavec=(1,\varrho)$, $\thetavec_0=(0,0)^\mathsf{T}$.}
    \label{fig:returnTimeCounterexample}
\end{figure}

For $\lambda\neq0$ and $\thetavec_0=(0,0)^\mathsf{T}$, let
\begin{equation}\label{eq:returnCounterexampleFrame}
    V(\thetavec)=\mathrm e^{-i\lambda(\sin\theta_1+\sin\theta_2)\sigma^x}, \qquad H_{\mathrm d}=\frac{\mathds{1}-\sigma^z}{4},
\end{equation}
\begin{equation}\label{eq:returnCounterexampleModel}
    H(\thetavec)=i(\omega^\mu\partial_\mu V)V^\dagger+VH_{\mathrm d}V^\dagger.
\end{equation}
Its average energy is $\mAE=VH_{\mathrm d}V^\dagger$, with $\mae_-=0$ and $\mae_+=1/2$. Here $\mae_-$ denotes the lower average energy. The two invariant projectors are
\begin{equation}
    \Pi_-[\thetavec]=V(\thetavec)\frac{\mathds{1}+\sigma^z}{2}V(\thetavec)^\dagger, \qquad \Pi_+=\mathds{1}-\Pi_-.
\end{equation}
The globally smooth frame $V$ makes the drive abelian and untwisted. It transports the spin states while $H_{\mathrm d}$ accumulates their relative phase, giving the exact evolution
\begin{equation}\label{eq:returnCounterexamplePropagator}
    U(t,0)=V(\thetavec_t)\mathrm e^{-iH_{\mathrm d}t}V(\thetavec_0)^\dagger.
\end{equation}
We start at $\thetavec_0=0$, where the ground-state projector is $\Pi_-^{(\infty,z)}=(\mathds{1}+\sigma^z)/2$.

Let $p^{(i)},q^{(i)}$ be consecutive Fibonacci numbers and $e^{(i)}=q^{(i)}\varrho-p^{(i)}$. The closest approaches to the initial drive phases occur at
\begin{equation}
\begin{aligned}
    T^{(i)}&=\frac{2\pi(q^{(i)}+\varrho p^{(i)})}{1+\varrho^2},\\
    \delta\thetavec^{(i)}&=\frac{2\pi e^{(i)}}{1+\varrho^2}(-\varrho,1)\longrightarrow0.
\end{aligned}
\end{equation}
For even $q^{(i)}$, the relative phase approaches zero modulo $2\pi$ and $U(T^{(i)},0)\to\mathds{1}$. The small remaining motion contains both an $x$-rotation from $V$ and a $z$-rotation from $H_{\mathrm d}$. Dropping an overall phase,
\begin{equation}
    U(T^{(i)},0)\propto\mathds{1}-ie^{(i)}B+\mathcal O((e^{(i)})^2),
\end{equation}
\begin{equation}\label{eq:returnTiltedGenerator}
    B=\frac{2\pi\lambda(1-\varrho)}{1+\varrho^2}\sigma^x +\frac{\pi\varrho}{2(1+\varrho^2)}\sigma^z.
\end{equation}
Both rotations vanish as $e^{(i)}\to0$, but their ratio stays finite: the return eigenvectors therefore approach the tilted spin axis of $B$, rather than the invariant $z$ axis. We follow the return eigenprojector with positive $z$-polarization as the candidate ground projector $\Pi_-^{(i)}$. Its even-return limit, $\Pi_-^{(\infty,B)}$, is the positive-eigenvalue projector of $B$.

Odd $q^{(i)}$ instead retain a relative phase of $\pi$, so $U(T^{(i)},0)\to\sigma^z$ and $\Pi_-^{(i)}\to\Pi_-^{(\infty,z)}$. These returns also expose the failure of the tilted candidate:
\begin{equation}
    \sigma^z\Pi_-^{(\infty,B)}\sigma^z\neq\Pi_-^{(\infty,B)}.
\end{equation}
The drive phases return, but this spin direction does not. Both selected sequences converge, as Fig.~\ref{fig:returnTimeCounterexample} shows, yet only one recovers the invariant splitting. The problem here is the vanishing relative evolution phase, rather than any lack of smooth invariant states.

\subsubsection{Floquet Eigenspaces of Rational Periodic Approximations}

We construct a smooth version of a quasiperiodically kicked spin for which the periodic approximations select the wrong invariant states. We first solve the resulting spin motion exactly and explain how the kicks produce this error. We then construct the smooth drive that keeps the Floquet states of every selected periodic approximation (Floquet approximants) well defined.

Keep $\omegavec=(1,\varrho)$ and start at $\thetavec_0=(0,0)^\mathsf{T}$. We follow the initial spin-$x$ projectors $(\mathds{1}\pm\sigma^x)/2$ under
\begin{equation}\label{eq:periodicApproxCounterexampleModel}
    H(\theta_1,\theta_2)=\frac12h_x(\theta_1)f(\theta_2)\sigma^x +\frac12h_z(\theta_1)\sigma^z.
\end{equation}
The pulse envelopes $h_x,h_z$ are nonnegative, smooth, $2\pi$ periodic in $\theta_1$, with disjoint supports and with areas of $\int_0^{T_1}h_x(s)\,\mathrm ds=1$ and $\int_0^{T_1}h_z(s)\,\mathrm ds=\pi$, where $T_1=2\pi$ since $\omega_1=1$.  Each cycle in $\theta_1$ therefore contains an $x$ kick followed by a $z$ kick that flips the spin. The function $f(\theta_2)$ modulates the $x$-kick strength. For either initial $x$-eigenstate $(\mathds{1}\pm\sigma^x)/2$, the $x$ kick contributes only a phase; the $z$-kick reverses the spin after every $\theta_1$-cycle, independent of $f$.

\begin{figure*}[t!]
    \centering
    \includegraphics[width=\textwidth]{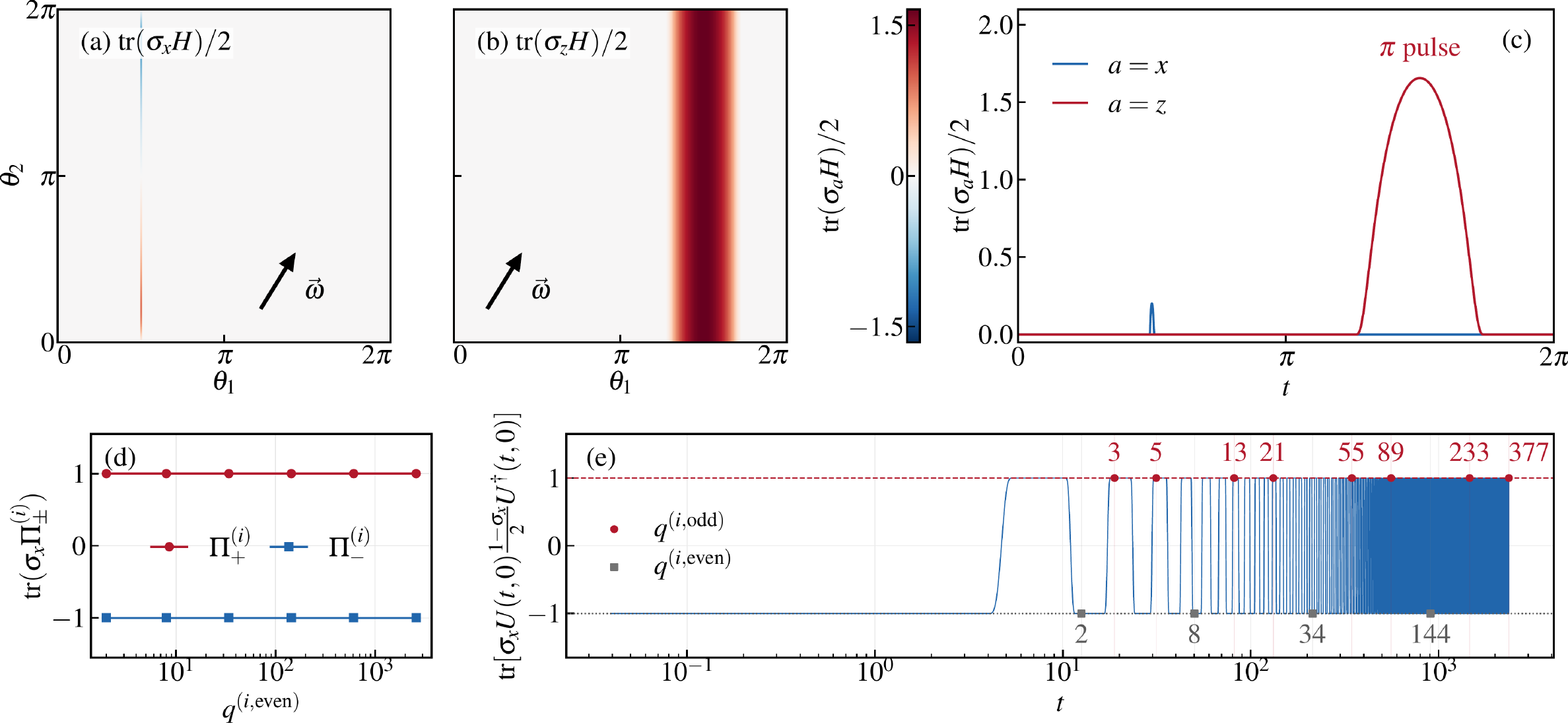}
    \caption{\textbf{Eigenspaces of Floquet approximants: converged limit along one Diophantine sequence and a different limit along another.} (a,b) Hamiltonian from Eq.~\eqref{eq:periodicApproxCounterexampleModel} defined on the drive-phase torus; (c) smooth pulse of quasiperiodic kick $H(\omegavec t+\thetavec_0)$ for $t\in[0,T_1]$, both showing a shape of $s\mapsto\exp(-1/[1-s^2])$; (d) $\mathrm{tr}(\sigma^x\Pi_\pm^{(i)})$ versus $q^{(i,\mathrm{even})}$, where $\Pi_\pm^{(i)}$ are eigenprojectors of Floquet approximants $U^{(i)}(T^{(i)},0)$; (e) the $x$ component of the initial projector $(\mathds{1}-\sigma^x)/2$ evolving in time $t$; red circles and gray squares mark odd- and even-denominator returns, labelled by their Fibonacci denominators. $\omegavec=(1,\varrho)$, $\thetavec_0=(0,0)^\mathsf{T}$.}
    \label{fig:periodicApproxCounterexample}
\end{figure*}

We use the same Fibonacci rational approximations as above, now separating the even- and odd-denominator subsequences as $p^{(i,\mathrm{even})}/q^{(i,\mathrm{even})}$ and $p^{(i,\mathrm{odd})}/q^{(i,\mathrm{odd})}$; $i$ counts each subsequence separately. We choose the periodic drives $H^{(i)}(t)$ with
\begin{equation}
    \omegavec^{(i)}=\left(1,\frac{p^{(i,\mathrm{even})}}{q^{(i,\mathrm{even})}}\right), \qquad T^{(i)}=q^{(i,\mathrm{even})}T_1.
\end{equation}
These Floquet periods $T^{(i)}$ always contain an even number of spin flips. Both initial $x$-spin projectors return to themselves at these Floquet periods. But the original irrational driving frequency $\omegavec$ also returns arbitrarily close to its initial phase $\thetavec_0$ after an odd number of $x+z$ kicks, exchanging the $x$-up and $x$-down projectors. Looking only at the chosen periodic drives misses this exchange and mistakes their Floquet states for invariant spaces of the irrational drive.

Figure~\ref{fig:periodicApproxCounterexample}(a,b) shows the Hamiltonian components $\mathrm{tr}(\sigma^{x,z}H)/2$ over the torus; panel (c) shows them during the first pulse cycle. Since the $x$ kicks only change the phase of the spin state, its direction under the original drive is determined entirely by the $z$ pulses:
\begin{equation}\label{eq:periodicExactSpin}
    \mathrm{tr}\!\left[\sigma^xU(t,0)\frac{\mathds{1}\pm\sigma^x}{2} U(t,0)^\dagger\right] =\pm\cos\!\left[\int_0^t h_z(s)\,\mathrm ds\right].
\end{equation}
This exact time trace is shown in Fig.~\ref{fig:periodicApproxCounterexample}(e).

For the periodic drives $H^{(i)}(t)=H(\omegavec^{(i)}t+\thetavec_0)$ with $\thetavec_0=\vec 0$, each $z$-flip reverses the sense of the next $x$-rotation. The net rotation angle in $x$ is therefore the alternating sum of the $x$-kick areas,
\begin{equation}
    \beta^{(i)}=\sum_{k=0}^{q^{(i,\mathrm{even})}-1}(-1)^k \int_0^{T_1}\mathrm ds\,h_x(s)\,f\!\left(\omega_2^{(i)}(s+k T_1)\right).
\end{equation}
The even number of $z$-kicks contributes only an overall phase, giving
\begin{equation}\label{eq:alternatingFilterProof}
    U^{(i)}(T^{(i)},0)=(-1)^{q^{(i,\mathrm{even})}/2} \mathrm e^{-i\beta^{(i)}\sigma^x/2}.
\end{equation}
The construction below keeps $0<|\beta^{(i)}|<\pi$, so the two Floquet eigenvalues are distinct. Their Floquet projectors are exactly $\Pi_-^{(i)}=(\mathds{1}-\sigma^x)/2$ and $\Pi_+^{(i)}=(\mathds{1}+\sigma^x)/2$ at every $T^{(i)}$, and hence so are their limits; see Fig.~\ref{fig:periodicApproxCounterexample}(d).

For the original irrational driving $\omegavec$, we take instead $T^{(i,\mathrm{odd})}=q^{(i,\mathrm{odd})}T_1$. The phase mismatch tends to zero because $q^{(i,\mathrm{odd})}\varrho-p^{(i,\mathrm{odd})}\to0$, but the odd number of $z$-kicks takes $(\mathds{1}\mp\sigma^x)/2$ to $(\mathds{1}\pm\sigma^x)/2$, shown in Fig.~\ref{fig:periodicApproxCounterexample}(e). Thus, the converged eigenspaces from the Floquet approximant sequence are not the true dynamically invariant spaces of the original irrational drive $H(\omegavec t+\thetavec_0)$ with $\thetavec_0=\vec 0$.

It remains to choose $f(\theta_2)$ in order to keep the Floquet eigenvalues distinct as the periods become longer. Equal successive $x$-kicks (e.g., $f(\theta_2)=\text{const.}$) cancel because the intervening $z$-pulse reverses the sense of rotation. A harmonic with wave number $q^{(i,\mathrm{even})}/2$ compensates this reversal: between kicks its phase advances by $\pi q^{(i,\mathrm{even})}\omega_2^{(i)} =\pi p^{(i,\mathrm{even})}$, so it changes sign because $p^{(i,\mathrm{even})}$ is odd. A single $\cos\theta_2$ only achieves this for $q^{(i,\mathrm{even})}=2$; for all later approximations its kicks still cancel exactly. Instead, take the simple smooth $f(\theta_2)$
\begin{equation}\label{eq:periodicSimpleModulation}
\begin{aligned}
    f(\theta_2)&=\frac{\sin\theta_2} {25-20\cos\theta_2}=\frac{1}{10}\sum_{k=1}^{\infty}2^{-k} \sin(k\theta_2).
\end{aligned}
\end{equation}
It contains the required harmonic for every approximation, with exponentially decreasing strength that makes $f$ smooth. The small overall amplitude keeps every net rotation below $\pi$, giving $0<|\beta^{(i)}|<\pi$ for a narrow $h_x(\theta_1)$.

With this definition of $f(\theta_2)$, we have defined a smooth quasiperiodic drive for which the Floquet approximants of a Diophantine sequence remain nondegenerate, with projectors converging at $\thetavec_0=(0,0)^\mathsf{T}$ to
\begin{equation}
    \Pi_\pm^{(\infty,\mathrm{even})}[\thetavec_0] =\frac{\mathds{1}\pm\sigma^x}{2}.
\end{equation}
However, from Fig.~\ref{fig:periodicApproxCounterexample}(e), we see that $\Pi_\pm^{(\infty,\mathrm{even})}$ are not the true invariant projectors. The true invariant spin directions actually lie close to $z$, with a tilt toward the $y$ direction. At $\theta_1=0$, we write this tilt angle as $u(\theta_2)$. The $x$-pulse rotates the spin and the $z$-pulse reverses its tilt; after one cycle it must match the direction $u(\theta_2+2\pi\varrho)$ at the new $\theta_2+2\pi\varrho$. Hence,
\begin{equation}\label{eq:periodicCounterexampleTilt}
    u(\theta_2+2\pi\varrho)-(-u(\theta_2)) =\int_0^{T_1}\mathrm ds\,h_x(s)f(\theta_2+\varrho s).
\end{equation}
The periodic solution, $u(\theta_2)=\sum_{k_2}u_{k_2}e^{ik_2\theta_2}$, follows by dividing each Fourier coefficient on the RHS by $1+\mathrm e^{2\pi i k_2\varrho}$. The exponentially decaying harmonics of the RHS keep this resolved $u(\theta_2)$ smooth despite golden-ratio-small denominators. These spin directions therefore match continuously at every near return of the irrational $\omegavec$.

At $\thetavec_0=(0,0)^\mathsf{T}$, the true projectors are
\begin{equation}\label{eq:periodicCounterexampleTrueProjectors}
    \Pi_\pm[\thetavec_0] =\frac{\mathds{1}\pm [\sin u(0)\sigma^y+\cos u(0)\sigma^z]}{2}.
\end{equation}
As long as $f(\theta_2)$ remains small, so does $u(\theta_2)$; then, the entire projector $\Pi_\pm$ still remains near $\pm z$. Thus, the Floquet-approximant counterexample Hamiltonian defined in Eq.~\eqref{eq:periodicApproxCounterexampleModel} has two abelian untwisted invariant spaces and an average energy operator of
\begin{equation}
\begin{aligned}
    \mAE[\thetavec]&=\mae_+ V(\thetavec) [\sin u(\theta_2-\varrho\theta_1)\sigma^y\\&\quad+\cos u(\theta_2-\varrho\theta_1)\sigma^z]V(\thetavec)^\dagger,\\
    \mae_\pm&=\pm\frac14 \int_0^{2\pi}\frac{\mathrm d\theta_2}{2\pi}\cos u(\theta_2),\\
       V(\thetavec)&=\exp\left[-i\frac{\sigma^z}{2}\int_0^{\frac{\theta_1}{\omega_1}}h_z(s)\mathrm{d}s\right]\\ &\quad \times \exp\left[-i\frac{\sigma^x}{2}\int_0^{\frac{\theta_1}{\omega_1}}h_x(s)f(\theta_2+\varrho s-\varrho\theta_1)\mathrm{d}s\right].
\end{aligned}
\end{equation}

\subsection{Frequency-lattice Schr\"odinger operator \texorpdfstring{$\mathcal K$}{K} and Liouvillian superoperator \texorpdfstring{$\LL$}{L}}
\label{app:avila-comparison}

The operator $\mathcal K$ defined in Eq.~\eqref{eq:Sambe} acting on the frequency-lattice Hilbert space looks similar to the Schr\"odinger operator on a lattice with quasiperiodic potential. The latter has been studied using Avila's global theory~\cite{avila2015global}. This theory organizes the localization regimes, with spectral properties distinguishing pure point, singular continuous, and absolutely continuous spectra. Since the frequency-lattice Schr\"odinger operator $\mathcal K=H-i\omega^\mu\partial_\mu$ also describes a lattice in frequency space, one might expect the same construction to determine the dynamics of quasiperiodic drives. We examine this possibility below.

\subsubsection{Comparison with the Mathematics Literature}
\label{app:subsec:math_lit}

In the mathematics literature, the question of whether a quasiperiodic drive admits a Floquet-type decomposition is known as the problem of reducibility of the associated quasiperiodic cocycle: whether a smooth change of basis on $\T^D$ conjugates the evolution to one generated by a constant operator. In the perturbative regime over Diophantine rotations, reducibility holds for a set of parameters of positive measure~\cite{krikorian1999presque,her2008full}, while non-reducible, uniquely ergodic cocycles are topologically generic~\cite{nerurkar1997construction,nerurkar2007construction,eliasson2002ergodic}; for $\mathbb{T}^d\times \text{SU}(2)$ a dichotomy between measurable reducibility and continuous fiber spectrum has been established~\cite{karaliolios2018continuous}; over Liouville rotations generic cocycles are not reducible. In our language, reducible cocycles form the abelian untwisted class and non-reducible ones comprise the other three; the coexistence of both behaviors at fixed Diophantine $\omegavec$ is the coexistence of four classes of invariant spaces in Sec.~\ref{sec:classification}, and unique ergodicity of the skew-system is the counterpart of complete nonabelian ergodicity in Sec.~\ref{sec:ergodicity}.

Our contribution is complementary to this body of work in three respects. First, we do not ask whether the evolution can be conjugated to a constant generator, but which spaces are invariant under it; the invariant projectors $\Pi_n[\thetavec]$ exist and are smooth whether or not the cocycle is reducible, and the decomposition~\eqref{eq:main} built from them requires no arithmetic condition on $\omegavec$. Second, we organize the non-reducible cases by two geometric properties of the invariant bundles, topology and rank, each of which has a physical signature. Third, the classification is constructive: the variational principle of Sec.~\ref{subsec:varl} produces $\Pi_n[\thetavec]$ for a given drive, whereas reducibility results are typically existence or perturbative. The spectral type of the frequency-lattice Schr\"odinger operator $\mathcal{K}$, central to the mathematical analysis~\cite{blekher1992floquet}, plays no role in our construction, see Appendix~\ref{app:subsec:spetral_types}.

\subsubsection{Comparison with Avila's Global Theory}

Avila's global theory investigates the following systems: for a 1D chain with nearest-neighbor hopping and quasiperiodic potential $v(x+2\pi k\varrho)$, the Schr\"odinger eigenvalue equation is
\begin{equation}\label{eq:avila-scalar-chain}
    u_{k+1}+u_{k-1}+v(x+2\pi k\varrho)u_k=E u_k,
\end{equation}
where $v$ is real analytic and $2\pi$ periodic, and $\varrho$ is irrational.

To solve this problem analytically, Avila's global theory constructs a $2\times 2$ transfer matrix:
\begin{equation}
    T_E(x)=\begin{pmatrix}E-v(x)&-1\\1&0\end{pmatrix}.
\end{equation}
This transfer matrix advances the two neighboring scalar amplitudes $(u_k,u_{k-1})^\mathsf T$ by one site to $(u_{k+1},u_{k})^\mathsf T$. Writing $T_E^{(L)}$ for $L$ successive transfers, as the number of transfers tends to infinity $L\to\infty$, its growth defines the Lyapunov exponent,
\begin{equation}\label{eq:avila-transfer-LE}
    \ell(E,y)=\lim_{L\to\infty}\frac{1}{L} \int_0^{2\pi}\frac{\mathrm dx}{2\pi} \log\|T_E^{(L)}(x+iy)\|.
\end{equation}
Complexifying the phase $x$ makes $\ell$ convex in $y$, with integer-valued right derivative in $y$~\cite{avila2015global}.

The value of $\ell$ and its slope in $y$ at $y=0$ determine the type of spectrum: pure-point, singular continuous, or absolutely continuous. For energies in the spectrum, the three regimes are
\begin{equation}\label{eq:avila-three-regimes}
\begin{array}{@{}ll@{}}
    \text{pure-point:}&\ell(E,0)>0,\\
    \text{singular continuous:}&\ell(E,0)=0,\quad \partial_y\ell(E,0)>0,\\
    \text{absolutely continuous:}&\ell(E,0)=\partial_y\ell(E,0)=0.
\end{array}
\end{equation}
A positive Lyapunov exponent corresponds to a locally pure-point $E$ regime where the lattice solutions have an exponential decay rate $\ell(E,0)$. When $\ell(E,0)=0$, its slope separates the singular continuous spectrum from the absolutely continuous regime in the vicinity of $E$. The absolutely continuous spectrum has a corresponding delocalized lattice solution; the singular continuous spectrum has a fractal-like delocalized solution.

This phenomenon resembles what we have found in Fig.~\ref{fig:SambeBad}; however, we find this transfer matrix technique and Lyapunov exponent are unsuitable to investigate quasiperiodic drives. A two-tone drive already gives a different and more complicated transfer problem. For $H(\thetavec)=H_0+V_1\cos\theta_1+V_2\cos\theta_2$, with noncommuting Hermitian matrices $H_0$, $V_1$ and $V_2$, $\mathcal K\psi=E\psi$ becomes
\begin{equation}\label{eq:avila-drive-lattice}
\begin{aligned}
    &(\kvec\cdot\omegavec+H_0)c_{\kvec}+\frac12\sum_{\mu=1}^2 V_\mu (c_{\kvec+\hat e_\mu}+c_{\kvec-\hat e_\mu}) =E c_{\kvec}.
\end{aligned}
\end{equation}
Along $k_1$, each layer still contains all $k_2$ amplitudes and their internal states. Advancing one layer therefore requires an entire transfer over the one-frequency-extended Hilbert space, instead of a $2\times 2$ transfer matrix from one $k_1$ site to the next. This is much more complicated than the above formulation of Avila's global theory. Thus, the simple $2\times 2$ transfer matrix method, in general, cannot compute the Lyapunov exponent or its slope, and thus cannot provide information on whether the different regimes of spectra of $\mathcal K$ are pure-point, singular continuous, or absolutely continuous.

Even when the Lyapunov exponents can be obtained directly via the evolution operator for some exactly solvable systems, a spectral-type label does not give the invariant projectors $\Pi_n$ satisfying $\LL(\Pi_n)=0$, their quantum geometry, or their internal dynamics. Continuous quasienergy spectrum alone cannot distinguish a topological obstruction from nonabelian mixing. Determining which obstruction is present requires solving the invariant spaces themselves.

\subsubsection{Properties of the Liouvillian operator}
\label{app:subsec:spetral_types}

The operator $\LL_0=-i\omega^\mu\partial_\mu$ acts in Fourier space as multiplication by the real sequence $\vec k\cdot\omegavec$ and is therefore self-adjoint on the domain of $\thetavec$-dependent operators in $\mathbb{H}^2\otimes L^2(\T^D)$ with square-integrable derivative along $\omegavec$. Since $[H,\cdot]$ is bounded (for bounded $H$) and symmetric w.r.t.~the inner product~\eqref{eq:inner_product}, the Liouvillian $\LL=\LL_0+[H,\cdot]$ is self-adjoint on the same domain by the Kato--Rellich theorem~\cite{ReedSimon2}. As a consequence, $\LL$ is closed, and hence $\ker\LL$ is a closed subspace. Thus, the kernel projection $P_{\ker\LL}$ is a bounded orthogonal projector. Therefore, the average energy $\mAE=P_{\ker\LL}H$ is a well-defined element of $\mathbb{H}^2\otimes L^2(\T^D)$ for any bounded drive Hamiltonian $H$.

Moreover, $i\LL$ generates the strongly continuous unitary group $(\mathrm e^{it\LL}O)(\thetavec_0)=U(t,0)^\dagger O(\omegavec t+\thetavec_0)\,U(t,0)$, whose fixed-point space is $\ker\LL$. By the mean ergodic theorem~\cite{vonNeumann1932proof}, the long-time average in Eq.~\eqref{eq:AE_varl} converges in the torus-averaged norm to $P_{\ker\LL}H$; this definition of $\mAE$ is a well-posed physical procedure. Since $\ker\LL$ is the eigenspace at the point $\lambda=0$, which is always an eigenvalue ($\LL(\mathds{1}_{N\times N})=0$) with a finite-dimensional eigenspace of dimension $\dim\ker\LL\leq N^2$, the kernel projection is agnostic to the spectral type (pure-point, continuous) of the rest of the spectrum $\sigma(\LL)$.

Note that none of these arguments requires a bounded pseudoinverse $\LL^+$. Indeed, observe that multiplication by $e^{i\vec l\cdot\thetavec}$ is unitary and preserves the domain; hence, the relation $e^{-i\vec l\cdot\thetavec}\,\LL\,e^{i\vec l\cdot\thetavec}=\LL+\vec l\cdot\omegavec$ implies $\sigma(\LL)=\sigma(\LL)+\vec l\cdot\omegavec$ for all $\vec l\in\mathbb Z^D$. Since $\sigma(\LL)$ is closed and nonempty and the set $\{\vec l\cdot\omegavec\}\subset\mathbb{R}$ is dense for incommensurate $\omegavec$ with $D\ge2$, it follows that $\sigma(\LL)=\mathbb R$; explicitly, if $\Pi_n\in\ker\LL$ then $e^{i\vec l\cdot\thetavec}\Pi_n$ is an eigenvector of $\LL$ with eigenvalue $\vec l\cdot\omegavec$, accumulating at zero. Hence zero is not an isolated eigenvalue. For a self-adjoint operator, the range is closed iff zero is either isolated in the spectrum or absent: thus, for $D\geq 2$, $\im\LL$ is not closed, and hence the pseudoinverse $\LL^+$ is unbounded.

Nonetheless, the gauge potential $A^\mAE=H-\mAE$ is always well-defined as the difference of two well-defined objects, and lies in $(\ker\LL)^\perp=\overline{\im\LL}$; it need not lie in $\im\LL$ itself. The latter is a stronger property, equivalent to $H\in\mathrm{dom}\,\LL^+$: this translates to a small-divisor condition, which is not needed for the decomposition $H=\mAE+A^\mAE$ or for the defining equation of the AGP, $\LL(A^\mAE)=\LL(H)$.

Compared to the analysis of quasiperiodically driven systems based on the frequency-lattice Schr\"odinger operator $\mathcal K=H-i\omega^\mu\partial_\mu$, we note that eigenkets of $\mathcal K$ exist only in the pure-point sector of its spectrum; twisted and nonabelian-irreducible invariant spaces come with a continuous spectrum and have no $L^2$ kets, and hence a frequency-lattice $\mathcal K$-based analysis cannot represent them. By contrast, the eigenprojectors of $\mAE$ exist for all four classes of quasiperiodic drives and are globally smooth on $\T^D$.

\newcommand{\HModeTableHeader}{%
\makebox[0.20\linewidth]{\textbf{Mode $\vec{k}$}} & \makebox[0.45\linewidth]{\textbf{Component $M_{\vec{k},i}$}} & \makebox[\dimexpr0.35\linewidth-0.7pt\relax]{\textbf{Value $c_{\vec{k},i}$}} \\
\hline
}

\newcommand{\HAbelianUntwistedRows}{%
\multirow[c]{4}{*}{ $\displaystyle
\begin{pmatrix}
0\\
0
\end{pmatrix}$
} & $\mathds{1}_{A}\otimes\sigma_{B}^{x}$ & $-0.0003855$ \\
& $\mathds{1}_{A}\otimes\sigma_{B}^{y}$ & $0.0002383$ \\
& $\mathds{1}_{A}\otimes\sigma_{B}^{z}$ & $3$ \\
& $\sigma_{A}^{z}\otimes\mathds{1}_{B}$ & $0.01$ \\
\hline

\multirow[c]{6}{*}{ $\displaystyle
\begin{pmatrix}
1\\
0
\end{pmatrix}$
} & $\mathds{1}_{A}\otimes\sigma_{B}^{x}$ & $-i\,0.5$ \\
& $\mathds{1}_{A}\otimes\sigma_{B}^{x}$ & $0.0001452$ \\
& $\mathds{1}_{A}\otimes\sigma_{B}^{y}$ & $0.0042593$ \\
& $\mathds{1}_{A}\otimes\sigma_{B}^{z}$ & $0.5$ \\
& $\mathds{1}_{A}\otimes\sigma_{B}^{z}$ & $i\,0.0008741$ \\
& $\sigma_{A}^{z}\otimes\mathds{1}_{B}$ & $0.5$ \\
\hline

\multirow[c]{6}{*}{ $\displaystyle
\begin{pmatrix}
0\\
1
\end{pmatrix}$
} & $\mathds{1}_{A}\otimes\sigma_{B}^{x}$ & $-0.0068916$ \\
& $\mathds{1}_{A}\otimes\sigma_{B}^{y}$ & $-i\,0.5$ \\
& $\mathds{1}_{A}\otimes\sigma_{B}^{y}$ & $-0.0000898$ \\
& $\mathds{1}_{A}\otimes\sigma_{B}^{z}$ & $0.5$ \\
& $\mathds{1}_{A}\otimes\sigma_{B}^{z}$ & $-i\,0.0005402$ \\
& $\sigma_{A}^{z}\otimes\mathds{1}_{B}$ & $0.5$ \\

}

\newcommand{\HAbelianTwistedRows}{%
\multirow[c]{4}{*}{ $\displaystyle
\begin{pmatrix}
0\\
0
\end{pmatrix}$
} & $\mathds{1}_{A}\otimes\sigma_{B}^{x}$ & $-0.0157656$ \\
& $\mathds{1}_{A}\otimes\sigma_{B}^{y}$ & $0.0097437$ \\
& $\mathds{1}_{A}\otimes\sigma_{B}^{z}$ & $1$ \\
& $\sigma_{A}^{z}\otimes\mathds{1}_{B}$ & $0.01$ \\
\hline

\multirow[c]{6}{*}{ $\displaystyle
\begin{pmatrix}
1\\
0
\end{pmatrix}$
} & $\mathds{1}_{A}\otimes\sigma_{B}^{x}$ & $-i\,0.5$ \\
& $\mathds{1}_{A}\otimes\sigma_{B}^{x}$ & $0.0095777$ \\
& $\mathds{1}_{A}\otimes\sigma_{B}^{y}$ & $0.0025350$ \\
& $\mathds{1}_{A}\otimes\sigma_{B}^{z}$ & $0.5$ \\
& $\mathds{1}_{A}\otimes\sigma_{B}^{z}$ & $i\,0.0033899$ \\
& $\sigma_{A}^{z}\otimes\mathds{1}_{B}$ & $0.5$ \\
\hline

\multirow[c]{6}{*}{ $\displaystyle
\begin{pmatrix}
0\\
1
\end{pmatrix}$
} & $\mathds{1}_{A}\otimes\sigma_{B}^{x}$ & $-0.0041017$ \\
& $\mathds{1}_{A}\otimes\sigma_{B}^{y}$ & $-i\,0.5$ \\
& $\mathds{1}_{A}\otimes\sigma_{B}^{y}$ & $-0.0059194$ \\
& $\mathds{1}_{A}\otimes\sigma_{B}^{z}$ & $0.5$ \\
& $\mathds{1}_{A}\otimes\sigma_{B}^{z}$ & $-i\,0.0020951$ \\
& $\sigma_{A}^{z}\otimes\mathds{1}_{B}$ & $0.5$ \\

}

\newcommand{\HNonAbelianUntwistedRows}{%
\multirow[c]{4}{*}{ $\displaystyle
\begin{pmatrix}
0\\
0
\end{pmatrix}$
} & $\mathds{1}_{A}\otimes\sigma_{B}^{x}$ & $-0.0003855$ \\
& $\mathds{1}_{A}\otimes\sigma_{B}^{y}$ & $0.0002383$ \\
& $\mathds{1}_{A}\otimes\sigma_{B}^{z}$ & $3$ \\
& $\sigma_{A}^{x}\otimes\mathds{1}_{B}$ & $0.01$ \\
\hline

\multirow[c]{6}{*}{ $\displaystyle
\begin{pmatrix}
1\\
0
\end{pmatrix}$
} & $\mathds{1}_{A}\otimes\sigma_{B}^{x}$ & $-i\,0.5$ \\
& $\mathds{1}_{A}\otimes\sigma_{B}^{x}$ & $0.0001452$ \\
& $\mathds{1}_{A}\otimes\sigma_{B}^{y}$ & $0.0042593$ \\
& $\mathds{1}_{A}\otimes\sigma_{B}^{z}$ & $0.5$ \\
& $\mathds{1}_{A}\otimes\sigma_{B}^{z}$ & $i\,0.0008741$ \\
& $\sigma_{A}^{z}\otimes\mathds{1}_{B}$ & $0.5$ \\
\hline

\multirow[c]{6}{*}{ $\displaystyle
\begin{pmatrix}
0\\
1
\end{pmatrix}$
} & $\mathds{1}_{A}\otimes\sigma_{B}^{x}$ & $-0.0068916$ \\
& $\mathds{1}_{A}\otimes\sigma_{B}^{y}$ & $-i\,0.5$ \\
& $\mathds{1}_{A}\otimes\sigma_{B}^{y}$ & $-0.0000898$ \\
& $\mathds{1}_{A}\otimes\sigma_{B}^{z}$ & $0.5$ \\
& $\mathds{1}_{A}\otimes\sigma_{B}^{z}$ & $-i\,0.0005402$ \\
& $\sigma_{A}^{z}\otimes\mathds{1}_{B}$ & $0.5$ \\

}

\newcommand{\HNonAbelianTwistedRows}{%
\multirow[c]{4}{*}{ $\displaystyle
\begin{pmatrix}
0\\
0
\end{pmatrix}$
} & $\mathds{1}_{A}\otimes\sigma_{B}^{x}$ & $-0.0157656$ \\
& $\mathds{1}_{A}\otimes\sigma_{B}^{y}$ & $0.0097437$ \\
& $\mathds{1}_{A}\otimes\sigma_{B}^{z}$ & $1$ \\
& $\sigma_{A}^{x}\otimes\mathds{1}_{B}$ & $0.01$ \\
\hline

\multirow[c]{6}{*}{ $\displaystyle
\begin{pmatrix}
1\\
0
\end{pmatrix}$
} & $\mathds{1}_{A}\otimes\sigma_{B}^{x}$ & $-i\,0.5$ \\
& $\mathds{1}_{A}\otimes\sigma_{B}^{x}$ & $0.0095777$ \\
& $\mathds{1}_{A}\otimes\sigma_{B}^{y}$ & $0.0025350$ \\
& $\mathds{1}_{A}\otimes\sigma_{B}^{z}$ & $0.5$ \\
& $\mathds{1}_{A}\otimes\sigma_{B}^{z}$ & $i\,0.0033899$ \\
& $\sigma_{A}^{z}\otimes\mathds{1}_{B}$ & $0.5$ \\
\hline

\multirow[c]{6}{*}{ $\displaystyle
\begin{pmatrix}
0\\
1
\end{pmatrix}$
} & $\mathds{1}_{A}\otimes\sigma_{B}^{x}$ & $-0.0041017$ \\
& $\mathds{1}_{A}\otimes\sigma_{B}^{y}$ & $-i\,0.5$ \\
& $\mathds{1}_{A}\otimes\sigma_{B}^{y}$ & $-0.0059194$ \\
& $\mathds{1}_{A}\otimes\sigma_{B}^{z}$ & $0.5$ \\
& $\mathds{1}_{A}\otimes\sigma_{B}^{z}$ & $-i\,0.0020951$ \\
& $\sigma_{A}^{z}\otimes\mathds{1}_{B}$ & $0.5$ \\

}

\begin{table*}[t!]
\centering
\small
\setlength{\tabcolsep}{0pt}
\setlength{\arrayrulewidth}{0.35pt}
\renewcommand{\arraystretch}{1.08}

\begin{tabular}{
    c !{\vrule width 0.8pt} c | c }
\noalign{\hrule height 0.8pt}

\makebox[1.65cm]{$(H^\alpha_{A}+\centernot{H}\!\!^{\,\tau}_{B})_{\vec{k}}$} & \textbf{Untwisted} & \textbf{Twisted} \\
\noalign{\hrule height 0.8pt}

\textbf{Abelian} &
\begin{minipage}[c]{\dimexpr0.50\textwidth-0.825cm-0.575pt\relax}
\begin{tabular}{@{}c|c|c@{}}
\HModeTableHeader
\HAbelianUntwistedRows
\end{tabular}
\end{minipage}
&
\begin{minipage}[c]{\dimexpr0.50\textwidth-0.825cm-0.575pt\relax}
\begin{tabular}{@{}c|c|c@{}}
\HModeTableHeader
\HAbelianTwistedRows
\end{tabular}
\end{minipage}
\\
\hline

\textbf{Nonabelian} &
\begin{minipage}[c]{\dimexpr0.50\textwidth-0.825cm-0.575pt\relax}
\begin{tabular}{@{}c|c|c@{}}
\HModeTableHeader
\HNonAbelianUntwistedRows
\end{tabular}
\end{minipage}
&
\begin{minipage}[c]{\dimexpr0.50\textwidth-0.825cm-0.575pt\relax}
\begin{tabular}{@{}c|c|c@{}}
\HModeTableHeader
\HNonAbelianTwistedRows
\end{tabular}
\end{minipage}
\\
\noalign{\hrule height 0.8pt}
\end{tabular}

\caption{\textbf{Table of Fourier modes and matrix components of $(H^\alpha_{A}+\centernot{H}\!\!^{\,\tau}_{B})$.} The four decompositions $(H^\alpha_{A}+\centernot{H}\!\!^{\,\tau}_{B})_{\vec{k}} =\sum_i c_{\vec{k},i}M_{\vec{k},i}$ are organized according to abelian versus nonabelian structure and untwisted versus twisted topology. Table obtained for perturbation strength $\epsilon=10^{-2}$ and drive parameters $|\omegavec|=0.1$, $\omega_2/\omega_1=(1+\sqrt 5)/2$. In all four cases, $(H^\alpha_{A}+\centernot{H}\!\!^{\,\tau}_{B})_{-\vec{k}} =(H^\alpha_{A}+\centernot{H}\!\!^{\,\tau}_{B})_{\vec{k}}^\dagger$. }
\label{tab:H-modes-four-classes}
\end{table*}

\end{appendix}

\FloatBarrier

\fi %

\bibliography{references}

\end{document}

%% file: preamble/macros.tex
\renewcommand{\i}{\text{i}}

\newcommand{\im}{\text{im}\,}

\newcommand{\mAE}{\text{\AE}}
\newcommand{\mae}{\text{\ae}}

\newcommand{\thetavec}{\vec{\theta}}
\newcommand{\omegavec}{\vec{\omega}}
\newcommand{\kvec}{\vec{k}}

\newcommand{\T}{\mathbb{T}}

\newcommand{\K}{\text{K}}
\newcommand{\F}{\text{F}}
\newcommand{\U}{\text{U}}
\newcommand{\D}{\mathrm{d}}
\newcommand{\LL}{\mathcal{L}}

\definecolor{lime}{HTML}{A6CE39}
\DeclareRobustCommand{\orcidicon}{%
  \hspace{-4pt}%
  \begin{tikzpicture}
    \draw[lime, fill=lime] (0,0) circle [radius=0.16]
      node[white] {\hspace{0.1mm}{\fontfamily{qag}\selectfont \tiny ID}};
    \draw[white, fill=white] (-0.06,0.1) circle [radius=0.005];
  \end{tikzpicture}%
  \hspace{-3.2mm}%
}
\newcommand\orcid[1]{\href{https://orcid.org/#1}{~\orcidicon~~}}